\documentclass[graybox]{svmult}

\usepackage[round]{natbib}
\usepackage{mathptmx}       
\usepackage{helvet}         
\usepackage{courier}        
\usepackage{type1cm}        
\usepackage{lmodern}

\usepackage[T1]{fontenc}
\usepackage{amsmath,amsbsy}     
\usepackage{makeidx}         
\usepackage{graphicx}        
\usepackage{multicol}        
\usepackage[bottom]{footmisc}
\usepackage{bm}        

\newcommand{\sig}{\lower0.6ex\hbox{$\stackrel{\textstyle >}{\sim}$}\:}
\newcommand{\sil}{\lower0.6ex\hbox{$\stackrel{\textstyle <}{\sim}$}\:}
\newcommand{\sigs}{\lower0.4ex\hbox{$\stackrel{\scriptstyle
      >}{\scriptstyle \sim}$}\,}
\newcommand{\sils}{\lower0.4ex\hbox{$\stackrel{\scriptstyle
      <}{\scriptstyle \sim}$}\,}

\newcommand{\hii}{H{\textsc{ii}}}

\newcommand*\Bell{\ensuremath{\boldsymbol\ell}}

\def\aj{{\sl AJ}}
\def\apj{{\sl ApJ}}
\def\apjl{{\sl ApJ}}
\def\apjs{{\sl ApJS}}
\def\apj{{\sl ApJ}}
\def\araa{{\sl ARA\&A}}
\def\aap{{\sl A\&A}}
\def\aaps{{\sl A\&AS}}
\def\aapr{{\sl A\&A Rev.}}
\def\an{{\sl AN}}

\def\apss{{\sl Astrophys.\ Space Sci.}}
\def\assl{{\sl Astrophys.\ Space Sc.\ Lib.}}
\def\aujp{{\sl Aust.\ J.\ Phys.,}}

\def\bain{{\sl Bull.\ Astron.\ Inst.\ Netherlands}}
\def\bjp{{\sl  Brazilian J. Phys.}}
\def\fcp{{\sl Fund.\ Cosm.\ Phys.}}

\def\jfm{{\sl J.\ Fluid Mech.}}
\def\jpb{{\sl J.\ Phys.\ B}}
\def\jpcs{{\sl J.\ Phys.\ Conf. Ser.}}
\def\mnras{{\sl MNRAS}}
\def\nat{{\sl Nature}}
\def\na{{\sl New.\ Astron.}}
\def\pta{{\sl Phil.\ Trans.\ A.}}

\def\pre{{\sl PRE}}
\def\prl{{\sl PRL}}
\def\physa{{\sl Physica A}}
\def\physrep{{\sl Phys.\ Reports}}
\def\prsa{{\sl Proc.\ R.\ Soc.\ London A}}
\def\pasj{{\sl PASJ}}
\def\pasa{{\sl Pub.\ Astron.\ Soc.\ Aust.}}
\def\pasp{{\sl PASP}}
\def\procps{{\sl Proceedings of the
  Cambridge Philosophical Society}}
\def\plss{{\sl Plan.\ Space Sci.}}

\def\rmp{{\sl Rev.\ Mod.\ Phys.}}
\def\rslpsa{{\sl Royal Society of London Proceedings Series A}}
\def\sci{{\sl Science}}
\def\ssr{{\sl Space Science Reviews}}
\def\sovastron{{\sl Sov.\ Astron.,}}

\def\za{{\sl Z.\ Astrophys.}}

\usepackage{fancyhdr}
\makeindex             

\begin{document}

\title*{Physical Processes in the Interstellar Medium}
\author{Ralf S.\ Klessen and Simon C.\ O.\ Glover}
\institute{Universit\"{a}t Heidelberg, Zentrum f\"{u}r Astronomie, Institut f\"{u}r Theoretische Astrophysik, \\Albert-Ueberle-Stra{\ss}e 2, 69120 Heidelberg, Germany\\ \email{klessen@uni-heidelberg.de, glover@uni-heidelberg.de}}

\maketitle


\abstract{\mbox{~}\\
Interstellar space is filled with a dilute mixture of charged particles, atoms, molecules and dust grains, called the interstellar medium (ISM). Understanding its physical properties and dynamical behavior is of pivotal importance to many areas of astronomy and astrophysics. Galaxy formation and evolution, the formation of stars, cosmic nucleosynthesis, the origin of large complex, prebiotic molecules and the abundance, structure and growth of dust grains which constitute the fundamental building blocks of planets, all these processes are intimately coupled to the physics of the interstellar medium. However, despite its importance, its structure and evolution is still not fully understood. Observations reveal that the interstellar medium is highly turbulent, consists of different chemical phases, and is characterized by complex structure on all resolvable spatial and temporal scales. Our current numerical and theoretical models describe it as a strongly coupled system that is far from equilibrium and where the different components are intricately linked together by complex feedback loops. Describing the interstellar medium is truly a multi-scale and multi-physics problem. In these lecture notes we introduce the microphysics necessary to better understand the interstellar medium. We review the relations between large-scale and small-scale dynamics, we consider turbulence as one of the key drivers of galactic evolution, and we review the physical processes that lead to the formation of dense molecular clouds and that govern stellar birth in their interior. 
}

\tableofcontents 

\pagestyle{fancy}
 \renewcommand{\sectionmark}[1]{%
\markboth{#1}{}}
\renewcommand{\chaptermark}[1]{%
\markboth{#1}{}}
\renewcommand{\sectionmark}[1]{\markright{\ #1}}

\section{Introduction}
\label{sec:introduction}

Understanding the physical processes that govern the dynamical behavior of the interstellar medium (ISM) is central to much of modern astronomy and astrophysics. The ISM is the primary galactic repository out of which stars are born and into which they deposit energy, momentum and enriched material as they die. It constitutes the  anchor point of the galactic matter cycle, and as such is the key to a consistent picture of galaxy formation and evolution. The dynamics of the ISM determines where and when stars form. Similarly, the properties of the planets and planetary systems around these stars are intimately connected to the properties of their host stars and the details of their formation process. 

When we look at the sky on a clear night, we can notice dark patches of obscuration along the band of the Milky Way. These are clouds of dust and gas that block the light from distant stars. With the current set of telescopes and satellites we can observe dark clouds at essentially all frequencies possible, ranging from low-energy radio waves all the way up to highly energetic $\gamma$-rays. We have learned that all star formation occurring in the Milky Way and other galaxies is associated with these dark clouds that mostly consist of cold molecular hydrogen and dust. In general, these dense clouds are embedded in and dynamically connected to the larger-scale and less dense atomic component. Once stellar birth sets in, feedback becomes important. Massive stars emit copious amounts of ionizing photons and create bubbles of hot ionized plasma, thus converting ISM material into a hot and very tenuous state. 

We shall see in this lecture that we cannot understand the large-scale dynamics of the ISM without profound knowledge of the underlying microphysics. And vice versa, we will argue that dynamical processes on large galactic scales determine the local properties of the different phases of the ISM, such as their ability to cool and collapse, and to give birth to new stars. ISM dynamics spans a wide range of spatial scales, from the extent of the galaxy as a whole down to the local blobs of gas that collapse to form individual stars or binary systems. Similarly, it covers many decades in temporal scales, from the hundreds of millions of years it takes to complete one galactic rotation down to the hundreds of years it takes an ionization front to travel through a star-forming cloud. This wide range of scales is intricately linked by a number of competing feedback loops. Altogether, characterizing the ISM is truly a multi-scale and multi-physics problem. It requires insights from quantum physics and chemistry, as well as knowledge of magnetohydrodynamics, plasma physics, and gravitational dynamics. It also demands a deep understanding of the coupling between matter and radiation, together with input from high-resolution multi-frequency and multi-messenger astronomical observations. 

By mass, the ISM consists of around 70\% hydrogen (H), 28\% helium (He), and 2\% heavier elements. The latter are generally termed metals in the sometimes very crude astronomical nomenclature. We give a detailed account of the composition of the ISM in Section \ref{sec:comp-ISM}. Because helium is chemically inert, it is customary, and indeed highly practical, to distinguish the different phases of the ISM by the chemical state of hydrogen. Ionized bubbles are called H{\textsc{ii}} regions, while atomic gas is often termed H{\sc i} gas, in both cases referring to the spectroscopic notation. H{\textsc{ii}} regions are best observed by looking at hydrogen recombination lines or the fine structure lines of ionized heavy atoms. The properties of H{\sc i} gas are best studied via the 21$\,$cm hyperfine structure line of hydrogen. Dark clouds are sufficiently dense and well-shielded against the dissociating effects of interstellar ultraviolet radiation to allow H atoms to bind together to form molecular hydrogen (H$_2$). They are therefore called molecular clouds. 

H$_2$ is a homonuclear molecule. Its dipole moment vanishes and it radiates extremely weakly under normal Galactic ISM conditions. Direct detection of H$_{2}$ is therefore generally possible only through ultraviolet absorption studies. Due to atmospheric opacity these studies can only be done from space, and are limited to pencil-beam measurements of the absorption of light from bright stars or active galactic nuclei (AGN). We note that rotational and ro-vibrational emission lines from H$_2$ have indeed been detected in the infrared, both in the Milky Way and in other galaxies. However, this emission comes from gas that has been strongly heated by shocks or radiation, and it traces only a small fraction of the overall amount of molecular hydrogen. Due to these limitations, the most common tool for studying the molecular ISM is radio and sub-millimeter emission either from dust grains or from other molecules that tend to be found in the same locations as H$_2$. By far the most commonly used molecular tracer is carbon monoxide with its various isotopologues. The most abundant, and hence easiest to observe is $^{12}$C$^{16}$O, usually referred to simply as $^{12}$CO or just CO. However, this isotopologue is often so abundant that its emission is optically thick, meaning that it only traces conditions reliably in the surface layers of the dense substructure found within most molecular clouds. The next most abundant isotopologues are $^{13}$C$^{16}$O (usually written simply as $^{13}$CO) and $^{12}$C$^{18}$O (usually just C$^{18}$O). Their emission is often optically thin and can freely escape the system. This allows us to trace the full volume of the cloud. As CO has a relatively low critical density and also freezes out on dust grains at very high densities, other tracers such as HCN or N$_{2}$H$^{+}$ need to be used to study conditions within high density regions such as prestellar cores. We discuss the microphysics of the interaction between radiation and matter and the various heating and cooling processes that determine the thermodynamic response of the various phases of the ISM in Section \ref{sec:heating-cooling}.

A key physical agent controlling the dynamical evolution of the ISM is turbulence. For a long time it was thought that supersonic turbulence in the interstellar gas could not produce significant compressions, since this would result in a rapid dissipation of the turbulent kinetic energy. In order to avoid this rapid dissipation of energy, appeal was made to the presence of strong magnetic fields in the clouds, which were thought to greatly reduce the dissipation rate. However, it was later shown in high-resolution numerical simulations and theoretical stability analyses that magnetized turbulence dissipates energy at roughly the same rate as hydrodynamic turbulence. In both cases, the resulting density structure is highly inhomogeneous and intermittent in time. Today, we think that ISM turbulence plays a dual role. It is energetic enough to counterbalance gravity on global scales, but at the same time it may provoke local collapse on small scales. This apparent paradox can be resolved when considering that supersonic turbulence establishes a complex network of interacting shocks, where converging flows generate regions of high density. These localized enhancements can be sufficiently large for gravitational instability to set in. The subsequent evolution now depends on the competition between collapse and dispersal. The same random flows that create high-density regions in the first place may also destroy them again. For local collapse to result in the formation of stars, it must happen rapidly enough for the region to decouple from the flow. Typical collapse timescales are found to be comparable to dispersal times of shock-generated density fluctuations in the turbulent gas. This makes the outcome highly unpredictable and theoretical models are based on stochastic theory. In addition, supersonic turbulence dissipates quickly and so needs to be continuously driven for the galaxy to reach an approximate steady state. Finding and investigating suitable astrophysical processes that can drive interstellar turbulence remains a major challenge. We review the current state of affairs in this field in Section \ref{sec:turbulence}.

We think that molecular clouds form by a combination of turbulent compression and global instabilities. This process connects large-scale dynamics in the galaxy with the localized transition from warm, tenuous, mostly atomic gas to a dense, cold, fully molecular phase. The thermodynamics of the gas, and thus its ability to respond to external compression and consequently to go into collapse, depends on the balance between heating and cooling processes. Magnetic fields and radiative processes also play an important role. The chemical reactions associated with the transition from H to H$_2$, the importance of dust shielding, and the relation between molecular cloud formation and the larger galactic context are discussed in Section \ref{sec:cloud-form}. 

These clouds constitute the environment where new stars are born. The location and the mass growth of young stars are therefore intimately coupled to the dynamical properties of their parental clouds. Stars form by gravitational collapse of shock-compressed density fluctuations generated from the supersonic turbulence ubiquitously observed in molecular clouds. Once a gas clump becomes gravitationally unstable, it begins to collapse and its central density increases considerably until a new stars is born. Altogether, star formation in molecular clouds can be seen as a two-phase process. First, supersonic turbulence creates a highly transient and inhomogeneous molecular cloud structure that is characterized by large density contrasts. Some of the high-density fluctuations are transient, but others exceed the critical mass for gravitational contraction, and hence begin to collapse. Second, the collapse of these unstable cores leads to the formation of individual stars and star clusters. In this phase, a nascent protostar grows in mass via accretion from the infalling envelope until the available gas reservoir is exhausted or stellar feedback effects become important and remove the parental cocoon. In Section \ref{sec:collapse-SF}, we discuss the properties of molecular cloud cores, the statistical characteristics of newly born stars and star clusters, and our current theoretical models of dynamical star formation including the distribution of stellar masses at birth.

Finally, we conclude these lecture notes with a short summary in Section \ref{sec:summary}.

\newpage
\section{Composition of the ISM}
\label{sec:comp-ISM}

\subsection{Gas}
\label{subsec:comp-ISM-gas}

The gas in the ISM is composed almost entirely of hydrogen and helium, with hydrogen accounting for around 70\% of the total mass, helium for 28\%, and all other elements for the remaining 2\%. The total gas mass in the Milky Way is difficult to estimate, but is probably close to $10^{10} \: {\rm M_{\odot}}$ \citep{kk09}. The majority of the volume of the ISM is occupied by ionized gas, but the total mass associated with this component is not more than around 25\% of the total gas mass. The majority of the mass is located in regions dominated by neutral atomic gas (H, He) or molecular gas (H$_{2}$). Much of the atomic gas and all of the molecular gas is found in the form of dense clouds that occupy only 1--2\% of the total volume of the ISM.

The thermal and chemical state of the ISM are conventionally described in terms of a number of distinct phases.
An early and highly influential model of the phase structure of the ISM was put forward by \citet{fgh69}, who 
showed that if one assumes that the atomic gas in the ISM is in thermal equilibrium, then there exists a wide
range of pressures for which there are two thermally stable solutions:  one corresponding to cold, dense gas
with $T \sim 100$~K  that we can identify with the phase now known as the Cold Neutral Medium (CNM), and a 
second corresponding to warm, diffuse gas with $T \sim 10^{4}$~K that we can identify with the phase now 
known as the Warm Neutral Medium (WNM). In the \citet{fgh69} model, gas at intermediate temperatures is 
thermally unstable and depending on its density will either cool down and increase its density until it joins the CNM, 
or heat up and reduce its density until it joins the WNM. 

This two-phase model of the ISM was extended by \citet{mo77}, who pointed out that supernovae exploding in
the ISM would create large, ionized bubbles filled with very hot gas ($T \sim 10^{6} \: {\rm K}$). Although this
gas would eventually cool, the temperature dependence of the atomic cooling curve at high temperatures is
such that the cooling time around $T \sim 10^{6}$~K is considerably longer than the cooling time in the
temperature range $10^{4} < T < 10^{6}$~K (see Section~\ref{atmolcool} below). Therefore, rather than this hot gas having
a wide range of temperatures, one would instead expect to find most of it close to $10^{6}$~K. This hot, ionized
phase of the ISM has subsequently become known as the Hot Ionized medium (HIM).

Evidence for an additional phase, the so-called Warm Ionized Medium (WIM), comes from a variety of
observations, including free-free absorption of the Galactic synchrotron background \citep{he63}, the dispersion of
radio signals from pulsars \citep{reynolds89,gae08}, and faint optical emission lines produced by ionized species such as O$^{+}$
and N$^{+}$ \citep{reynolds73,mie06}. This ionized phase has a density comparable to that of the WNM, and has a scale-height
of the order of 1~kpc \citep[see e.g.][]{reynolds89}. Its volume filling factor is relatively small in regions close to the 
Galactic midplane, but increases significantly as one moves away from the midplane \citep[see e.g.][]{gae08}.
Overall, 90\% or more of the total ionized gas within the ISM is located in the WIM \citep{haff09}.
It should be noted that the gas in classical H{\textsc{ii}} regions surrounding O stars is generally not considered to be part of the WIM.

Finally, a distinction is often drawn between the dense, molecular phase of the ISM, observed to be distributed 
in the form of discrete molecular clouds of various masses and sizes \citep[see e.g.][]{blitz07a} and the lower density, cold
atomic gas surrounding these clouds, which is part of the CNM. The distribution of this molecular gas in our
Galaxy is of particular interest, as star formation is observed to correlate closely with the presence of molecular
gas. The distribution of molecular gas with
Galactocentric radius can be measured by combining data from CO observations, which trace clouds with high
concentrations of both H$_{2}$ and CO, and C$^{+}$ observations, which trace so-called ``dark molecular gas'',
i.e.\ clouds with high H$_{2}$ fractions but little CO \citep[see e.g.][]{pineda13}. The molecular gas surface density 
shows a pronounced peak within the central 500~pc of the Galaxy, a region known as the Central Molecular Zone (CMZ). 
It then falls off sharply between 0.5 and 3~kpc, possibly owing to the influence of the Milky Way's central stellar bar 
\citep{ms96}, before peaking again at a Galactocentric radius of around 4--6~kpc in a structure known as the Molecular 
Ring. Outside of the Molecular Ring, the surface density of molecular gas declines exponentially, but it can still be 
traced out to distances of at least 12--13~kpc \citep{heyer98}.

\begin{table}[th]
\begin{center}
\caption{{Phases of the ISM} \label{ism-phases}}
\vspace{.1in}
\begin{tabular}{c @{\hspace{0.5cm}}  c @{\hspace{0.5cm}} c @{\hspace{0.5cm}} c}
\hline
Component  & Temperature (K) & Density (${\rm cm^{-3}}$) & Fractional ionization \\
\hline
Molecular gas & 10--20 & $> 10^{2}$ & $< 10^{-6}$ \\
Cold neutral medium (CNM) & 50--100 & 20--50 & $\sim 10^{-4}$ \\
Warm neutral medium (WNM) & 6000--10000 & 0.2--0.5 & $\sim$0.1 \\
Warm ionized medium (WIM) & $\sim 8000$ & 0.2--0.5 & 1.0 \\
Hot ionized medium (HIM) & $\sim 10^{6}$ & $\sim 10^{-2}$ & 1.0 \\
\hline\\[-0.2cm]
\end{tabular}
{\footnotesize Adapted from \citet{ferriere01}, \citet{cas98},
\citet{wolf03}, and \citet{jenk13}.}
\end{center}
\end{table}

An overview of the main physical properties of these different phases is given in Table~\ref{ism-phases}.
The information on the typical density and temperature ranges was taken from the review by \citet{ferriere01}, 
while the information on the typical fractional ionization of the various phases is based on \citet{cas98},
\citet{wolf03}, and \citet{jenk13}.

Although gas in the ISM is often classified purely in terms of these five distinct phases, the question
of how distinct these phases truly are remains open. For example,
in the classical \citet{fgh69} model and the many
subsequent models inspired by it,  the CNM and WNM are two completely distinct phases in pressure
equilibrium with each other, and all neutral atomic hydrogen in the ISM belongs to one phase or the 
other. However, observations of H{\sc i} in the ISM suggest that the true picture is more complicated,
as there is good evidence that a significant fraction of the atomic gas has a temperature intermediate
between the CNM and WNM solutions, in the thermally unstable regime \citep{ht03,roy13}. This gas cannot
be in equilibrium, and cannot easily be assigned to either the CNM or the WNM.

One important reason why this picture of the ISM appears to be an oversimplification is that the
ISM is a highly turbulent medium. Turbulence in the ISM is driven by a number of different physical
processes, including thermal instability \citep{kritsuk02a}, supernova feedback \citep[see e.g.][]{maclow04}, and 
the inflow of gas onto the disk \citep{klessen10,eb10}, and acts to mix together what would otherwise be 
distinct phases of the ISM \citep[see e.g.][]{joung09,ssn11}.
We discuss the role that turbulence plays in structuring the ISM together with the various driving mechanisms proposed at much greater length in Section \ref{sec:turbulence}.

Finally, it is useful to briefly summarize what we know about the metallicity of the ISM, i.e.\ of the fractional abundance of elements heavier than helium, since this plays an important role in regulating the thermal behavior of the ISM. In the Milky Way, the metallicity can be measured using a variety of methods \citep{mc10}.
Measurements of the optical emission lines of O{\textsc{ii}} and O{\textsc{iii}} together
with H$\alpha$ and H$\beta$ can be used to constrain the oxygen abundance
in Galactic H{\textsc{ii}} regions \citep[see e.g.][]{deh00}, from which the total metallicity
$Z$ follows if we assume that the oxygen abundance scales linearly with $Z$. 
Alternatively, the abundances of carbon, nitrogen, oxygen and many other elements
can be measured using ultraviolet (UV)  absorption lines in the spectra of bright background stars
\citep[see e.g.][]{cs86,ss96,sofia04}. The metallicity can also be measured using stars, specifically
by studying the spectra of young, massive B-type stars \citep[see e.g.][]{roll00}.
Technically, stellar measurements constrain the metallicity at the time that the star 
formed, rather than at the present day, but since B stars have short lifetimes, this 
distinction does not turn out to be particularly important in practice. 

None of these techniques gives us a completely unbiased picture of the metallicity
of the ISM. Emission line measurements are sensitive to the temperature distribution
within the H{\textsc{ii}} regions, which is difficult to constrain accurately. UV absorption line
measurements are much less sensitive to excitation effects, but can only be carried
out from space, and also require the presence of a UV-bright background source.
Therefore, although they can give us information on the composition of the more
diffuse phases of the ISM, including the WNM and CNM, they cannot be used to
probe denser regions, such as molecular clouds, as the extinction in these regions
is typically far too high for us to be able to detect the required background sources
in the UV. In addition, these measurements tell us only about the gas-phase metals
and not about the metals that are locked up in dust grains (see Section~\ref{dust} below).
Finally, stellar measurements provide us with good tracers of the total metallicity,
but do not tell us how much of this was formerly in the gas phase, and how much
was in dust.

Nevertheless, by combining the information provided by these different methods,
we can put together a pretty good picture of the metallicity distribution of the gas
in the ISM. Measurements of the metallicities of B stars and of H{\textsc{ii}} regions  both show 
that there is a large-scale radial metallicity gradient in the ISM, with a value of
around $-0.04 \:\: {\rm dex} \: {\rm kpc^{-1}}$ \citep{mc10}. The metallicity of the gas 
in the CMZ is therefore around twice the solar value \citep{fgj07}, while in the outer 
Galaxy, metallicities are typically somewhat sub-solar \citep{rud06}. 

Comparison of the abundances of individual elements derived using B stars 
and those derived using UV absorption lines shows that most elements are
depleted from the gas phase to some extent, a finding that we can explain
if we suppose that these elements are locked up in interstellar dust grains.
Support for this interpretation comes from the fact that the degree to which
elements are depleted generally correlates well with their condensation
temperature, i.e.\ the critical gas temperature below which a solid form is
the favored equilibrium state for the elements \citep{lodders03}. Elements
with high condensation temperatures are more easily incorporated into
dust grains than those with low condensation temperatures, and so if
the observed depletions are due to dust formation, one expects the
degree of depletion to increase with increasing condensation temperature,
as observed \citep[see e.g.\ Figure 15 in][]{jenk09}. In addition, depletions 
of high condensation temperature elements such as iron, nickel or silicon,
also seem to correlate with the mean gas density \citep{jenk09}, and so are
higher in the CNM than in the WNM \citep{welty99}. A plausible 
explanation of this fact is that dust growth in the ISM is an ongoing process
that occurs more rapidly in cold, dense gas than in warm, diffuse gas
\citep[see e.g.][]{zgt08}.

\subsection{Dust}
\label{dust}
The reddening of starlight in the ISM, and the fact that this effect correlates closely with the
hydrogen column density rather than with distance, points towards there being an additional
component of the ISM, responsible for absorbing light over a wide range of frequencies.
Measurements of the strength of the absorption at different frequencies show that when
there are distinct features in the extinction curve -- e.g.\ the 217.5\ {nm} bump --  they tend
to be extremely broad, quite unlike what we expect from atoms or small molecules. In
addition, measurement of elemental abundances in the local ISM show that a number of
elements, notably silicon and iron, are considerably less abundant in the gas-phase than in the
Sun. Finally, mid-infrared and far-infrared observations show that there is widespread continuum
emission, with a spectrum close to that of a black-body, and an intensity that once again
correlates well with the hydrogen column density. Putting all of these separate pieces of
evidence together, we are lead to the conclusion that in addition to the ionized, atomic and
molecular constituents of the ISM, there must also be a particulate component, commonly
referred to simply as dust.

Our best evidence for the nature of this dust comes from detailed measurements of the
spectral shape of the extinction curve that it produces. To a first approximation, individual
dust grains absorb only those photons with wavelengths smaller than the physical size of
the grain. Therefore, the fact that we see a large amount of absorption in the ultraviolet,
somewhat less in the optical and even less at infrared wavelengths tells us immediately
that there are many more small dust grains than there are large ones. In addition, we can
often associate particular spectral features in the extinction curve, such as the  217.5\ {nm}~bump
or the infrared bands at $9.7\ \umu$m and $18\ \umu$m, with particular types of dust grain:
graphite in the case of the 217.5\ {nm}~bump \citep{mrn77} and amorphous silicates in the
case of the infrared bands \citep[e.g.][]{dl84,draine07}; see also Figure \ref{fig:ISRF}. 

This argument can be made more quantitative, and has been used to derive detailed
constraints on the size distribution of interstellar dust grains. One of the earlier and still highly influential attempts
to do this was made by \citet{mrn77}. They were able
to reproduce the then-extant measurements of the ISM extinction curve between 0.1--1$\,\umu$m
with a mixture of  spherical graphite and silicate grains with a size distribution
\begin{equation}
N(a) da \propto a^{-3.5} da\,,
\label{eqn:MRN}
\end{equation}
where $a$ is the grain radius, and where the distribution extends over a range of
radii from $a_{\rm min} = 50\;$nm to  $a_{\rm max} = 0.25\;\umu$m. Subsequent
studies have improved on this simple description \citep[see e.g.][]{dl84,wd01a}, but
it remains a useful guide to the properties of interstellar dust. In particular, it is easy
to see that for grains with the size distribution given by equation (\ref{eqn:MRN}) -- commonly known as the MRN distribution
-- the total mass of dust is dominated by the contribution made by large grains, while
the total surface area is dominated by the contribution made by small grains. This
general behavior remains true in more recent models \cite[see the detailed discussion by][]{draine11}.

The total mass in dust is difficult to constrain purely with absorption measurements,
but if we combine these with measurements of elemental depletion patterns in the
cold ISM, then we can put fairly good constraints on how much dust there is. In the local
ISM, we find that the total mass of metals locked up in grains is roughly the same
as the total mass in the gas phase. The dust therefore accounts for around 1\% of
the total mass of the ISM. Therefore, when we attempt to model the behavior of the
ISM -- particularly its thermal and chemical behavior -- the dust can play a role that
is as important or more important than the gas-phase metals (see e.g.\ Section \ref{subsec:gas-grain-transfer}).

\subsection{Interstellar radiation field}
\label{subsec:ISRF}
The chemical and thermal state of the gas in the ISM is determined in large part by the interaction of the gas and the dust with the interstellar radiation field (ISRF). 
Several processes are important. First, the chemical state of the gas (the ionization fraction, the balance between atomic and molecular gas,
etc.) depends on the rate at which molecules are photodissociated and atoms are photoionized by the
radiation field. Second, the thermal state of the gas depends on the photoionization rate, and also on the rate of a process
known as  photoelectric heating: the ejection of an energetic electron from a dust grain due to the
absorption of a UV photon by the grain. And finally, the thermal state of the dust is almost entirely determined by the balance between the absorption
by the grains of radiation from the ISRF and the re-emission of this energy in the form of thermal radiation.

\begin{figure}
\center{\includegraphics[width=0.90\textwidth]{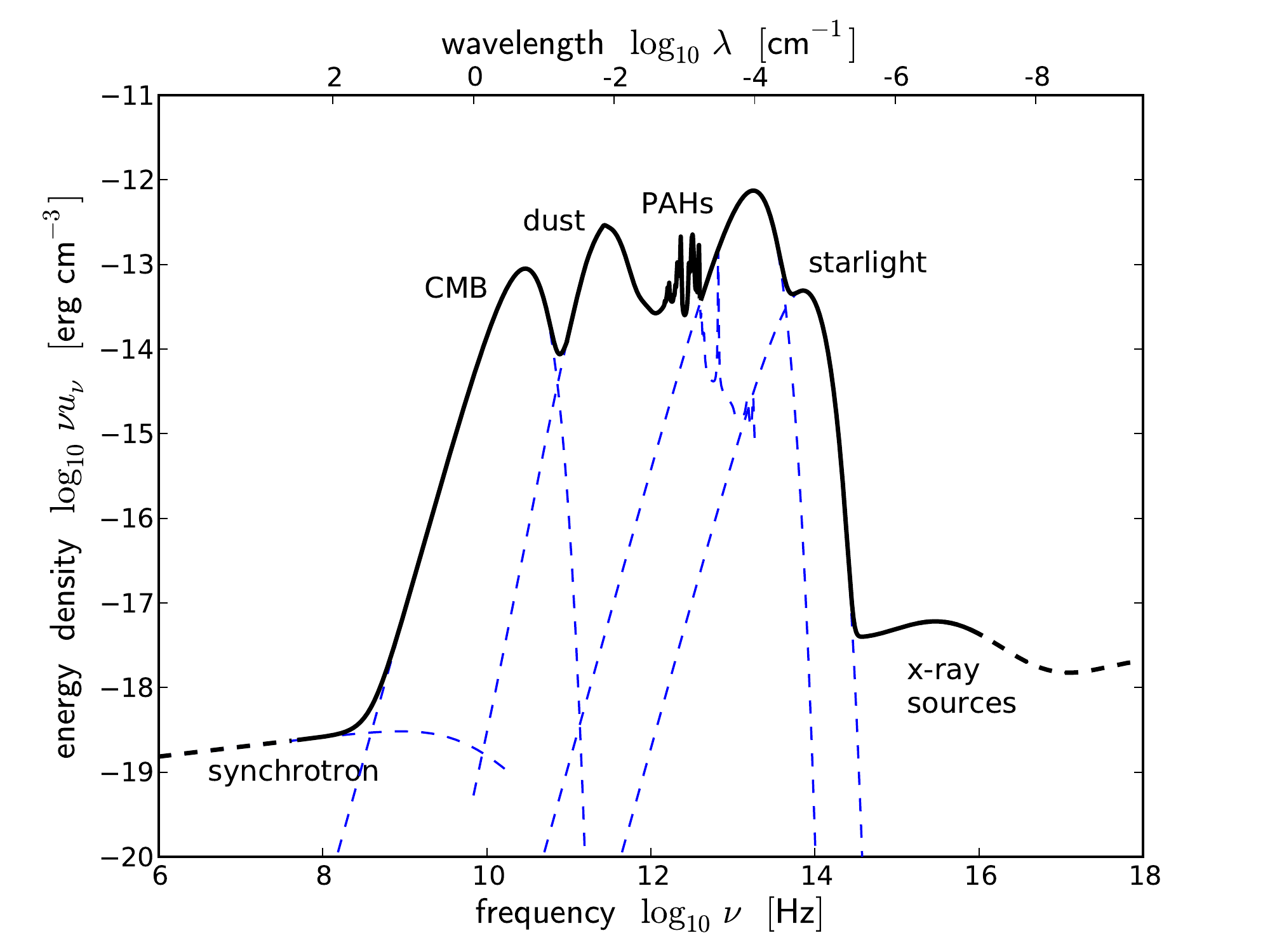}}
\caption{\label{fig:ISRF}
Schematic sketch of the energy density of the interstellar radiation field at different frequencies. The contributions of the cosmic microwave background (CMB) as well as of old, low-mass and young, high-mass stars are taken to be perfect blackbodies with temperatures  2.73$\,$K,  3500$\,$K, and 18000$\,$K, respectively \cite[see][]{chakreborty13}. The contributions from dust and PAHs are obtained from \citet{draine07}. The estimate for the Galactic synchrotron emission is taken from \citet{draine11} and the one for the X-ray flux from \citet{snowden97}. Note that in the vicinity of massive star clusters, the contributions from massive stars can be orders of magnitude larger than the numbers provided here. For further discussions, see for example \citet{draine11}.  }
\end{figure}

In the solar neighborhood, the ISRF is dominated by six components, {\em (1) } galactic synchrotron emission from relativistic electrons, {\em (2) } the cosmic microwave background (CMB), {\em (3) } infrared and far-infrared emission from dust grains heated by starlight, {\em (4) } bound-bound (bb), bound-free (bf) and free-free (ff) emission from $10^{4} \: {\rm K}$ ionized plasma (sometimes referred to as nebular emission), {\em (5) } starlight, and finally {\em (6) } X-rays from hot ($10^{5}$--$10^{8}$~K) plasma. 
The energy densities of each of these components are summarized in Table~\ref{isrf}
(adapted from \citealt{draine11}); see also Figure \ref{fig:ISRF}.

\begin{table}
\begin{center}
\caption{{ Energy densities in different components of the ISRF} \label{isrf}}
\vspace{.1in}
\begin{tabular}{l @{\hspace{0.5cm}} c}
\hline
Component  of ISRF & Energy density (erg~cm$^{-3}$) \\
\hline
Synchrotron & $2.7 \times 10^{-18}$ \\
CMB & $4.19 \times 10^{-13}$ \\
Dust emission & $5.0 \times 10^{-13}$ \\
Nebular emission (bf, ff) & $4.5 \times 10^{-15}$ \\
Nebular emission (H$\alpha$) & $8 \times 10^{-16}$ \\
Nebular emission (other bb) & $10^{-15}$ \\
Starlight, $T_{1} = 3000$~K & $4.29 \times 10^{-13}$ \\
Starlight, $T_{2} = 4000$~K & $3.19 \times 10^{-13}$ \\
Starlight, $T_{3} = 7000$~K & $2.29 \times 10^{-13}$ \\
Starlight, power-law & $7.11 \times 10^{-14}$ \\
Starlight, total & $1.05 \times 10^{-12}$ \\
Soft X-rays & $10^{-17}$ \\
\hline\\[-0.2cm]
{\footnotesize Adapted from \citet{draine11}.}\\
\end{tabular}
\end{center}
\end{table}
We see that most of the energy density of the ISRF is in the infrared,
where thermal dust emission and the CMB dominate, and in the optical and UV,
where starlight dominates. It is these components that play the main role in 
regulating the properties of the ISM, and so we focus on them below.

\subsubsection{Cosmic microwave background}
At wavelengths between $\lambda = 600\;\umu$m and $\lambda = 30\;$cm, the energy budget of the
ISRF is dominated by the CMB. This has an almost perfect black-body spectrum with a temperature
$T_{\rm CMB} = 2.725\;$K \citep{fixsen02}. 
This temperature is significantly lower than
the typical temperatures of the gas and the dust in the local ISM, and so despite the high energy density
of the CMB, energy exchange between it and these components does not substantially affect their
temperature \citep{black94}. The CMB therefore does not play a major role in the overall energy balance
of the ISM in the Milky Way or in other local galaxies. In high-redshift galaxies, however, the CMB temperature
and energy density are both much larger, with a redshift dependence of  $T_{\rm CMB} \propto (1+z)$ and $u_{\rm CMB} \propto (1+z)^{4}$.  The CMB can therefore play a much more significant role in regulating the thermal evolution of the
gas and the dust. The extent to which this affects the outcome of
the star formation process in high-redshift galaxies, and in particular the stellar initial mass function (IMF)
remains very unclear. Some authors have suggested that as the CMB essentially imposes a temperature
floor at $T_{\rm CMB}$, it can potentially affect the form of the IMF by suppressing low-mass star formation
when the CMB temperature is large \citep[see e.g.][]{cb03,so10}. However, the observational evidence for a 
systematic change in the IMF as one moves to higher redshift remains weak \citep{Bastian10,Offner14},
and although some simulations find evidence that a high CMB temperature can suppress low-mass star 
formation \citep[see e.g.][]{smith09}, other work suggests that low-mass stars can form even at very high 
redshifts \citep[see e.g.][]{clark11a,greif11b,greif12, dopcke13} by the fragmentation of the accretion disk surrounding the central star (see also Section \ref{subsec:IMF-models}).

\subsubsection{Infrared and and far-infrared emission from dust}
\label{isrf_mir}
Infrared emission from dust grains dominates the spectrum of the
ISRF between $\lambda = 5 \,\umu$m and $\lambda = 600 \,\umu$m.
About two-thirds of the total power is radiated in the mid and far-infrared,
at $\lambda > 50 \,\umu$m. This emission is largely in the form of thermal emission
from dust grains: the spectrum is that of a modified black-body
\begin{equation}
J_{\nu} \propto B_{\nu}(T_{0}) \left(\frac{\nu}{\nu_{0}}\right)^{\beta},
\end{equation}
where $J_{\nu}$ is the mean specific intensity of the radiation field,
$B_{\nu}(T_{0})$ is the Planck function, $T_{0}$ is the mean temperature
of the dust grains, and $\beta$ is the spectral index. In the Milky Way,
the mean dust temperature $\langle T_{\rm d} \rangle \approx 20$~K, and the
spectral index is typically around $\beta \approx 1.7$ \citep{planck13}.
The question of whether $\beta$ depends on temperature is somewhat
controversial. Many studies find an apparent anti-correlation
between $\beta$ and $T_{\rm d}$ \citep[see e.g.][]{dupac03,des08}. However, these
studies generally fit the spectral energy distribution (SED) with the $\chi^2$ linear regression method, which is known to produce an artificial anti-correlation from uncorrelated data  in some cases simply due to the presence of noise in the observations \citep{shetty09a,shetty09b}. When using hierarchical Bayesian methods for  determining $\beta$ and $T_{\rm d}$,  \citet{shetty09b} and the \citet{planck13} instead find a slight positive correlation between the two
parameters.

The remaining one-third of the dust emission is largely concentrated in a series of
distinct peaks at wavelengths $\lambda = 3.3, 6.2, 7.7, 8.6, 11.3$ and $12.7 \,\umu$m. These
peaks correspond to vibrational emission bands produced by so-called {polycyclic
aromatic hydrocarbons}, or PAHs for short. These are large organic molecules, containing one or
more benzene rings (hence `aromatic').

Although dust grains are large enough that we can usually treat them as macroscopic
objects without distinct radiative transitions, the same is not true for the much smaller
PAH molecules. The rate at which individual PAH molecules absorb photons is small, 
but each photon causes a significant change in the internal energy of the molecule. 
Their ``temperature'' therefore varies greatly with time -- they are very hot (i.e.\ have 
a large internal energy) immediately after they absorb a photon, but spend much of 
their time being very cold. Physically, what happens is actually a form of fluoresence 
-- the PAHs absorb UV photons, putting them into a highly excited state, and then cascade 
back to the ground state via a large number of infrared transitions.
An important implication of this is that PAH emission dominates at short wavelengths (i.e.\ in 
the near and mid-infrared) unless the other grains are also very hot. Since the strength of the PAH 
emission depends on the strength of the UV radiation field, it is therefore a useful tracer of the 
formation of massive stars. 

\subsubsection{Starlight}
Stars produce energy primarily at near infrared, visible and soft ultraviolet wavelengths. However,
in neutral regions of the ISM, stellar photons with energies greater than the ionization energy of
hydrogen, 13.6$\,$eV, are largely absent -- they are absorbed by hydrogen atoms, ionizing them,
and hence cannot penetrate deeply into neutral regions.

\cite{mmp83} showed that in the solar neighborhood, the starlight component 
of the ISRF could be represented at long wavelengths as the sum of three diluted black-body spectra.
At wavelengths $\lambda > 245\,${nm}, the radiation energy density is
\begin{equation}
\nu u_\nu = \sum_{i=1}^{3} \frac{8\pi h \nu^{4}}{c^{3}} \frac{W_{i}}{e^{h\nu / k_{\rm B}T_{i}} - 1}  \;\; {\rm erg \: cm^{-3}}\;.
\end{equation}
As usual $h= 6.626 \times10^{-27}\;$erg$\,$s and $k_{\rm B}= 1.381\times 10^{-16}\,$erg$\,$K$^{-1}$ are  Planck's and Boltzmann's constants. The quantities $W_{i}$ and $T_{i}$ are the dilution factor and temperature of each component, with
\begin{eqnarray}
T_{1} = 3000 \: {\rm K}, & \hfill & W_{1} = 7.0 \times 10^{-13} \;,\\
T_{2} = 4000 \: {\rm K}, & \hfill & W_{2} = 1.65 \times 10^{-13} \;,\\
T_{3} = 7500 \: {\rm K}, & \hfill & W_{3} = 1.0 \times 10^{-14} \;.
\end{eqnarray}
At wavelengths $\lambda < 245\,${nm}, the starlight contribution to the ISRF has been estimated
by a number of authors. The earliest widely-cited estimate was made by \citet{habing68}. He estimated
that $\nu u_{\nu} \approx 4 \times 10^{-14} \: {\rm erg \: cm^{-3}}$ at $\lambda = 100\ ${nm}, corresponding
to a photon energy of 12.4\ eV. It is often convenient to reference other estimates to this value, which
we do via the dimensionless parameter
\begin{equation}
\chi \equiv \frac{(\nu u_{\nu})_{100 {\mbox{\small {nm}}}}}{4 \times 10^{-14} \: {\rm erg \: cm^{-3}}}.
\end{equation}
Alternatively, we can reference other estimates to the \citet{habing68} field by comparing the total energy 
density in the range 6\ --\ 13.6$\,$eV. In this case, we define a different dimensionless parameter
\begin{equation}
G_{0} \equiv \frac{u(6\; \mbox{--}\, 13.6 \: {\rm eV})}{5.29 \times 10^{-14} \: {\rm erg \: cm^{-3}}}\;.
\label{eqn:Habing}
\end{equation}
If we are interested in e.g.\ the photodissociation of H$_{2}$ or CO, which requires photons with
energies above $10$~eV, then $\chi$ is the appropriate parameter to use. On the other hand, if we are
interested in e.g.\ the photoelectric heating rate, which is sensitive to a wider range of photon energies,
then $G_{0}$ is more appropriate.

Two other estimates of the UV portion of the ISRF are in widespread use: one due to \citet{dr78}
and the other due to \citet{mmp83}. \citet{dr78} fit the field with a polynomial function:
\begin{equation}
\lambda u_{\lambda} = 6.84 \times 10^{-14} \;\lambda_{2}^{-5} \left(31.016 \;\lambda_{2}^{2} - 49.913 \,\lambda_{2}
+ 19.897 \right) \: {\rm erg \: cm^{-3}},
\end{equation}
where $\lambda_{2} \equiv \lambda / 100\ ${nm}. This field has a normalization, relative to the Habing
field, of $\chi = 1.71$ and $G_{0} = 1.69$. 

\citet{mmp83} instead used a broken power-law fit:
\begin{equation}
\nu u_{\nu} = \left \{ \begin{array}{ll}
2.373 \times 10^{-14} \lambda^{-0.6678} \hspace{.3in} & \mbox{for~} 0.134\;\umu\mbox{{m}} < \lambda < 0.245\;\umu\mbox{{m}} \\
6.825 \times 10^{-13} \lambda & \mbox{for~} 0.110\;\umu\mbox{{m}} < \lambda \le 0.134\;\umu\mbox{{m}} \\
1.287 \times 10^{-9} \lambda^{4.4172} & \mbox{for~} 0.091\;\umu\mbox{{m}} < \lambda \le 0.110\;\umu\mbox{{m}} \\ 
\end{array} \right.
\end{equation}
Here, all wavelengths are in units of $\umu$m, and the energy densities are in units of
${\rm erg \: cm^{-3}}$. This estimate has $\chi = 1.23$ and $G_{0} = 1.14$. The available
observational evidence \citep[see e.g.][]{henry80,gond80} is better fit by the \citet{mmp83} field
than by the \citet{dr78} field, but the latter estimate is probably in wider use in models of the ISM.

\subsection{Cosmic rays}
\label{cr}
The final part of our inventory of the ISM are the cosmic rays. These are high energy, relativistic particles, mostly being nuclei ($\sim 99$\%) with a small fraction of electrons ($\sim 1$\%). The nuclei are primarily protons, but with about 10\% being alpha particles and $\sim 1$\% being metal nuclei. 
Their energy spans a wide range, from 100 MeV up to more than 1 TeV. The total energy
density in cosmic rays is approximately $2 \: {\rm eV} \: {\rm cm^{-3}}$, within a factor of a few
of the mean thermal energy density of the ISM. Cosmic rays therefore play an important role
in the overall energy balance of the gas.

\begin{figure}
\center{\includegraphics[width=0.55\textwidth]{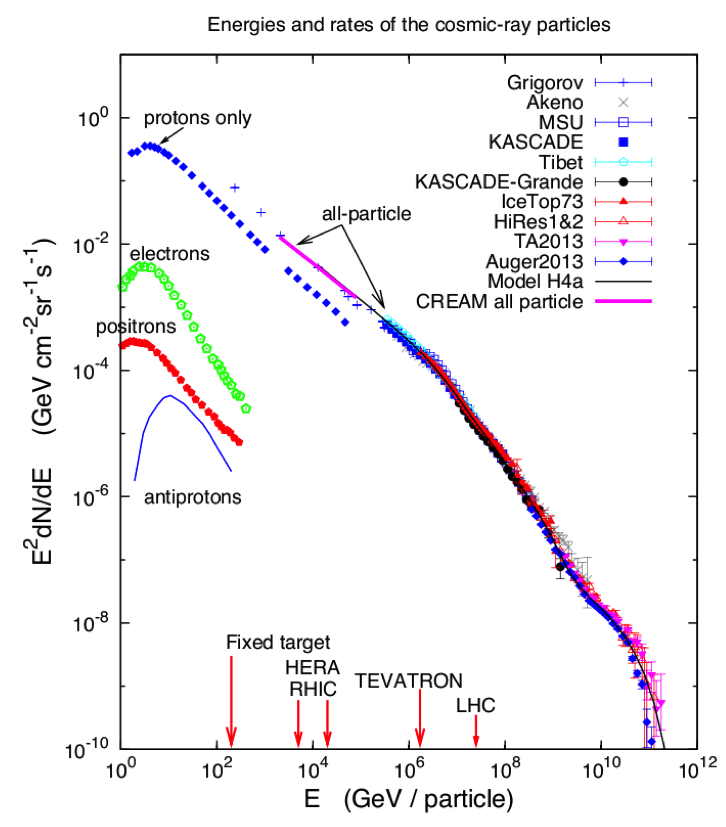}}
\parbox[b]{0.36\textwidth}{\vspace{2.0cm}
\caption{\label{fig:CR}
Energy spectrum of cosmic rays as observed with different instruments and telescopes. Plot from \citet{Blasi14}; see also  \citet{gaisser06} for further information. }\vspace*{0.7cm}} 
\end{figure}

All but the most highly energetic cosmic rays are tied to the magnetic field of the galaxy
and therefore scatter repeatedly within the disk. The expectation is therefore that the local
energy density in cosmic rays should be relatively uniform. Observations of cosmic rays
with TeV energies, which are not significantly affected by interactions with the solar wind,
find that their intensity in the solar rest-frame is almost isotropic, consistent with this picture
of a uniform energy density \citep{amen06}.

The spectrum of the cosmic rays (i.e.\ the flux per unit energy) decreases sharply with
increasing energy, and so the majority of the heating and ionization that they provide
comes from the least energetic cosmic rays, with energies of $\sim 100$~MeV or below.
Unfortunately, it is precisely 
this part of the cosmic ray energy spectrum that we know the least about. At this energy, cosmic 
rays are unable to penetrate within the heliosphere, owing to interactions with the solar wind. 
Our determination of the cosmic ray ionization rate is therefore indirect, based on chemical 
constraints.

An important example of this kind of constraint is provided by the abundance of the H$_{3}^{+}$
ion. This is formed in the diffuse ISM via the reaction chain
\begin{eqnarray}
{\rm H_{2}} + {\rm CR} & \rightarrow & {\rm H_{2}^{+}} + {\rm e^{-}}, \\
{\rm H_{2}^{+}} + {\rm H_{2}} & \rightarrow & {\rm H_{3}^{+}} + {\rm H},
\end{eqnarray}
where the first reaction is the rate-limiting step. It is destroyed by dissociative recombination
\begin{equation}
{\rm H_{3}^{+}} + {\rm e^{-}} \rightarrow {\rm various ~ products}.
\end{equation}
In a diffuse cloud, with $n_{\rm H_{2}} > n_{\rm H}$,  the equilibrium number density of H$_{3}^{+}$ that 
results from these reactions is given approximately by
\begin{equation}
n_{\rm H_{3}^{+}} = \frac{\zeta_{\rm H_{2}}}{k_{\rm dr}} \frac{n_{\rm H_{2}}}{n_{\rm e}},
\label{eqn:CR-density}
\end{equation}
where $k_{\rm dr}$ is the dissociative recombination rate coefficient and $\zeta_{\rm H_{2}}$
is the cosmic ray ionization rate of H$_{2}$, and where $n_{\rm H}$, $n_{\rm H_{2}}$, $n_{\rm H_{3}^+}$, and $n_{\rm e}$ are the number densities of H, H$_2$, H$_3^+$, and  free electrons, respectively.  

If we assume that the temperature and the values of the H$_{2}$-to-electron ratio do not vary
greatly along the line of sight, then we can convert  equation (\ref{eqn:CR-density}) into an expression relating the
column densities of H$_{2}$, H$_{3}^{+}$ and electrons,
\begin{equation}
N_{\rm H_{3}^{+}} = \frac{\zeta_{\rm H_{2}}}{k_{\rm dr}} \frac{N_{\rm H_{2}}}{N_{\rm e}}. \label{nh3p}
\end{equation}
Next, we note that within diffuse molecular clouds, the main source of electrons is ionized
carbon, C$^{+}$. We therefore assume that $N_{\rm C^{+}} = N_{\rm e}$ and  write
\begin{equation}
N_{\rm H_{3}^{+}} = \frac{\zeta_{\rm H_{2}}}{k_{\rm dr}} \frac{N_{\rm H_{2}}}{N_{\rm C^{+}}}. 
\end{equation}
Finally, this can be rearranged to give
\begin{equation}
\zeta_{\rm H_{2}} = \frac{N_{\rm H_{3}^{+}} N_{\rm C^{+}} k_{\rm dr}}{N_{\rm H_{2}}}.
\end{equation}
Since all of the column densities on the right-hand side of this expression can be measured
observationally \citep[see e.g.][]{savage77,cardelli96,mccall02}, and $k_{\rm dr}$ can be measured 
experimentally \citep{mccall04}, we can use this expression to constrain $\zeta_{\rm H_{2}}$.

In practice, a slightly more sophisticated version of this technique is used that accounts
for the fact that not all of the H$_{2}^{+}$ ions produced by cosmic ray ionization of H$_{2}$
survive for long enough to form H$_{3}^{+}$, and that also includes several additional
destruction processes for H$_{3}^{+}$ \citep{ind12}. Applying this technique to many lines-of-sight 
in the diffuse ISM, one finds that the resulting mean value for the cosmic ray ionization 
rate of H$_2$ is $\zeta_{\rm H_{2}} = 3.5 \times 10^{-16} \: {\rm s^{-1}}$, but also that there is
a very substantial scatter around this mean, with some lines-of-sight having $\zeta_{\rm H_{2}}
\sim 10^{-15} \: {\rm s^{-1}}$ or more, and others having $\zeta_{\rm H_{2}} \sim 
10^{-17} \: {\rm s^{-1}}$ \citep{ind12}.

In dense clouds of gas, the chemical abundances of other observable species such as OH or HCO$^{+}$
are also sensitive to the cosmic ray ionization rate, and measurements of the abundances of
these species relative to CO can therefore be used to provide additional constraints on 
$\zeta_{\rm H_{2}}$. These other techniques typically find that in dense gas, $\zeta_{\rm H_{2}}
\sim 10^{-17}\: {\rm s^{-1}}$ \citep[see e.g.][]{will98,vv00}. This is consistent with the low end of the range of values found using
H$_{3}^{+}$, but not with the higher values found in many diffuse clouds. This difference between
diffuse and dense clouds may indicate that the cosmic rays that dominate the heating and ionization
of the local ISM have energies of only a few MeV, allowing them to penetrate low column density,
diffuse atomic or molecular gas, but not high column density clouds \citep{pad09}. Alternatively, 
purely magnetic effects, such as the interaction between low energy cosmic rays and their self-generated
MHD waves \citep{ps05} may explain the apparent inability of low energy cosmic rays to travel into dense 
molecular cloud regions. In either case, these mechanisms do not explain the large scatter in $\zeta_{\rm H_{2}}$ seen in purely 
diffuse clouds, which may indicate that the energy density in very low energy cosmic rays is significantly less 
uniform than has been previously supposed \citep{ind12}.

\newpage
\section{Heating and cooling of interstellar gas}
\label{sec:heating-cooling}

\subsection{Optically-thin two-level atom}
\label{subsec:opt-thin-2-level-atom}
A convenient starting point for understanding how radiative cooling operates in the ISM is the 
two-level atom. This toy model allows us to illustrate many of the most important concepts 
involved without unnecessarily complicating the mathematical details. 

Picture an atomic system with two bound states, a lower level $l$ and an upper level $u$,
with statistical weights $g_{l}$ and $g_{u}$, separated by an energy $E_{ul}$. We will
write the number density of the atoms in the lower level as $n_{l}$ and the number density
in the upper level as $n_{u}$. The total number density of the atoms then  follows
as $n_{\rm atom} = n_{l} + n_{u}$. For simplicity, we consider for the time being monoatomic gases at rest, so that $dn_{\rm atom}/dt = 0$. Even if the number density $n_{\rm atom}$ remains constant, the values of $n_{l}$ and $n_{u}$ will change over time, as individual atoms transition from the lower to the upper 
level due to collisional excitation or the absorption of a photon, and transition from the 
upper to the lower level due to collisional de-excitation or the emission of a photon. For the time being, we consider optically thin conditions, which means that the emitted photon can leave the region of interest unimpeded. In the opposite, optically thick case, it would very likely be  absorbed by a neighboring atom.  

We note that in general, the atomic species under consideration is just one amongst many. Typically, the medium is a mixture of  different atomic or molecular species $i$, with the total number density being $n= \sum n_i$. In this case our atoms will not only collide with each other, but also with particles of the other species. Chemical reactions lead to further complications as the total number (and consequently the number density) of our atoms is no longer conserved, but instead may change with time. If we also consider gas motions, even for monoatomic gases  the local density will vary with time, so that again $dn_{\rm atom}/dt \ne 0$. 

For a static system without chemical reactions,  we can write the rates of change of $n_{l}$ and $n_{u}$ at fixed $n_{\rm atom}$ as
\begin{eqnarray}
\frac{{d}n_{u}}{{d}t} & = & C_{lu} n_{l} - C_{ul} n_{u} - A_{ul} n_{u} - B_{ul} I_{ul} n_{u} + B_{lu} I_{ul} n_{l} \;, \\
\frac{{d}n_{l}}{{d}t}  & = & - C_{lu} n_{l} + C_{ul} n_{u} + A_{ul} n_{u} + B_{ul} I_{ul} n_{u} - B_{lu} I_{ul} n_{l}\;.
\end{eqnarray}
Here, $C_{lu}$ and $C_{ul}$ are the collisional excitation and de-excitation rates, which we discuss
in more detail below, $A_{ul}$, $B_{ul}$ and $B_{lu}$ are the three Einstein
coefficients for the transition (describing spontaneous emission, stimulated emission and
absorption, respectively), and $I_{ul}$ is the specific intensity of the local radiation field at a
frequency $\nu_{ul} = E_{ul} / h$. For a detailed derivation of this equation and the parameters involved, see e.g.\ the textbook by \citet{rybicki86}. 

Typically, the radiative and/or collisional transitions occur rapidly compared to any of the other timescales of 
interest in the ISM, and so it is usually reasonable to assume that the level populations have reached a state
of statistical equilibrium in which
\begin{equation}
\frac{{d}n_{u}}{{d}t} = \frac{{d}n_{l}}{{d}t} = 0.
\end{equation}
In this case, $n_{l}$ and $n_{u}$ are linked by a single algebraic equation,
\begin{equation}
(C_{lu} + B_{lu} I_{ul}) n_{l} = \left(C_{ul} + A_{ul} + B_{ul} I_{ul} \right) n_{u}.
\end{equation}
A further simplification that we can often make is to ignore the effects of the incident radiation field.
This is justified if the gas is optically thin and the strength of the interstellar radiation field at 
the frequency $\nu_{ul}$ is small. In this regime, we have
\begin{equation}
C_{lu} n_{l} = \left(C_{ul} + A_{ul} \right) n_{u}\;,
\end{equation}
which we can rearrange to give
\begin{equation}
\frac{n_{u}}{n_{l}} = \frac{C_{lu}}{C_{ul} + A_{ul}} \;.
\end{equation}

The collisional excitation rate $C_{lu}$ describes the rate per atom at which collisions with
other gas particles cause the atom to change its quantum state from level $l$ to level $u$.
In principle, collisions with any of the many different chemical species present
in the ISM will contribute towards $C_{lu}$, but in practice, the main contributions come
from only a few key species: H, H$_{2}$, H$^{+}$, He, and free electrons. We can write
$C_{lu}$ as a sum of the collisional excitation rates due to these species,
\begin{equation}
C_{lu} = \sum_{i} q_{lu}^{i} n_{i}\;,
\end{equation}
where $i = {\rm H, H_{2}, H^{+}, He, e^{-}}$ and $q_{lu}^{i}$ is the collisional excitation rate
coefficient for collisions between our atom of interest and species $i$. The collisional
excitation rate coefficients themselves can be computed using the tools of quantum chemistry
or measured in  laboratory experiments. Values for many atoms and molecules of astrophysical interest can
be found in the LAMDA database\footnote{http://home.strw.leidenuniv.nl/$\sim$moldata/} \citep{sch05}.

Given the excitation rate $C_{lu}$, it is straightforward to obtain the de-excitation rate $C_{ul}$
by making use of the principle of detailed balance. This states that in local thermal equilibrium (LTE), the 
rate at which collisions cause transitions from level $l$ to level $u$ must be the same as the rate at 
which they cause transitions from level $u$ to level $l$. This is a consequence of microscopic 
reversibility, i.e.\ the fact that the microscopic dynamics of particles and fields are time-reversible, 
because the  governing equations are symmetric in time. 

In thermal equilibrium, we know that the ratio of atoms in level $u$ to those in level $l$
is simply given by the Boltzmann distribution
\begin{equation}
\frac{n_{u}}{n_{l}} = \frac{g_{u}}{g_{l}} e^{-E_{ul} / k_{\rm B}T}\;.
\end{equation}
The principle of detailed balance tells us that for any collisional transition the equilibrium condition reads as
\begin{equation}
q_{lu}^{i} n_{l} n_{i} = q_{ul}^{i} n_{u} n_{i}\;.
\end{equation}
It therefore follows that
\begin{equation}
C_{lu} n_{l} = C_{ul} n_{u}\;,
\end{equation}
and consequently, we can write 
\begin{equation}
\frac{C_{lu}}{C_{ul}}  = \frac{n_{u}}{n_{l}} = \frac{g_{u}}{g_{l}} e^{-E_{ul} / k_{\rm B}T} 
\label{det_bal}
\end{equation}
for a system in thermal equilibrium. The true power of the principle of detailed balance becomes
clear once we realize that the values of $C_{lu}$ and $C_{ul}$ depend only on the quantum 
mechanical properties of our atoms, and not on whether our collection of atoms actually is in thermal equilibrium
or not. Therefore, although we have assumed thermal equilibrium in deriving equation~(\ref{det_bal}),
we find that the final result holds even when the system is not in equilibrium.

We can use this relation between $C_{lu}$ and $C_{ul}$ to write our expression for
$n_{u}/n_{l}$ in the form
\begin{equation}
\frac{n_{u}}{n_{l}} = \frac{(g_{u} / g_{l}) e^{-E_{ul} / k_{\rm B}T}}{1 + A_{ul} / C_{ul}}\;.
\end{equation}
In the limit that collisions dominate the behavior, i.e.\ for $C_{ul} \gg A_{ul}$, we recover the Boltzmann distribution. On the other hand, if radiative de-excitation is more important than the collisional one, that is in the limit $C_{ul} \ll A_{ul}$, we find that
\begin{equation}
\frac{n_{u}}{n_{l}} \approx \frac{C_{ul}}{A_{ul}} \frac{g_{u}}{g_{l}} e^{-E_{ul} / k_{\rm B}T}\;.
\end{equation} 
Together with  equation~(\ref{det_bal}) we arrive at
\begin{equation}
\frac{n_{u}}{n_{l}} \approx \frac{C_{lu}}{A_{ul}}\;.
\end{equation}
We see therefore that when collisions dominate over radiative decays, the level populations approach
their LTE values, while in the other limit, collisional excitations are balanced by radiative de-excitations, 
and collisional de-excitations are unimportant.

In the simple case in which collisions with a single species dominate $C_{ul}$, we can write the collisional
de-excitation rate as $C_{ul} = q_{ul}^{i} n_{i}$, where $n_{i}$ is the number density of the dominant collision
partner. Since the key parameter that determines whether collisions or radiative decays dominate is the ratio
$A_{ul} / C_{ul}$, we can define a critical density for the collision partner, such that this ratio is one,
\begin{equation}
n_{{\rm cr}, i} \equiv \frac{A_{ul}}{q_{ul}^{i}}\;.  
\label{crit_c}
\end{equation}
When $n_{i} \gg n_{{\rm cr}, i}$, collisions dominate and the level populations tend to their LTE values.
On the other hand, when $n_{i} \ll n_{{\rm cr}, i}$, radiative decay dominates and most atoms are in their
ground states.

In the more general case in which collisions with several different species make comparably large
contributions to $C_{ul}$, we can define the critical density in a more general fashion. If we take $n$
to be some reference number density (e.g.\ the number density of H nuclei, which has the benefit that it
is invariant to changes in the ratio of atomic to molecular hydrogen), then we can define a critical density
with the following expression,
\begin{equation}
\frac{A_{ul}}{C_{ul}} \equiv \frac{n_{\rm cr}}{n}\;.
\end{equation}
Here, $n_{\rm cr}$ is the critical value of our reference density, rather than that of a specific collision
partner. In terms of the individual fractional abundances and collisional de-excitation rates, we have
\begin{equation}
n_{\rm cr} = \frac{A_{ul}}{\sum_{c} q_{ul}^{i} x_{i}}\;
\label{eqn:ncrit}
\end{equation}
where $x_{i} \equiv n_{i} / n$ is the relative abundance of the species $i$. Alternatively, if we divide through by $A_{ul}$, we can easily show that
\begin{equation}
n_{\rm cr} = \left[ \sum_{i} \frac{x_{i}}{n_{{\rm cr}, i}} \right]^{-1}\;,
\end{equation}
where the critical densities for the individual colliders are given by equation~(\ref{crit_c}) above.

Using our general definition of the critical density, we can write the ratio of the level populations
of our two level atom as
\begin{equation}
\frac{n_{u}}{n_{l}} = \frac{(g_{u} / g_{l}) e^{-E_{ul} / k_{\rm B}T}}{1 + n_{\rm cr} / n}\;.
\end{equation}
We can now use the fact that for our species of interest in the two-level approximation the density  $n_{\rm atom} = n_{l} + n_{u}$, and rewrite this equation as
\begin{equation}
\frac{n_{u}}{n_{\rm atom} - n_{u}} = \frac{(g_{u} / g_{l}) e^{-E_{ul} / k_{\rm B}T}}{1 + n_{\rm cr} / n}\;. 
\end{equation}
Further rearrangement gives
\begin{equation}
\frac{n_{u}}{n_{\rm atom}} = \frac{(g_{u} / g_{l}) e^{-E_{ul} / k_{\rm B}T}}{1 +  n_{\rm cr} / n + (g_{u} / g_{l}) e^{-E_{ul} / k_{\rm B}T}}\;. 
\label{levelp}
\end{equation}
The radiative cooling rate $\Lambda_{ul}$ of our collection of atoms is simply the rate at which they emit photons multiplied 
by the energy of the photons, i.e.\
\begin{equation}
\Lambda_{ul}  = A_{ul} E_{ul} n_{u}\;.
\end{equation}
If we make use of the expression derived above for $n_{u}$,  this becomes
\begin{equation}
\Lambda_{ul} = A_{ul} E_{ul} n_{\rm atom} \frac{(g_{u} / g_{l}) e^{-E_{ul} / k_{\rm B} T}}{1 +  n_{\rm cr} / n + (g_{u} / g_{l}) e^{-E_{ul} / k_{\rm B}T}}\;.
\label{cool_atom}
\end{equation}
It is informative to examine the behavior of this expression in the limits of very low and very
high density. At low densities, $n \ll n_{\rm cr}$, equation~(\ref{cool_atom}) reduces to
\begin{equation}
\Lambda_{ul, \, n \rightarrow 0} = A_{ul} E_{ul} n_{\rm atom} \frac{(g_{u} / g_{l}) e^{-E_{ul} / k_{\rm B}T}}{n_{\rm cr} / n}\;.
\label{eqn:Lambda01}
\end{equation}
We can use the equation of detailed balance in the form (\ref{det_bal}) together with the definition of the critical density as given by  equation (\ref{eqn:ncrit})  to derive the more useful expression 
\begin{equation}
\Lambda_{ul, \, n \rightarrow 0} = E_{ul} \left( \sum_{i} q_{lu}^{i} n_{i} \right) n_{\rm atom} = E_{ul} C_{lu} n_{\rm atom}\;\;.
\end{equation}
Physically, this expression has a simple interpretation. At low densities, every collisional excitation is followed by 
radiative de-excitation and hence by the loss of a photon's worth of energy from the gas. The cooling
rate in this limit therefore depends only on the excitation rate of the atom, and is independent of the
radiative de-excitation rate. Moreover, this rate is proportional to the total number density of the gas, $C_{lu} \propto n$, and in addition $n_{\rm atom} \propto n$, if the fractional abundance of our atomic coolant is
independent of density. As a consequence,  the cooling rate scales with the density squared in the low-density regime, 
\begin{equation}
\Lambda_{ul, \, n \rightarrow 0} \propto n^{2}\;.
\end{equation}
The behavior is different at high densities, $n \gg n_{\rm cr}$. The expression for the cooling rate now becomes 
\begin{equation}
\Lambda_{ul, \, \rm LTE}  = A_{ul} E_{ul} n_{\rm atom} \left[ \frac{(g_{u} / g_{l}) e^{-E_{ul} / k_{\rm B}T}}{1 + (g_{u} / g_{l}) e^{-E_{ul} / k_{\rm B} T}} \right].
\end{equation}
The term in square brackets is simply the fraction of all of the atoms that are in the
upper level $u$ when the system is in LTE, a quantity that we will refer to as $f_{u, \, \rm LTE}$.
In this limit, we write 
\begin{equation}
\Lambda_{ul, \, \rm LTE}  = A_{ul} E_{ul} f_{u, \, \rm LTE} n_{\rm atom}\;.
\end{equation}
This is known as the LTE limit. In this limit,  the {\em mean} cooling rate per atom depends only on
the temperature, and not on the collisional excitation rate. Consequently,  the cooling
rate scales linearly with the density,
\begin{equation}
\Lambda_{ul, \, \rm LTE} \propto n\;.
\end{equation}

\subsection{Effects of line opacity}
So far, we have assumed that the strength of the local radiation field at the frequency of the
atomic transition is negligible, allowing us to ignore the effects of absorption and stimulated
emission. This is a reasonable approximation when the gas is optically thin, provided that
the ISRF is not too strong, but becomes a poor approximation once the gas becomes optically
thick. Therefore, we now generalize our analysis to handle the effects of absorption and stimulated 
emission.

Consider once again our two-level atom, with level populations that are in statistical equilibrium. 
In this case, we have
\begin{equation}
(C_{lu} + B_{lu} J_{ul}) n_{l} = (A_{ul} + B_{ul} J_{ul} + C_{ul}) n_{u}, \label{full_lp}
\end{equation}
where $J_{ul}$ is the mean specific intensity
\begin{equation}
J_{ul} = \frac{1}{4\pi} \oint I_{ul}(\vec{n}) {d}\Omega,
\end{equation}
where the integral is over all directions $\vec{n}$ and ${d}\Omega$ is the  solid angle element.

In general, to solve this equation throughout our medium, 
we need to know $J_{ul}$ at every point, and since $J_{ul}$ depends on the level populations, we 
end up with a tightly coupled problem that is difficult to solve even for highly symmetric systems, 
and that in general requires a numerical treatment. A detailed discussion of the different 
numerical methods that can be used to solve this optically-thick line transfer problem is outside the scope of
our lecture notes. Instead, we refer the reader to the paper by \citet{vz02} and the references therein. 

Here, we restrict our attention to a simple but important limiting case. We start by assuming that any incident 
radiation field is negligible, and hence that the only important contribution to $J_{ul}$ comes from the 
emission of the atoms themselves. We also assume that there are only three possible fates for the emitted
photons: 
\begin{enumerate}
\setlength{\itemindent}{0.4cm}
\item[\em (1)] Local absorption, followed by collisional de-excitation of the atom\footnote{By local, we generally mean within a small volume around the emission site, within which we can assume that physical conditions such 
as density and temperature do not  vary appreciably.}
\item[\em (2)] Local absorption, followed by re-emission (i.e.\ scattering)
\item[\em (3)] Escape from the gas
\end{enumerate}
Photons which scatter may do so once or many times, before either escaping from the gas, or being removed
by absorption followed by collisional de-excitation. The probability that the photon eventually escapes from
the gas is termed the escape probability. In its most general form, this can be written as
\begin{equation}
\beta(\vec{x}) = \frac{1}{4\pi} \oint \int e^{-\tau_{\nu}(\vec{x, n})}
\phi(\nu) {d}\nu {d}\Omega,
\label{beta}
\end{equation}
where $\beta(\vec{x})$ is the escape probability at a position $\vec{x}$, $\phi(\nu)$ is the line profile function, a function normalized to unity that describes the shape of the line, and $\tau_{\nu}(\vec{x, n})$ is the optical depth at frequency $\nu$ at position $\vec{x}$ in the direction $\vec{n}$. 

We note that the net number of absorptions (i.e.\ the number of photons absorbed minus the number produced 
by stimulated emission) must equal the number of photons emitted that do not escape from
the gas, i.e.\
\begin{equation}
(n_{l} B_{lu} - n_{u} B_{ul}) J_{ul} = n_{u} (1 - \beta) A_{ul}.
\end{equation}
Using this, we can rewrite equation (\ref{full_lp}) for the statistical equilibrium level populations as
\begin{equation}
C_{lu} n_{l} = (C_{ul} + \beta A_{ul}) n_{u}.
\end{equation}
Local absorptions reduce the effective radiative de-excitation rate by a factor determined by the  escape probability $\beta$, i.e.\ we go from $A_{ul}$ in the 
optically thin case to $A^{\prime}_{ul} = \beta A_{ul}$ in the optically thick case.
Therefore, all of our previously derived results still hold provided that we make the substitution 
$A_{ul} \rightarrow A^{\prime}_{ul}$.
One important consequence of this is that the critical density {decreases}. 
Since $n_{\rm cr} \propto A_{ul}$, we see that when the gas is optically thick, 
$n_{\rm cr} \propto \beta$. This means that the effect of local absorption
(also known as {photon trapping}) is to lower the density at which LTE is reached.
The higher the optical depth, the more pronounced this effect becomes.

In order for the escape probability approach to be useful in practice, we need to be able to 
calculate $\beta$ in a computationally efficient fashion. Unfortunately, the expression for
$\beta$ given in equation~(\ref{beta}) is not well suited for this.
The reason for this is the dependence of $\beta$ on the direction-dependent optical
depth $\tau_{\nu}(\vec{x, n})$. This can be written in terms of the absorption coefficient 
$\alpha_{\nu}$ as
\begin{equation}
\tau_{\nu}(\vec{x, n}) = \int_{0}^{\infty} \alpha_{\nu}(\vec{x} + s\vec{n}, \vec{n}) {d}s,
\end{equation}
where $\alpha_{\nu}(\vec{x} + s\vec{n}, \vec{n})$ is the absorption coefficient at position $\vec{x} + s\vec{n}$ for photons propagating in the direction $\vec{n}$. To compute the integral over solid angle in equation~(\ref{beta}), we need to integrate for
$\tau_{\nu}$ along many rays between the point of interest and the edge of the cloud. 
This can be done, but if we want to properly sample the spatial distribution of the gas, 
then the computational cost of performing these integrals will typically
scale as $N^{2/3}$, where $N$ is the number of fluid elements in our model cloud. 
If we then need to repeat this calculation for every fluid element (e.g.\ in order to calculate
the cooling rate at every point in the cloud), the result is a calculation that scales as 
$N^{5/3}$. For comparison, modeling the hydrodynamical or chemical evolution of the
cloud has a cost that scales as $N$. 

Because of the high computational cost involved in computing $\beta$ accurately, most
applications of the escape probability formalism make further simplifications to allow
$\beta$ to be estimated more easily. One common approach is to simplify the geometry
of the cloud model under consideration. For example, if we adopt a spherically symmetric
or slab-symmetric geometry, the inherent dimensionality of the problem can be reduced
from three to one, greatly speeding up our calculation of $\beta$. This approach can work
very well in objects such as prestellar cores that are quasi-spherical, but becomes less
applicable as we move to larger scales in the ISM, since real molecular clouds exhibit complex and highly inhomogeneous density and velocity structure (Section \ref{par:prop-MC}) and are not 
particularly well described as either slabs or spheres.

A more useful approximation for treating cooling in interstellar clouds is the Large Velocity Gradient 
(LVG) approximation, also known as the Sobolev approximation \citep{sob57}. The basic idea here
is that when there are large differences in the velocities of adjacent fluid elements, photons can
more easily escape from the gas. Suppose a photon is emitted at position $\vec{x}$ from gas
moving with velocity $\vec{v}$, and propagates a distance $\Delta \vec{x}$, to a point in the gas
where the velocity is $\vec{v} + \Delta \vec{v}$. The probability of the photon being absorbed
at this point depends on the frequency of the photon in the rest frame of the gas at that point.
If the change in velocity is sufficient to have Doppler-shifted the photon out of the core of the
line, the probability of it being absorbed is small. For lines dominated by thermal broadening,
the required change in velocity is roughly equal to the thermal velocity of the absorber, $v_{\rm th}$. If the photon can successfully propagate a distance
\begin{equation}
L_{\rm s} \equiv \frac{v_{\rm th}}{| {d}v/{d}x|},  \label{sobolev}
\end{equation}
where ${d}v/{d}x$ is the velocity gradient, then it is extremely likely that it will 
escape from the gas. The length-scale defined by equation~(\ref{sobolev}) is known as the
Sobolev length, and the LVG approximation can be used successfully when this length-scale
is significantly shorter than the length-scales corresponding to variations in the density,
temperature or velocity of the gas, or in the fractional abundance of the absorbing species.

More quantitatively, when the Sobolev approximation applies, the integral over frequency
can be solved analytically, yielding
\begin{equation}
\int e^{-\tau_{\nu}(\vec{x, n})} \phi(\nu) {d}\nu = \frac{1 - e^{-\tau_{\rm LVG}(\vec{x, n})}}{\tau_{\rm LVG}(\vec{x, n})},
\end{equation}
where $\tau_{\rm LVG}(\vec{x, n})$ is the direction-dependent LVG optical depth,
\begin{equation}
\tau_{\rm LVG}(\vec{x, n}) = \frac{A_{ul}c^{3}}{8 \pi \nu_{ul}^{3}} \frac{1}{\left| \vec{n} \cdot \vec{\nabla} v \right|}
\left(\frac{g_{l}}{g_{u}} n_{u} - n_{l} \right).
\end{equation}
In this case, $\beta$ is given by the expression
\begin{equation}
\beta(\vec{x}) = \frac{1}{4\pi} \oint \frac{1 - e^{-\tau_{\rm LVG}(\vec{x, n})}}{\tau_{\rm LVG}(\vec{x, n})} {d}\Omega.
\end{equation}
The validity of the Sobolev approximation for line transfer in turbulent molecular clouds was examined by
\citet{ossen97}, who showed that most of the fluid elements contributing significantly to the $^{13}$CO
line emission produced by a typical turbulent cloud had short Sobolev lengths, justifying the use
of the LVG approximation for modeling the emission \citep[see also][]{oss02}. 

\subsection{Multi-level systems}
So far, we have restricted our discussion to the case of a simple two-level system.
However, most of the important coolants in the ISM have more than two energy levels that need
to be taken into account when computing the cooling rate, and so in this section, we briefly look
at how we can generalize our analysis to the case of multiple levels.

When we are dealing with more than two levels, and hence more than a single transition 
contributing to the cooling rate, then the net cooling rate can be written in terms of the level
populations as
\begin{equation}
 \Lambda = \sum_{u} \sum_{l < u} E_{ul} \left[ (A_{ul} + B_{ul} J_{ul}) n_{u} - B_{lu} J_{ul} n_{l} \right],
\end{equation}
where the second sum is over all states $l$ that have energies $E_{l} < E_{u}$. 
A major difficulty here comes from the need to compute the level populations $n_{u}$.
If the levels are in statistical equilibrium, then the level populations satisfy the  
equation
\begin{eqnarray}
& & \sum_{j > i} \left[n_{j} A_{ji} + (n_{j} B_{ji} - n_{i} B_{ij}) J_{ij} \right] - \sum_{j < i} \left[n_{i} A_{ij} 
+ (n_{i} B_{ij} - n_{j} B_{ji}) J_{ij} \right] \nonumber \\
& & \mbox{} + \sum_{j \neq i} \left[n_{j} C_{ji} - n_{i} C_{ij} \right] = 0.
\end{eqnarray}
If we have $N$ different levels, then this equation can also be written in the form of $N$
coupled linear equations. These are straightforward to solve numerically if the mean
specific intensities $J_{ij}$ are known, but just as in the two-level case, these specific intensities 
will in general depend on the level populations at every point in our gas, meaning that the 
general form of the non-LTE statistical equilibrium equation is very challenging to solve
numerically in an efficient manner. 

Consequently, when computing cooling rates for complicated multi-level systems,
we often make use of  simplifying assumptions similar to those we have already 
discussed in the case of the two-level system. For example, when the gas
density is very low, it is reasonable to assume that essentially all of our coolant
atoms or molecules will be in the ground state, and that every collisional
excitation from the ground state will be followed by the loss of a photon from the
gas (possible after one or more scattering events, if the gas is optically thick). In
this limit, the cooling rate simplifies to
\begin{equation}
\Lambda_{n \rightarrow 0} = \sum_{u} E_{u0} C_{u0} n_{0},
\end{equation}
where $n_{0}$ is the number density of coolant atoms/molecules in the ground state,
$C_{u0}$ is the collisional excitation rate from the ground state to state $u$, and 
$E_{u0}$ is the difference in energy between state $u$ and the ground state.

In the LTE limit, the cooling rate is also easy to calculate, as the level populations
will simply have the values implied by the Boltzmann distribution,
\begin{equation}
n_{u} = \frac{(g_{u} / g_{0}) e^{-E_{u0} / k_{\rm B}T}}{Z(T)} n_{\rm atom},
\end{equation}
where $n_{\rm atom}$ is the total number density of the coolant of interest,
$g_{0}$ is the statistical weight of the ground state, $g_{u}$ is the statistical
weight of level $u$, $E_{u0}$ is the energy difference between level $u$ and the
ground state, and $Z$ is the partition function. It is given by the expression
\begin{equation}
Z(T) = \sum_{i} \frac{g_{i}}{g_{0}} e^{-E_{i0}/k_{\rm B}T},
\end{equation}
where we sum over all of the states of the coolant, including the ground state.
In this limit, we still need to know the mean specific intensities of the various lines in
order to calculate the total cooling rate. In principle, these are straightforward to compute 
when the level populations are fixed, although for reasons of computational efficiency,
a further simplifying assumption such as the LVG approximation is often adopted.

In the case of our simple two-level system, we have already seen that at densities
$n \ll n_{\rm crit}$, it is safe to use the low-density limit of the cooling rate, while at
densities $n \gg n_{\rm crit}$, the LTE limit applies. In the multi-level case, the
situation is slightly more complicated, as in principle there is a critical density
$n_{{\rm crit}, ul}$ associated with every possible transition that can occur. 
Moreover, since there can be large differences in the value of $A_{ul}$ between
one transition and another, these individual critical densities can differ by orders
of magnitude. We can therefore often be in the situation where some of the energy
levels of our coolant are in LTE, while others are not. In practice, what is often done
if we are interested in the total cooling rate and not in the strengths of the individual
lines is to define an effective critical density \citep{hm79}
\begin{equation}
n_{\rm crit, eff} = \frac{\Lambda_{\rm LTE}}{\Lambda_{n \rightarrow 0}} n\;,
\end{equation}
and  write the density-dependent cooling rate as
\begin{equation}
\Lambda = \frac{\Lambda_{\rm LTE}}{1 + n_{\rm crit, eff} / n}.
\end{equation}
This expression can be somewhat approximate at densities close to $n_{\rm crit, eff}$,
but becomes very accurate in the limit of low density or high density.

\subsection{Atomic and molecular coolants in the ISM}
\label{atmolcool}
Having briefly outlined the basic physical principles of line cooling, we now go on to examine which of the many possible forms of
line emission are most important for the cooling of interstellar gas.

\subsubsection{\bf Permitted transitions}
At high temperatures, in regions dominated by atomic or ionized gas, the cooling of
the ISM takes place largely via the permitted (i.e.\ dipole-allowed) electronic transitions
of various atoms and ions. At temperatures close to $10^{4}$~K, excitation of the Lyman
series lines of atomic hydrogen is the dominant process\footnote{Note that this is often 
referred to in the literature simply as ``Lyman-$\alpha$'' cooling}, giving rise to a cooling rate 
per unit volume \citep{black81,cen92} of
\begin{equation}
\Lambda_{\rm H} = 7.5 \times 10^{-19} \frac{1}{1 + (T / 10^{5})^{1/2}} \exp \left(\frac{-118348}{T} \right)
n_{\rm e} n_{\rm H},
\end{equation}
where $n_{\rm e}$ and $n_{\rm H}$ are the number densities of free electrons and atomic
hydrogen, respectively. At temperatures $T \sim 3 \times 10^{4}$~K and above, however, the 
abundance of atomic hydrogen generally becomes very small, and other elements, particularly 
C, O, Ne and Fe, start to dominate the cooling \citep[see e.g.][]{gf12}. 

\begin{figure}
\center{\includegraphics[width=0.9\textwidth]{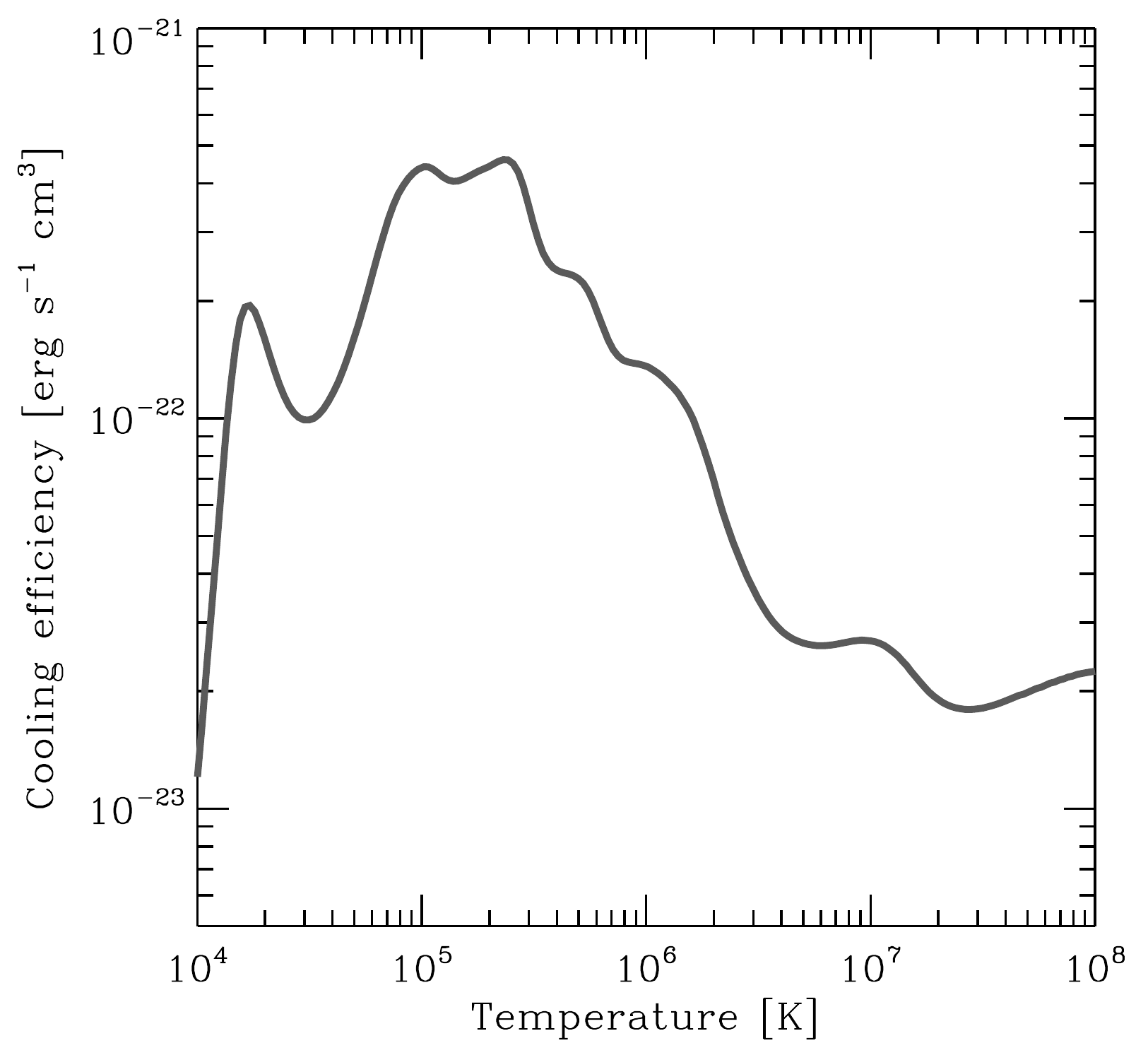}}
\caption{Cooling efficiency of solar-metallicity gas in collisional ionization equilibrium,
plotted as a function of temperature. This plot is based on data taken from \citet{gf12}.
\label{high-T-cool}}
\end{figure}

In conditions where collisional ionization equilibrium (CIE) applies, and where the fractional
abundance of each ion or neutral atom is set by the balance between collisional ionization
and radiative recombination, the total cooling function is relatively straightforward to
compute and only depends on the temperature and metallicity. As an example, we show
in Figure~\ref{high-T-cool} the CIE cooling efficiency, ${\cal L}_{\rm CIE}$, 
as a function of temperature, computed for a solar 
metallicity  gas using the data given in \citet{gf12}. The cooling efficiency plotted in the 
figure has units of erg~cm$^{3}$~s$^{-1}$ and is related to the cooling rate per unit volume
by the expression
\begin{equation}
\Lambda_{\rm CIE} = {\cal L}_{\rm CIE} n_{\rm e} n,
\end{equation}
where $n_{\rm e}$ is the number density of free electrons and $n$ is the number density of
hydrogen nuclei.

Unfortunately, there are many cases in the real ISM in which the CIE assumption does 
not apply. As an obvious example, consider gas in the H{\textsc{ii}} regions around massive
stars, where the ionization state of the various elements is determined primarily by 
photoionization rather than collisional ionization. The CIE assumption also breaks
down whenever the gas cools rapidly. Whenever the cooling time becomes shorter
than the recombination time the  gas cannot adjust its ionization state rapidly
enough to remain in equilibrium. Instead, it becomes over-ionized compared to the
CIE case \citep[see e.g.][]{kaf73,sd93,gs07}. Similarly, if gas is heated more rapidly than it can 
collisionally ionize itself, such as in a very strong shock, then it can be under-ionized compared
to the CIE case. These non-equilibrium effects are particularly important around
$10^{4}$~K \citep[see e.g.][]{micic13,rso14}, but can also significantly affect the cooling rate
at higher temperatures. 

Efforts have been made to account for these non-equilibrium effects, either by explicitly
solving for the non-equilibrium ionization state of the main elements contributing to the
high temperature cooling \citep[see e.g.][]{cf06,db12,os13,rso14}, or by pre-computing and
tabulating rates appropriate for gas cooling at constant pressure or constant density
\citep{gs07}, or with an ionization state dominated by photoionization rather than
collisional ionization \citep[e.g.][]{wiersma09,gh12}. In any case, there is inevitably a trade-off
 between accuracy and speed -- full non-equilibrium calculations best represent
the behavior of the real ISM but have a considerably larger computational cost than
simple CIE-based calculations.

\subsubsection{\bf Fine structure lines}
\label{subsubsec:fine-structure-lines}
At temperatures below around $10^{4}$~K, it becomes extremely difficult for the gas to
cool via radiation from permitted atomic transitions, such as the Lyman series lines of
atomic hydrogen, as the number of electrons available with sufficient energy to excite
these transitions declines exponentially with decreasing temperature.  Atomic cooling
continues to play a role in this low temperature regime, but the focus now shifts from
permitted transitions between atomic energy levels with different principal quantum
numbers to forbidden transitions between different fine structure energy levels. 

Fine structure splitting is a phenomenon caused by the interaction between the
orbital and spin angular momenta of the electrons in an atom, an effect known as
spin-orbit coupling \citep[see e.g.][]{atkins}. Each electron within an atom has a magnetic moment 
due to its orbital motion and also an intrinsic magnetic moment due to its spin. States where 
these magnetic moments are parallel have higher energy than states where they are 
anti-parallel, which in the right conditions can lead to a splitting of energy levels that 
would otherwise remain degenerate. In order for an atom or ion to display fine structure
splitting in its ground state, the electrons in the outermost shell must have both non-zero total 
orbital angular momentum (i.e.\ $L > 0$) and non-zero total spin angular momentum
(i.e.\ $S > 0$), or else the spin-orbit coupling term in the Hamiltonian, which is proportional
to $\vec{L} \cdot \vec{S}$, will vanish. For example, the ground state of the hydrogen atom
has $S = 1/2$ but $L = 0$, and hence has no fine structure. On the other hand, the ground 
state of neutral atomic carbon has $L = 1$ and $S = 1$ and hence does have fine structure.

As the size of the spin-orbit term in the Hamiltonian is typically quite small compared to the
other terms, the resulting splitting of the energy levels is also small, with energy separations
typically of the order of $10^{-2}$~eV. This corresponds to a temperature of the order of
100~K, meaning that it is possible to excite these transitions even at relatively low gas
temperatures. 

The radiative transition rates associated with these fine structure transitions are very
small in comparison to the rates of the permitted atomic transitions discussed above,
for a couple of reasons. First, they are typically magnetic dipole transitions, with transition
matrix elements that are of the order of $\alpha^{2} \approx 5 \times 10^{-5}$ times smaller
than for electric dipole transitions, where $\alpha$ is the fine structure constant.
Second, it is easy to show that transitions with similar transition matrix elements 
but different frequencies have spontaneous transition rates that scale as 
$A_{ij} \propto \nu_{ij}^{3}$. Since the frequencies associated with the fine structure 
transitions are of the order of a thousand times smaller than those associated with 
the most important permitted electronic transitions, such as Lyman-$\alpha$, one 
expects the spontaneous transition rates to be a factor of $10^{9}$ smaller. 

Together, these two effects mean that we expect the size of the spontaneous transition 
rates associated with the fine structure transitions to be of the order of $10^{14}$ or more
times smaller than those associated with the most important permitted atomic transitions.
Consequently, the critical densities associated with many of the important fine structure
transitions are relatively low: $n_{\rm crit} \sim 10^{2}$--$10^{6} \: {\rm cm^{-3}}$ in conditions when collisions with H or H$_{2}$ dominate, 
and up to two to three orders of magnitude smaller when collisions with electrons dominate
\citep{hm89}. We therefore would expect cooling from fine structure emission to be effective
at moderate densities, e.g.\ in the WNM or CNM, but to become much less effective at the
much higher densities found in gravitationally collapsing regions within molecular clouds.

As hydrogen and helium have no fine structure in their ground states, fine structure
cooling in the ISM is dominated by the contribution made by the next most abundant
elements, carbon and oxygen \citep{wolf95}. In the diffuse ISM, carbon is found mainly 
in the form of C$^{+}$, as neutral atomic carbon can be photoionized by photons with
energies $E > 11.26\;$eV, below the Lyman limit. Singly ionized carbon has two fine
structure levels in its ground state, an upper level with total angular momentum
$J = 3/2$ and a lower level with total angular momentum $J = 1/2$. The energy separation 
between these two levels is approximately $E/k_{\rm B} = 92\;$K, and so this transition remains
easy to excite down to temperatures of around $20\;$K (see Figure~\ref{cool-rates}).

In denser regions of the ISM, where dust provides some shielding from the effects
of the ISRF, C$^{+}$ recombines to yield C. Atomic carbon has three fine structure levels, with total angular momenta 
$J = 0, 1, 2$ and energies relative to the ground state $\Delta E / k_{\rm B} = 0.0, 23.6, 62.4$~K, 
respectively. The small energy separations of these levels mean that neutral atomic
carbon remains an effective coolant down to very low temperatures. Indeed, in the
low density limit, neutral atomic carbon is a more effective coolant than CO
(Figure~\ref{cool-rates}), although it becomes ineffective at densities $n \gg 100 \:
{\rm cm^{-3}}$, owing to its low critical density.

The ionization energy of neutral oxygen is very similar to that of hydrogen, and so in 
the WNM and CNM, oxygen is present largely in neutral atomic form. Neutral oxygen
also has fine structure in its ground state. In this case, there are three fine structure 
levels, with total angular momenta $J = 0, 1, 2$ and energies relative to the ground
state $\Delta E / k_{\rm B} = 0.0, 227.7, 326.6$~K, respectively. The larger energy separation of 
these levels compared to the C$^{+}$ fine structure levels means that in the CNM, 
C$^{+}$ cooling is considerably more effective than cooling from oxygen, despite
the larger abundance of oxygen relative to carbon \citep[see e.g.][]{wolf95}. In warmer
gas, however, carbon and oxygen are both important coolants.

In gas with standard solar metallicity, other metals such as N,
Ne, Si, Fe and S also have relatively high abundances, but in practice they do
not contribute significantly to the cooling of the ISM. Nitrogen and neon are
present in the WNM and CNM primarily in neutral form, and have no fine structure
in their ground state in this form. Silicon and iron do have ground state fine
structure, but are strongly depleted in the ISM, particularly in the colder and 
denser phases \citep{jenk09}. Finally, sulfur is present primarily in the form of
S$^{+}$, which has no fine structure in its ground state. 

Data on the collisional excitation rates of the fine structure transitions of C$^{+}$, C and
O can be found in a number of places in the literature. Compilations of excitation rate
data are given in \citet{hm89}, \citet{gj07} and \citet{maio07}, as well as in the 
LAMDA database \citep{sch05}.

\subsubsection{\bf Molecular hydrogen}
\label{subsubsec:H2}
Molecular hydrogen is the dominant molecular species in the ISM and can have 
an abundance that is orders of magnitude larger than that of any other molecule 
or that of any of the elements responsible for fine structure cooling. Because of this,
it is natural to expect H$_{2}$ to play an important role in the cooling of the ISM. In 
practice, however, at metallicities close to solar,  H$_{2}$ cooling is important only  
in shocks \citep[see e.g.][]{hm79,hm89} and not in more quiescent regions of the diffuse ISM 
\citep{gc14}. The reason for this is that H$_{2}$ is not a particularly effective coolant
at low temperatures. 

There are several reasons for that. To a first approximation, we can treat H$_{2}$ as a linear rigid rotor, with rotational
energy levels separated by energies 
\begin{equation}
\Delta E = 2 B J,
\end{equation}
where $J$ is the rotational quantum number of the upper level and
$B$ is the rotational constant
\begin{equation}
B \equiv \frac{\hbar^{2}}{2I_{m}},
\end{equation}
and $I_{m}$ is the moment of inertia of the molecule \citep[see e.g.][]{atkins}.
Since H$_{2}$ is a light molecule, it has
a small moment of inertia, and hence a large rotational constant, leading to widely
spaced energy levels: $\Delta E / k_{\rm B} \approx 170 J$~K when the rotational quantum number 
$J$ is small. In addition, radiative transitions between states with odd $J$ and even $J$
are strongly forbidden. The reason for this is that the hydrogen molecule has two distinct forms, 
distinguished by the value of the nuclear spin quantum number $I$. If the two protons have 
anti-parallel spins, so that $I = 0$, then the total wave-function is anti-symmetric with respect to 
exchange of the two protons (as required by the Pauli exclusion principle) if and only if the 
rotational quantum number $J$ is even. Molecular hydrogen in this form is known as para-hydrogen.
On the other hand, if the protons have parallel spins,  then the Pauli principle requires that $J$ be odd.
H$_2$ in this form is known as ortho-hydrogen. Radiative transitions between the ortho and para states (or vice versa) therefore require a change in the nuclear spin, which is highly unlikely,
and hence the associated transition rates are very small. The first accessible rotational transition
is therefore the $J = 2 \rightarrow 0$ transition, which has an associated energy separation of
around 510~K. At low temperatures, it becomes extremely hard to excite this transition, and
therefore H$_{2}$ cooling becomes extremely ineffective.

\begin{figure}
\includegraphics[width=\textwidth]{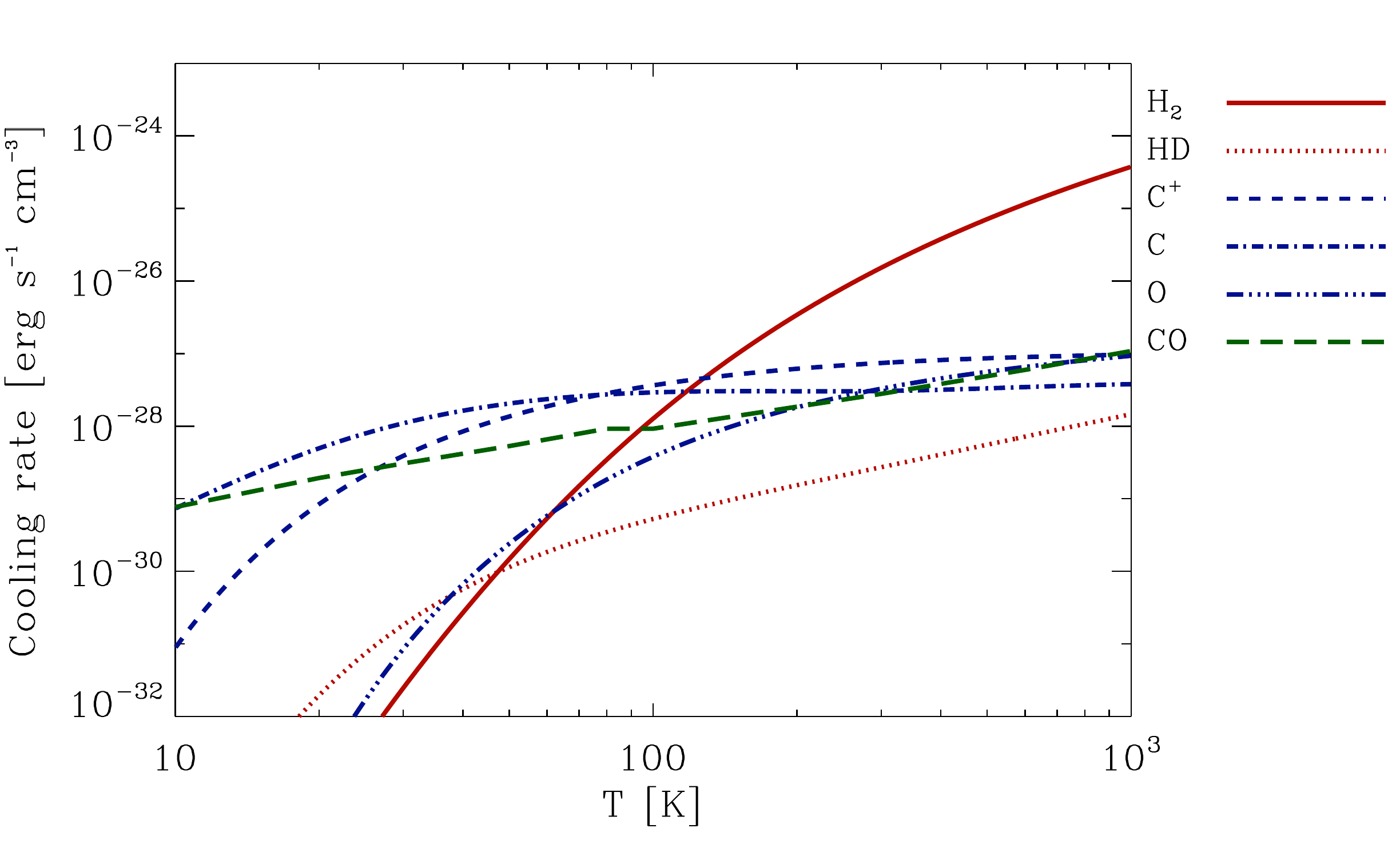}
\caption{Cooling rates for selected ISM coolants. The values plotted were computed assuming
that $n = 1 \: {\rm cm^{-3}}$ and are weighted by the fractional abundance (relative to the total 
number of hydrogen nuclei) 
of the coolant in question. For H$_{2}$ and HD, we assume that the gas is fully molecular, so 
that $x_{\rm H_{2}} = 0.5$ and $x_{\rm HD} = 2.5 \times 10^{-5}$. For the metals, we adopt
total C and O abundances from \citet{sem00}, and show the cooling rate that we would have if
all of the relevant element were in the form of the indicated species. In the case of CO, the 
adopted abundance is the total C abundance, since this is smaller than the total abundance
of oxygen \citep{sem00}. For cooling from H$_{2}$ and HD, we account for
collisions with both H$_{2}$ and He, while for the other species, we account only for collisions
with H$_{2}$, owing to a lack of data on the collision rates with He. The error introduced
by omitting He is unlikely to exceed 10--20\%. The data on which these cooling rates are 
based is taken from \citet{fr98}, \citet{fr99a}, and \citet{frz98}
for H$_{2}$, \citet{fr99b} and \citet{rz99} for HD, \citet{wg14} for C$^{+}$, 
\citet{sch91} and \citet{warin96a} for C, \citet{j92} for O, and \citet{nk93}, \citet{nlm95}, \citet{fl01}
and \citet{we06} for CO.
\label{cool-rates}}
\end{figure}

This is illustrated in Figure~\ref{cool-rates}, where we compare the cooling rate due to H$_{2}$
with the cooling rates of a number of other potentially important coolants, discussed in more
detail below. All of the cooling rates are computed in the low density limit and assume that the
hydrogen is fully molecular and that the fractional ionization is zero. 

From the figure, we see that in these conditions, H$_{2}$ cooling can be important at
temperatures $T > 100$~K, but becomes insignificant in comparison to fine structure
line cooling or CO rotational emission at $T < 100$~K, owing to the exponential fall-off
in the H$_{2}$ cooling rate. Changing the composition of the gas will change the relative
contributions of the different coolants, but in practice will typically make H$_{2}$ cooling
less important. For example, reducing the fractional abundance of H$_{2}$ causes the
H$_{2}$ cooling rate to drop significantly, because not only does one have fewer H$_{2}$
molecules to provide the cooling, but their collisional excitation rates also decrease,
since collisions with H atoms are much less effective at exciting the rotational 
levels of H$_{2}$ than collisions with other H$_{2}$ molecules 
\citep[see e.g.][and references therein]{ga08}. We therefore see that at solar metallicity, 
H$_{2}$ cooling is important only in gas with a high H$_{2}$ fraction {\em and} a 
temperature $T > 100$~K. In practice, it is difficult to satisfy both of these conditions at
the same time in quiescent gas. Temperatures of 100~K or more are easy to reach in 
low density CNM clouds, but the equilibrium H$_{2}$ abundance in these clouds is 
small. Increasing the density and/or column density of the clouds increases the equilibrium
H$_{2}$ abundance, but at the same time decreases the typical gas temperature.
For this reason, high H$_{2}$ fractions tend to be found only in cold gas \citep{klm11},
and therefore in conditions where H$_{2}$ cooling is ineffective. 

The combination of high H$_{2}$ fraction and a gas temperature $T > 100$~K can occur
in shocked molecular gas, provided that the shock is not so strong as to completely dissociate 
the H$_{2}$, and H$_{2}$ has long been known to be a significant coolant in these 
conditions \citep{hm79,hm89,pon12}. 

\subsubsection{\bf Hydrogen deuteride}
Although H$_{2}$ is an ineffective low temperature coolant, the same is not true for
its deuterated analogue, HD. Unlike H$_{2}$, HD does not have distinct ortho and
para forms, and hence for HD molecules in the $J = 0$ ground state, the first accessible
excited level is the $J = 1$ rotational level. In addition, HD is 50\%
heavier than H$_{2}$, and hence has a smaller rotational constant and more narrowly
spaced energy levels. The energy separation of its $J=0$ and $J=1$ rotational levels
is $\Delta E_{10} / k_{\rm B} = 128$~K, around a factor of four smaller than the separation of the
$J = 0$ and $J = 2$ levels of H$_{2}$. We would therefore expect HD cooling to remain
effective down to much lower temperatures than H$_{2}$. 

One important factor working against HD is the fact that the deuterium abundance is
only a small fraction of the hydrogen abundance, meaning that in general H$_{2}$
is orders of magnitude more abundant than HD. However, in cold gas that is not yet fully molecular,
the HD abundance can be significantly enhanced by a process known as chemical fractionation. 
HD is formed from H$_{2}$ by the reaction
\begin{equation}
{\rm H_{2} + D^{+}} \rightarrow {\rm HD + H^{+}}  \label{hdform}
\end{equation}
and is destroyed by the inverse reaction
\begin{equation}
{\rm HD + H^{+}} \rightarrow {\rm H_{2} + D^{+}}.  \label{hddest}
\end{equation}
The formation of HD via reaction~(\ref{hdform}) is exothermic and can take place at all temperatures,
but the destruction of HD via reaction~(\ref{hddest}) is mildly endothermic and becomes very
slow at low temperatures. As a result, the equilibrium HD/H$_{2}$ ratio is enhanced by
a factor \citep{gp02}
\begin{equation}
f_{\rm en} = 2 \exp \left( \frac{462}{T} \right) 
\end{equation}
over the elemental D/H ratio. At temperatures $T < 100$~K, characteristic of the CNM,
this corresponds to an enhancement in the equilibrium abundance by a factor of hundreds
to thousands. This fractionation effect helps HD to be a more effective low temperature
coolant than one might initially suspect. Nevertheless, there is a limit to how effective
HD can become, since the HD abundance obviously cannot exceed the total deuterium
abundance. The total abundance of deuterium relative to hydrogen in primordial gas
is \citep{cooke14}
\begin{equation}
({\rm D / H}) = (2.53 \pm 0.04) \times 10^{-5}.
\end{equation}
In the local ISM, the ratio of D/H is even smaller \citep[see e.g.][]{lin06,prod10},  as some 
of the primordial deuterium  has been destroyed by stellar processing.

From the comparison of cooling rates in Figure~\ref{cool-rates}, we see that in fully molecular
gas, HD becomes a more effective coolant than H$_{2}$ once the temperature drops below
50~K, despite the fact that in these conditions, the abundance of HD is more than $10^{4}$ times
smaller than that of H$_{2}$.  However, we also see that at these low temperatures, the amount
of cooling provided by HD is a factor of a hundred or more smaller than the cooling provided by
C$^{+}$, C or CO. It is therefore safe to conclude that HD cooling is negligible in low density,
solar metallicity gas. At higher densities, HD cooling could potentially become more important,
as HD has a larger critical density than C or C$^{+}$, but at the relevant densities ($n \sim
10^{6} \: {\rm cm^{-3}}$), dust cooling generally dominates.
  
\subsubsection{\bf Carbon monoxide} 
\label{subsubsec:CO}
Heavier molecules can also contribute significantly to the cooling of interstellar gas. In particular,
carbon monoxide (CO), the second most abundant molecular species in the local ISM, can play 
an important role in regulating the temperature within giant molecular clouds (GMCs). As we can see
from Figure~\ref{cool-rates}, CO is a particularly important coolant at very low gas temperatures, 
$T < 20$~K, owing to the very small energy separations between its excited rotational levels.
However, we also see from the figure that at low densities, fine structure cooling from neutral
atomic carbon is more effective than CO cooling, and that at $T \sim 20$~K and above, the 
contribution from C$^{+}$ also becomes significant. The overall importance of CO therefore
depends strongly on the chemical state of the gas. If the gas-phase carbon is primarily in the
form of C or C$^{+}$, then fine structure emission from these species will dominate, implying
that CO becomes important only once the fraction of carbon in CO becomes large. As we
will discuss in more detail later, this only occurs in dense, well-shielded gas, and so in
typical GMCs, CO cooling only dominates once the gas density exceeds 
$n \sim 1000 \: {\rm cm^{-3}}$. 

In practice, CO is able to dominate the cooling only over a restricted range in densities, as
it becomes  ineffective at densities $n \gg 1000 \: {\rm cm^{-3}}$. In part, this is
because CO has only a small dipole moment and hence the CO rotational transitions
have low critical densities. For example, in optically thin gas, the relative populations of the 
$J=0$ and $J=1$ rotational levels reach their LTE values at a density 
$n_{\rm crit} \sim 2200 \: {\rm cm^{-3}}$, while the $J=2$ level reaches LTE at 
$n_{\rm crit} \sim 23000  \: {\rm cm^{-3}}$. In addition, the low $J$ transitions of $^{12}$CO
rapidly become optically thick in these conditions, further lowering their effective critical
densities and significantly limiting their contribution to the cooling rate of the gas. 
This behavior has a couple of interesting implications. First, it means that cooling from
isotopic variants of CO, such as $^{13}$CO or C$^{18}$O can become important, despite
the low abundances of these species relative to $^{12}$CO, since they will often remain
optically thin even if $^{12}$CO is optically thick. Second, it means that the freeze-out
of CO onto the surface of dust grains, which is thought to occur in the cold, dense gas
at the center of many prestellar cores, has very little effect on the overall CO cooling rate.
This was demonstrated in striking fashion by \citet{Goldsmith01}, who showed that at densities
of order $10^{4}$--$10^{5} \: {\rm cm^{-3}}$ within a typical prestellar core, reducing the
CO abundance by a factor of a hundred reduces the CO cooling rate by only a factor of
a few.

\subsubsection{\bf Other heavy molecules}
Other molecular species can become important coolants in comparison to H$_{2}$ and CO
in the right conditions. An interesting example is water. H$_{2}$O molecules have
a very large number of accessible rotational and vibrational transitions and also have high
critical densities. Therefore, over a wide range of temperatures and densities, the amount
of cooling that one gets per water molecule can be much larger than the amount that one
gets per CO molecule \citep[see e.g.\ the comparison in][]{nk93}. Despite this, water does not
contribute significantly to the thermal balance of cold gas in molecular clouds, because the
fractional abundance of water in these regions is very small \citep[see e.g.][]{snell00}. This is, because most of the water molecules that form 
rapidly freeze out onto the surface of dust grains, forming a significant part of the ice mantles
that surround these grains \citep{bergin00,holl09}. On the other hand, in warm regions, such
as the shocked gas in molecular outflows, H$_2$O can be a very important coolant \citep{nisi10}.

The other molecules and molecular ions present in interstellar gas also provide some cooling,
but at low gas densities, their total contribution is relatively small compared to that of CO, 
since the latter generally has a much larger abundance. In very dense gas, however,
their contributions become much more important, owing to the high optical depth of the 
CO rotational lines. Of particular importance in this high density regime are species that
have large dipole moments, such as HCN or N$_{2}$H$^{+}$, as these species have high
critical densities and hence remain effective coolants up to very high densities. That said,
in typical molecular cloud conditions, dust cooling takes over from molecular line cooling
as the main energy loss route well before these species start to dominate the line cooling,
and so their overall influence on the thermal balance of the cloud remains small.

\subsection{Gas-grain energy transfer}
\label{subsec:gas-grain-transfer}
Dust can also play an important role in the cooling of the ISM \citep{gk74,l75}.
Individual dust grains are extremely
efficient radiators, and so the mean temperature of the population of dust grains very quickly relaxes
to an equilibrium value given by the balance between radiative heating caused by the absorption of 
photons from the ISRF and radiative cooling via the thermal emission from the grains.\footnote{The 
chemical energy released when H$_2$ molecules form on grain surfaces and the direct interaction 
between dust grains and cosmic rays also affect the grain temperature, but their influence on the
mean grain temperature is relatively minor \citep{ljo85}.} If the resulting dust temperature, $T_{\rm d}$, 
differs from the gas temperature, $T_{\rm K}$, then collisions between gas particles and dust grains 
lead to a net flow of energy from one component to the other, potentially changing both $T_{\rm K}$ 
and $T_{\rm d}$. 

The mean energy transferred from the gas to the dust by a single collision is given by:
\begin{equation}
\Delta E = \frac{1}{2} k_{\rm B} (T_{\rm K} - T_{\rm d}) \alpha,   \label{gd1}\;,
\end{equation}
where $\alpha$ is the thermal energy accommodation coefficient, which describes how efficiently
energy is shared between the dust and the gas \citep{burke83}. This efficiency typically varies 
stochastically from collision to collision, but since we are always dealing with a large number of
collisions, it is common to work in terms of the mean value of $\alpha$, which we denote as
$\bar{\alpha}$. However, even then, the treatment of the accommodation coefficient can be 
complicated, as $\bar{\alpha}$ depends in a complicated fashion on the nature of the dust grain, 
the nature of the collider (e.g.\ whether it is a proton, a hydrogen atom or an H$_{2}$ molecule), and 
the gas and grain temperatures \citep{burke83}.

The total rate at which energy flows from the gas to the dust is the product of the mean energy
per collision and the total collision rate. The latter can be written as
\begin{equation}
R_{\rm coll} = 4\pi \sigma_{\rm d} \bar{v} n_{\rm tot} n_{\rm d},  \label{gd2}
\end{equation}
where $\sigma_{\rm d}$ is the mean cross-sectional area of a dust grain, $n_{\rm d}$ is
the number density of dust grains, $n_{\rm tot}$ is the number density of particles, and
$\bar{v}$ is the mean thermal velocity of the particles in the gas. Note that both $n_{\rm tot}$
and $\bar{v}$ are functions of the composition of the gas -- in a fully atomic gas, $n_{\rm tot}$
and $\bar{v}$ are both larger than in a fully molecular gas. 

Combining equations~(\ref{gd1}) and (\ref{gd2}), we can write the cooling rate per unit volume
due to energy transfer from the gas to the dust as
\begin{equation}
\Lambda_{\rm gd} = \pi \sigma_{\rm d} \bar{v} \bar{\alpha} (2kT_{\rm K} - 2kT_{\rm d}) n_{\rm tot} n_{\rm d}.
\end{equation}
Note that although it is common to talk about this in terms of cooling, if $T_{\rm d} > T_{\rm K}$ then energy
will flow from the dust to the gas, i.e.\ this will become a heating rate.

Expressions given in the astrophysical literature for $\Lambda_{\rm gd}$ are typically written in the form:
\begin{equation}
\Lambda_{\rm gd} = C_{\rm gd} T_{\rm K}^{1/2} \left( T_{\rm K} - T_{\rm d} \right) n^{2}
\: {\rm erg \: s^{-1} \: cm^{-3}},
\end{equation}
where $n$ is the number density of hydrogen nuclei and $C_{\rm gd}$ is a cooling rate coefficient  given by
\begin{equation}
C_{\rm gd} = 2 \pi k \sigma_{\rm d} \left(\frac{\bar{v}}{T_{\rm K}^{1/2}} \right) \bar{\alpha} \frac{n_{\rm tot} n_{\rm d}}{n^{2}}. 
\end{equation}
The value of $C_{\rm gd}$ is largely determined by the assumptions that we make regarding the 
chemical state of the gas and the nature of the dust grain population, but in principle it also depends on 
temperature, through the temperature dependence of the mean accommodation coefficient, $\bar{\alpha}$.

Different authors introduce different assumptions about various of these issues, leading  to 
a wide spread of values for $C_{\rm gd}$ being quoted in the literature for Milky Way dust. For example, 
\citet{hm89} write $C_{\rm gd}$ as
\begin{equation}
C_{\rm gd} = 3.8 \times 10^{-33} \left[1 - 0.8 \exp \left(-\frac{75}{T_{\rm K}}\right) \right] 
\left(\frac{10 \, \mbox{nm}}{a_{\rm min}} \right)^{1/2} \: {\rm erg \: s^{-1} \: cm^{3} \: K^{-3/2}},
\end{equation}
where $a_{\rm min}$ is the minimum radius of a dust grain, often taken to be simply $a_{\rm min} = 10\,${nm}.
However, \citet{th85} quote a value for the same process that is almost an order of magnitude smaller:
\begin{equation}
C_{\rm gd} = 3.5 \times 10^{-34} \: {\rm erg \: s^{-1} \: cm^{3} \: K^{-3/2}},
\end{equation}
while \citet{Goldsmith01} quotes a value that is smaller still
\begin{equation}
C_{\rm gd} = 1.6 \times 10^{-34} \: {\rm erg \: s^{-1} \: cm^{3} \: K^{-3/2}}.
\end{equation}
Finally, Evans (private communication) argues for a rate
\begin{equation}
C_{\rm gd} = 1.8 \times 10^{-33}  \left[1 - 0.8 \exp \left(-\frac{75}{T_{\rm K}}\right) \right]  \: {\rm erg \: s^{-1} \: cm^{3} \: K^{-3/2}},
\end{equation}
close to the \citet{hm89} rate. Although it is not always clearly stated, all of these rates seem to be intended for use in 
H$_{2}$-dominated regions. In regions dominated by atomic hydrogen, one would expect the cooling rate to vary, owing to the difference in the value of $\bar{\alpha}$ appropriate for H atoms and that appropriate for H$_{2}$
molecules \citep{burke83}. In a recent study, \citet{klm11} attempted to distinguish between the molecular-dominated and
atomic-dominated cases, using 
\begin{equation}
C_{\rm gd} = 3.8 \times 10^{-33} \: {\rm erg \: s^{-1} \: cm^{3} \: K^{-3/2}}
\end{equation}
for molecular gas and
\begin{equation}
C_{\rm gd} = 1.0 \times 10^{-33} \: {\rm erg \: s^{-1} \: cm^{3} \: K^{-3/2}}
\end{equation}
for atomic gas.

\begin{figure}
\includegraphics[width=\textwidth]{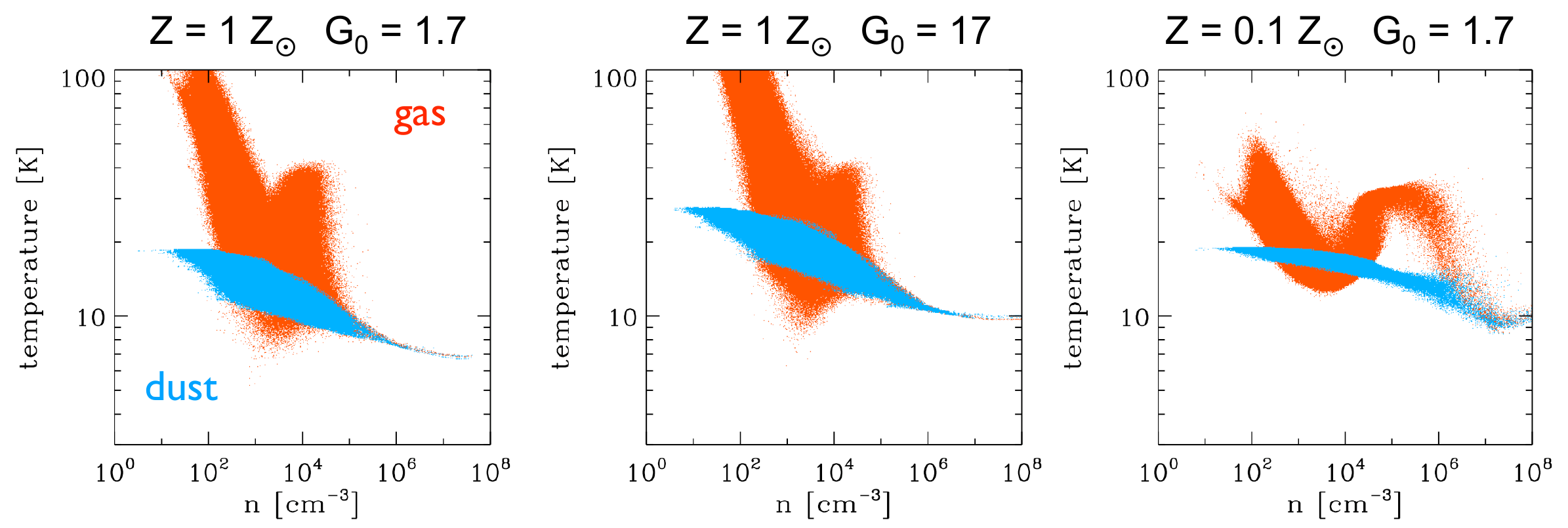}
\caption{Gas  and dust temperatures, $T_{\rm K}$ and $T_{\rm d}$, as a function of the hydrogen nuclei number density $n$ in the turbulent ISM for two different metallicities (the solar value, ${\rm Z = 1 \,Z_\odot}$, and a much smaller value typical of metal-poor dwarf galaxies, ${\rm Z = 0.1 \,Z_\odot}$),  and for two different strength of the ISRF (the solar neighborhood value with $G_{0} = 1.7$, and a value ten times larger with $G_{0} = 17$). We assume here that the grain size distribution is the same in each case, and that the dust-to-gas ratio scales linearly with the metallicity. The distribution of dust temperatures varies only weakly with changes in Z or $G_{0}$.
For ${\rm Z = 1 \,Z_\odot}$, the gas becomes thermally coupled to the gas at densities larger than $n \sim 10^5\;$cm$^{-3}$, so that $T_{\rm K} \approx T_{\rm d}$. For ${\rm Z= 0.1\, Z_\odot}$, this happens instead at densities above $n \sim 10^6\;$cm$^{-3}$. In these models, which started with atomic initial conditions, H$_2$ formation on dust grains releases latent heat and leads to a bump in the gas temperature at densities $ n \sim 10^4\;$cm$^{-3}$ for ${\rm Z = 1\,Z_\odot}$, and at $n \sim 10^5\;$cm$^{-3}$ for ${\rm Z = 0.1\,Z_\odot}$. This feature is absent in clouds that  have already converted all of their hydrogen to molecular form.
\label{fig:dust-gas-coupling}}
\end{figure}

The uncertainty in $C_{\rm gd}$ becomes even greater as we move to lower
metallicity, as less is known about the properties of the dust. It is often assumed that 
$C_{\rm gd}$ scales linearly with metallicity \citep[e.g.][]{gc12c}, but this is at
best a crude approximation, particularly as the dust abundance appears to scale 
non-linearly with metallicity in metal-poor galaxies \citep{gala11,hc12}.

The importance of gas-grain energy transfer as a cooling mechanism depends
strongly on the gas density, regardless of which value of $C_{\rm gd}$ we
adopt. At low densities, the cooling rate is low in comparison to that provided
by atomic fine structure lines or molecular transitions, as can be seen by comparing
the values quoted above with the atomic and molecular cooling rates plotted in
Figure~\ref{cool-rates}. This remains true for as long as we remain in the low-density line
cooling regime, where the cooling rate due to line emission scales as $n^{2}$.
However, as the density increases, we will eventually pass the critical densities
of our main atomic and molecular coolants. Once we do so, the line cooling rate
will begin to scale linearly with the density, while the gas-grain rate will continue 
to scale as $n^{2}$. We therefore see that eventually gas-grain energy transfer
will come to dominate the cooling rate of the gas. The considerable optical
depths that can build up in the main coolant lines simply act to hasten this process.

Once gas-grain energy transfer dominates the cooling rate, the gas
temperature is quickly driven towards the dust temperature. This is illustrated in Figure \ref{fig:dust-gas-coupling}, which shows the evolution of dust and gas temperature as function of number density in the turbulent ISM for different combinations of metallicity and strength of the ISRF 
\cite[based on numerical simulations similar to those described in][]{gc12a, gc12c}. The gas density at 
which the two temperatures become strongly coupled depends on the value of 
$C_{\rm gd}$. In a quiescent pre-stellar
core, coupling occurs once the cooling due to dust becomes larger than the
cosmic ray heating rate of the gas. In solar metallicity gas, this takes place at a density 
$n \sim 10^{5} \: {\rm cm^{-3}}$ if one uses the \citet{hm89} prescription for 
$C_{\rm gd}$, but not until $n \sim 10^{6} \: {\rm cm^{-3}}$ if one uses the 
\citet{Goldsmith01} prescription, provided that we assume a standard value for the cosmic
ray heating rate. In regions where the cosmic ray flux is highly enhanced or where the metallicity is low,
however, the coupling between gas and dust can be delayed until much higher densities \citep[see e.g.][]{papa10,clark13}.

\subsection{Computing the dust temperature}
As we saw in the previous subsection, energy transfer between gas and grains
acts to couple the gas temperature to the dust temperature in dense gas. It is
therefore important to understand the physics responsible for determining $T_{\rm d}$.
Because dust grains are extremely efficient radiators, it is usually a good approximation
to treat them as being in thermal equilibrium, with a temperature set by the balance 
between three processes: heating by photon absorption and by collisions with warmer
gas particles, and cooling by photon emission. 

As noted above, we typically assume that the dust is in thermal equilibrium, in which case
the dust temperature is the value that satisfies the following equation: 
\begin{equation}
\Gamma_{\rm ext} - \Lambda_{\rm dust} + \Lambda_{\rm gd} = 0\;.
\end{equation}
Here $\Gamma_{\rm ext}$ is the dust heating rate per unit volume due to the
absorption of radiation,  $\Lambda_{\rm dust}$ is the radiative cooling rate of the dust, 
and $\Lambda_{\rm gd}$, as we have already seen, is the net rate at which energy is 
transferred from the gas to the dust by collisions. 

In the simple case in which the main contribution to $\Gamma_{\rm ext}$ comes from the interstellar 
radiation field, we can write this term  as the product of a optically thin heating rate, 
$\Gamma_{\rm ext, 0}$, and a dimensionless factor, $\chi$, that represents the attenuation 
of the interstellar radiation field by dust absorption \citep{Goldsmith01},
\begin{equation}
\Gamma_{\rm ext} = \chi \Gamma_{\rm ext, 0}.
\end{equation}
The optically thin heating rate is given by
\begin{equation}
\Gamma_{\rm ext, 0} = 4\pi {\cal D} \rho \int_{0}^{\infty} J_{\nu} \kappa_{\nu}  \, {d}\nu,
\end{equation}
where ${\cal D}$ is the dust-to-gas ratio, $\rho$ is the gas density, $J_{\nu}$ is
the mean specific intensity of the incident interstellar radiation field (ISRF), 
and $\kappa_{\nu}$ is the dust opacity in units of ${\rm cm^{2}} \: {\rm g^{-1}}$.
To determine the attenuation factor $\chi$ at a specified point in the cloud, we
can use the following expression:
\begin{equation}
\chi = \frac{\oint \int_{0}^{\infty} J_{\nu} \kappa_{\nu} \exp \left[-\kappa_{\nu} \Sigma(\vec{n}) \right] 
\, {d}\nu \, {d}\Omega}{4\pi \int_{0}^{\infty} J_{\nu} \kappa_{\nu} \, {d}\nu}\;, \label{chi_func}
\end{equation}
where $\Sigma(\vec{n})$ is the column density of the gas between the point
in question and the edge of the cloud in the direction $\vec{n}$.

The values of both $\Gamma_{\rm ext, 0}$ and $\chi$ depend on the 
parameterization we use for the ISRF and on our choice of dust opacities. 
For example, \citet{Goldsmith01} uses values for the radiation field from \citet{mmp83}
plus a highly simplified dust grain model and derives an optically thin heating rate
\begin{equation}
\Gamma_{\rm ext, 0} = 1.95 \times 10^{-24} n \: {\rm erg} \: {\rm s^{-1}} \: {\rm cm^{-3}}.
\end{equation}
On the other hand, \citet{gc12b} make use of a more complicated model,
involving values for the ISRF taken from \citet{black94} and dust opacities 
taken from \citet{oh94} at long wavelengths and \citet{mmp83} at short
wavelengths, but their resulting value for $\Gamma_{\rm ext, 0}$ is relatively similar:
\begin{equation}
\Gamma_{\rm ext, 0} = 5.6 \times 10^{-24} n \: {\rm erg} \: {\rm s^{-1}} \: {\rm cm^{-3}}.
\end{equation}

The dust cooling rate, $\Lambda_{\rm dust}$, is given by
\begin{equation}
\Lambda_{\rm dust}(T_{\rm d}) =  4 \pi {\cal D} \rho \int_{0}^{\infty} B_{\nu}(T_{\rm d}) \kappa_{\nu} 
\, {d}\nu,
\end{equation}
where $B_{\nu}(T_{\rm d})$ is the Planck function for a temperature $T_{\rm d}$. 
Again, the resulting rate is sensitive to our choice of opacities. Using values
from \citet{oh94} yields a cooling rate that is well fit by the expression \citep{gc12b}
\begin{equation}
\Lambda_{\rm dust}(T_{\rm d}) = 4.68 \times 10^{-31} T_{\rm d}^{6} n \: {\rm erg} \: {\rm s^{-1}} \: {\rm cm^{-3}}
\end{equation}
for dust temperatures $5 < T_{\rm d} < 100 \: {\rm K}$. 

Comparing the expressions given above for $\Gamma_{\rm ext, 0}$ and $\Lambda_{\rm dust}(T_{\rm d})$,
we see that in optically thin, quiescent gas illuminated by a standard ISRF, the equilibrium dust 
temperature is $T_{\rm d} \sim 15$~K.\footnote{This is somewhat smaller than the mean value of
$\sim 20$~K that we quote in Section~\ref{isrf_mir}, but this discrepancy is most likely due to our use of the
\citet{oh94} opacities here, as these are intended to represent the behavior of dust in dense molecular
clouds and not in the diffuse WNM and CNM.}  Moreover, $T_{\rm d}$ decreases only very slowly as $\chi$
increases, since $T_{\rm d} \propto \chi^{1/6}$, and so substantial attenuation of the ISRF is
required in order to significantly alter the dust temperature. Note also, however, that once the attenuation
becomes very large, this simple prescription for computing $T_{\rm d}$ breaks down, as the re-emitted
far infrared radiation from the grains themselves starts to make a significant contribution to the overall
heating rate \citep[see e.g.][]{mmp83}.

Comparison of  $\Gamma_{\rm ext}$ and $\Lambda_{\rm gd}$ allows us to explore the role played by
the gas in heating the grains. In optically thin gas, $\Gamma_{\rm ext} \gg \Lambda_{\rm gd}$ even when
$T_{\rm K} \gg T_{\rm d}$ unless the gas density is very high, of the order of $10^{5} \: {\rm cm^{-3}}$
or higher, and so in these conditions, dust is heated primarily by the ISRF, with energy transfer from
the gas becoming important only in extreme conditions, such as in supernova blast waves. In dense
cores, where $n$ is large and $\chi$ is small, the importance of gas-grain energy transfer for heating
the dust depends on the difference between the gas temperature and the dust temperature, which in 
these conditions will generally be small. However, if we assume that we are at densities where the
gas and dust temperatures are strongly coupled, we can get a good idea of the importance of the
$\Lambda_{\rm gd}$ term by comparing the heating rate of the gas (e.g.\ by cosmic rays or compressional
heating) with  $\Gamma_{\rm ext}$, since most of the energy deposited in the gas will be quickly
transferred to the dust grains. 

If cosmic ray heating of the gas dominates, and we adopt the prescription
for cosmic ray heating given in \citet{goldsmith78}, then gas-grain energy transfer and heating from the ISRF
become comparable once $\chi \sim 10^{-4} \zeta_{17}$, where $\zeta_{17}$ is the cosmic ray
ionization rate of atomic hydrogen in units of $10^{-17} \: {\rm s^{-1}}$, and where we have adopted
the \citet{hm89} form for $\Lambda_{\rm gd}$.
In dense cores, $\zeta_{17}
\sim 1$ \citep{vv00}, and so in this scenario, gas-grain energy transfer only becomes important for
heating the grains once $\chi \sim 10^{-4}$, corresponding to an extremely high dust extinction
\citep[see e.g.\ Figure A1 in][]{gc12b}. On the other hand, if compressional heating or turbulent 
dissipation dominate, as appears to be the case in gravitationally collapsing cores \citep{gc12a}, 
then the heating rate can be considerably larger.  One important consequence of this is that once 
dynamical effects dominate the heating of the dust grains, the dust (and hence the gas) will start to 
heat up with increasing density, evolving with an effective adiabatic index $\gamma_{\rm eff}
\approx 1.1$ \citep{larson05,ban06}.

\subsection{Photoelectric heating}
\label{subsec:photoelectric-heating}
One of the most important forms of radiative heating in the diffuse ISM is the photoelectric
heating caused by the interaction between dust grains and UV photons. If a dust grain is hit by a 
suitably energetic photon, it can emit a photo-electron. The energy of this photo-electron is equal to 
the difference between the energy of the photon and the energy  barrier that needs to be overcome 
in order to detach the electron from the grain, a quantity often known as the work function. This 
difference in energies can often be substantial (of the order of an eV or more), and this energy is 
rapidly redistributed amongst the other particles in the gas in the form of heat.

For a dust grain with radius $a$, photon absorption cross-section $\sigma_{\rm d}(a, \nu)$, and
charge $Z_{\rm d} e$, the rate at which photo-electrons are ejected can be written as 
\begin{equation}
R_{\rm pe}(a, Z_{\rm d}) = 4 \pi \int_{\nu_{Z_{\rm d}}}^{\nu_{\rm H}} \frac{J_{\nu}}{h\nu} \sigma_{\rm d}(a, \nu)
Y_{\rm ion}(Z_{\rm d}, a, \nu) {d}\nu.
\end{equation}
Here, $J_{\nu}$ is the mean specific intensity of the ISRF,
$h\nu_{Z_{\rm d}}$ is the ionization potential of the grain (i.e.\ the energy required to remove a
single electron), and $h\nu_{\rm H} = 13.6\:$eV is the ionization potential of atomic hydrogen; we
assume that photons with $\nu > \nu_{\rm H}$ are absorbed by the neutral atomic hydrogen in the
ISM and do not reach the grains.

The $Y_{\rm ion}$ term is the photo-ionization yield. It can be written as 
\begin{equation}
Y_{\rm ion} = Y_{\infty} \left(1 - \frac{\nu_{Z_{\rm d}}}{\nu}\right) f_{y}(a),
\end{equation}
where $Y_{\infty}$ is the yield for very large grains in the limit $\nu \gg \nu_{Z_{\rm d}}$, and
$f_{y}(a)$ is a yield correction factor. This correction factor accounts for the fact that in large grains, 
the photon attenuation depth, $l_{a}$, i.e.\ the distance that a photon can penetrate into the material 
before it is absorbed, can be larger than the electron mean free path $l_{e}$.  We typically normalize
$Y_{\infty}$ such that $f_{y}(a) = 1$ for large grains, in which case our values for small grains
are enhanced by a factor $f_{y} = (l_{e} + l_{a}) / l_{e}$. Typical values for these yield-related parameters
are $Y_{\infty} = 0.15$, $l_{e} = 1\,${nm}~and $l_{a} = 10\,${nm}, respectively.

We therefore see that there are three main parameters that influence the size of the photoelectric heating
rate. These are (1)  the strength of the ISRF at the relevant frequencies, as quantified by the mean specific
intensity $J_{\nu}$, (2) the size distribution of the grains, often taken to be given by the simple MRN distribution (equation \ref{eqn:MRN}), and (3) the charge of the grains. The charge is important because it influences how easy it is to eject electrons
from the grains. When the grains are highly negatively charged, electron ejection is easy, the work function
is small, and the photo-ionization yield is high. As the grains become more neutral or even positively charged,
it becomes much harder to detach electrons from the grains: the work function increases and the photo-ionization
yield drops.

The photoelectric heating rate is therefore much larger in conditions when most grains are neutral, or 
negatively charged, than when most grains are positively charged. The main processes determining the charge 
on a typical grain are photo-ionization -- i.e.\ the same process that gives us the photoelectric heating -- together 
with the accretion of free electrons and the recombination of gas-phase ions with surface electrons. Schematically, 
we can write these reactions as
\begin{eqnarray}
{\rm GR}^{+n} + \gamma & \rightarrow  & {\rm GR}^{+n+1} + {\rm e^{-}}, \\
{\rm GR}^{+n} + {\rm e^{-}} & \rightarrow  & {\rm GR}^{+n-1}, \\ 
{\rm GR}^{+n} + {\rm A^{+}} & \rightarrow  & {\rm GR}^{+n+1} + {\rm A}.
\end{eqnarray}
In general, collisions with electrons are more important than collisions with positive ions, since the
electron thermal velocity is much larger than the thermal velocity of the ions. The level of charge on
the grains is therefore set primarily by the balance between photo-ionization and recombination with
free electrons. 

\begin{figure}
\center{\includegraphics[width=0.9\textwidth]{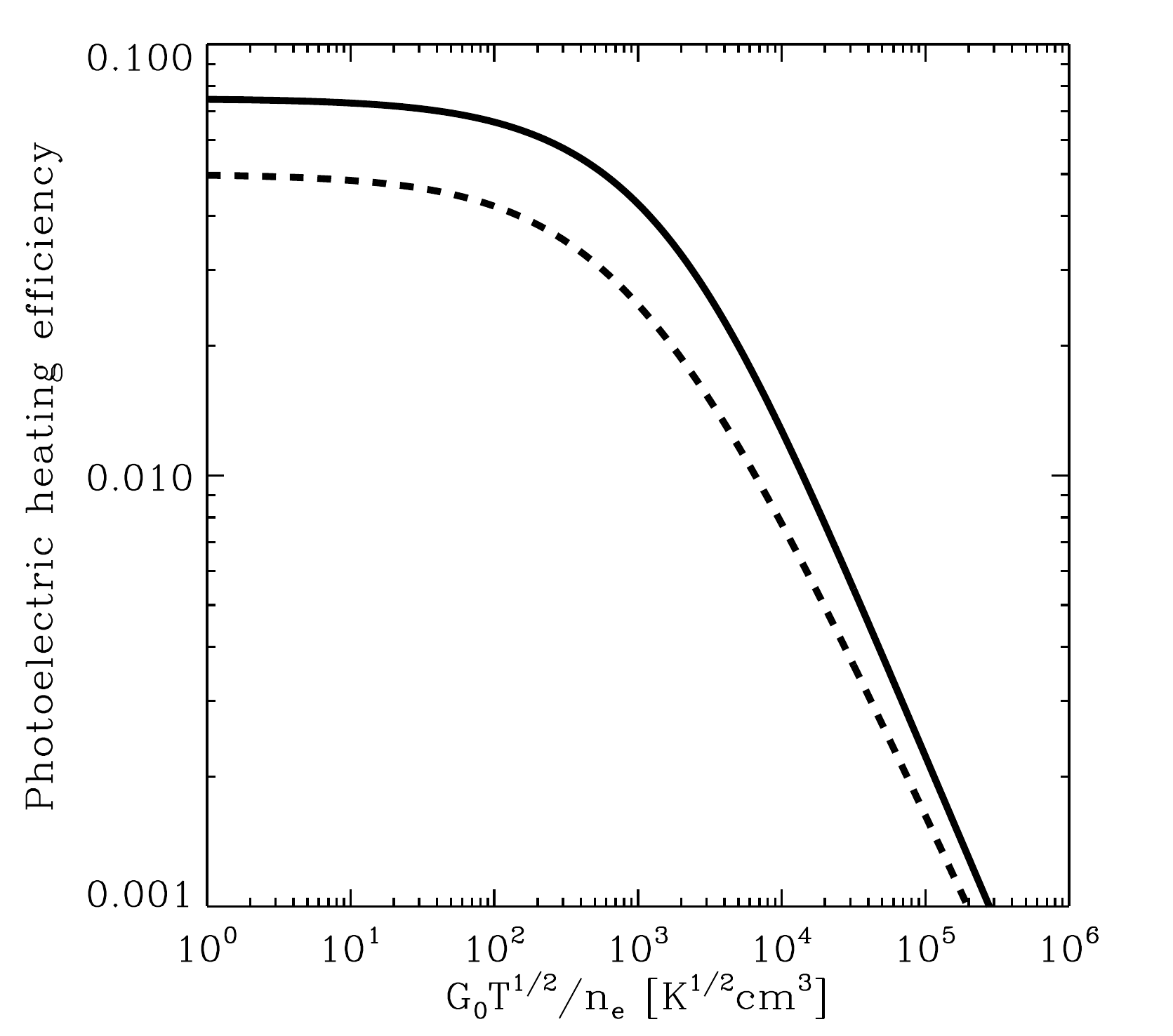}}
\caption{Photoelectric heating efficiency $\epsilon$ as a function of 
$\psi \equiv G_{0} T^{1/2} n_{\rm e}^{-1}$. Values are plotted for gas
temperatures of 6000~K (solid line), characteristic of the WNM, and
60~K (dashed line), characteristic of the CNM. The values shown here
are based on the work of \citet{bt94}, as modified by \citet{wolf03}.
In the local ISM, $\psi \sim 2 \times 10^{4}$ in the WNM and
$\sim 2000$ in the CNM \citep{wolf03}.
\label{PE-eff}}
\end{figure}

Although a detailed analysis of grain charging is rather complex, and beyond the scope of these lecture notes,
in practice one finds that the dependence of the photoelectric heating rate on the physical conditions in
the gas is fairly accurately described as a function of a single parameter, the combination
\begin{equation}
\psi \equiv \frac{G_{0} T^{1/2}}{n_{\rm e}},
\end{equation}
where $G_{0}$ is the strength of the ISRF, $n_{\rm e}$ is the number density of gas-phase electrons
and $T$ is the gas temperature \citep{ds87,bt94,wd01b}. Physically, this behavior makes sense: a strong 
ISRF or a paucity of free electrons will tend to lead to the grains being more positively charged, while the
converse will lead to grains being more negatively charged. The $T^{1/2}$ dependence simply reflects the
temperature dependence of  the rate coefficient for electron recombination with the grains.

For standard interstellar dust, the photoelectric heating rate has been parameterized as a function
of this $\psi$ parameter by \citet{bt94}. Their prescription for the heating rate per unit
volume can be written as
\begin{equation}
\Gamma_{\rm pe} = 1.3 \times 10^{-24} \epsilon G_{0} n  \: {\rm erg} \: {\rm s^{-1}} \: {\rm cm^{-3}},
\end{equation}
where $\epsilon$ is the photoelectric heating efficiency, given by
\begin{equation}
\epsilon = \frac{0.049}{1 + (\psi / 1925)^{0.73}} + \frac{0.037 (T / 10000)^{0.7}}{1 + (\psi / 5000)},
\end{equation}
and $G_{0}$ is the strength of the interstellar radiation field in units of the \citet{habing68} field
(see Equation \ref{eqn:Habing}).
In the limit of small $\psi$, we have $\epsilon \approx 0.05$ when the temperature is low, and
$\epsilon \approx 0.09$ when the temperature is high (see Figure~\ref{PE-eff}).
A more recent treatment by \citet{wd01b} gives similar values for $\epsilon$ for small $\psi$, but
predicts a more rapid fall-off in $\epsilon$ and $\Gamma_{\rm pe}$ with increasing $\psi$ for
$\psi > 10^{4}$.

A final important point to note is that because the photons required to eject photo-electrons must be energetic, 
with minimum energies typically around 6~eV, the photoelectric heating rate is highly sensitive to the dust 
extinction. This sensitivity can be approximately represented by a scaling factor $f_{\rm thick} = \exp(-2.5 A_{V})$
\citep{bergin04}. From this, we see that photoelectric heating will become ineffective once the visual extinction of 
the gas exceeds $A_{V} \sim 1$--2.

\subsection{Other processes responsible for heating}

\paragraph{\bf Ultraviolet radiation}
As well as heating the gas via the photoelectric effect, the ultraviolet component of the ISRF also heats
the gas in two other important ways. First, the photodissociation of H$_{2}$ by UV photons results in
a transfer of energy to the gas, as the hydrogen atoms produced in this process have kinetic energies
that on average are greater than the mean kinetic energy of the gas particles. The amount of energy
released varies depending upon which rovibrational level of the excited electronic state was involved
in the dissociation \citep{sd73,abg00}. Averaged over all the available dissociative transitions, the
total heating rate is around 0.4~eV per dissociation \citep{bd77}.

Second, UV irradiation of molecular hydrogen can lead to heating via a process known as
UV pumping. The absorption of a UV photon by H$_{2}$ leads to photodissociation only 
around 15\% of the time \citep{db96}. The rest of the time, the H$_{2}$ molecule decays from
its electronically excited state back into a bound rovibrational level in the electronic ground 
state. Although the molecule will occasionally  decay directly back into the $v=0$ vibrational
ground state, it is far more likely to end up in a vibrationally excited level. In low density gas,
it then radiatively decays back to the rovibrational ground state, producing a number of near
infrared photons in the process. In high density gas, on the other hand, collisional de-excitation
occurs more rapidly than radiative de-excitation, and so most of the excitation energy is converted
into heat. In this case, the resulting heating rate is around 2~eV per pumping event, corresponding
to around 10--11~eV per photodissociation \citep[see e.g.][]{bht90}.
The density at which this process becomes important is simply the critical density of
H$_{2}$, $n_{\rm crit} \sim 10^{4} \: {\rm cm^{-3}}$. This process is therefore not a major heat source
at typical molecular cloud densities, but can become important in dense cores exposed to strong
UV radiation fields.

\paragraph{\bf Cosmic rays}

In gas that is well shielded from the ISRF, both of these processes become unimportant, as does photoelectric 
heating. In this regime, cosmic rays provide one of the main sources of heat. When a cosmic ray proton
ionizes a hydrogen or helium atom, or an H$_{2}$ molecule, the energy lost by the cosmic ray is typically
considerably larger than the ionization energy of the atom or molecule \citep{gl73}. The excess
energy is shared between the resulting ion and electron as kinetic energy, and the collisions of these
particles with other atoms or molecules can lead to further ionizations, known as secondary ionizations.
Alternatively, the excess kinetic energy can be redistributed in collisions as heat. The amount of heat
transferred to the gas per cosmic ray ionization depends upon the composition of the gas
\citep{dyl99,ggp12}, but is typically around 10--20~eV. Most models of thermal balance in dark
clouds adopt a heating rate that is a fixed multiple of the cosmic ray ionization rate, rather than trying
to account for the dependence of the heating rate on the local composition of the gas 
\citep[see e.g.][]{goldsmith78,Goldsmith01,g10,klm11}. A commonly adopted parameterization is
\begin{equation}
\Gamma_{\rm cr} \sim 3.2 \times 10^{-28} (\zeta_{\rm H} / 10^{-17} \: {\rm s^{-1}}) \,n \:\:\: {\rm erg} \: {\rm cm^{-3} s^{-1}}\;,
\label{eqn:CR-rate}
\end{equation}
where the cosmic ray ionization rate of atomic hydrogen $\zeta_{\rm H}$ is scaled by its typical value of $10^{-17}\,$s$^{-1}$, and where $n$ is the number density of hydrogen nuclei. Note that the uncertainty introduced by averaging procedure is typically much smaller than the current uncertainty in the actual cosmic ray ionization rate in the considered region  (see Section~\ref{cr}).

\paragraph{\bf X-rays}

X-rays can also heat interstellar gas, and indeed in this case the chain of events is very similar to that
in the case of cosmic ray heating: X-ray ionization produces an energetic electron that can cause a
significant number of secondary ionizations, with some fraction of the excess energy also going into
heat. Unlike cosmic rays, X-rays are rather more sensitive to the effects of absorption, since their
mean free paths are typically much smaller. Therefore, although X-ray heating can be important in
the diffuse ISM \citep[see e.g.][]{wolf95}, it is generally not important in the dense gas inside molecular
clouds, unless these clouds are located close to a strong X-ray source such as an AGN
\citep[see e.g.][]{hs10}.

\paragraph{\bf Chemical reactions}

Another way in which the gas can gain energy is through changes in its chemical composition.
The formation of a new chemical bond, such as that between the two hydrogen nuclei in an
H$_{2}$ molecule, leads to a release of energy. Much of this energy will be in the form of
rotational and/or vibrational excitation of the newly-formed molecule, and in low density
environments, this will rapidly be radiated away. At high densities, however, collisional
de-excitation can convert this energy into heat before it can be lost via radiation. Some
of the energy released in a reaction may also be in the form of translational energy of the
newly-formed molecule, and this will also be rapidly converted into heat via collisions.
Many of the reactions occurring in interstellar gas lead to heating in this way, but for the most
part, their effects are unimportant, as the quantities of the reactants involved are too small
to do much. The one case in which this process can become significant,
however, is the formation of H$_{2}$. 
In the local ISM, the H$_{2}$ formation rate is approximately \citep{jura75}
\begin{equation}
R_{\rm H_{2}} \sim 3 \times 10^{-17} n n_{\rm H} \: {\rm cm^{-3} s^{-1}},
\end{equation}
and a total of 4.48~eV of energy is released for each H$_{2}$ molecule that is formed. If
this energy is converted to heat with an efficiency $\epsilon_{\rm H_{2}}$, then the resulting
heating rate is
\begin{equation}
\Gamma_{\rm H_{2} form} \sim 2 \times 10^{-28} \epsilon_{\rm H_{2}}  n n_{\rm H} \: {\rm erg} \: {\rm cm^{-3} s^{-1}}.
\end{equation}
Comparing this with the heating rate (\ref{eqn:CR-rate}) due to cosmic ray ionization,   we see 
that H$_{2}$ formation heating will dominate whenever $\epsilon_{\rm H_{2}}  n_{\rm H} > 
(\zeta_{\rm H} / 10^{-17} \: {\rm s^{-1}})$. In principle, therefore, H$_{2}$ formation heating can be
an important process, provided that the efficiency factor $\epsilon_{\rm H_{2}}$ is not too small.
Unfortunately, the value of $\epsilon_{\rm H_{2}}$ remains a matter of debate within the astrochemical
community. Some studies \citep[see e.g.][and references therein]{leb12} indicate that a significant fraction of the H$_{2}$ binding energy
should be available for heating the gas, while others \citep[e.g.][]{roser03,con09} predict that $\epsilon_{\rm H_{2}}$
should be small.

\begin{figure}
\includegraphics[width=\textwidth]{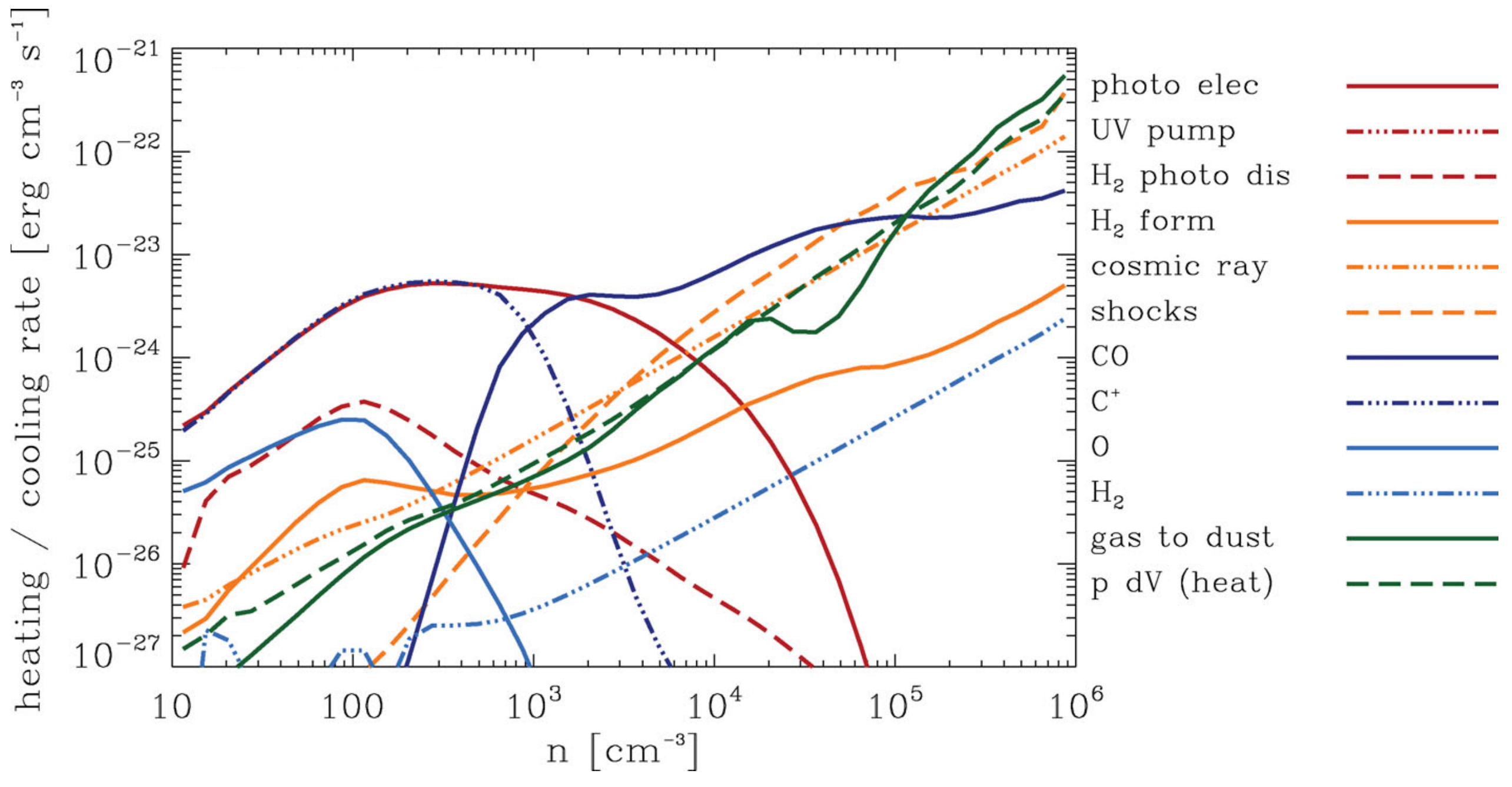}
\caption{Overview of the main  heating and cooling processes plotted as a function of the hydrogen nuclei number density $n$ calculated from a simulation of molecular cloud formation from initially atomic gas in the solar neighborhood. Adopted from \citet{gc12a}.
\label{fig:heating-cooling}}
\end{figure}

\paragraph{\bf Dynamical processes}

Finally, hydrodynamical and magnetohydrodynamical effects can also lead to significant heating. In subsonic, gravitationally collapsing regions, such as low mass prestellar cores, adiabatic compression ($PdV$ heating) can be a major source of heat and can actually be more important in the overall thermal balance of the core than cosmic ray heating. In less quiescent regions, where the gas flow is supersonic, turbulent dissipation in  shocks is another major heat source. Figure \ref{fig:heating-cooling} provides an overview of the most important heating and cooling processes for the solar neighborhood ISM.  Unlike in Figure \ref{cool-rates}, the rates are here plotted as a function of the hydrogen nuclei number density $n$. The figure shows that  initially atomic gas exhibits three different regimes. At densities $n < 2000\,$cm$^{-3}$, the gas heating is dominated by photoelectric emission from dust grains (Section \ref{subsec:photoelectric-heating}), while cooling is provided by fine structure emission from C$^+$. In the density regime $2000 < n < 10^5\;$cm$^{-3}$, rotational line emission from CO becomes the main coolant. Photoelectric heating remains the main heat source initially, but steadily becomes less effective, owing to the larger visual extinction of the cloud at these densities, and other processes -- adiabatic compression of the gas, dissipation of turbulent kinetic energy in shocks and cosmic ray ionization heating -- become more important at $n \sim 6000 \: {\rm cm^{-3}}$ and above. Finally, at densities above about $10^5\;$cm$^{-3}$, the gas couples to the dust (Section \ref{subsec:gas-grain-transfer}), which acts as a thermostat and provides most of the cooling power. Weak shocks and adiabatic compressions together dominate the heating of the gas in this regime, each contributing close to half of the total heating rate \cite[for a more detailed discussion, see][]{gc12a}.

The rate at which turbulent kinetic energy is dissipated in regions where the turbulence is supersonic
is well established \citep{maclow98, Stone98,maclow99}. The energy dissipation rate within a cloud of mass $M$ and velocity dispersion $\sigma$ can be written to within a factor of order unity as \citep{maclow99}
\begin{equation}
\dot{E}_{\rm kin} \sim - M k_{\rm d} \sigma^{3},
\end{equation}
where $k_{\rm d}$ is the wavenumber on which energy is injected into the system. If we assume that
this is comparable to the size of the cloud \citep[see e.g.][]{brunt09}, and adopt Larson's relations between 
the size of the cloud and its velocity dispersion and number density \citep{larson81}, then we arrive at an average 
turbulent  heating rate \citep{pp09}
\begin{equation}
\Gamma_{\rm turb} = 3 \times 10^{-27} \left(\frac{L}{1 \: {\rm pc}} \right)^{0.2} n \: {\rm erg \: s^{-1} \: cm^{-3}}.
\end{equation}
This heating rate is of a similar order of magnitude to the cosmic ray heating rate. Unlike cosmic ray
heating, however, turbulent heating is highly intermittent \citep{pp09}. This means that in much of the cloud,
the influence of the turbulent dissipation is small, while in small, localized regions, very high heating rates
can be produced \citep[see e.g.][]{fal95,god09}. We provide a more detailed account of ISM turbulence in the next Section. 

Finally, note that the physical nature of the heating process depends upon the strength of the magnetic
field within the gas. If the field is weak, energy dissipation occurs mostly through shocks, whereas if
the field is strong, a substantial amount of energy is dissipated via ambipolar diffusion
\citep{pzn00,lmm12}.

\newpage
\section{ISM Turbulence}
\label{sec:turbulence}
The dynamical evolution of the ISM and many of its observational parameters cannot be understood without acknowledging the importance of supersonic turbulence. Here, we summarize some of the key measurements that point towards the presence of strong supersonic turbulent motions in the various phases of the ISM on a wide range of spatial scales. We introduce the most important theoretical concepts behind our current understanding of ISM turbulence, and discuss some statistical properties of compressible turbulent flows. Finally, we speculate about the physical origin of the observed turbulence in the ISM. For an overview of ISM turbulence we refer the reader to the review articles by \cite{elmegreen04} and \cite{Scalo04}, and for a discussion of the relation between turbulence and star formation on local as well as galactic scales, we point to the reviews by \citet{maclow04} and \cite{ballesteros07b}.  More recent discussions on the topic of ISM dynamics can be found in \citet{Hennebelle:2012p72333} as well as in {\em Protostars and Planets VI}, in particular in the chapters by \citet{Padoan:2013p88872} or \citet{dobbs14}.

\subsection{Observations}
\label{subsubsec:turb-obs}

\subsubsection{\bf Observational tracers of ISM dynamics}
The best approach to learn more about the dynamical and kinematic state of the ISM is to look for the line emission (or sometimes absorption) of various atomic and molecular species. We take a spectrum, and once we have identified the line, we can compare the observed frequency with the rest-frame frequency in order to obtain information about the velocity distribution of gas along the line of sight (LOS). Ideally, we obtain spectra at multiple positions and fully cover the projected area of the object of interest on the sky. By doing so, we obtain a three-dimensional data cube containing the line intensity at different positions on the sky and different LOS velocities. Such position-position-velocity (PPV) cubes form the basis of most kinematic studies of ISM dynamics.

For the warm neutral medium (Section \ref{subsec:comp-ISM-gas}), most studies focus on the $21\,$cm hyperfine structure line of atomic hydrogen (H{\sc i}). It occurs with a spin flip from the excited state ($S=1$), where the spins of proton and electron are parallel, to the ground state ($S=0$), where the two spins are anti-parallel. The energy difference is $\Delta E = 5.87 \times 10^{-6}\,$eV or $\Delta E / k_{\rm B} = 0.06816\,$K, corresponding to the emission of a photon with the wavelength $\lambda = 21.106\,$cm or the frequency $\nu = 1.4204\,$GHz (for further details, see e.g.\ chapter 8 in \citealt{draine11}).

Molecular hydrogen is much more difficult to observe. It is a homonuclear molecule, and as a consequence its dipole moment vanishes. The quadrupole radiation requires high excitation temperatures and is extremely weak under normal molecular cloud conditions (Section \ref{subsubsec:H2}). Direct detection of cold interstellar H$_2$ requires ultraviolet absorption studies. However, due to the atmospheric absorption properties, this is only possible from space and limited to pencil-beam measurements of the absorption of light from bright stars or from AGN.\footnote{Note that rotational and ro-vibrational emission lines from H$_2$ have also been detected in the infrared, both in the Milky Way and in other galaxies. However, this emission comes from gas that has been strongly heated by shocks or radiation, and it traces only a small fraction of the total H$_2$ mass (e.g.\ \citealt{vanderwerf00}).} Studies of the molecular ISM therefore typically rely on measuring the radio and sub-millimeter emission either from dust grains or from other molecules that tend to be found in the same locations as H$_2$. 

The most prominent of these tracer molecules is CO and its various isotopologues. As previously mentioned, the most abundant of these isotopologues is $^{12}$C$^{16}$O, often referred to just as $^{12}$CO or simply CO. However, the high abundance of this tracer can actually become problematic, as it is often optically thick, and hence we cannot use it to trace the properties of the turbulence in the whole of the cloud. For example, numerical studies have shown that many of the smaller-scale structures identified in PPV cubes of $^{12}$CO emission are actually blends of multiple unrelated features along the LOS \citep{bpm02,beau13},  and that the statistical properties of the velocity field that can be derived using $^{12}$CO emission are not the same as those derived using the $^{12}$CO number density \citep{bertram14}. For this reason, studies of the properties of the turbulence within molecular clouds often focus on less abundant isotopologues, such as $^{13}$C$^{16}$O (usually written simply as $^{13}$CO) or  $^{12}$C$^{18}$O (often written just as C$^{18}$O). The optical depths of these tracers are much lower, and we therefore expect them to provide a less biased view of the properties of the turbulent velocity field. Nevertheless, problems still remain. The lowest rotational transition of CO, the $J = 1-0$ transition, has a critical density of only $n_{\rm cr} = 1.1 \times 10^3 \,$ particles per cm$^3$, only a factor of a few larger than the typical mean density of a molecular cloud. Observations of this transition are therefore useful at providing us with information on the properties of the cloud at densities close to the mean density, but provide little information on highly underdense or highly overdense regions. This is exacerbated by the chemical inhomogeneity of the CO distribution within molecular clouds. In low density, low extinction regions, much of the CO is photodissociated (see Section~\ref{cco}), and most of the available carbon is found instead in the form of C$^{+}$, while in high density cores, CO freezes out onto the surface of dust grains. 

\begin{figure}[t]
\center{\includegraphics[width=0.9\textwidth]{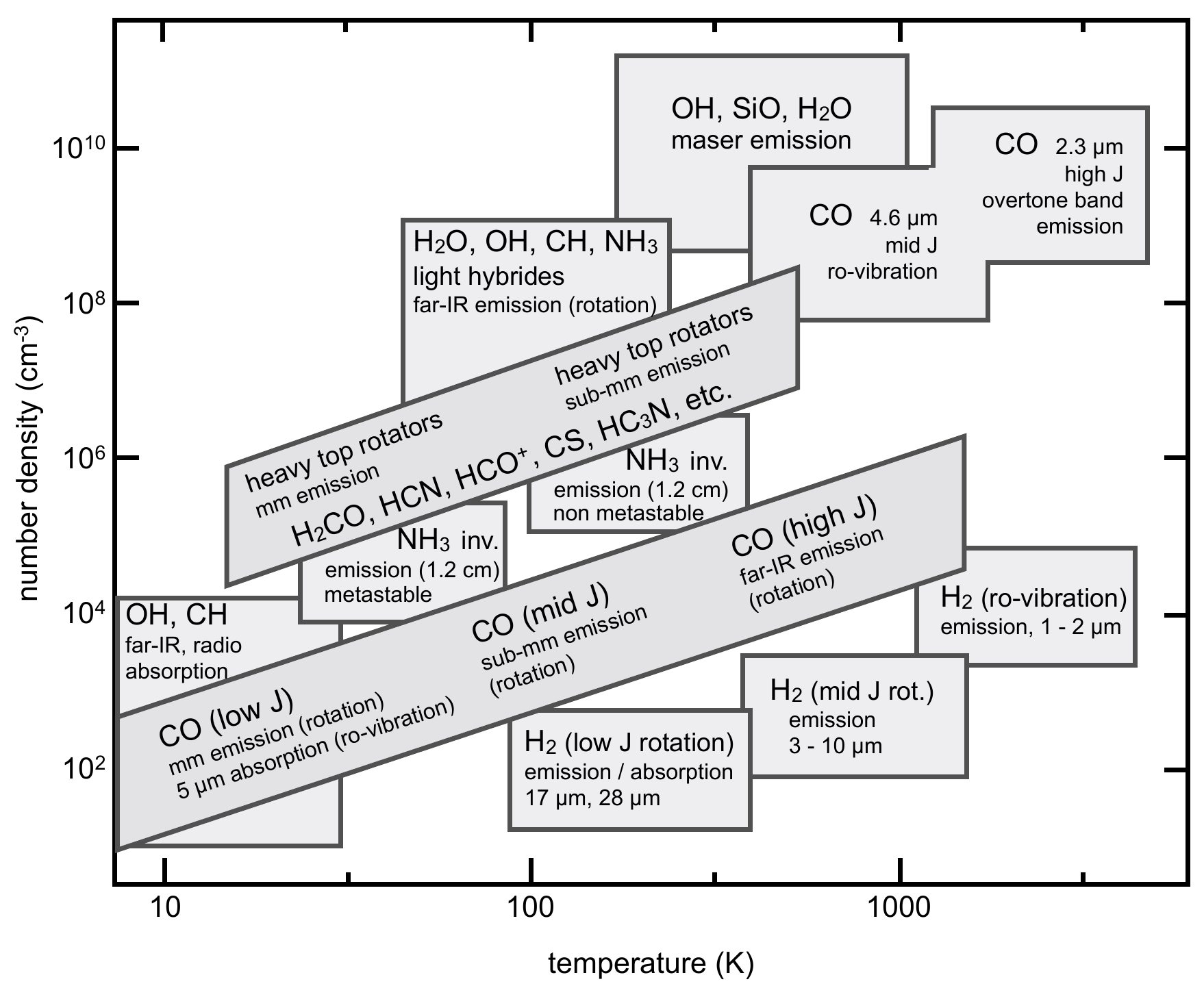}}
\caption{Temperature and density range of various observational tracers of molecular cloud structure and dynamics. Adapted from \citet{genzel91}.}
\label{fig:genzel-plot}
\end{figure}

To trace the properties of the turbulence in these regions, different observational tracers are required. In low density regions, this is difficult, as C$^{+}$ emits at a wavelength of $158 \, \umu$m which cannot be observed from the ground owing to the effects of atmospheric attenuation. It has been observed from the stratosphere by the Kuiper Airborne Observatory \citep[see e.g.][]{chokshi88} and more recently by the Stratospheric Observatory for Infrared Astronomy (SOFIA; see e.g.\ \citealt{simon12}), and from space by ISO and by the Herschel space telescope \citep[e.g.][]{pineda13}, but efforts to map the large-scale distribution of C$^{+}$ emission within molecular clouds are still in their infancy. In addition, they are hampered by the fact that 
the energy separation of the ground state and first excited state of C$^{+}$ corresponds to a temperature of around 92~K, higher than one expects to find in the low density regions of most molecular clouds, making the properties of the observed C$^{+}$ emission highly sensitive to the temperature distribution of the gas in the cloud. In high density regions, the situation is much simpler, as a number of different observational tracers are readily available, with the most popular ones being HCN, NH$_{3}$, HCO$^{+}$ and N$_{2}$H$^{+}$. A summary of the most relevant tracers, together with the range of temperature and density they are most suitable for, is depicted in Figure \ref{fig:genzel-plot}. 

For studying the properties of H{\textsc{ii}} regions, atomic recombination lines are the best available tool.  These are electronic transitions that occur when the recombination event leaves the electron in an excited state, which consequently decays down towards the ground state by emitting photons. The classic example is line emission from the hydrogen atom itself in the Lyman, Balmer, Paschen, etc.\ series  \cite[e.g.][]{spitzer78, osterbrock89}. For low quantum numbers these photons typically have UV or optical wavelengths, but if highly excited Rydberg states are involved, the emission can be detected at radio or sub-mm wavelengths. Besides hydrogen (and in part helium) recombination lines,  H{\textsc{ii}} regions also show a large number of metal lines, both at optical wavelengths, where they result from the recombination of (multiply) ionized atoms (such as O$^{++}$ or N$^+$), and in
the infrared, where they result from fine structure transitions of ions or atoms with high ionization potentials.

\begin{figure}[t]
\unitlength1.0cm
\begin{picture}(10,7)
\put(0.0,0.0){\includegraphics[height=6.7cm]{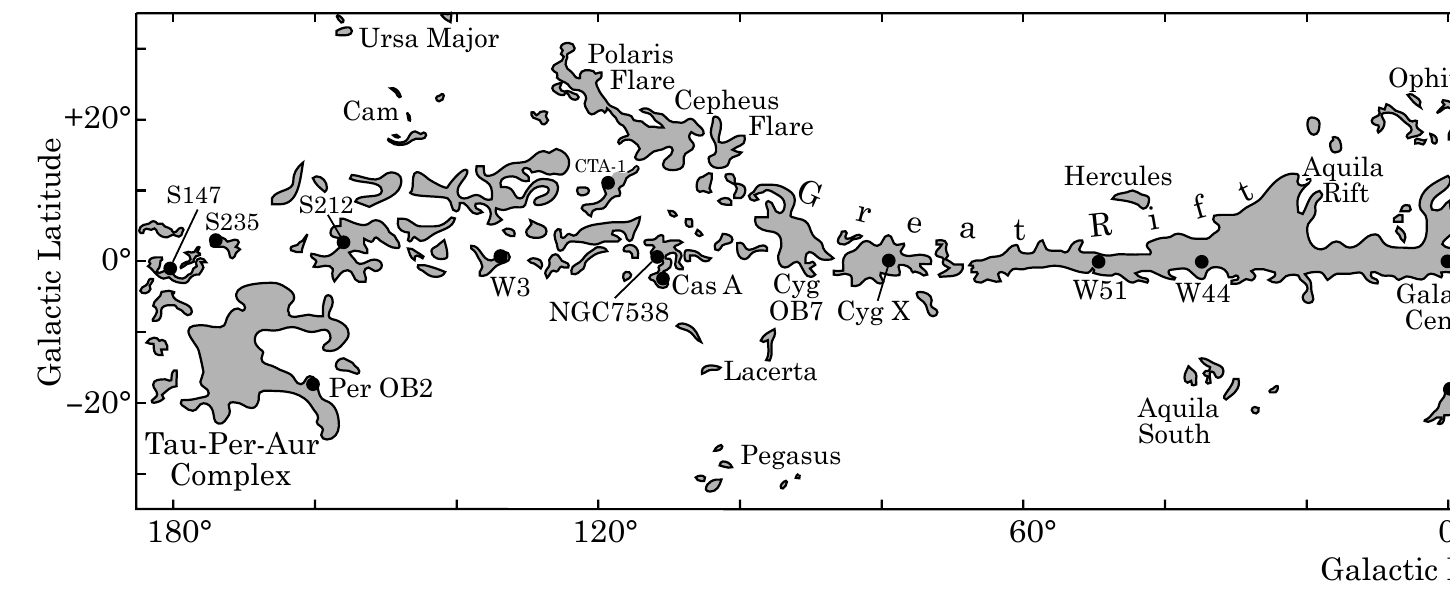}}
\end{picture}
\caption{Schematic distribution of molecular cloud complexes in the disk of the Milky Way. Data from  \citet{dame01}. {\textcolor{red}{}}}
\label{fig:gal-plane-survey}
\end{figure}

\subsubsection{Properties of molecular clouds}
\label{par:prop-MC}
In the following, we focus on the properties of the molecular component of the ISM. The all-sky survey conducted by \citet{dame01} shows that molecular gas is mostly confined to a thin layer in the Galactic disk, and that the gas in this layer is organized into cloud complexes of various sizes and masses. Figure \ref{fig:gal-plane-survey} illustrates the distribution of Galactic H$_2$ as traced by the $J=1-0$ line of  $^{12}$CO with some of the most prominent molecular cloud complexes indicated by name. One of the best studied complexes in the northern sky contains the the two giant molecular clouds Orion A and B, lying between a Galactic longitude $200^{\circ} < \ell < 220^{\circ}$ and a latitude  $-20^{\circ} < b < -10^{\circ}$. A detailed map of the total CO intensity from this region is shown in Figure \ref{fig:mon-orion-total-intensity}, taken from the study of \citet{Wilson:2005p80793}. Their observations reveal a complex hierarchy of filaments and clumps on all resolved scales. 

Studies of the structure of the molecular gas show that molecular clouds appear to display self-similar behavior over a wide distribution of spatial scales \citep[see e.g.\ the review by][]{williams00}, ranging from scales comparable to the disk thickness down to the size of individual prestellar cores, where thermal pressure starts to dominate the dynamics. The molecular cloud mass spectrum is well described by a power law of the form
\begin{equation}
\label{eqn:mass-spectrum}
\frac{dN}{dm} \propto m^{-\alpha}\:,
\end{equation}
with the exponent being somewhere in the range $3/2 < \alpha < 2$. Consequently there is no natural mass or size scale for molecular clouds between the observed lower and upper limits.  The largest molecular structures are giant molecular clouds (GMCs). They have  masses of typically  $10^5$ to $10^6\,{\rm M}_{\odot}$ and extend over a few tens of parsecs. On the other hand, the smallest observed entities are protostellar cores with masses of a few solar masses or less and sizes of $\sil 10^{-2}\,$pc. The volume filling factor of dense clumps, even denser subclumps and so on, is very low. It is of the order of 10\% or less. In the following, we distinguish between molecular cloud complexes, cluster-forming clumps (often called infrared dark clouds, IRDCs, in the phases prior to the onset of massive star formation), and protostellar cores (which give rise to individual stars or binary systems). Table \ref{tab:mol-clouds} summarizes their basic parameters.

\begin{figure}[t]
\unitlength1.0cm
\begin{picture}(10,7)
\put(-3.0,0.1){\includegraphics[height=7.2cm]{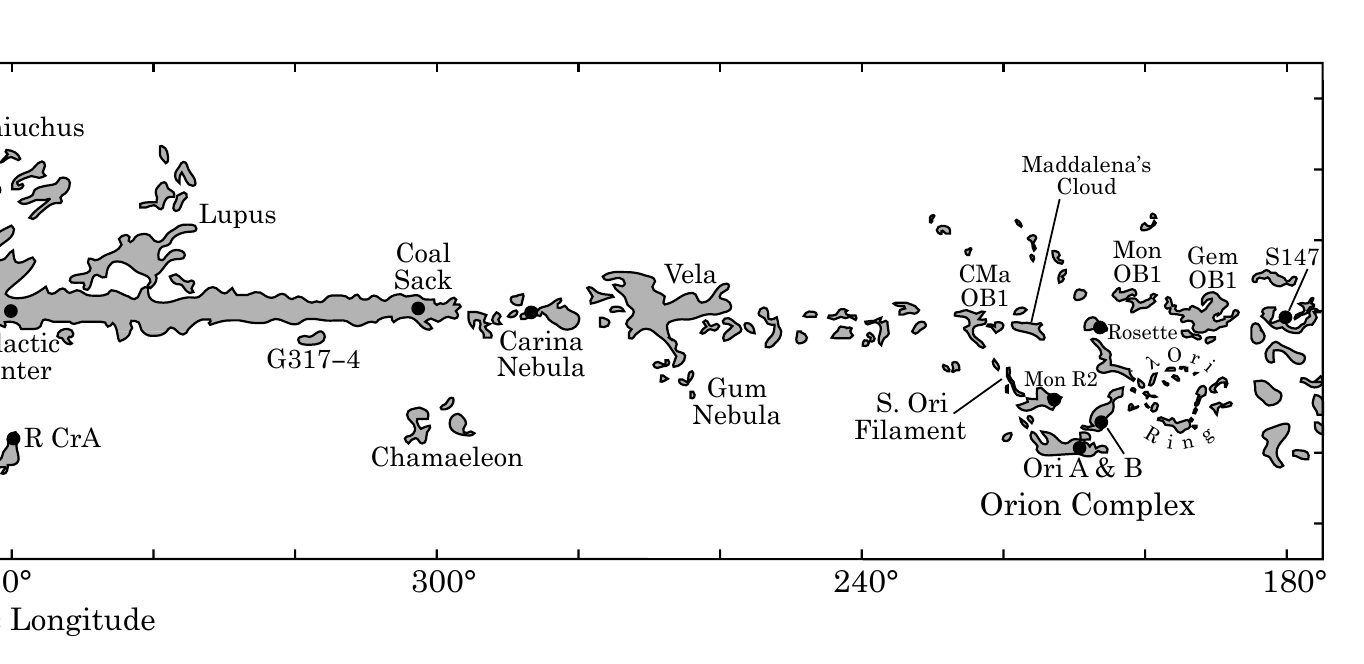}}
\end{picture}
\end{figure}

The fact that all studies obtain a similar power law is remarkable, and we argue below that it is the result of turbulent motions acting on self-gravitating gas \cite[see also][]{maclow04,ballesteros07b}. This result holds for clouds over a wide range of cloud masses and densities, and is based on data obtained with different reduction and analysis techniques. Furthermore, the result seems to be independent of whether it was derived for very actively star-forming clouds or very cold and quiescent ones. Given the uncertainties in determining the slope, it appears reasonable to conclude that there is a universal mass spectrum, and it appears plausible that the physical processes at work are rather similar from cloud to cloud. And vice versa, clouds that show significant deviation from this universal distribution most likely have different dynamical histories or live in different environments \cite[for a discussion of molecular cloud properties in the spiral galaxy M51 based on probability distribution functions of $^{12}$CO integrated intensity, see][]{hughes13b}.

\begin{table}[t]
\begin{center}
{\caption{\label{tab:mol-clouds}
 Physical properties of molecular clouds, clumps, and cores$^a$}}
\begin{tabular}[t]{p{3.5cm} @{\hspace{0.5cm}} p{1.9cm} @{\hspace{0.5cm}} p{1.9cm} @{\hspace{0.5cm}} p{1.9cm}}
\hline
& molecular clouds & cluster-forming clumps & protostellar cores \\
\hline
Size (pc)                            &$2 - 20$   & $0.1-2$    &$\sil 0.1$\\
Mean density (H$_2\, {\rm cm}^{-3}$)   &$10^2-10^3$& $10^3-10^5$&$>10^5$ \\
Mass (M$_{\odot}$)                  &$10^2-10^6$& $10 - 10^3$&$0.1-10$ \\
Temperature (K)                      &$10-30$    & $10-20$    &$7-12$ \\
Line width (km$\,$s$^{-1}$)           &$1 - 10$   & $0.5-3$    &$0.2-0.5$\\
Turbulent Mach number            &$5 - 50$   & $2 - 15$    &$0 - 2$\\
Column density (g cm$^{-2}$) & $0.03$ & $0.03-1.0$ & $0.3-3$ \\
Crossing time (Myr)           & $2 - 10$   & $\sil 1$    &$0.1-0.5$\\
Free-fall time (Myr)           & $0.3 - 3$   & $0.1 - 1$    &$\sil 0.1$\\
Examples                             & Orion, Perseus & L1641, L1709 & B68, L1544\\
\hline
\end{tabular}
{\footnotesize \vspace{0.2cm} $^a$~Adapted from \citet{Cernicharo1991} and \citet{bergin07}.}
\end{center}
\end{table}


Temperatures within molecular clouds are generally lower than in other phases of the ISM. Simulations suggest that the low density portions of molecular clouds that are CO-dark (i.e.\ that are H$_{2}$-rich, but CO-poor; see Section \ref{dust-shield-import} below) have temperatures ranging from around 20~K up to the temperatures of 50--100~K that are typical of the CNM \citep[see e.g.][]{gc12c,glo14}. Unfortunately, observational verification of this prediction is difficult, as the main observational tracer of the gas in these regions, C$^{+}$, has only a single fine structure emission line and hence does not directly constrain the temperature of the gas. In the denser, CO-emitting gas, both observations and simulations find temperatures of around 10--20~K for dark, quiescent clouds, and somewhat higher values in clouds close to sites of ongoing high-mass star formation. It is notable that within this dense, well-shielded gas, the temperature remains remarkably constant over several orders of magnitude in density \citep[see e.g.][]{gold88,gc12a, glo14}. This has important consequences for theoretical and numerical models of molecular cloud dynamics and evolution, because to a good approximation the gas can be described by a simple isothermal equation of state, where pressure $P$ and density $\rho$ are linearly related, 
\begin{equation}
P = c_{\rm s}^2 \,\rho,
\end{equation}
with the sound speed $c_{\rm s}$ being the proportionality factor. The assumption of isothermality breaks down when the gas becomes optically thick and heat can no longer be radiated away efficiently. In the local ISM, this occurs when the number density exceeds values of $n({\rm H}_2) \approx 10^{10}$~cm$^{-3}$.

\begin{figure}[htbp]
\begin{center}
\includegraphics[width=0.85\textwidth]{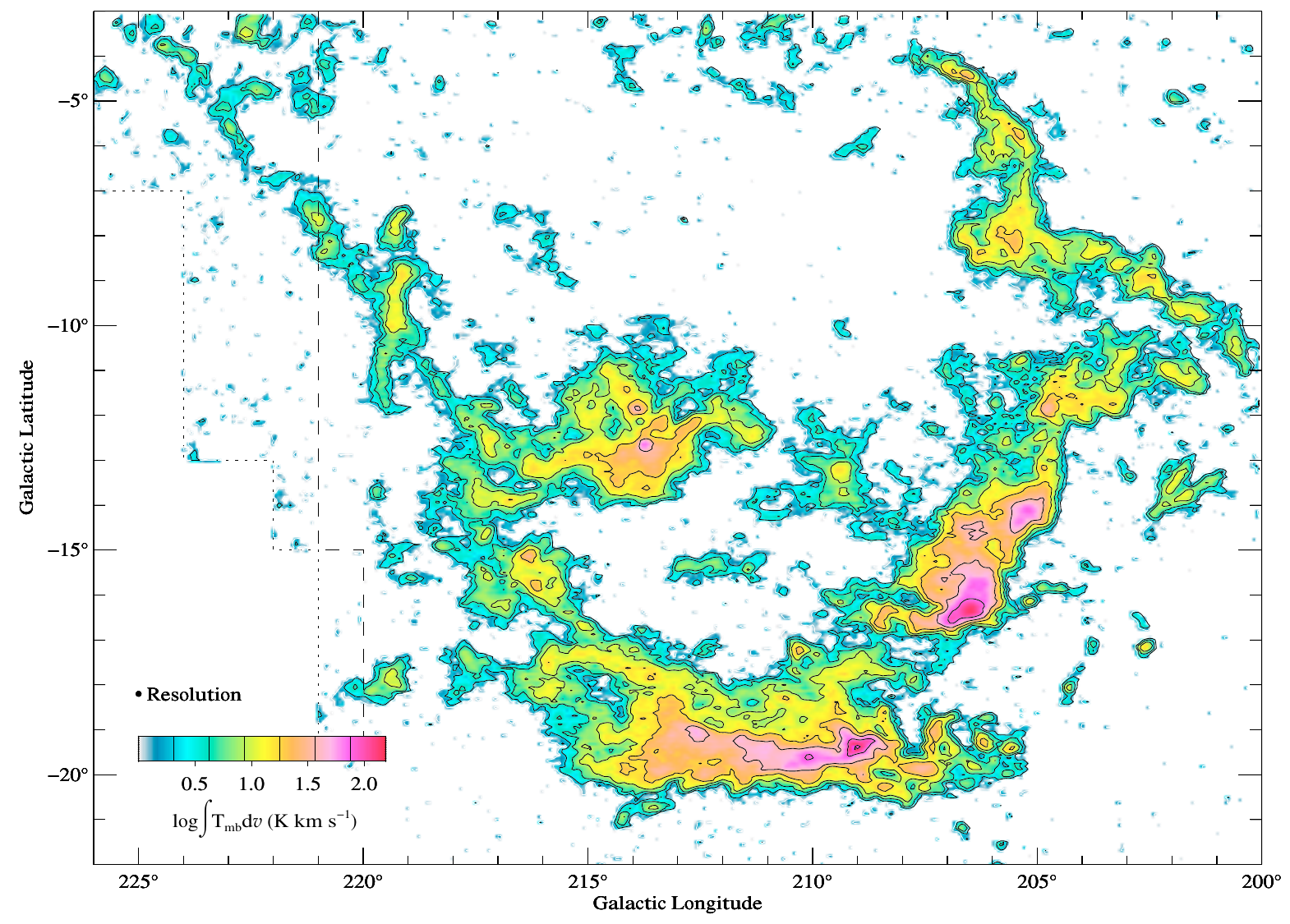}
\caption{Map of the velocity-integrated $J=1-0$ rotational line emission of the $^{12}$CO molecule as tracer of the total molecular hydrogen gas in the Orion / Monoceros region. The image shows the complex spatial structure of H$_2$ gas in a typical molecular cloud complex. The figure is taken from \citet{Wilson:2005p80793}.}
\label{fig:mon-orion-total-intensity}
\end{center}
\end{figure}

The masses of molecular clouds are orders of magnitude larger than the critical mass for gravitational collapse computed from the average density and temperature (see Section \ref{sec:collapse-SF}).  If we assume that only thermal pressure opposes gravitational attraction they should collapse and quickly form stars  on timescales comparable to the free-fall time. However, this is {\sl not} observed (for an early discussion, see \citealt{zuckerman74}; for more recent discussions, consult \citealt{Kennicutt:2012p70781}). The typical lifetime of giant molecular clouds is about $10^7\,$years \cite[][]{blitz07a,dobbs14}, and the average star formation efficiency is low, with values ranging between 1\% and 10\% \citep{blitz80,krumholz07e}. This tells us that there must be additional physical agents that provide stability against large-scale cloud collapse. 

\begin{figure}[htbp]
\begin{center}
\includegraphics[width=0.9\textwidth]{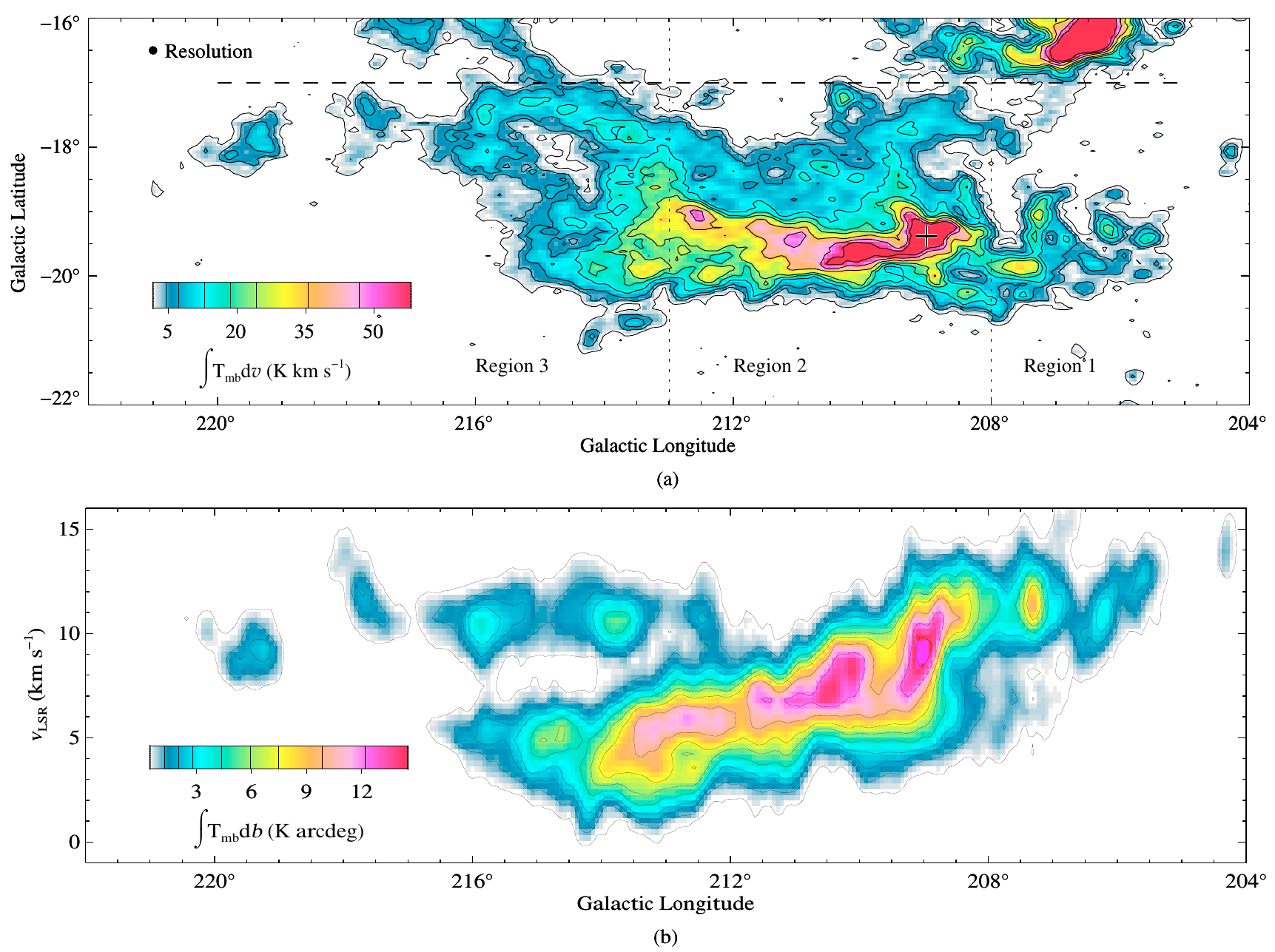}
\caption{{\it Top}: Integrated intensities of the $J = 1-0$ transition of $^{12}$CO of the Orion A cloud (enlargement of the lower central parts of Figure \ref{fig:mon-orion-total-intensity}). {\it Bottom}: Distribution of line-of-sight velocities as a function of Galactic longitude for the same region, based on data integrated in stripes of Galactic latitude from $-22^\circ < b< -17^\circ$. The width of the velocity distribution gives a good indication of the turbulent velocity dispersion in the region. (For more information on both panels, see \citealt{Wilson:2005p80793}).}
\label{fig:orion-A-vel-disp}
\end{center}
\end{figure}

For a long time, magnetic fields have been proposed as the main agent responsible for preventing collapse \cite[e.g.][]{shu87}. However, it appears that the typical field strengths observed in molecular clouds are not sufficient to stabilize the clouds as a whole \citep{ver95a,ver95b,tro96,padoan99,Lunttila:2009p33749,crutcher09b,crutcher10,Bertram:2012p55309}. This is the point at which ISM turbulence comes into play \citep{elmegreen04, Scalo04}. Virtually all observations of molecular cloud dynamics reveal highly supersonic gas motions on scales above a few tenths of a parsec. The observed linewidths are always wider than what is implied by the excitation temperature of the molecules. This is illustrated in Figure \ref{fig:orion-A-vel-disp}, which shows the $^{12}$CO $J = 1-0$ integrated intensity from the Orion A cloud in the top panel together with the distribution of Doppler velocities of the line peak as a function of the cloud's major axis in the bottom panel, each entry sampled in strips across the face of the cloud parallel to the minor axis. The width of the velocity distribution along the ordinate is a good indicator of the one-dimensional velocity dispersion $\sigma_{\rm 1D}$ of the cloud. We see that $\sigma_{\rm 1D}$ reaches values of a few km$\,$s$^{-1}$, about an order of magnitude larger than the sound speed of the dense molecular gas, $c_{\rm s} \approx 0.2 \,$km$\,$s$^{-1}$. 

\begin{figure}[htbp]
\begin{center}
\includegraphics[width=0.9\textwidth]{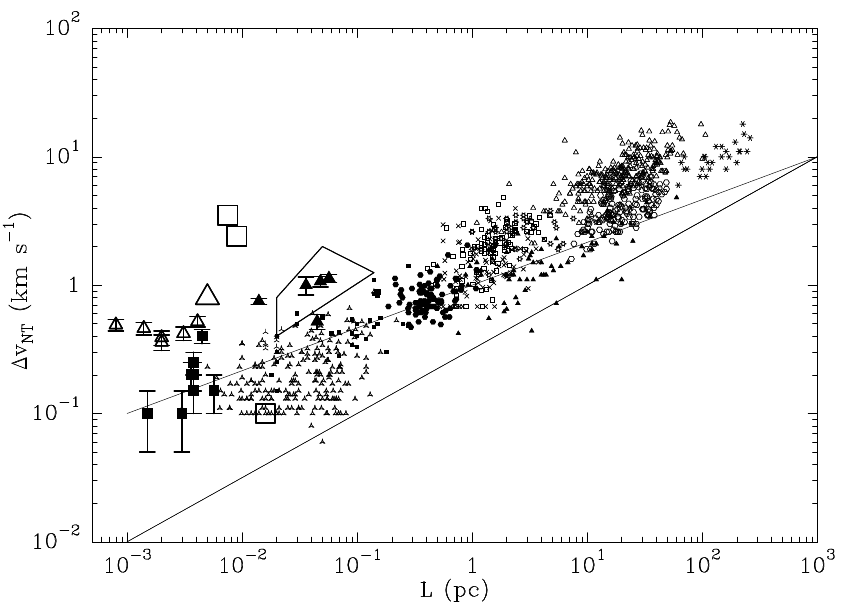}
\caption{Relation between velocity dispersion (as measured by the width of the $J=1 - 0$ rotational line transition of $^{12}$CO)  and spatial scale in the Galactic ISM. The data points come from a wide range of observations  that trace different structures and physical conditions (in terms of density, temperature, excitation parameter, etc.). This in part contributes to the large scatter in the data. However,  altogether the observations reveal a clear power-law relation of the form $\Delta v_{\rm NT} \propto L^\alpha$. To guide the eyes, the solid lines illustrate the slopes $\alpha = 1/3$ and $\alpha = 1/2$.  The lower limit of $\Delta v_{\rm NT} \approx 0.1\,$km$\,$s$^{-1}$ is due to the spectral resolution in the data and corresponding noise level. The figure is taken from \citet{Falgarone:2009p13509}, where further details and a full list of references can be found. }
\label{fig:Larson}
\end{center}
\end{figure}

More detailed analysis reveals that the observed velocity dispersion $\sigma_{\rm 1D}$ is related to the size $L$ of the cloud by
\begin{equation}
\label{eqn:larson}
\sigma_{\rm 1D} \approx 0.5 \left(\frac{L}{1.0\mbox{ pc}}\right)^{1/2}\mbox{ km s}^{-1}.
\end{equation}
This goes back to the seminal work by \citet{larson81}, who compared measurements of different clouds available at that time, and it has been confirmed by many follow-up studies both in our Milky Way as well as neighboring satellite galaxies \cite[e.g.][]{solomon87, heyer04a, bolatto08a, Falgarone:2009p13509, RomanDuval:2011p52934, CalduPrimo:2013p80430}. There is still some debate about  the normalization and about slight variations in the slope \citep{Heyer:2009p40203, Shetty:2012p68723, hughes13}, but in general the relation (\ref{eqn:larson}) is thought to reflect a more or less universal  property of the ISM. The most common interpretation is the presence of turbulent gas motions. On scales above $\sim 0.1\,$pc (which corresponds to the typical sizes of prestellar cores; see Section \ref{subsec:cores}), the velocities inferred from equation (\ref{eqn:larson}) exceed values of the thermal line broadening (where the one-dimensional velocity dispersion $\sigma_{\rm 1D}$ is comparable to the sound speed $c_{\rm s}$). On scales of molecular cloud complexes, we measure root mean square Mach numbers of 10 or larger, clearly indicative of highly supersonic turbulence. We also note that these motions seem to exceed the typical Alfv\'{e}n velocities in molecular clouds,
\begin{equation}
v_{\rm A} = \left( \frac{B^2}{4 \pi \rho} \right)^{1/2}\;,
\label{eqn:Alfven-velocity}
\end{equation}
with $B$ and $\rho$ being the magnetic field strength and the mass density of the gas, respectively. The observed turbulence is not only supersonic but also super-Alfv\'{e}nic \cite[e.g.][]{padoan99, Heyer:2012p74096}. In essence, this means that the energy density associated with turbulent gas motions dominates over both the thermal energy density as well as the magnetic energy density. We also note that the observed linewidths generally are not due to large-scale collapse as inferred from the generally rather low star formation rates and the absence of inverse P-Cygni line profiles.

\begin{figure}[th]
\begin{center}
\includegraphics[width=0.9\columnwidth]{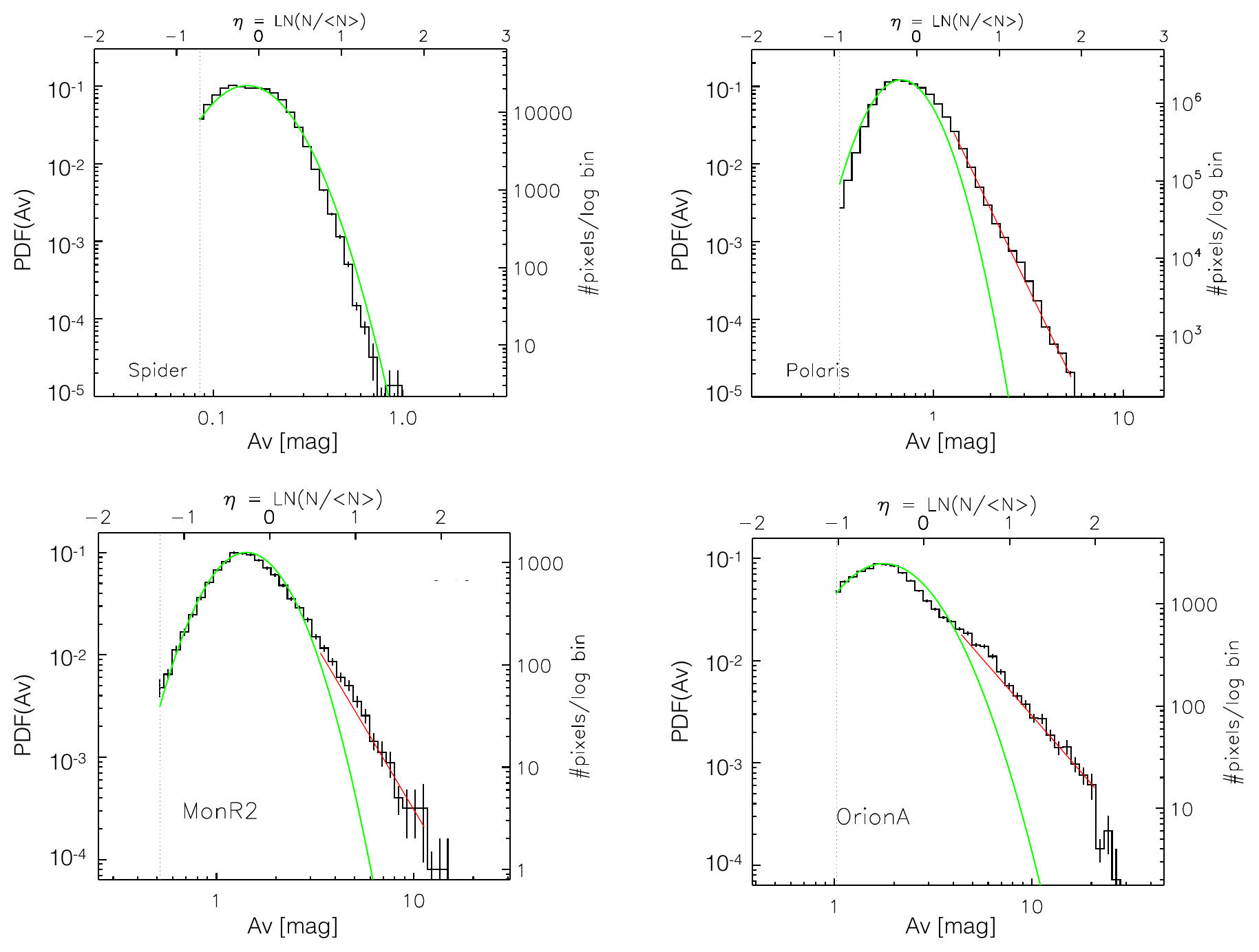}
\caption{Column density PDFs derived from {\sl Herschel} observations \cite[see e.g.][]{Schneider2013} for four different nearby molecular clouds:  two tenuous high-lattitude clouds, Spider (top left) and Polaris (top right), and two star dense star-forming regions, Monoceros R2 (bottom left) and Orion A (bottom right). For a map of the latter two clouds, see Figure \ref{fig:mon-orion-total-intensity}.  The lower abscissa gives the visual extinction, $A_{\rm V}$, which we take as a proxy of the column density $N$. The upper axis indicates the natural logarithm of the column density normalized to the mean value, $\eta = \ln(N/\langle N \rangle)$. The left ordinate is the PDF of the extinction, and to the right, we provide the corresponding total number of pixels to indicate the statistical significance of the observation. The green curve indicates the fitted PDF, and the red line shows a possible power-law fit to the high $A_{\rm V}$ tail. 
The plots are adopted from \citet{Schneider2014a}. }
\label{fig:column-density-PDF}
\end{center}
\end{figure}

The analysis of extinction or dust emission maps in nearby molecular clouds reveals a roughly log-normal distribution of column densities in tenuous cirrus-like clouds with no or little star formation, and they show the development of a power-law tail at high column density that becomes more pronounced for more massive and more vigorously star-forming clouds \citep{lada10, Kainulainen2011, Kainulainen2013, Schneider2012, Schneider2014a, Alves14}. Typical examples are provided in Figure \ref{fig:column-density-PDF}, where we take the visual extinction, $A_{\rm V}$, as a proxy for the column density. Spider (top left) and Polaris (top right) are high latitude clouds located in the North Celestial Loop \citep{meyerdierks91}. Spider shows no signs of star formation and is a prototypical example of a cloud with a log-normal PDF, while Polaris seems to be forming some low-mass stars and exhibits a weak power-law tail. Monoceros and Orion A (see also Figure \ref{fig:mon-orion-total-intensity}) have much higher average densities and are forming clusters containing intermediate to high-mass stars. They exhibit a clear power-law tail at high extinctions. These observations are essential, because the characteristics of the (column) density distribution function are essential input to our current theoretical star formation models. This is discussed in detail in Section \ref{subsubsec:turb-IMF}.

\subsection{Simple theoretical considerations}
\label{subsubsec:turb-theory}

At this point, we need to digress from our discussion of molecular cloud properties and turn our attention to the theoretical models introduced to describe turbulent flows. We begin by introducing the classical picture of incompressible turbulence. This is a good description for turbulent flows with velocities that are significantly smaller than the speed of sound, such as those we typically encounter on Earth. For very subsonic flows we can infer from the continuity equation (\ref{eqn:continuity}) that density fluctuations are negligible. We note that for typical ISM conditions, however, the turbulence is highly supersonic, and we need to go beyond this simple picture as the compressibility of the medium becomes important, for instance when we want to understand the formation of stars as discussed in Section \ref{subsec:gravoturb-SF}. In any case, we assume that kinetic energy is inserted into the system on some well-defined, large scale $L_{\rm D}$, and that it cascades down through a sequence of eddies of decreasing size until the size of the eddies becomes comparable to the mean free path $\lambda$. The kinetic energy associated with eddy motion turns into heat (random thermal motion) and is  dissipated away. This picture of the turbulent cascade goes back to \cite{Richardson1920}. 

We first follow \citet{Kolmogorov1941a} and derive the corresponding power spectrum of the turbulent kinetic energy, which describes terrestrial flows, such as the motion of air in the Earth's atmosphere or the flow of water in rivers and oceans. Then we turn to supersonic motions and focus on additional aspects that are characteristic of ISM turbulence. For the level of our discussion, it is sufficient to think of turbulence as the gas flow resulting from random motions on many scales, consistent with the simple scaling relations discussed above. For a more detailed discussion of the complex statistical characteristics of turbulence, we refer the reader to the excellent textbooks by \cite{Frisch96}, \citet{Lesieur97}, or \cite{Pope2000}. For a thorough account of ISM turbulence, we point again to the reviews by \cite{elmegreen04} and \cite{Scalo04}, and for the relation to star formation to \citet{maclow04} and \cite{ballesteros07b}.

\subsubsection{\bf Energy cascade in stationary subsonic turbulence}
Hydrodynamical flows exhibit two fundamentally different states. For small velocities, they tend to be laminar and smooth. If the  velocity increases, however, the flow becomes unstable. It becomes turbulent and highly chaotic. This transition occurs when  advection strongly dominates over dissipation. To see this, let us consider the equations describing the motion of a fluid element. From the continuity equation,
\begin{equation}
\frac{\partial \vec{\rho}}{\partial t} + \vec v \cdot \vec \nabla \rho   = - \rho  \vec \nabla \cdot \vec{v}\;,\label{eqn:continuity}
\end{equation}
we can infer for incompressible flows ($\rho = {\rm const.}$) that $\vec{\nabla} \cdot \vec{v} = 0$. The equation of motion, also called the Navier-Stokes equation, then simplifies to 
\begin{equation}
\frac{\partial \vec{v}}{\partial t} + (\vec v \cdot \vec \nabla) \vec{v}  -  \nu\,\vec\nabla^2 \vec v = 
-\frac{1}{\rho} \vec\nabla P\;,
\label{eqn:Navier-Stokes}
\end{equation}
where $\vec{v}$ is the fluid velocity, $P$ and $\rho$ are pressure and density, and $\nu$ is the kinetic viscosity with the units cm$^2\,$s$^{-1}$.  At any spatial scale $\ell$, we can compare the advection term $(\vec v \cdot \vec \nabla) \vec{v}$ with the dissipation term $\nu\,\vec\nabla^2 \vec v$. To get an estimate of their relative importance for the flow dynamics we approximate $\vec{\nabla}$ by $1/\ell$ and obtain
\begin{equation}
(\vec v \cdot \vec \nabla) \vec{v} \,\approx \frac{v_\ell^2}{\ell} \mathrm{~~and~~} \nu\,\vec\nabla^2 \vec v \,\approx \frac{\nu v_\ell}{\ell^2}\;,
\end{equation}
with $v_\ell$ being the typical velocity on scale $\ell$. This ratio defines the dimensionless Reynolds number on that scale,
\begin{equation}
  \mathrm{Re}_{\ell}=\frac{v_\ell \ell}{\nu}\;.
\label{eqn:reynolds-number}
\end{equation}
It turns out that a flow becomes unstable and changes from being laminar to turbulent if the Reynolds number exceeds a  critical value $\mathrm{Re}_\mathrm{cr} \approx \mathrm{few} \times 10^3$. The exact value depends on the flow characteristics. For example, pipe flows  with $\mathrm{Re} <  2 \times 10^3$ are usually laminar, while flows with $\mathrm{Re}>  4 \times 10^3$ are most certainly turbulent. For intermediate Reynolds numbers both laminar and turbulent flows are possible, depending on other factors, such as pipe roughness and flow uniformity. These flows  are often called transition flows. In the ISM, the Reynolds number can easily exceed values of $10^9$ or more, indicating that the ISM is highly turbulent. 

A full analysis of the turbulent instability is very difficult and in general an unsolved problem. In the classical picture turbulence causes the formation of eddies on a wide range of different length scales. In this picture, some driving mechanism creates eddies on some large scale. These live for about one crossing time, and then fragment into smaller eddies, which again break up into even smaller eddies, and so forth. Most of the kinetic energy of the turbulent motion is contained on large scales. It cascades down to smaller and smaller ones in an inertial and essentially inviscid way. This holds as long as the advection term dominates over dissipation, i.e. as long as $\mathrm{Re} \gg \mathrm{Re}_\mathrm{cr}$. Eventually this hierarchy creates structures on scales that are small enough so that molecular diffusion or other forms of dissipation become important. The turbulent eddies become so tiny that they essentially turn into random thermal motion, the kinetic energy they carry becomes heat, and may be radiated away.

To obtain an estimate of the scaling behavior of turbulent flows let us look at the specific kinetic energy $\epsilon_\ell$ carried by eddies of size $\ell$. With $v_\ell$ being the typical rotational velocity across the eddy and with $t_\ell \approx \ell/ v_\ell$ being the typical eddy turnover time, we can estimate the energy flow rate through eddies of size $\ell$ as 
\begin{equation}
  \dot\epsilon \approx \frac{\epsilon_\ell}{t_\ell} 
  \approx \left(\frac{v_\ell^2}{2}\right)\left(\frac{\ell}{v_\ell}\right)^{-1}
  \approx\frac{v_\ell^3}{\ell}\;.
\label{eq:kolmogorov-1}
\end{equation}
As long as Re$_\ell \gg 1$,  dissipation is negligible. The rate $\dot\epsilon$ is conserved, and the kinetic energy simply flows across $\ell$ from larger scales down to smaller ones. This defines the inertial range of the turbulent cascade. It ends when Re$_\ell$ approaches unity at the dissipation scale  $\lambda_\nu$. Assume now that kinetic energy is inserted into the system on some large scale $L$ with a typical velocity  $v_L$. Then, the inertial range covers the scales
\begin{equation}
L > \ell >  \lambda_\nu.
\label{eq:intertial-range}
\end{equation}
In this regime, the energy flow $\dot\epsilon$ is independent of scale, as kinetic energy cannot be accumulated along the turbulent cascade. This implies that the typical eddy velocity $v_\ell$ changes with  eddy scale $\ell$ as $ v_\ell = \epsilon_\ell \ell^{1/3}$. 
As a consequence, the largest eddies  carry the highest velocities,  
\begin{equation}
  v_\ell \approx v_L \left(\frac{\ell}{L}\right)^{1/3}\;,
\label{eq:eddy-vel}
\end{equation}
but the smallest ones have the highest vorticity,
\begin{equation}
  \Omega_\ell = \vec\nabla \times \vec{v}_\ell \approx\frac{v_\ell}{\ell}\approx
  \frac{v_L}{(\ell^2L)^{1/3}} \approx \left(\frac{L}{\ell}\right)^{2/3} \Omega_L\;.
\label{eq:10-99}
\end{equation}
Indeed, in agreement with this picture of the turbulent cascade, observations of nearby molecular clouds reveal that the energy is carried by large-scale modes, indicating that the turbulent velocity field in these clouds is driven by external sources (see Section  \ref{subsub:ISM-driving-external}). 

We can obtain an estimate for the size of the inertial range (\ref{eq:intertial-range}) based on the requirement that $\mathrm{Re} \approx 1$ on the dissipation scale $\lambda_\nu$. In combination with (\ref{eq:eddy-vel}),  this leads to 
\begin{equation}
\frac{L}{\lambda_\nu} \approx \mathrm{Re}^{3/4}\;.
\end{equation}
With Reynolds numbers $\mathrm{Re} \approx 10^9$ and above, the turbulent cascade in the ISM extends over more than six orders of magnitude in spatial scale. 

We now look at the autocorrelation function of the velocity fluctuations on the scale $\ell$ defined as 
\begin{equation}
  \xi_{v} ({\Bell}) = \langle [ \vec{v}(\vec{x} + {\Bell}) - \vec{v}(\vec{x})]^2 \rangle  \;,
\label{eq:vel-correlation}
\end{equation}
which is the average of the square of all velocity differences between any two points in space separated by a lag ${\Bell}$. We have assumed 
that the system has zero net velocity, $\langle \vec{v}(\vec{x}) \rangle = 0$. If turbulence is isotropic, the autocorrelation function depends only on the separation $\ell = |{\Bell}|$ and not on the direction, and so $ \xi_{v} ({\Bell}) =  \xi_{v} (\ell)$. From (\ref{eq:kolmogorov-1}) we obtain 
\begin{equation}
  \xi_{v}(\ell) \propto \langle v_\ell^2\rangle  
  \propto(\dot\epsilon\ell)^{2/3}\;.
\label{eq:kolmogorov-2}
\end{equation}
In particular, we are interested in the Fourier transform of $ \xi_{v} ({\Bell})$, the power spectrum $P(k)$ of the velocity fluctuations. For random Gaussian fluctuations, both are related via
\begin{equation}
\frac{1}{2} \int_0^\infty \langle v_\ell'^2 \rangle d^3\ell'   = \int_\infty^{0} P_v(k)d^3k\;,
\label{eq:energy}
\end{equation}
which simply is the specific kinetic energy in the system. On each scale $\ell = 2\pi / k$ equation (\ref{eq:energy}) can be approximated by  
\begin{equation}
  P_{v}(k) \propto\ell^3\xi_{v}\propto
  k^{-3}\left(\dot\epsilon k^{-1}\right)^{2/3}\propto
  \dot\epsilon^{2/3}k^{-11/3}\;.
\label{eq:turbulent-power-spectrum}
\end{equation}
If we consider isotropic turbulence  with $d^3{k} \rightarrow 4\pi k^2 dk$, then the power in modes in the wave number range $k$ to $k+dk$  is 
\begin{equation}
  P_{v} \,k^2 dk\propto\dot\epsilon^{2/3}k^{-5/3}dk\;.
\label{eq:kolmogorov}
\end{equation}
This  is the  Kolmogorov spectrum of isotropic incompressible turbulence in the inertial range.
 
\subsubsection{\bf Energy cascade in stationary supersonic turbulence}
We now turn to the opposite limit of highly compressible turbulence, where the flow can be described as a network of interacting shock fronts. To simplify our discussion, we neglect the effects of pressure forces. This leads to the so-called \citet{Burgers:1939p70817} turbulence. He introduced a simplified non-linear partial differential equation of the form 
\begin{equation}
\frac{\partial v}{\partial t} + v \frac{\partial v}{ \partial x} = \nu \frac{\partial^2 v}{ \partial x^2}\;,
\label{eqn:Burgers}
\end{equation}
as approximation to the full Navier-Stokes equation, in order to study the mathematical properties of turbulent flows. Indeed, in the high Mach number regime, where the velocities $v$ are much larger than the sound speed $c_{\rm s}$, we can neglect the pressure term, since $P \propto c^2_{\rm s}$, and Burgers' equation (\ref{eqn:Burgers})  is identical to equation (\ref{eqn:Navier-Stokes}) in one dimension. If we consider the width of the shock transition to be infinitely thin, then the density or the velocity jump in the shock can be mathematically described by a Heaviside step function. For isotropic turbulence, there is always a shock that runs parallel to the considered line-of-sight, and we can naively estimate the power in the wavenumber range $k$ to $k+dk$ as 
\begin{equation}
  P_{v} dk \propto v_k^2 dk \propto k^{-2} dk\;.
\label{eq:Burgers-power-spectrum}
\end{equation}
In a more general sense, this follows from the fact that the energy spectrum of a field is determined by its strongest singularity. If a function $u(x)$ has a discontinuity in the $(m-1)$th order derivative, then its energy spectrum has the form of a power-law $P(k) \propto k^{-2m}$. In a shock the velocity itself is discontiuous, and with $m=1$ we obtain equation (\ref{eq:Burgers-power-spectrum}).  A less handwaving derivation is very involved \cite[see e.g.][]{Boldyrev:1998p82360,Verma:2000p82635,Boldyrev:2004p26595,Bec:2007p81884}, but in general leads to a similar conclusion.

\subsubsection{\bf Decay rate of turbulent energy}
\label{par:decay-turb}
So far, we have considered the case of stationary turbulence. This requires an energy source that continuously excites large-scales modes to compensate for the loss of energy at the dissipation scale. For a long time, it was thought that the rate of energy dissipation in a  magnetized gas differs significantly from purely hydrodynamic turbulence. \citet{arons75}, for example, suggested that  presence of strong magnetic fields could explain the supersonic motions ubiquitously observed in molecular clouds (Section \ref{subsubsec:turb-obs}). In their view,  interstellar turbulence is a superposition of Alfv\'{e}n waves, propagating in many different directions with different wavelengths and amplitudes. These are transverse waves traveling with the Alfv\'{e}n velocity (equation \ref{eqn:Alfven-velocity}), and they are dissipationless in the linear regime under the assumption of ideal magnetohydrodynamics. However, if ambipolar diffusion is taken into account, i.e.\ the drift between charged and neutral particles in the partially ionized ISM, these waves can be dissipated away at a rate substantial enough to require energy input from a driving source to maintain the observed motions \cite[e.g.][]{Zweibel:1983p84059,Zweibel:2002p84077}. Furthermore, if one includes higher order effects, then the dissipation becomes comparable to the purely hydrodynamic case \cite[e.g.][]{cho2003}.

\begin{figure}[bt]
\begin{center}
\includegraphics[width=.9\textwidth]{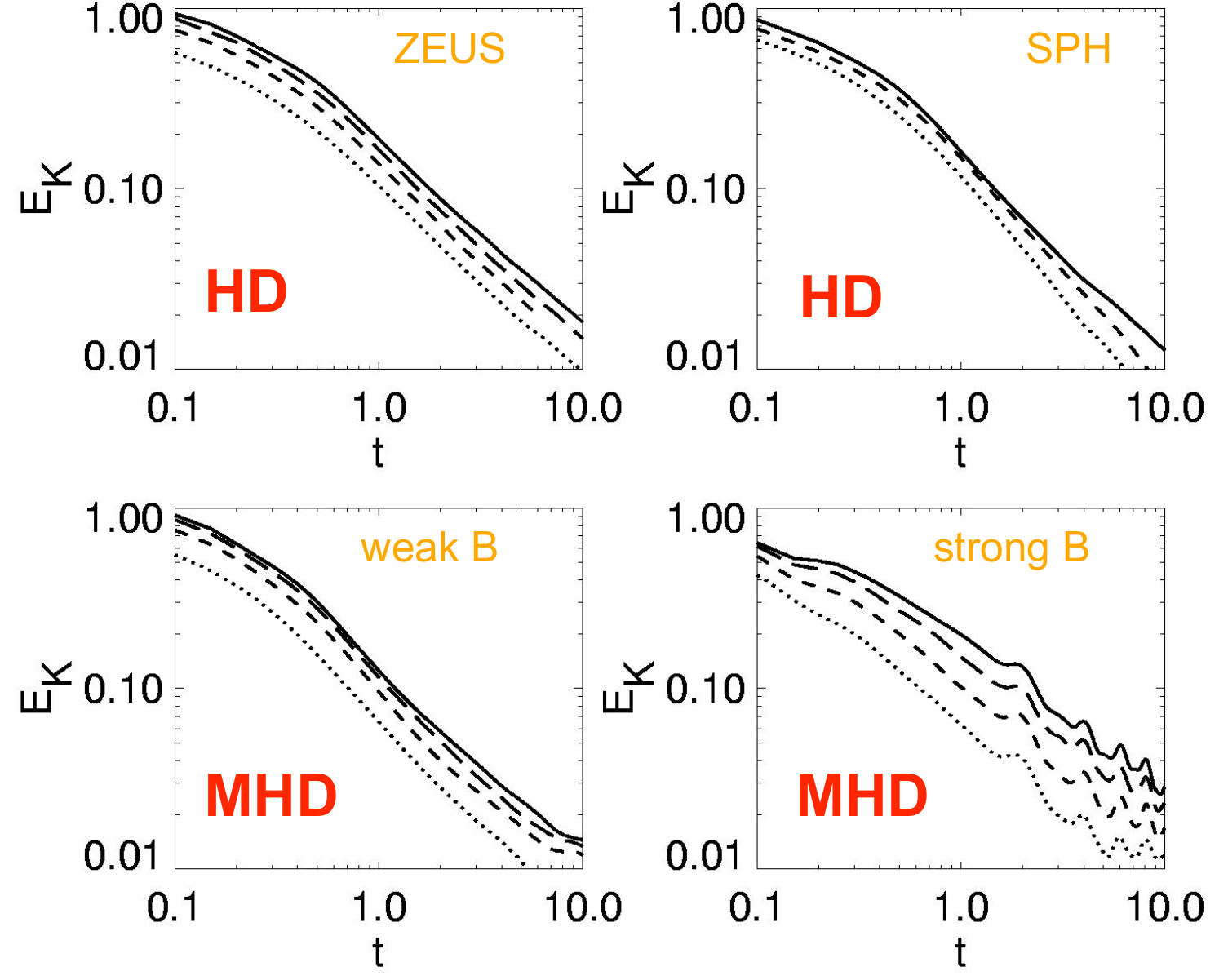}
\end{center}
\caption[Kinetic energy vs.\ time in decaying turbulence]{Decay of supersonic turbulence. The plots show the time evolution of the total kinetic energy $E_K$ in a variety of three-dimensional numerical calculations of decaying supersonic turbulence in isothermal ideal gas with initial root mean square Mach number of 5, calculated with two different numerical codes. ZEUS is a Eulerian grid code that solves the equations of magnetohydrodynamics \citep{Stone92a,Stone92b}, while SPH follows a particle-based approach \citep{benz90,Springel:2010p63410}. The top panels depict the decay of purely hydrodynamic turbulence, the bottom panels show the decay properties with weak and strong magnetic fields. For more details see \citet{maclow98}.}
\label{fig:prlres}
\end{figure}

This analysis is supported by numerical simulations. One-dimensional calculations of non-driven, compressible, isothermal, magnetized turbulence by \citet{Gammie:1996p84111} indicate a very efficient dissipation of kinetic energy. They also found that the decay rate depends strongly on the adopted initial and boundary conditions. \citet{maclow98}, \citet{Stone98}, and \citet{padoan99} determined the decay rate in direct numerical simulations in three dimensions, using a range of different numerical methods. They uniformly report very rapid decay rates and propose a power-law behavior for the decay of the specific kinetic energy of the form $\dot{\epsilon} \propto t^{-\eta}$, with $0.85 < \eta < 1.1$. A typical result is shown in Figure~\ref{fig:prlres}. Magnetic fields with strengths ranging up to equipartition with the turbulent motions seem to reduce $\eta$ to the lower end of this range, while unmagnetized supersonic turbulence shows values closer to  $\eta \,\sils\, 1.1$.

Besides directly measuring the decay of the kinetic energy in the absence of driving sources in a closed system, we can also continuously insert energy and determine the resulting velocity dispersion. \citet{maclow99} and \cite{elmegreen00} argue that the dissipation time, $t_{\rm d} = \epsilon /\dot{\epsilon}$, is comparable to the turbulent crossing time in the system,
\begin{equation} 
\label{eqn:decay-time}
t_{\rm d}\approx \frac{L}{\sigma}\;,
\end{equation}
where $L$ is again the driving scale and $\sigma$ the velocity dispersion. This holds regardless of whether the gas is magnetized or not and also extends into the subsonic regime. The loss of the specific turbulent kinetic energy, $\epsilon = 1/2 \sigma^2$ is then,
\begin{equation} 
\label{eqn:dissip}
\dot{\epsilon} =  \frac{\epsilon}{t_{\rm d}} =  \xi \frac{1}{2}  \frac{\sigma^3}{L}\;.
\end{equation}
We have  recovered Kolmogorov's formula for the energy decay rate (\ref{eq:kolmogorov-1}), modulo an efficiency coefficient $\xi/2$ that depends on the physical parameters of the system or on the details of the numerical method employed.

\subsection{Scales of ISM Turbulence}
\label{subsubsec:ISM-scales}

As introduced in Section \ref{par:decay-turb}, interstellar turbulence decays on roughly a crossing time. It needs to be continuously driven in order to maintain a steady state. Here we compare the various astrophysical processes that have been proposed as the origin of ISM turbulence, and mostly follow the discussion in \citet{maclow04} and \citet{klessen10}. 

We begin with an analysis of the typical scales of ISM turbulence in our Galaxy, then calculate the corresponding turbulent energy loss, and finally turn our attention to the various astrophysical driving mechanisms that have been proposed to compensate for the decay of turbulent energy. We point out that a key assumption is that the ISM in the Milky Way evolves in a quasi steady state, so that energy input and energy loss rates roughly balance when being averaged over  secular timescales and over large enough volumes of the Galactic disk. We caution the reader that this need not necessarily be the case. 

The self-similar behavior of turbulent flows only hold in the inertial range, i.e.\ on scales between the driving and dissipation scales. We now want to address the question of what these scales are in the Galactic ISM. The answer clearly depends on the different physical processes that stir the turbulence and that provide dissipation. There is a wide range of driving mechanisms proposed in the literature, ranging from stellar feedback acting only on very local scales  up to accretion onto the Galaxy as a whole, inserting energy on very large scales. As we discuss below, we favor the latter idea.

Regardless of the detailed driving process, a firm outer limit to the turbulent cascade in disk galaxies is given by the disk scale height. If molecular clouds are created at least in part by converging large-scale flows triggered by accretion, or by spiral shocks, or by the collective influence of recurring supernovae explosions, then the extent of the Galactic disk is indeed the true upper scale of turbulence in the Milky Way. For individual molecular clouds this means that turbulent energy is fed in at scales well above the size of the cloud itself. This picture is supported by the observation that the clouds' density and velocity structure exhibits a power-law scaling behavior extending all the way up to the largest scales observed in today's surveys {\citep{Ossenkopf02, Brunt:2003p17014, brunt09}.

One could argue that the outer scale of the ISM turbulence actually corresponds to the diameter of the Galaxy as a whole (rather than the disk scale height) and that the largest turbulent eddy is the rotational motion of the disk itself. However, because the disk scale height $H$ is typically less than 10\% of the disk radius $R$, this motion is intrinsically two-dimensional and if we restrict our discussion to three-dimensional turbulence then $H$ is the maximum outer scale. In addition, we note that the decay time (\ref{eqn:decay-time}) is comparable for both approaches. At the solar radius, $R = 8.5\,$kpc, the rotational speed is $v_{\rm rot} = 220\,$km$\,$s$^{-1}$, leading to $t_{\rm d} \approx R/v_{\rm rot} \approx 38\,$Myr. If we adopt an average H{\sc i} disk scale height of $H = 0.5\,$kpc and a typical velocity dispersion of $\sigma = 12\,$km$\,$s$^{-1}$ \citep{ferriere01,Kalberla:2003p2981}, then our estimate of the turbulent decay time, $t_{\rm d} \approx L/\sigma \approx 40\,$Myr, is essentially the same. In conclusion, for the estimate of the energy decay rate in typical disk galaxies, and by the same token, for the calculation of the required turbulent driving rate, it does not really matter what we assume for the outer scale of the turbulence \cite[for further discussions, see e.g.][]{klessen10}. This follows, because for typical disk galaxies, the disk scale height and radius, as well as the velocity dispersion and rotational velocity scale similarly, that is  $H/R \approx 0.1$ and  $\sigma/v_{\rm rot} \approx 0.1$. 

The estimate of the dissipation scale is also difficult. For purely hydrodynamic turbulence, dissipation sets in when molecular viscosity becomes important. The corresponding spatial scales are tiny. In the ISM the situation is more complex because we are dealing with a magnetized, partially ionized, dusty plasma. \citet{Zweibel:1983p84059} argue that ambipolar diffusion (i.e.\ the drift between charged and neutral species in this plasma) is the most important dissipation mechanism in typical molecular clouds with very low ionization fractions $x = \rho_i/\rho_n$, where $\rho_i$  and $\rho_n$  are is the densities of ions and neutrals, respectively, with $\rho = \rho_i + \rho_n$ being the total density. We can then replace the kinetic viscosity in the Navier-Stokes equation (\ref{eqn:Navier-Stokes}) by the ambipolar diffusion coefficient 
\begin{equation}
\nu_{\rm AD} = v_{\rm A}^2 / \zeta_{ni},
\end{equation}
where $v_{\rm A}^2 = B^2/4\pi\rho_n$ approximates the effective Alfv\'en speed for the coupled neutrals and ions if $\rho_n \gg
\rho_i$, and $\zeta_{ni} = \alpha \rho_i$ is the rate at which each neutral is hit by ions. The coupling constant $\alpha$ is given by
\begin{equation} \label{eqn:couple}
\alpha = \langle \sigma v \rangle/(m_i + m_n) \approx  9.2 \times
10^{13}\,\mbox{cm}^3\,\mbox{s}^{-1}\,\mbox{g}^{-1}\;,
\end{equation}
with $m_i$ and $m_n$ being the mean mass per particle for the ions and neutrals, respectively. Typical values in molecular clouds are $m_i = 10 \,m_{\rm H}$ and $m_n = 2.35 m_{\rm H}$.  It turns out that $\alpha$ is roughly independent of the mean velocity, as the ion-neutral cross-section $\sigma$ scales inversely with velocity in the regime of interest. For further details on the microphysics, consult the excellent textbooks by \citet{osterbrock89}, \citet{tielens2010}, or \citet{draine11}. 

We can define an ambipolar diffusion Reynolds number in analogy to equation (\ref{eqn:reynolds-number}) as
\begin{equation}
{\rm Re}_{{\rm AD},\ell} = \ell v_{\ell} / \nu_{\rm AD} = {\cal M}_{\rm A} \ell \,\zeta_{ni} / v_{\rm A},
\end{equation}
which must fall below unity on scales where  ambipolar diffusion becomes important. As before, $v_\ell$ is the characteristic velocity at scale $\ell$,  and we define ${\cal M}_{\rm A} = v_\ell/v_{\rm A}$ as  the characteristic Alfv\'en Mach number at that scale. From the condition Re$_{{\rm AD},\lambda} = 1$, we derive the dissipation scale due to ambipolar diffusion as  
\begin{equation} 
\label{eqn:AD-scale}
\lambda_{\rm AD} = \frac{v_{\rm A}}{{\cal M}_{\rm A} \zeta_{ni}} 
         \approx 0.041 \mbox{ pc}\left(\frac{B}{10\,\mu{\rm G}}\right) {\cal M_{\rm
         A}}^{-1} \left(\frac{x}{10^{-6}}\right)^{-1}
         \left(\frac{n_n}{10^3\,{\rm cm}^{-3}}\right)^{-3/2},
\end{equation}
with the magnetic field strength $B$, the ionization fraction $x$, the neutral number density $n_n$, and where we have taken $\rho_n = \mu n_n$, with a mean particle mass $\mu = 2.35\,m_{\rm H} = 3.92 \times 10^{-24}\,$g typical for molecular clouds. It is interesting to note, that the resulting value for $\lambda_{\rm AD}$ is comparable to the typical sizes of protostellar cores \cite[e.g.][]{bergin07}. Indeed, the velocity dispersion in these objects is dominated by thermal motions \citep{goodman98}.

We note that there are wave families that can survive below $\lambda_{\rm AD}$ that resemble gas dynamic sound waves. Consequently, even on scales where the magnetic field becomes uniform, the gas dynamical turbulent cascade could continue. This is determined by the dimensionless magnetic Prandtl number, 
\begin{equation}
\label{eqn:Prandtl-number}
{\rm Pr}_{\rm mag} = \frac{{\rm Re}_{\rm AD}}{{\rm Re}} = \frac{\nu}{\nu_{\rm AD}}\;,
\end{equation}
which compares the relative importance of viscous and magnetic diffusion processes. For small magnetic Prandtl numbers the hydrodynamic inertial range extends beyond the magnetic one, and vice versa, for ${\rm Pr}_{\rm mag} \gg 1$ fluctuations in the magnetic field can occur on scales much smaller than the hydrodynamic diffusion limit \cite[for further discussions see e.g.][]{schekochihin04a,schekochihin04b,Schober:2012p70633,Schober:2012p54505}.

\begin{figure}[tbp]
\begin{center}
\includegraphics[width=1.0\columnwidth]{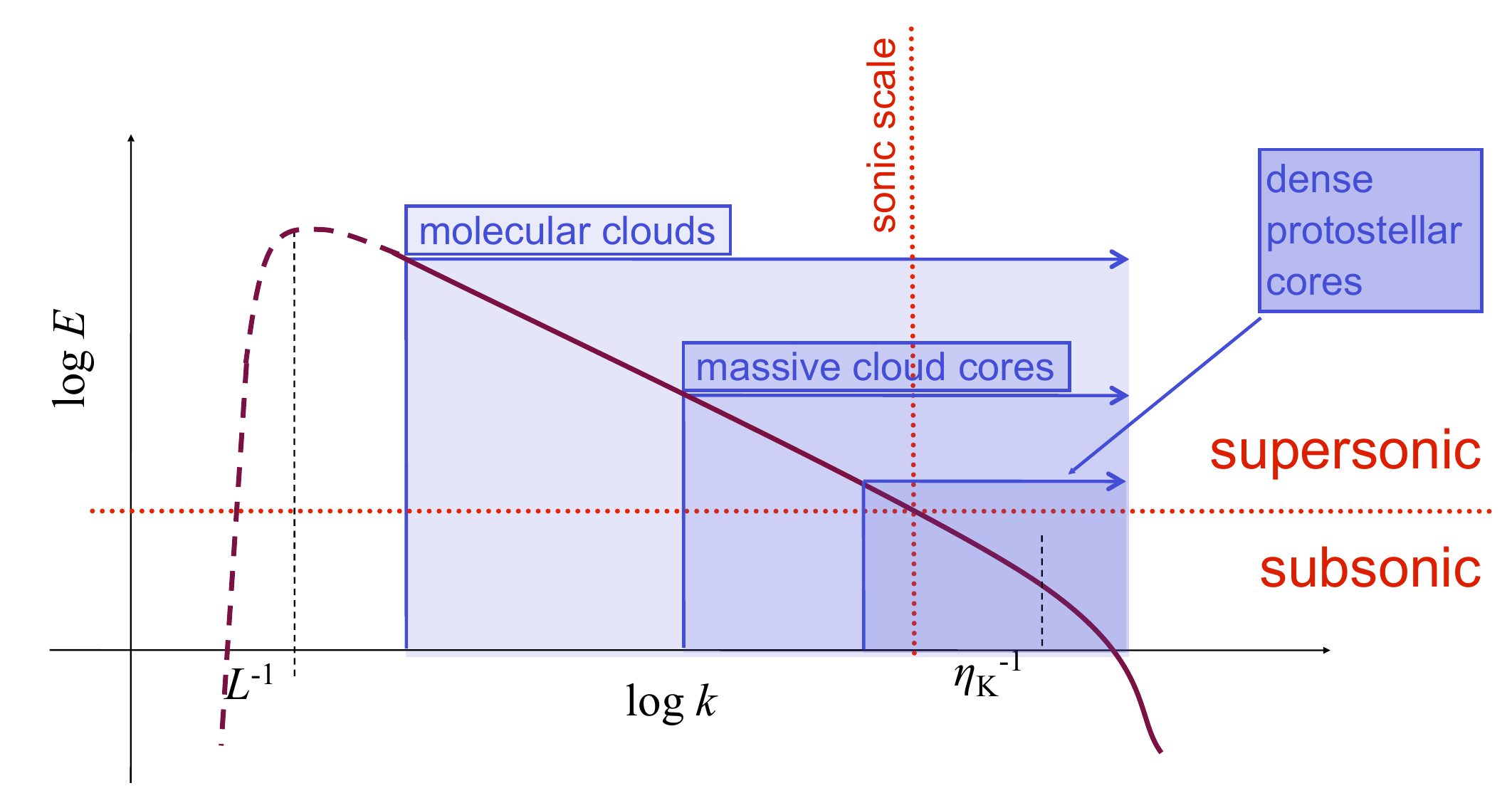}
\caption{Simple cartoon picture of the turbulent energy spectrum, i.e.\ of the kinetic energy carried by modes of different wave numbers $k$, and their relation to different cloud structures (see also Table \ref{tab:mol-clouds}). Turbulence is driven on large scales  comparable to the size $L$ of the cloud and is dissipated on very small scales $\eta_{\rm K}$. Adopted from \cite{klessen11b}.}
\label{fig:turb-spectrum}
\end{center}
\end{figure}

Combining these findings with the molecular cloud properties discussed in Section \ref{par:prop-MC} we arrive at the following picture, as illustrated in Figure \ref{fig:turb-spectrum}.  On the scales of individual molecular clouds and large molecular cloud complexes, the observed turbulence is highly supersonic. We know that the density contrast created by shocks in isothermal gas scales with the Mach number $\cal M$ to the second power, $\Delta \rho / \rho \propto {\cal M}^2$ \cite[e.g.][]{landaulifshitzfluiddynamics}. Consequently, for ${\cal M} \approx 10$ we expect density contrasts of roughy 100. This is indeed observed in molecular clouds, where the mean density is around $100$ particles per cubic centimeter and where the high-density cores exceed values of $10^4\,$cm$^{-3}$ and more (see Table \ref{tab:mol-clouds}). When we focus on cluster-forming cloud cores (or their not-yet-star-forming counterparts, the so-called infrared dark clouds) we still measure ${\cal M} \approx 5$ leading to localized density fluctuations of $\Delta \rho / \rho \approx 25$, on average. As we discuss further in Section \ref{sec:collapse-SF}, some of these fluctuations may exceed the critical mass for gravitational collapse to set in. The presence of turbulence thus leads to the break-up into smaller units. The core fragments to build up a cluster of stars with a wide range of masses rather than forming one single high-mass star. We call this process gravoturbulent fragmentation, because turbulence generates the distribution of clumps in the first place, and then gravity selects a subset of them for subsequent star formation (see also Section \ref{subsec:gravoturb-SF}). Finally, when focusing on low-mass cores, the velocity field becomes more coherent and the turbulence subsonic (see Section \ref{subsec:cores}). This defines the sonic scale at around $0.1\,$pc. Such structures are no longer subject to gravoturbulent fragmentation and are the direct progenitors of individual stars or binary systems. We note, however, that gravitational fragmentation may still occur within the protostellar accretion disk that builds up in the center of the core due to angular momentum conservation (Section \ref{subsec:collapse}). This process is likely to produce close binaries \cite[see e.g.][]{bodenheimer95,machida08}. The fact that the observed velocity dispersion approaches the thermal value as one zooms in on smaller and smaller scales is a direct consequence of the turbulent cascade, as expressed in the observed Larson relation (equation \ref{eqn:larson}).

\subsection{Decay of  ISM turbulence}
\label{subsubsec:ISM-decay}
With the above considerations, we are now in a position to calculate the rate of energy loss in the Galactic ISM due to the decay of turbulence. Our Milky Way is a typical $L_\star$ galaxy with a total mass including dark matter of about $10^{12}\,$M$_\odot$ out to the virial radius at $\sim 250\,$kpc \cite[e.g.][]{Xue:2008p3090}. The resulting rotation curve is $220\,$km$\,$s$^{-1}$ at the solar radius, $R_\odot \approx 8.5\,$kpc, and it declines to values slightly below $200\,$km$\,$s$^{-1}$ at a radius of $60\,$kpc \citep{Xue:2008p3090}.  Star formation occurs out to a radius of about $15\,$kpc \citep{Rix:2013p88908}. The total mass in the disk in stars is about $2.7 \times 10^{10}\,$M$_\odot$, and in gas it is about $8 \times 10^{9}\,$M$_\odot$ \cite[see Table \ref{tab:MW}, as well as][]{Naab:2006p2645}. Assuming a global baryon fraction of 17\%, this corresponds to 40\% of all the baryonic mass within the virial radius and  implies that roughly the same amount of baryons is in an extended halo in form of hot and tenuous gas.  The gaseous disk of the Milky Way can be decomposed into a number of different phases. We follow \citet{ferriere01} and  \citet{Kalberla:2003p2981}, and consider molecular gas (H$_2$ as traced e.g.\ via CO emission) as well as atomic hydrogen gas (H{\sc i} as observed e.g.\ by its $21\,$cm emission). The H{\sc i} component can be separated into a cold ($T \approx \mbox{few} \times 10^2\,$K) and a hot ($T \approx \mbox{few} \times 10^3\,$K) component. Because they have similar overall distributions we consider them together. The scale height of H{\sc i} ranges from $\sim 230\,$pc within $4\,$kpc up to values of $\sim 3\,$kpc at the outer Galactic boundaries. The H{\sc i} disk therefore is strongly flared. We take 500$\,$pc as a reasonable mean value, but note that this introduces a high degree of uncertainty. Also, we neglect the warm and the hot ionized medium in our analysis, since the ionized gas within H{\textsc{ii}} regions or supernova remnants carries little of the turbulent kinetic energy compared to the other components. Indeed, roughly 95\% of the turbulent kinetic energy is carried by the atomic component. The adopted values are summarized in Table \ref{tab:MW}. 

\begin{table}[ht]
\caption{Properties of gas components of the Milky Way.}
{\begin{center}
\begin{tabular}{r@{\hspace{0.5cm}}c@{\hspace{0.4cm}}c@{\hspace{0.4cm}}c@{\hspace{0.4cm}}c@{\hspace{0.4cm}}c}
\hline
component  & $M$ $(10^9\,$M$_\odot)^a$ & $n$  $($cm$^{-3})^b$  & $L$ $($pc$)^c$  &  $\sigma$ $($km$\,$s$^{-1})^d$ & $E_{\rm kin}$ ($10^{55}\,$erg)$^e$ \\
\hline\\[-0.3cm]
molecular gas & 2 & 0.7 &150 & 5 & 0.5 \\
atomic gas  & 6 & 0.4 &1000 & 12 & 8.6   \\
 \hline
\end{tabular}
\end{center}
}

{\footnotesize
$^a$~Total mass of the component. Values from \citet{ferriere01} and \citet{Kalberla:2003p2981}.\\
$^b$~Estimate of volume-averaged midplane number density. Values from \citet{ferriere01} and \citet{Kalberla:2003p2981}. Note that the value for H$_2$ gas is so low compared to Table \ref{tab:mol-clouds}, because the volume filling factor of molecular clouds in the Galactic disk is very small. \\
$^c$~We consider the disk thickness as being twice the observed scale height. \\
$^d$~The parameter $\sigma$ is the three-dimensional velocity dispersion.  \\
$^e$~Total kinetic energy of the component, $E_{\rm kin} = 1/2 \, M \sigma^2$.\\
}
\label{tab:MW}
\end{table}%

One of the remarkable features of spiral galaxies is the nearly constant velocity dispersion $\sigma$, e.g.\ as measured using the H{\sc i} 21\ cm emission line, seemingly independent of galaxy mass and type \citep{Dickey:1990p4035,vanZee:1999p4021,Tamburro:2009p3039}. The inferred  values of $\sigma$ typically fall in a range between $10\,$km$\,$s$^{-1}$ and $20\,$km$\,$s$^{-1}$ \citep{bigiel08,Walter08a} and extend well beyond the optical radius of the galaxy with only moderate fall-off as one goes outwards. Quite similar behavior is found in the molecular gas, when increasing the sensitivity in the outer regions by stacking the data \citep{CalduPrimo:2013p80430}. It is interesting in this context that the transition from the star-forming parts of the galaxy to the non-star-forming outer disk seems not to cause significant changes in the velocity dispersion. This approximate indifference to the presence of stellar feedback sources sets severe constraints on the physical processes that can drive the observed level of turbulence (see Sections \ref{subsub:ISM-driving-external} and \ref{subsub:ISM-driving-feedback} below).

Using equation (\ref{eqn:dissip}), we can calculate the average loss of kinetic energy density $e$ (erg/cm$^{3}$) per unit time in the ISM. With $e = \rho \epsilon$, where $\epsilon$ is the specific energy (in units of erg/g) and $\rho = \mu n$ is the mass density, we obtain 
\begin{eqnarray} 
\label{eqn:decay-law}
\dot{e} &=& - \frac{1}{2}\frac{\mu n \sigma^3}{L} \\
& \approx &  -3.5 \times 10^{-27} \,\mbox{erg}\,\mbox{cm}^{-3}\,\mbox{s}^{-1} 
\left(\frac{n}{1\,\mbox{cm}^{-3}}\right)
\left(\frac{\sigma}{10\,\mbox{km}\,\mbox{s}^{-1}}\right)^3 
\left(\frac{L}{100\,\mbox{pc}}\right)^{-1}\;,   \nonumber
\end{eqnarray}
where we have set the efficiency parameter $\xi$ in equation (\ref{eqn:dissip}) to unity, and where again $n$, $\sigma$, and $L$ are the number density, the velocity dispersion, and the turbulent driving scale, respectively. For simplicity, we have assumed a mean mass per particle $\mu = 1.26\, m_{\rm H} = 2.11\times 10^{-24}\;$g typical for purely atomic gas. If we consider different ISM phases, this number needs to be adapted. If we plug in the values from Table \ref{tab:MW}, we obtain the following estimate for  our Galaxy:
\begin{equation}
\label{eqn:decay-MW}
\dot{e}_{\rm ISM} = \dot{e}_{{\rm H}_2} + \dot{e}_{{\rm HI}}  \approx  - 4.5 \times 10^{-28} \,\mbox{erg}\,\mbox{cm}^{-3}\,\mbox{s}^{-1}\;.
\end{equation}
According to (\ref{eqn:decay-time}) the decay timescale can be computed as 
\begin{equation} 
\label{eqn:disstime}
t_{\rm d} = \frac{e}{\dot{e}} =  \frac{L}{\sigma} \approx 10\, \mbox{Myr}
\left(\frac{L}{100\,\mbox{pc}}\right)
\left(\frac{\sigma}{10\,\mbox{km}\,\mbox{s}^{-1}}\right)^{-1}\;,
\end{equation}
which is simply the turbulent crossing time on the driving scale.

\subsection{Sources of ISM turbulence: gravity and rotation}
\label{subsub:ISM-driving-external}
There is a wide range of physical processes that could potentially drive the observed turbulent flows in the ISM. We will introduce and discuss the most important ones that have been proposed in the literature. We identify two main categories of sources. In this Section we focus on processes that convert a fraction of the potential energy available in the galaxy into turbulent gas motions. We first look at the process of accretion-driven turbulence, and then turn to rotation. As the rotational motion of the Galaxy is ultimately driven by gravity, we include all mechanisms that can tap the rotational energy here as well. In  Section \ref{subsub:ISM-driving-feedback} we then assess the influence of stellar feedback on the large-scale dynamics of the ISM. Our list is sorted in such a way that the processes which seem most important to us are introduced first.

\subsubsection{Accretion onto the galaxy}
\label{par:input-accretion}

We argue that it is the accretion process that inevitably accompanies any astrophysical structure formation, whether it is the formation of galaxies or the birth of stars, that drives the observed turbulent motions. We propose that this process is universal and makes significant contributions to the turbulent energy on all scales \cite[see also][]{Field:2008p17384}. When cosmic structures grow, they gain mass via accretion. This transport of matter is associated with kinetic energy and provides an ubiquitous source for the internal turbulence on smaller scales. We follow the analysis of \citet{klessen10} and ask whether the accretion flow onto galaxies provides enough energy to account for the observed ISM turbulence. 

We begin with a summary of what we know about accretion onto spiral galaxies like our Milky Way. Our Galaxy forms new stars at a rate of $\dot{M}_{\rm SF} \sim 2 - 4\,$M$_\odot\,$yr$^{-1}$ \cite[e.g.][]{Naab:2006p2645, Adams:2013p85330}. Its gas mass is about $M_{\rm gas} \approx 8 \times 10^9\,$M$_\odot\,$ (see Table \ref{tab:MW}, and also \citealt{ferriere01} and \citealt{Kalberla:2003p2981}). If we assume a constant star formation rate, then the remaining gas should be converted into stars within about $2 - 4$ Gyr. Similar gas depletion timescales of the order of a few billion years are reported for many nearby spiral galaxies \citep{bigiel08}. We note, however, that there is a debate in the community as to whether the depletion timescale is constant or whether it varies with the surface gas density. While \citet{Leroy:2008p4217,Leroy:2012p65552, leroy13} argue in favor of a more or less universal gas depletion time of $\sim 2\,$Gyr, \citet{shetty13} and \citet{shetty14} find that the depletion time varies from galaxy to galaxy, and as a general trend, increases with surface density. Nevertheless, it is fair to say that the inferred overall gas depletion times are shorter by a factor of a few than the ages of these galaxies of   $\sim 10\;$Gyr. If we discard the possibility that most spiral galaxies are observed at an evolutionary phase close to running out of gas, and instead assume that they evolve in quasi steady state, then this requires a supply of fresh gas at a rate roughly equal to the star formation rate. 

There is additional support for this picture. \citet{Dekel:2009p1173} and \citet{Ceverino:2009p1271}, for example, argue that massive galaxies are continuously fed by steady, narrow, cold gas streams that penetrate through the accretion shock associated with the dark matter halo down to the central galaxy. This is a natural outcome of cosmological structure formation calculations if baryonic physics is considered consistently. In this case, roughly three quarters of all galaxies forming stars at a given rate are fed by smooth streams \cite[see also][]{Agertz:2009p1291}. The details of this process, however, seem to depend on the numerical method employed and on the way gas cooling is implemented  \cite[e.g.][]{Bird:2013p70458}. Further evidence for accretion onto galaxies comes from the observation that the total amount of atomic gas in the universe appears to be roughly constant for redshifts $z\, \sils \,3$. This holds despite the continuous transformation of  gas into stars, and it  suggests that H{\sc i} is continuously replenished \citep{Hopkins:2008p9891,Prochaska:2009p9829}. In our Galaxy, the presence of deuterium at the solar neighborhood \citep{Linsky:2003p3845} as well as in the Galactic Center \citep{Lubowich:2000p3908} also points towards a continuous inflow of low-metallicity material. As deuterium is destroyed in stars and as there is no other known source of deuterium in the Milky Way, it must be of cosmological and extragalactic origin \citep{Ostriker:1975p3929,Chiappini:2002p3920}. 

In order to calculate the energy input rate from gas accretion we need to know the velocity $v_{\rm in}$ with which this gas falls onto the disk of the Galaxy and the efficiency with which the kinetic energy of the infalling gas is converted into ISM turbulence. As the cold accretion flow originates from the outer reaches of the halo and beyond, and because it lies in the nature of these cold streams that gas comes in almost in free fall, $v_{\rm in}$ can in principle be as high as the escape velocity $v_{\rm esc}$ of the halo. For the Milky Way in the solar neighborhood we find $v_{\rm esc} \sim 550\,$km$\,$s$^{-1}$ \citep{Fich:1991p3643,Smith:2007p3637}. However, numerical experiments indicate that the inflow velocity of cold streams is of order of the virialization velocity of the halo \citep{Dekel:2009p1173}, which typically is $\sim 200\,$km$\,$s$^{-1}$. The actual impact velocity with which this gas interacts with disk material will also depend on the sense of rotation. Streams which come in co-rotating with the disk will have smaller impact velocities  than material that comes in counter-rotating. To relate to quantities that are easily observable and to within the limits of our approximations, we adopt $v_{\rm in} = v_{\rm rot}$ as our fiducial value, but note that considerable deviations are possible. We also note that even gas that shocks at the virial radius and thus heats up to $10^5 - 10^6\,$K, may cool down again and some fraction of it may be available for disk accretion. This gas can condense into higher-density clumps that sink towards the center and replenish the disk  \citep{Peek:2009p9660}. Again, $v_{\rm in} \approx v_{\rm rot}$ is a reasonable estimate.

Putting everything together, we can now calculate the energy input rate associated with gas accretion onto the Milky Way as 
\begin{eqnarray}
\label{eqn:input-MW}
\dot{e} & = &\rho \dot{\epsilon} = 
\frac{1}{2} \rho \frac{\dot{M}_{\rm in}}{M_{\rm gas}}  v^2_{\rm in} \\
 & = & 6.2 \times 10^{-27} \,\mbox{erg}\,\mbox{cm}^{-3}\,\mbox{s}^{-1} \,  \left(\frac{n}{1\,\mbox{cm}^{-3}}\right) \left(\frac{\dot{M}_{\rm in}}{3\,\mbox{M}_{\odot}\,\mbox{yr}^{-1}}\right)
\left(\frac{v_{\rm in}}{220\,\mbox{km}\,\mbox{s}^{-1}}\right)^2\,,\nonumber
\end{eqnarray}
where again $\rho = \mu \,n$ with the mean particle mass $\mu= 2.11\times 10^{-24}\;$g suitable for atomic gas. We take an average number density $n$ in the Galactic disk from Table \ref{tab:MW}, set the mass infall rate $\dot{M}_{\rm in}$ to the average star formation rate of $3\,\mbox{M}_{\odot}\,\mbox{yr}^{-1}$, and approximate the infall velocity $v_{\rm in}$ by the circular velocity of $220\,\mbox{km}\,\mbox{s}^{-1}$.

If we compare the input rate (\ref{eqn:input-MW}) with the decay rate (\ref{eqn:decay-MW}), we note that only about 7\% of the infall energy is needed to explain the observed ISM turbulence. However, the fraction of the infall energy that actually is converted into turbulent motions is very difficult to estimate. Some fraction will turn into heat and is radiated away. In addition, if the system is highly inhomogeneous with most of the  mass residing in high density clumps with a low volume filling factor, most of the incoming flux will feed the tenuous interclump medium rather than the dense clumps, and again, will not contribute directly to driving turbulence in the dense ISM. Numerical experiments indicate that the efficiency of converting infall energy into turbulence  scales linearly with the density contrast between the infalling gas and the ISM \citep{klessen10}. For the Milky Way, this means that the infalling material should have average densities of $\sils \,0.1\,$cm$^{-3}$.  

It seems attractive to speculate that the population of high-velocity clouds observed around the Milky Way is the visible signpost for high-density peaks in this accretion flow. Indeed the inferred infall rates of high-velocity clouds are in the range  $0.5 - 5\,$M$_\odot$yr$^{-1}$  \citep{Wakker:1999p3760,Blitz:1999p4002,Braun:2004p4007,Putman:2006p9534}, in good agreement with the Galactic star formation rate or with chemical enrichment models \cite[see e.g.][and references therein]{Casuso:2004p3934}. An important question in this context is where and in what form the gas reaches the Galaxy. This is not known well. Recent numerical simulations indicate that small clouds (with masses less then a few $10^4\,$M$_\odot$) most likely will dissolve, heat up and merge with the hot halo gas, while larger complexes will be able to deliver cold atomic gas even to the inner disk \citep{Heitsch:2009p8345}. In any case, it is likely that the gas is predominantly accreted onto the outer disk of the Milky Way. However, it is consumed by star formation mostly in the inner regions. To keep the Galaxy in a steady state there must be an inwards gas motion of the order of $v_{\rm R} \approx \dot{M}_{\rm in} / (2\pi R \Sigma) \approx  3 \,$km$\,$s$^{-1}$, where we adopt a gas surface density at the solar radius $R_\odot = 8.5 \;$kpc of $\Sigma = 15\,$M$_\odot\;$pc$^{-1}$ \citep{Naab:2006p2645}. Whether this net inward flow exists is not known, given our viewpoint from within the Galaxy and given a typical velocity dispersion of $\sim 10\,$km$\,$s$^{-1}$, which exceeds the strength of the signal we are interested in. In other galaxies, where we have an outside view onto the disk, we could in principle try to decompose the observed line-of-sight motions and find signs of the proposed inward mass transport.

\subsubsection{Spiral arms}
\label{par:input-spiral-arms}

Spiral galaxies such as our Milky Way are rotationally supported. This means that the gas is prevented from freely flowing towards the Galactic Center by angular momentum conservation, which forces the gas into circular orbits about a common origin. The process is very similar to the formation of protostellar accretion disks during the collapse of rotating cloud cores, which are a natural part of the process of stellar birth (see Section \ref{subsec:collapse}). Parcels of gas can only change their radial distance from the center of the disk by exchanging angular momentum with neighboring gas. Fluid elements that lose angular momentum move inwards, while those that gain angular momentum move outwards. Exchange of angular momentum between fluid elements may be due to dynamical friction or to the influence of some effective viscosity. In the ISM, molecular viscosity is far too small to explain the observed gas motions. However, we can resort to either spiral arms (which reflect the onset of gravitational instability) or to magnetic fields in the disk. In both cases, some of the energy stored in Galactic rotation can be converted into turbulent kinetic energy as the gas moves inwards. 

Indeed, the spiral structure that is almost ubiquitously observed in disk galaxies has long been proposed as an important source of ISM turbulence. \citet{roberts69} argued that the gas that flows through spiral arms  forming in marginally stable disks \citep{toomre64,ls64,lin69} may shock and so distribute energy throughout different scales. \citet{Gomez:2002p86006} and \citet{Martos:1998p85818}, for example, found that some fraction of the gas will be lifted up in a sudden vertical jump at the position of the shock. Some portion of this flow will contribute to interstellar turbulence. However, we note that the observed presence of interstellar turbulence in irregular galaxies without spiral arms as well as in the outer regions of disk galaxies beyond the extent of the spiral arm structure suggests that there must be additional physical mechanisms driving turbulence.  

For purely hydrodynamic turbulence in the absence of magnetic fields or for flows with weak Maxwell stresses (see equation \ref{eqn:stress} below), purely gravitational stress terms may become important. \citet{Wada:2002p86350} estimated the energy input resulting from these Newton stress terms. They result from correlations in the different components of the flow velocity $\vec{v}$ as $T_{R\Phi} = \langle \rho v_{R} v_{\Phi} \rangle$ \citep{LyndenBell:1972p87362}, and will only add energy for  a positive correlation between radial and azimuthal gravitational forces. It is not clear, however, whether this is always the case. Despite this fact, we can use this approach to get an upper limit to the energy input. As an order of magnitude estimate, we obtain 
\begin{eqnarray}
\dot{e} & \approx & G (\Sigma_g/H)^2 L^2 \Omega  \\
        & \approx & 4 \times 10^{-29} \mbox{ erg cm$^{-3}$ s}^{-1} \times \nonumber \\
        & & \times \left(\frac{\Sigma_g}{10 \mbox{ M$_{\odot}$ pc}^{-2}} \right)^2
\left(\frac{H}{100 \mbox{ pc}} \right)^{-2}
\left(\frac{L}{100 \mbox{ pc}} \right)^{2}
\left(\frac{\Omega}{(220 \mbox{ Myr})^{-1}}\right)\;,\nonumber
\end{eqnarray}
where $G$ is the gravitational constant, $\Sigma_g$ is the density of gas, $H$ is the scale height of the disk, $L$ is the length scale of turbulent perturbations, and $\Omega$ is the angular velocity.  The normalization is appropriate for the Milky Way.  This is about an order of magnitude less than the value required to maintain the observed ISM turbulence (equation \ref{eqn:decay-MW}).

We note that the fact that spiral arms are curved adds another pathway to driving ISM turbulence. Curved shocks are able to generate vortex motions as the gas flows through the discontinuity. \citet{Kevlahan:2009p86402} argue that this process is able to produce a Kolmogorov-type energy spectrum in successive shock passages \cite[see also][]{Wada:2008p86257}. However, further investigations are needed to determine whether this process is able to produce the observed energy density in the Galactic ISM.

\subsubsection{Magnetorotational instabilities}
\label{par:ISM-driving-MRI}

\citet{Sellwood:1999p5200} proposed that the magnetorotational instability \citep{balbushawley98} could efficiently couple large scale rotation with small-scale turbulence. The instability generates Maxwell stresses, which lead to a positive correlation between radial $B_R$ and azimuthal $B_{\Phi}$ components of the magnetic field, transfering energy from shear into turbulent motions at a rate 
\begin{equation}
\label{eqn:stress}
\dot{e} = - T_{R\Phi} (d\Omega / d \ln R) =  T_{R\Phi} \Omega\;,
\end{equation}
where the last equality holds for a flat rotation curve.
Typical values are $T_{R\Phi} \approx 0.6 \,B^2/(8\pi)$ \citep{Hawley:1995p86443}. At the radius of the Sun, $R_\odot = 8.5\;$kpc and a circular velocity of $v_{\rm rot} = 220 \, $km$\,$s$^{-1}$, we obtain an angular velocity of  
\begin{equation}
\Omega = \frac{v_{\rm rot}}{2 \pi R_\odot} = \frac{1}{220\, \rm{Myr}} \approx 1.4 \times 10^{-16} \mbox{ rad s}^{-1}\;.
\end{equation}
If we put both together, we conclude that the magnetorotational instability could contribute energy at a rate
\begin{equation}
\label{eqn:source-MRI}
\dot{e} = 3 \times 10^{-29}\,\mbox{erg}\,\mbox{cm}^{-3}\,\mbox{s}^{-1}
\left(\frac{B}{3 \mu\mbox{G}}\right)^2 \left(\frac{\Omega}{(220\,\mbox{Myr})^{-1}}\right)\;.
\end{equation}
\citet{Sellwood:1999p5200} tested this hypothesis for the small galaxy NGC~1058 and concluded that the  magnetic field required to produce the observed velocity dispersion of 6$\,$km$\,$s$^{-1}$ is roughly 3 $\mu$G which is reasonable value for such a galaxy. Whether the process is efficient enough to explain ISM turbulence in large spiral galaxies such as our Milky Way remains an open question. The typical values derived from equation (\ref{eqn:source-MRI}) are considerably lower than the energy required to compensate for the loss of turbulent energy (equation \ref{eqn:decay-MW}). Numerical simulations geared towards the Galactic disk \cite[e.g.][]{Dziourkevitch:2004p5202, piontek04, Piontek:2005p34248} are not fully conclusive and in general deliver values of $\dot{e}$ that are too small. Overall, the magnetorotational instability may provide a base value for the velocity dispersion below which no galaxy will fall, but it seems likely that additional processes are needed to explain the observations.

\subsection{Sources of ISM turbulence: stellar feedback}
\label{subsub:ISM-driving-feedback}

There are various stellar feedback processes that could act as potential sources of ISM turbulence. In general, we can distinguish between mechanical and radiative energy input. Supernova explosions that accompany the death of massive stars, line-driven winds in the late phases of stellar evolution, as well as the protostellar jets and outflows that are associated with stellar birth belong to the first category. The ionizing and non-ionizing radiation that stars emit during all of their life belongs to the latter one. As before, we discuss these various feedback processes in decreasing order of importance. 

\subsubsection{Supernovae}
\label{par:SN}

The largest contribution from massive stars to interstellar turbulence most likely comes from supernova explosions. In order to understand their impact on ISM dynamics in our Galaxy, we first need to determine the supernova rate $\sigma_{\rm SN}$. The exact number is quite uncertain, but typical estimates fall in the range of 2 -- 5  supernovae per century \cite[e.g.][]{Mckee:1989p84991,mckee97, Adams:2013p85330}. Note that we do not distinguish between core collapse supernovae from massive stars and type Ia explosions which are triggered by accretion onto white dwarfs. In addition, we assume for simplicity that each event releases the same energy of $E_{\rm SN} = 10^{51}\,$erg. Next, we need to obtain an estimate for the volume of the star forming disk of the Galaxy. Following the values discussed in Section \ref{subsubsec:ISM-decay}, we take the star forming radius to be $R = 15\,$kpc and the disk thickness to be $H = 100\,$pc. The corresponding energy input rate normalized to Milky Way values becomes 
\begin{eqnarray}
\dot{e} & = &\frac{\sigma_{\rm SN} \xi_{\rm SN} E_{\rm SN}}{\pi R_{\rm sf}^2 H}  \\
       &  = & 3 \times 10^{-26} \mbox{ erg s$^{-1}$ cm}^{-3} \times \nonumber \\
&& \times \left(\frac{\xi_{\rm SN}}{0.1} \right)
\left(\frac{\sigma_{\rm SN}}{(100\,{\rm yr})^{-1}} \right) 
\left(\frac{H}{100 \mbox{ pc}} \right)^{-1}
\left(\frac{R}{15 \mbox{ kpc}} \right)^{-2}
\left(\frac{E_{\rm SN}}{10^{51} {\rm erg}} \right)\;. \nonumber
\end{eqnarray}
The efficiency of energy transfer from supernova blast waves to the interstellar gas $\xi_{\rm SN}$ depends on many factors, including the strength of radiative cooling in the initial shock, or whether the explosion occurs within a hot and tenuous H{\textsc{ii}} region or in dense gas. Substantial amounts of energy can escape in the vertical direction in galactic fountain flows. The scaling factor $\xi_{\rm SN} \approx 0.1$ used here was derived by \citet{Thornton:1998p86655} from detailed one-dimensional numerical simulations of supernovae expanding in a uniform medium. The efficiency can also be estimated analytically \citep{Norman:1996p86663}, mostly easily by assuming momentum conservation, comparing the typical expansion velocity of 100$\,$km$\,$s$^{-1}$ to the typical velocity of ISM turbulence of 10$\,$km$\,$s$^{-1}$. Clearly, fully three-dimensional models, describing the interaction of multiple supernovae in the multi-phase ISM are needed to better constrain the efficiency factor $\xi_{\rm SN}$.

Supernova driving appears to be powerful enough to maintain ISM turbulence at the observed levels and to compensate for the energy loss estimated in equation (\ref{eqn:dissip}).  In the star-forming parts of the Galactic disk, it provides a large-scale self-regulation mechanism. As the disk becomes more unstable, the star formation rate goes up. Consequently, the number of OB stars increases which leads to a higher  supernova rate. As the velocity dispersion increases, the disk becomes more stable again and the star formation rate goes down again. However, this process does not explain the large velocity dispersion observed in the outer parts of disk galaxies, which show little signs of star formation, and hence, will not have much energy input from supernovae. Here other processes, such as those described in Section  \ref{subsub:ISM-driving-external}, appear to be required.

\subsubsection{Stellar winds}
\label{par:winds}
The total energy input from a line-driven stellar wind over the main-sequence lifetime of an early O~star can equal the energy from its supernova explosion, and the Wolf-Rayet wind can be even more powerful \citep{Nugis:2000p87766}.  The wind mass-loss rate scales somewhat less than quadratically with the stellar luminosity \cite[e.g.][]{Pauldrach:1990p87697, Puls:1996p87734, Vink:2000p87669, Vink:2001p87644}, and as the luminosity $L$ itself is a very steep function of stellar mass $M$, with $L\propto M^{3.5}$ providing a reasonable approximation \cite[e.g.][]{kippenhahn12}, only the  most massive stars contribute substantial energy input \cite[for a review, see][]{Lamers:1999p87970}. We also note that stellar rotation can dramatically change the derived stellar mass loss rates and the energy and momentum inserted by line-driven winds (for recent reviews, see \citealt{Meynet:2009p88254} or \citealt{Maeder:2012p63503}, or for a grid of evolutionary tracks, see \citealt{Ekstrom:2012p63511} and \citealt{Georgy:2012p63508}). \citet{Krumholz:2014p85764} concluded that even the most optimistic wind models lead to momentum and energy input rates comparable to the radiation field (see below, Section \ref{par:radiation}). In comparison, the energy from supernova explosions remains nearly constant down to the least massive star that can explode.  Because there are far more low-mass stars than massive stars in the Milky Way and other nearby galaxies (for a discussion of the stellar initial mass function, see Section \ref{par:IMF}), supernova explosions inevitably dominate over stellar winds after the first few million years of the lifetime of an OB association. Nevertheless, realistic three-dimensional numerical models of  the momentum and kinetic energy input into the ISM and its effects on molecular cloud evolution and on interstellar turbulence are needed. At the moment too little is known about this process \cite[see also][]{Krumholz:2006p18956,Yeh:2012p88293}.

\subsubsection{Protostellar jets and outflows}
\label{par:jets}

Protostellar jets and outflows are another very popular potential energy source for the observed ISM turbulence. They propagate with velocities of about $300\,$km$\,$s$^{-1}$ as seen in the radial velocity shift of forbidden emission lines, but also in proper motion of jet knots.  Many of these jets remain highly collimated with opening angles less than $5^{\circ}$ over a distance up to several parsec  \cite[e.g.][]{Mundt:1990p86759,Mundt:1991p86713}.

Protostellar jets and outflows are launched by magnetic forces \cite[for a summary, see][]{Pudritz07}. The scenario of magnetohydrodynamic jet formation has been studied with stationary models \citep{camenzind90, Shu94, Fendt:1996p87007, Ferreira97} as well as by time-dependent MHD simulations \citep[e.g.][]{Ouyed97b, Ouyed03, Krasnopolsky:1999p87137}.

Essentially, the MHD jet formation process works by transferring magnetic energy (Poynting flux) into kinetic energy.  As a consequence, the asymptotic, collimated jet flow is in energy equipartition between magnetic and kinetic energy.  The general characteristics of jet propagation can be summarized as follows.  Along the interface between the propagating jet and the surrounding material at rest, Kelvin-Helmholtz instabilities develop and lead to the entrainment of  matter from this region into the jet. This slows down the outward propagation while roughly conserving the overall momentum of the flow. At the front of the jet two leading shocks build up, a bow shock at the  interface between the jet and the ambient medium and a Mach shock where the  propagating matter is decelerated to low velocities. It is diverted  into a cocoon of back-flowing material which is highly turbulent and heated up to high temperatures, leading to emission from the fine structure lines of carbon, nitrogen, oxygen, or sulfur atoms and to some degree from their ions (Section \ref{subsubsec:fine-structure-lines}). Eventually the outflow dissolves as it reaches a speed that is comparable to the typical velocity dispersion in the ISM. 

\citet{ns80} estimated the amount of energy injected into the ISM by protostellar outflows, and showed that they could be an important energy source for turbulent motions in molecular clouds. They suggested that this in turn may influence the structure of the clouds and regulate the rate of gravitational collapse and star formation  \cite[see also][]{Li:2006p13226,Banerjee:2007p6178,Nakamura:2008p13041,Wang:2010p23176}. The existence of a kinematic interrelation between outflows and their ambient medium has been inferred from high resolution CO observations, e.g.\ of the PV Cephei outflow HH\,315 \citep{Arce:2002p87219,Arce:2002p87224}. Optical observations surveying nearby molecular clouds furthermore indicate a similar influence of the outflows on the ionization state and energetics of the inter-cloud medium that surrounds low-mass star forming regions (for Perseus, see \citealt{Bally:1997p87208} or \citealt{Arce:2010p62101}; for Orion~A, see \citealt{Stanke02}).

We begin with an estimate of the protostellar jet kinetic luminosity. It can be described as
\begin{equation}
L_{\rm jet} =  \frac{1}{2}\dot{M}_{\rm jet} \,
v^2_{\rm jet} = 1.3\times10^{32}\,{\rm erg\,s}^{-1}
\left(\frac{\dot{M}_{\rm jet}}{10^{-8}\,{\rm M}_{\odot} {\rm
yr}^{-1}}\right)
\left(\frac{v_{\rm jet}}{200\,{\rm km\,s}^{-1}}\right)^2\;,
\end{equation}
with $\dot{M}_{\rm jet}\approx 10^{-8}\,{\rm M}_{\odot} {\rm yr}^{-1}$ being the mass loss associated with the jet material that departs from the protostellar disk system at typical velocities of ${v_{\rm jet}}\approx 200\,{\rm km\,s}^{-1}$. This outflow rate is closely coupled to the accretion rate $\dot{M}_{\rm acc}$ onto the central star by $\dot{M}_{\rm jet} = f_{\rm jet}  \dot{M}_{\rm acc}$, with the efficiency factor typically being in the range $0.1 \,\sils\, f_{\rm jet} \,\sils\, 0.4$ (see e.g.\ \citealt{shu00}, \citealt{Ouyed97b}, \citealt{bontemps96a};  or consult the reviews by \citealt{Bally07}, \citealt{Pudritz07}, \citealt{frank2014}, or \citealt{li2014} for further details). 

A simple estimate of the jet lifetime in this phase is $t_{\rm jet} \approx 2\,{\rm pc}/{200\,{\rm km\,s}^{-1}} \approx 10^4\,{\rm yr}$. This coincides to within factors of a few with the typical duration of the class 0 and early class 1 phases of protostellar evolution (see Section \ref{subsec:collapse}). During these phases, we expect the strongest outflow activity (see the review by \citealt{andre00}). The total amount of energy provided by the jet is therefore
\begin{equation}
E_{\rm jet} =  L_{\rm jet}\,t_{\rm jet} \approx 8 \times 10^{43} {\rm erg}\;.
\end{equation}
This kinetic luminosity is smaller than but comparable to the radiative luminosity of protostars.  The outflow-ISM coupling is more direct and, thus, supposedly more efficient than the energy exchange between the protostellar radiation and the ISM. However, determinations of the coupling strength are controversial and require further investigation  \citep{Banerjee:2007p6178, Nakamura:2008p13041, Cunningham08a, Wang:2010p23176, Carroll:2010p85799}.  

The total energy input from protostellar winds will substantially exceed the amount that can be transferred to the turbulence, because of radiative cooling at the wind termination shock. This introduces another efficiency factor $\xi_{\rm jet}$.  A reasonable upper limit to the energy loss can be obtained by assuming that this cooling process is very efficient so that only momentum conservation holds, 
\begin{equation}
\label{eqn:mntm-cons}
\xi_{\rm jet}\, \,\sils\,\, \frac{\sigma}{v_{\rm jet}} = 0.05 \left(\frac{\sigma}{10\,\mbox{km}\,{\rm s}^{-1}}\right) \left(\frac{v_{\rm jet}}{200\,\mbox{km}\,{\rm s}^{-1}}\right)^{-1},
\end{equation}
where $\sigma$ as before is the velocity dispersion of ISM turbulence.  If we assumed that most of the energy went into driving dense gas, the efficiency would be lower, as typical velocities for CO outflows are only $1 - 2\,$km$\,$s$^{-1}$. The energy injection rate per unit volume then follows as 
\begin{eqnarray}
 \dot{e}& = &  \frac{1}{2} \xi_{\rm jet} f_{\rm jet} \frac{\dot{M}_{\rm SF} v_{\rm jet}^2 }  {\pi R^2 H}    =   \frac{1}{2}  f_{\rm jet} \frac{\dot{M}_{\rm SF} v_{\rm jet} \sigma }  {\pi R^2 H}\\
 & = & 1.4 \times 10^{-28}  \,\mbox{erg}\,\mbox{cm}^{-3}\,\mbox{s}^{-1} \times \nonumber \\
 && \!\!\! \times 
  \left(\frac{f_{\rm jet}}{0.2}\right)\!\!  \left(\frac{\dot{M}_{\rm SF}}{ 3\,{\rm M}_\odot \, {\rm yr}^{-1}} \right)\!  \left(\frac{v_{\rm jet}}{200\,\mbox{km}\,\mbox{s}^{-1}}\right)\!  \left(\frac{\sigma}{10\, \mbox{km}\,{\rm s}^{-1}}\right)\!   \left(\frac{H}{100 \mbox{ pc}} \right)^{-1}\!\! \left(\frac{R}{15 \mbox{ kpc}} \right)^{-2}\!, \nonumber 
\end{eqnarray}
where we again normalize to  the Galactic star formation  rate $\dot{M}_{\rm SF} = 3\,$M$_\odot \, $yr$^{-1}$ and take the volume of the star forming disk as $V = \pi R^2 H$, with radius $R = 15\,$kpc and disk thickness $H = 100\,$pc.

Although protostellar jets and outflows are very energetic, they are likely to deposit most of their energy into low density gas \citep{henning89}, as is shown by the observation of multi-parsec long jets extending completely out of molecular clouds \citep{bd94}. Furthermore, observed motions of molecular gas show increasing power on scales all the way up to and perhaps beyond the largest scale of molecular cloud complexes \citep{Ossenkopf02}. It is hard to see how such large scales could be driven by protostars embedded in the clouds.

\subsubsection{Radiation}
\label{par:radiation}

Next, we consider the radiation from massive stars. We focus our attention on ionizing radiation, because H{\textsc{ii}} regions can affect large volumes of interstellar gas and their expansion converts thermal energy into kinetic energy. To a much lesser degree, the same holds for radiation in the spectral bands that can lead to the dissociation of molecular hydrogen into atomic gas. The thermal radiation mostly from low-mass stars will not be able to trigger large gas motions in the ISM, and we will not concern ourselves with it here (but see Section \ref{subsec:ISRF}).

The total energy density carried by photons at frequencies high enough to ionize hydrogen is very large. We use the information provided, e.g. by \citet{tielens2010} or \citet{draine11}, and estimate the integrated luminosity of ionizing radiation in the disk of the Milky Way to be
\begin{equation}
\dot{e} = 1.5 \times 10^{-24} \mbox{ erg s$^{-1}$ cm}^{-3}\;.
\end{equation}
\citep[See also earlier work by][]{Abbott:1982p87255}. We note, however, that only a small fraction of this energy is converted into turbulent gas motions. There are two main pathways for this to happen. First, ionizing radiation will produce free electrons with relatively large velocities. This process heats up the resulting plasma to $7000 -10000\,$K. The ionized regions are over pressured compared to the ambient gas and start to expand. They cool adiabatically and convert thermal energy into kinetic energy. Second, the medium can also cool radiatively and possibly contract. The ISM in this regime is thermally unstable \citep{Field:1965p8403,mo77}. This instability can excite turbulent motions \cite[e.g.][]{VazquezSemadeni:2000p13329, kritsuk02a, Piontek:2005p34248, Hennebelle:2007p12522} with typical conversion factors from thermal to kinetic energy of less than 10\%. 

We begin with the first process, and look at the supersonic expansion of H{\textsc{ii}} regions after photoionization heating raises their pressures above that of the surrounding neutral gas. By integrating over the H{\textsc{ii}} region luminosity function derived by \citet{mckee97},  \citet{matzner02} estimates the average momentum input from expanding H{\textsc{ii}} regions as
\begin{equation}
\label{eqn:momentum}
  p_{\rm H{\textsc{ii}}} \,\approx\, 260 \,\mbox{km}\,\mbox{s}^{-1}
\left(\frac{N}{1.5 \times 10^{22}\,\mbox{cm}^{-2}}\right)^{-3/14}
\left(\frac{M_{\rm cloud}}{10^6\,\mbox{M}_{\odot}}\right)^{1/14}  M_* \;,
\end{equation}
where the column density $N$ is scaled to the mean value for Galactic molecular clouds \citep{solomon87}, $M_{\rm cloud}$ is a typical molecular cloud mass, and $M_* = 440 \,\mbox{M}_{\odot}$ is the mean stellar mass per cluster in the Galaxy \citep{matzner02, ladalada03}.  

We focus our attention on clusters and OB associations producing more than $10^{49}$ ionizing photons per second, because these are responsible for most of the available ionizing photons. From the luminosity function presented by \citet{mckee97}, we estimate that there are about $N_{49} = 650$ such clusters in the Milky Way. To derive an energy input rate per unit volume from the mean momentum input per cluster (Eq.~\ref{eqn:momentum}), we need to obtain an estimate for the typical expansion velocity $v_{{\rm H}{\textsc{ii}}}$  of the H{\textsc{ii}} regions as well as for the duration of this process. While expansion is supersonic with respect to the ambient gas, it is by definition subsonic with respect to the hot interior. The age spread in massive star clusters and OB associations can be several million years \citep{PreibischZinnecker1999, pozwetal10, Longmore:2014p85628}. We take a value of $t_{*} = 10\;$Myr. With the stellar mass -- luminosity relation on the main sequence being $L / {\rm L}_{\odot} \approx 1.5 \,(M/{\rm M}_\odot)^{3.5}$ for stars with masses up to $M \approx 20\,$M$_{\odot}$ and  $L / {\rm L}_{\odot} \approx 3200 \,(M/{\rm M}_\odot)$ for stars with $M \;\sigs\; 20\,$M$_{\odot}$, the energy output in a cluster is dominated by the most massive stars. The main sequence lifetime can be estimated as $t_{\rm MS} \approx 10^{10} \;{\rm yr}\;  (M/{\rm M}_\odot) (L/{\rm L}_\odot)^{-1} \approx 10^{10} \;{\rm yr}\;  (M/{\rm M}_\odot)^{-2.5}$ for stars with $M < 20\,$M$_{\odot}$, asymptoting to a value of around $t_{\rm MS} \approx 3$~Myr for more massive stars \cite[e.g.][]{Hansen:1994p85789}. For typical O-type stars, $t_{\rm MS} < t_*$, and as a consequence we can take $t_*$ as a good order of magnitude estimate for the duration of strong ionizing feedback from the clusters of interest. Once again, we estimate the volume of the star forming disk of the Galaxy as $V = \pi R^2 H$, with radius $R = 15\,$kpc and disk thickness $H = 100\,$pc. Putting this all together, the estimated energy input rate from expanding H{\textsc{ii}} regions is then
\begin{eqnarray}
\dot{e}& =& \frac{ N_{49} p_{\rm H{\textsc{ii}}} v_{\rm H{\textsc{ii}}} }{\pi R^2 H \,t_{*}}  \\
       & = &3\times 10^{-30} \mbox{ erg s$^{-1}$ cm}^{-3}
\left(\frac{N}{1.5\times 10^{22} \mbox{ cm}^{-2}}\right)^{-3/14}\!
\left(\frac{M_{\rm cloud}}{10^6 \mbox{ M}_{\odot}}\right)^{1/14}
\times \nonumber \\
  & \times  & 
  \left(\frac{M_*}{440 \mbox{ M}_{\odot}}\right)
\left(\frac{{N_{49}}}{650}\right)      
\left(\frac{v_{\rm H{\textsc{ii}}}}{10 \mbox{ km s}^{-1}}\right)
\left(\frac{H}{100 \mbox{ pc}}\right)^{-1}\!
\left(\frac{R}{15 \mbox{ kpc}}\right)^{-2}\!
\left(\frac{t_{*}}{10 \mbox{ Myr}}\right)^{-1}. \nonumber
\end{eqnarray}
Nearly all of the energy in ionizing radiation goes towards maintaining the ionization and temperature of the diffuse  medium, and hardly any towards driving turbulence.  Flows of ionized gas may be important very close to young clusters and may terminate star formation locally \cite[for the difference between two- and three-dimensional simulations, see][]{yorke02, krumholz07a, krumholz09a,  Peters2010b, Peters2011, kuiper11}. They can also influence the molecular cloud material that surrounds the young star cluster \cite[e.g.][]{Dale:2005p27348,Dale:2011p48360,Walch:2012p77141,Walch:2013p81833}. However, they appear not to contribute significantly on a global scale.

Now we turn our attention to the second process, to the thermal instability. \citet{kritsuk02b} find that the thermal energy released can be converted into turbulent kinetic energy,  $e_{\rm kin} = \xi_{\rm ion} e_{\rm th}$, with an efficiency $\xi_{\rm ion} \approx 0.07$. \citet{parravano03} study the time dependence of the local UV radiation field. They find that the corresponding  photoelectric heating rate increases by a factor of $2 - 3$ due to the formation of a nearby OB association every $100 - 200\;$Myr. Note, however, that substantial motions only lasted about $1\;$Myr after a heating event \citep{kritsuk02b,deavillez04,deavillez05,2007ApJ...665L..35D}. We follow this line of reasoning and estimate the resulting average energy input by taking the kinetic energy input from the heating event and dividing by the typical time $t_{\rm OB}$ between heating events. We determine the thermal energy for gas at a number density of $n=1\;$cm$^{-3}$  at a temperature of $T =10^4\;$K and find that
\begin{eqnarray}
\dot{e} &= &\frac{3}{2} \frac{n k T \xi_{\rm ion} }{t_{\rm OB}} \\
     &=&   5\times 10^{-29} \,\mbox{erg}\,\mbox{cm}^{-3}\,\mbox{s}^{-1}
       \left(\frac{n}{1\,\mbox{cm}^{-3}}\right)
       \left(\frac{T}{10^4\,\mbox{K}}\right)
       \left(\frac{\xi_{\rm ion}}{0.07}\right)
       \left(\frac{t_{\rm OB}}{100\,\mbox{Myr}}\right)^{-1}\;.\nonumber 
\end{eqnarray}
In comparison to some other proposed energy sources discussed here, this mechanism appears unlikely to be as important as the supernova explosions from the same OB stars discussed before.

\pagebreak 
\section{Formation of molecular clouds}
\label{sec:cloud-form}

\subsection{Transition from atomic to molecular gas}
\label{subsec:transition-H-to-H2}
Our starting point for considering the physics of molecular cloud formation is the chemistry
of the gas. After all, molecular clouds are, by definition, dominated by molecular gas, while
the gas in the more diffuse neutral phases of the ISM is almost entirely atomic. Cloud
formation must therefore involve, at some stage of the process, a chemical transition from 
gas which is mainly atomic to gas which is mainly molecular.

There are two main chemical transitions, occurring at different points in the assembly of 
a molecular cloud, that we could use to identify the point at which our assembling cloud
becomes ``molecular''. The first and most obvious of these is the transition between atomic
and molecular hydrogen: once most of the hydrogen in the cloud is in the form of H$_{2}$
rather than H, it is obviously reasonable to talk of the cloud as being molecular. However,
this transition has the disadvantage that it is extremely difficult to observe, since H$_{2}$
does not emit radiation at typical molecular cloud temperatures. Therefore, it is common
to use a different, observationally-motivated definition of the point when a cloud becomes
molecular, which is the moment it becomes visible in CO emission. Understanding when
this occurs requires understanding the chemical transition from C$^{+}$ to C to CO that
occurs within the assembling cloud. 

Below, we discuss the chemistry involved in both of these transitions in more detail,
and then examine some of the approximations used to model the atomic-to-molecular
transition in numerical studies of molecular cloud formation.

\subsubsection{Transition from H to H$_{2}$}
\label{hh2}
The simplest way to form H$_{2}$ in the ISM is via the radiative association of
two hydrogen atoms, i.e.\
\begin{equation}
{\rm H + H} \rightarrow {\rm H_{2}} + \gamma.
\end{equation}
However, in practice the rate coefficient for this reaction is so small that only a very small
amount of H$_2$ forms in this way. Somewhat more can form via the ion-neutral
reaction pathways
\begin{eqnarray}
{\rm H + e^{-}} & \rightarrow & {\rm H^{-} + \gamma}, \\
{\rm H^{-} + H} & \rightarrow & {\rm H_{2} + e^{-}},
\end{eqnarray}
and
\begin{eqnarray}
{\rm H + H^{+}} & \rightarrow & {\rm H_{2}^{+} + \gamma}, \\
{\rm H_{2}^{+} + H} & \rightarrow & {\rm H_{2} + H^{+}},
\end{eqnarray}
but it is difficult to produce H$_{2}$ fractional abundances larger
than around $f_{\rm H_{2}} \sim 10^{-2}$ with these reactions, even in the most
optimal conditions \citep[see e.g.][]{teg97}. Moreover, in the local ISM, photodetachment of 
H$^{-}$ and photodissociation of H$_{2}^{+}$ by the ISRF render these pathways
considerably less effective \citep{glover03}. We are therefore forced to conclude that 
gas-phase formation of H$_{2}$ is extremely inefficient in typical ISM conditions. Nevertheless, 
we do observe large quantities of H$_{2}$ in Galactic molecular clouds. 

The resolution to this apparent puzzle comes when we realize that most of the
H$_{2}$ in the ISM does not form in the gas-phase, but instead forms on the
surface of dust grains \citep{gs63}. Association reactions between adsorbed hydrogen 
atoms occur readily on grain surfaces, and the rate at which H$_2$ forms there is
limited primarily by the rate at which H atoms are adsorbed onto the surface.
For typical Milky Way conditions, the resulting H$_{2}$ formation rate
is approximately \citep{jura75}
\begin{equation}
R_{\rm H_{2}} \sim 3 \times 10^{-17} n n_{\rm H} \: {\rm cm^{-3} s^{-1}}.
\label{eqn:H2-form-rate}
\end{equation}
Here, $n$ is the total number density of  gas particles, while $n_{\rm H}$ is the number density of atomic hydrogen. For atomic hydrogen gas, both quantities are identical if we neglect contributions from helium and possibly metals. Note that $n_{\rm H}$ goes down as the molecular fraction increases, while $n$ remains the same in the absence of compression or expansion.  The H$_{2}$ formation timescale corresponding to the formation rate (\ref{eqn:H2-form-rate}) is
approximately
\begin{equation}
t_{\rm form} = \frac{n_{\rm H}}{R_{\rm H_{2}}} \sim 10^{9} n^{-1} \: {\rm yr}.
\end{equation}
When the gas density is low, this timescale can be considerably longer than
the most important dynamical timescales, such as the turbulent crossing time
or the gravitational free-fall time. Accounting for the effects of the small-scale
transient density structures produced by supersonic turbulence does shorten
the timescale somewhat \citep{gm07b,micic12}, but typically not by more than
an order of magnitude. 

Molecular hydrogen in the ISM can be collisionally dissociated by
\begin{eqnarray}
{\rm H_{2} + H} & \rightarrow & {\rm H + H + H}, \\
{\rm H_{2} + H_{2}} & \rightarrow & {\rm H + H + H_{2}}.
\end{eqnarray}
However, these reactions are effective at destroying H$_{2}$ only in
warm, dense gas, and so although they are important in certain circumstances,
such as in molecular outflows \citep[see e.g.][]{flower03}, they do not play a major role in regulating
the molecular content of the ISM. Instead, the dominant process responsible for 
destroying H$_{2}$ in the local ISM is photodissociation.

Photodissociation of H$_{2}$ occurs via a process known as spontaneous
radiative dissociation \citep{sw67,vd87}. The H$_{2}$ molecule first absorbs a
UV photon with energy $E > 11.2$~eV, placing it in an excited electronic state.
The excited H$_{2}$ molecule then undergoes a radiative transition back to the
electronic ground state. This transition can occur either into a  bound ro-vibrational 
level in the ground state, in which case the molecule survives, or into the vibrational
continuum, in which case it dissociates. The dissociation probability depends strongly
on the rotational and vibrational quantum numbers that the molecule has while in the
excited electronic state, but on average, it is around 15\% \citep{db96}. The discrete
set of UV absorption lines produced by this process are known as the Lyman and
Werner bands, and hence it has become common to refer to the energetic photons
responsible for destroying H$_{2}$ as Lyman-Werner photons.

Because H$_2$ photodissociation is line-based, rather than continuum-based, 
the H$_2$ photodissociation rate in the ISM is highly sensitive to an effect known
as self-shielding. This term refers to the fact that in a region with a high H$_{2}$
column density, the Lyman-Werner photons with energies corresponding to the
main absorption lines are mostly absorbed by H$_{2}$ on the outskirts of the region,
with only a few surviving to reach the center. Consequently, the H$_{2}$ 
photodissociation rate in the gas at the center of the region is reduced by a large
factor compared to the rate in the unshielded, optically thin gas. Detailed studies
of this process show that it starts to significantly affect the H$_{2}$ photodissociation
rate once the H$_{2}$ column density exceeds $N_{\rm H_{2}} \sim 10^{14}
\: {\rm cm^{-2}}$ \citep{db96}. The corresponding total column density of hydrogen
depends on the strength of the ISRF and the density of the gas. In unshielded gas
illuminated by an ISRF with a strength $G_{0}$ in Habing units, the equilibrium
number density of H$_2$ is given approximately by
\begin{equation}
n_{\rm H_{2}} \sim \frac{3 \times 10^{-17} n n_{\rm H}}{3 \times 10^{-10} G_{0}} 
 = 10^{-6} n n_{\rm H} G_{0}^{-1}.
\end{equation}
The resulting H$_{2}$ column density, $N_{\rm H_{2}}$, is therefore related to the total 
hydrogen column density $N$ by
\begin{equation}
N_{\rm H_{2}}  = 10^{-6} n_{\rm H} G_{0}^{-1} N.
\end{equation}
From this, we see that in order to produce an H$_{2}$ column density of 
$10^{14} \: {\rm cm^{-2}}$, we need a total column density
\begin{equation}
N = 10^{20} G_{0} n^{-1} \: {\rm cm^{-2}}.
\end{equation}
For comparison, the visual extinction required to reduce the H$_{2}$ photodissociation
rate by a factor of ten is  approximately $A_{\rm V} \approx 0.65$, which in the diffuse ISM 
corresponds to a total hydrogen column density $N \sim 10^{21} \: {\rm cm^{-2}}$.
Therefore, H$_2$ self-shielding becomes important earlier, at lower total column densities,
than dust shielding in conditions when $G_{0} / n$ is small, such as in CNM clouds far
from regions of massive star formation. On the other hand, if $G_{0} / n$ is large, such as
can be the case in photodissociation regions close to massive stars, then dust extinction
typically dominates.

\subsubsection{Transition from C$^{+}$ to C to CO}
\label{cco}
The chemistry involved in the transition from C$^{+}$ to C is very simple: atomic carbon
forms via the radiative recombination of C$^{+}$,
\begin{equation}
{\rm C^{+} + e^{-}} \rightarrow {\rm C + \gamma},
\end{equation}
and is destroyed by photoionization,
\begin{equation}
{\rm C + \gamma} \rightarrow {\rm C^{+} + e^{-}}.
\end{equation}
However, the formation of CO is considerably more complicated, as in this case there is not a single
dominant process responsible for CO formation, but rather a variety of different pathways that one can
follow to get to CO. In this section, we give a very brief introduction to the basics of CO formation 
chemistry, but we refer readers in search of a more detailed and comprehensive treatment to the
classic papers by \citet{gl75}, \citet{l76}, \citet{db76}, \citet{th85} and \citet{sd95}.

\paragraph{{\bf CO formation}}
The majority of the CO found in molecular clouds forms via one or the other of two main sets of
chemical intermediates. One set of intermediates includes hydroxyl (OH), its positive ion (OH$^{+}$)
and their products, while the other set includes the simple hydrocarbons CH and CH$_{2}$ and
their positive ions.

The formation of CO from OH occurs rapidly via the neutral-neutral reaction
\begin{equation}
{\rm C + OH} \rightarrow {\rm CO + H}.
\end{equation}
Unlike many neutral-neutral gas-phase reactions, this reaction has no activation energy
and hence remains effective even at the very low temperatures found within 
molecular clouds. In addition, in gas with a high C$^{+}$ to C ratio, CO$^{+}$ ions are produced  by
\begin{equation}
{\rm C^{+} + OH} \rightarrow {\rm CO^{+} + H},
\end{equation}
which then form CO either directly,
\begin{equation}
{\rm CO^{+} + H} \rightarrow {\rm CO + H^{+}}, 
\end{equation}
or indirectly, via HCO$^{+}$ in the reactions
\begin{eqnarray}
{\rm CO^{+} + H_{2}} & \rightarrow & {\rm HCO^{+} + H}, \\
{\rm HCO^{+} + e^{-}} & \rightarrow & {\rm CO + H}. 
\end{eqnarray}

We therefore see that once OH forms, CO follows rapidly. However, forming the necessary
OH radical is not so straightforward. One obvious pathway to OH involves the reaction of
atomic oxygen with H$_{2}$,
\begin{equation}
{\rm O + H_{2}} \rightarrow {\rm OH + H}.  \label{oh_nn}
\end{equation}
However, this reaction has an activation energy of 0.26~eV, and so although it is an important source 
of OH in hot gas \citep[see e.g.][]{hm79}, in the cold gas found in CNM clouds and molecular clouds, other
less-direct routes to OH dominate. One of these involves the reaction of atomic oxygen with
H$_{3}^{+}$:
\begin{equation}
{\rm O + H_{3}^{+}} \rightarrow {\rm OH^{+} + H_{2}}.
\end{equation}
The OH$^{+}$ ions formed in this reaction react rapidly with H$_{2}$, forming H$_{2}$O$^{+}$
and H$_{3}$O$^{+}$ via
\begin{eqnarray}
{\rm OH^{+} + H_{2}} & \rightarrow & {\rm H_{2}O^{+} + H}, \\
{\rm H_{2}O^{+} + H_{2}} & \rightarrow & {\rm H_{3}O^{+} + H}.
\end{eqnarray}
H$_{3}$O$^{+}$ does not readily react further with H$_{2}$, but instead is removed
from the gas via dissociative recombination, yielding a variety of products that include OH and
water \citep[see e.g.][]{jensen00}
\begin{eqnarray}
{\rm H_{3}O^{+} + e^{-}} & \rightarrow & {\rm H_{2}O + H}, \\
{\rm H_{3}O^{+} + e^{-}} & \rightarrow & {\rm OH + H_{2}}, \\
{\rm H_{3}O^{+} + e^{-}} & \rightarrow & {\rm OH + H + H}, \\
{\rm H_{3}O^{+} + e^{-}} & \rightarrow & {\rm O + H_{2} + H}.
\end{eqnarray}
The other main route to OH involves O$^{+}$. This can be produced by cosmic ray ionization
of neutral oxygen, or by charge transfer from H$^{+}$, and can react with H$_{2}$ to yield
OH$^{+}$,
\begin{equation}
{\rm O^{+} + H_{2}} \rightarrow {\rm OH^{+} + H}.
\end{equation}
The OH$^{+}$ ions produced in this reaction then  follow the same chain of reactions as outlined 
above.  

An important point to note here is that in every case, the rate-limiting step is the formation of the
initial OH$^{+}$ ion. Although the reactions between O and H$_{3}^{+}$ and between
O$^{+}$ and H$_{2}$ are rapid, the fractional abundances of O$^{+}$ and H$_{3}^{+}$
are small, and so the overall rate of OH$^{+}$ formation is relatively small. Once the OH$^{+}$
ions have formed, however, the remainder of the reactions in the chain leading to CO are
rapid. Since all of the reactions involved in the formation of OH$^{+}$ depend on H$_{2}$,
either directly or as a source for the H$_{3}^{+}$ ions, one consequence of this is that CO
formation via the OH pathway is sensitive to the molecular hydrogen abundance.

The other main route to CO involves the simple hydrocarbons CH and CH$_{2}$ and their
ions. In gas with a high C$^{+}$ fraction, CH$^{+}$ can be formed via the reaction
with H$_{2}$,
\begin{equation}
{\rm C^{+} + H_{2}} \rightarrow {\rm CH^{+} + H},
\end{equation}
or by radiative association with atomic hydrogen,
\begin{equation}
{\rm C^{+} + H} \rightarrow {\rm CH^{+} + \gamma}.  \label{chp_ra}
\end{equation}
As radiative association is a slow process, one might expect that the reaction with H$_{2}$
would dominate. However, this suffers from the same problem as reaction~(\ref{oh_nn}). It has a
substantial energy barrier, in this case 0.4~eV, and therefore proceeds at a very slow rate
at the temperatures typical of the CNM or of molecular clouds. Indeed, this presents something
of a problem, as the resulting CH$^{+}$ formation rate is too slow to explain the observed CH$^{+}$
abundance in the diffuse atomic ISM, possibly indicating that some form of non-thermal chemistry
is active there \citep[see e.g.][]{shef08,god09}.

In gas with significant fractions of C$^{+}$ and H$_{2}$, the CH$_{2}^{+}$ ion can be formed by
radiative association,
\begin{eqnarray}
{\rm C^{+} + H_{2}}  \rightarrow  {\rm CH_{2}^{+} + \gamma}.
\end{eqnarray}
The rate coefficient for this reaction is significantly larger than the rate coefficient for 
reaction~(\ref{chp_ra}) as discussed by \citet{udfa12}, and so in regions with $n_{\rm H_{2}} \geq n_{\rm H}$, this reaction 
is usually the main starting point for the formation of CO via the hydrocarbon pathway.

Once CH$^{+}$ or CH$_{2}^{+}$ has formed via one of the above reactions, it quickly 
reacts further with H$_{2}$,
\begin{eqnarray}
{\rm CH^{+} + H_{2}} & \rightarrow & {\rm CH_{2}^{+} + H}, \\
{\rm CH_{2}^{+} + H_{2}} & \rightarrow & {\rm CH_{3}^{+} + H}.
\end{eqnarray}
Although the CH$_{3}^{+}$ ions formed by this reaction chain can react further with H$_{2}$, they do so 
via a slow radiative association reaction,
\begin{equation}
{\rm CH_{3}^{+} + H_{2}} \rightarrow {\rm CH_{5}^{+} + \gamma}.
\end{equation}
Therefore, most of the CH$_{3}^{+}$ is destroyed instead by dissociative recombination,
\begin{eqnarray}
{\rm CH_{3}^{+} + {\rm e^{-}}} & \rightarrow & {\rm CH + H_{2}}, \nonumber \\
{\rm CH_{3}^{+} + {\rm e^{-}}} & \rightarrow & {\rm CH + H + H}, \nonumber \\
{\rm CH_{3}^{+} + {\rm e^{-}}} & \rightarrow & {\rm CH_{2} + H}.
\end{eqnarray}
The CH and CH$_{2}$ radicals produced by this process react readily with atomic oxygen,
forming CO via
\begin{eqnarray}
{\rm CH + O} & \rightarrow & {\rm CO + H}, \\
{\rm CH_{2} + O} & \rightarrow & {\rm CO + H_{2}}, \\
{\rm CH_{2} + O} & \rightarrow & {\rm CO + H + H}.
\end{eqnarray}
The CH$_{2}$ radicals are also destroyed rapidly in a reaction with atomic hydrogen,
\begin{equation}
{\rm CH_{2} + H}  \rightarrow {\rm CH + H_{2}},
\end{equation}
but in the case of CH, the analogous reaction has a significant energy barrier and hence
is negligible at the temperatures of interest. The end result is therefore that a large fraction
of the carbon incorporated into CH$^{+}$ or CH$_{2}^{+}$ by the reactions described 
above ultimately ends up in the form of CO.

In gas with a high abundance of neutral atomic carbon, a few other processes contribute
significantly to the formation rate of the CH and CH$_{2}$ radicals that are the precursor of
CO. They can be formed directly via radiative association of C with H or H$_{2}$,
\begin{eqnarray}
{\rm C + H} & \rightarrow & {\rm CH + \gamma}, \\
{\rm C + H_{2}} & \rightarrow & {\rm CH_{2} + \gamma},
\end{eqnarray}
although these reactions are relatively slow. Alternatively, atomic carbon can react with
H$_{3}^{+}$,
\begin{equation}
{\rm C + H_{3}^{+}} \rightarrow {\rm CH^{+} + H_{2}},
\end{equation}
forming a CH$^{+}$ ion that the reacts further as described above

Looking at the hydrocarbon pathway as a whole, we see that it shares some common
features with the OH pathway. In each case, the rate-limiting step is the initiating reaction,
whether this is the formation of CH, CH$_{2}$, CH$^{+}$ or CH$_{2}^{+}$ by radiative
association, or the formation of H$_{3}^{+}$ as a consequence of the cosmic ray ionization
of H$_2$. Once the initial molecular ion or radical has formed, the remainder of the reactions
that lead to CO proceed relatively quickly. This behavior forms the basis of several simplified
methods for treating CO formation that we discuss in Section~\ref{chem-model} below.
In addition, we also see that all of the different ways that we can proceed from C$^{+}$ or C to CO 
rely on the presence of molecular hydrogen. This is important, as it implies that substantial quantities 
of CO will form only in regions that already have high H$_{2}$ fractions. Therefore, although the
characteristic timescales of the chemical reactions involved in CO formation are generally shorter than
the H$_{2}$ formation time, non-equilibrium, time-dependent behavior can nevertheless still be important, 
owing to the dependence on the H$_{2}$ fraction.

\paragraph{{\bf CO destruction}}
In gas with a low visual extinction, the destruction of CO is dominated by photodissociation:
\begin{equation}
{\rm CO} + \gamma \rightarrow {\rm C + O}.
\end{equation}
The photodissociation of the CO molecule occurs via a process known as predissociation
\citep{vd87}. The molecule first absorbs a UV photon with energy $E > 11.09 \: {\rm eV}$, 
placing it in an excited electronic state. 
From here, it can either return to the ground state via radiative decay, or it can undergo a transition 
to a repulsive electronic state via a radiationless process. In the latter case, the molecule very
rapidly dissociates. In the case of CO, dissociation is typically far more likely than decay back to
the ground state \citep{vdb88}. Consequently, the lifetimes of the excited electronic states are
very short. This is important, as Heisenberg's uncertainty principle then implies that their energy
is comparatively uncertain. The UV absorption lines associated with the photodissociation of CO
are therefore much broader than the lines associated with H$_{2}$ photodissociation. As a
result, CO self-shielding is less effective than the analogous process for H$_{2}$.

The classic work on the photodissociation of CO in an astrophysical context is the paper by
\citet{vdb88}. However, in the two decades since this paper was published, improved 
experimental data on the properties of the CO molecule has become available, and a revised
treatment of CO photodissociation was recently given by \citet{visser09}.

Once the visual extinction of the gas becomes large, CO photodissociation becomes unimportant.
In these circumstances, two other processes take over as the main routes by which CO is destroyed.
First, cosmic ray ionization of hydrogen molecules or hydrogen atoms produces energetic  
photo-electrons. If these collide with other hydrogen molecules before dissipating their energy,
they can excite the H$_{2}$ molecules into excited electronic states. The subsequent
radiative decay of the molecules back to the ground state produces UV photons that can
produce localized photodissociation of CO and other molecules \citep{pt83,gld87,gredel89}. 
Second, CO is also destroyed via dissociative charge transfer with He$^{+}$ ions
\begin{equation}
{\rm CO + He^{+}} \rightarrow {\rm C^{+} + O + He}.
\end{equation}
The He$^{+}$ ions required by this reaction are produced by cosmic ray ionization of neutral 
helium. We therefore see that the rate at which CO molecules are destroyed in high $A_{\rm V}$
gas is controlled by the cosmic ray ionization rate in these high $A_{\rm V}$ regions. In local
molecular clouds, this is relatively small \citep{vv00}, and so almost all of the carbon 
in these high $A_{\rm V}$ regions is found in the form of CO. In clouds illuminated by a much higher
cosmic ray flux, however, such as those in the Central Molecular Zone of the Galaxy, the
destruction of CO by these processes in high extinction gas is considerably more important, and
the CO fraction can be significantly suppressed even in well-shielded gas \citep[see e.g.][]{clark13}.

\subsubsection{Modeling the atomic-to-molecular transition}
\label{chem-model}
There are a number of different approaches that one can use in order to numerically model 
the transition from atomic to molecular gas that occurs as one builds up a molecular cloud,
each with their own strengths and weaknesses. 

One of the most obvious approaches is to build a model that incorporates all of the main
chemical reactions occurring in the gas. The forty or so reactions discussed in 
Sections~\ref{hh2} \& \ref{cco} above represent only a small fraction of the full range of possible 
reactions that can occur, particular once one accounts for the role played by additional chemical 
elements such as nitrogen or sulfur. An example of the degree of chemical complexity that is
possible is given by the UMIST Database for Astrochemistry \citep{udfa12}. The latest release
of this database contains details of 6173 different gas-phase reactions of astrophysical interest,
involving 467 different chemical species. If one also attempts to account for the full range of
possible grain-surface chemistry and also for important isotopic variants of the main chemical
species (e.g.\ molecules with one or more deuterium atoms in place of a hydrogen atom), then
the size of the resulting chemical network can easily be an order of magnitude larger still
\citep[see e.g.][for a recent example]{albert13}. By coupling a comprehensive chemical model such
as this to a detailed model for the penetration of UV radiation through the gas and in addition 
a treatment of its magnetohydrodynamical and thermal evolution, we can in 
principle model the chemical evolution of the gas with a very high degree of accuracy. 

Unfortunately, the computational requirements of such an approach are currently prohibitive.
The chemistry of the ISM evolves on a wide range of different timescales, and hence the
set of coupled ordinary differential equations (ODEs) that describe the chemical evolution
of the gas are what is known as ``stiff''. To ensure stability, these equations must be solved
implicitly, and the cost of doing so scales as the cube of the number of ODEs. Consequently,
solving for the chemical evolution of the gas using a comprehensive chemical model is rather
time-consuming, owing to the large number of ODEs involved. This is not necessarily a problem
if one is interested in solving for the chemical evolution of only a small number of fluid elements,
but becomes a major difficulty once one tries to solve for the chemical evolution of the gas within
a high-resolution three-dimensional simulation, when one is dealing with tens or hundreds of
millions of fluid elements. Chemical networks involving $\sim$10--20 different species can be
used within such models \citep[see e.g.][]{g10}, although this is already computationally
demanding, but scaling up to $\sim$400--500 species requires approximately $10^{4}$ times
more computational power, rendering it completely impractical at the present time.

Because of this, any attempt to model the atomic-to-molecular transition numerically must
make some simplifications. If one is interested in a time-dependent, non-equilibrium description
of the transition, then there are two main strategies that can be used to make the problem
simpler. First, we can simplify the chemistry while continuing to use a detailed model of
the hydrodynamical evolution of the gas. The basic idea here is to strip the chemical model
down to its bare essentials, i.e.\ only those reactions that most directly affect the abundances
of H, H$_{2}$, C$^{+}$, C and CO. In the case of H and H$_{2}$, the simplicity of the chemistry
makes this relatively straightforward, and a number of different implementations of H$_{2}$
formation chemistry within large hydrodynamical simulations are now available
\citep[see e.g.][]{ann97,gm07a,dobbs08,gnedin09,christ12}. The only real difficulty in this
case is how to handle the effects of H$_{2}$ self-shielding and dust shielding. Several
different approaches have been used in the literature, ranging from simple Sobolev-like
approximations \citep{gnedin09}, to more sophisticated approximations based on computing
the column density of dust and H$_{2}$ along a limited number of sight-lines 
\citep[see e.g.][]{clark12a,clark12b,hartwig14a}.

Modeling the chemistry involved in the transition from C$^{+}$ to C to CO is rather harder,
owing to the significantly greater complexity of the required chemical network. Nevertheless,
several different possibilities have been put forward in the literature \citep{nl97,nl99,kc08,kc10,g10}. 
Typically, these approximate treatments ignore any reactions involving elements other than
H, He, C and O, ignore those parts of the carbon chemistry not directly involved in the formation
or destruction of CO, and greatly simplify the treatment of the main pathways from C$^{+}$
to CO. As we have seen, the rate limiting step in these pathways is typically the initiating 
reaction, and so a decent estimate of the CO formation rate can be arrived at by computing
how rapidly carbon is incorporated into any one of CH, CH$_{2}$, CH$^{+}$ and CH$_{2}^{+}$,
and how rapidly oxygen is incorporated into OH$^{+}$, without the need to follow all of the
details of the subsequent chemistry. A number of these approximate treatments were compared
with each other by \citet{gc12b}, who showed that although very simple treatments such as
that of \citet{nl97} tend to over-produce CO, more detailed models such as those of \citet{nl99}
and \citet{g10} produced results that agreed well with each other.

The other main non-equilibrium approach retains far more of the chemical complexity of the full 
network, choosing to simplify instead the treatment of the gas dynamics and often also the geometry 
of the gas. This is the strategy used, for example, in most PDR codes.\footnote{The acronym PDR stands for photodissociation region or photon dominated region.} For a long time, the standard
approach has been to ignore the effects of dynamics completely, and to adopt either spherical
symmetry or one-dimensional slab symmetry in order to model the clouds. Neglecting the hydrodynamical
evolution of the gas is often justified, if the chemical species one is interested in have characteristic
evolutionary timescales that are much shorter than a representative dynamical timescale such as
the turbulent crossing time or gravitational free-fall time. However, it is probably not a good approximation
for treating species such as H or H$_{2}$ that have long chemical timescales and whose abundances
at any given time in the evolution of a molecular cloud are therefore sensitive to the previous dynamical
history of the gas \citep[see e.g.][]{bergin04,gm07b}. The assumption of one-dimensional symmetry, although computationally 
convenient, is less easy to justify, as the resulting models are unable to explain some notable features of real
molecular clouds such as the widespread distribution of atomic carbon \citep{fr89,little94,sch95}. Accounting for
clumping within the cloud greatly alleviates this issue \citep[see e.g.][]{kr08}, and although it is possible to model
a clumpy cloud using a one-dimensional PDR code by representing the cloud as an ensemble of spherically-symmetric
clumps \citep[see e.g.][]{stut88}, ideally one would use a full three-dimensional approach. Recently, three-dimensional PDR codes are starting to become available \citep[see e.g.][]{levrier12,offner13}, although they are not yet as fully-featured as their one-dimensional cousins.

An even simpler approach to modeling the atomic-to-molecular transition involves relaxing the
assumption that the chemistry is out of equilibrium. If we assume that the gas is in chemical equilibrium,
then instead of solving a set of coupled ordinary differential equations in order to obtain the current
values of the chemical abundances, we have the simpler task of solving a set of linear equations.
In particular, note the studies by \citet{kmt08,kmt09} and \citet{mk10}
in which they solve for the equilibrium H and H$_{2}$ abundances as a function of the gas surface
density, the UV field strength, and the metallicity. They also derive simple analytical approximations
to their numerical results, suitable for implementing in large-scale numerical simulations that do not
have the resolution to model the structure of individual molecular clouds \citep[see e.g.][]{kuh12,thom14}.

The validity of the equilibrium approach depends upon the extent to which the equilibrium abundances
reflect the true chemical abundances in the ISM, and hence on the relative sizes of the H$_{2}$
formation timescale, $t_{\rm form}$, and the dynamical time, $t_{\rm dyn}$. At solar metallicity,
the two timescales are roughly equal in dense GMCs, and so it is reasonable to expect 
equilibrium models to be a good guide to the behavior of the H$_{2}$ fraction in these clouds.
Indeed, recent observations of the H{\sc i} and H$_{2}$ content of the Perseus molecular cloud 
made by \citep{lee12} yield results that are well fit by the \citet{kmt08} model. However, as
$t_{\rm form} \propto n^{-1}$, the formation time can be significantly longer than the dynamical
time in lower density clouds, such as the diffuse H$_{2}$ clouds observed in UV absorption
line studies \citep{snow06}, and it is therefore unclear whether the H$_{2}$ content of these clouds has yet
reached equilibrium \citep[see also][]{mg12}. Furthermore, since H$_{2}$ formation timescale 
scales inversely with the dust-to-gas ratio, the equilibrium approximation can fail badly in very
low metallicity, dust-poor systems. In these conditions, cloud formation, gravitational collapse and 
star  formation can all take place before the gas has had a chance to reach chemical equilibrium 
\citep{gc12c,krum12}. At very low metallicities, this can even lead to star formation occurring in regions 
that are primarily atomic rather than molecular.

\subsection{Importance of dust shielding}
\label{dust-shield-import}
In our discussion of the atomic to molecular transition in the previous Section, we saw that 
the column density of the gas in the ISM plays an important role in regulating its chemical 
state. Regions with high column densities have large visual extinctions, and hence can shield
themselves effectively from the UV portion of the ISRF (see Section \ref{subsec:ISRF}). In these regions, the gas is primarily
molecular once it reaches chemical equilibrium, although the approach to equilibrium can
take a long time when the volume density of the gas is small. On the other hand, regions with
a low column density have low visual extinctions and so are unable to resist the dissociating effects of
the ISRF. These regions are generally dominated by atomic gas. 

The precise value of the visual extinction corresponding to the transition between
mostly-atomic and mostly-molecular gas depends on a number of factors: the strength
of the ISRF, the volume density of the gas, and the effectiveness of self-shielding. However, 
the equilibrium molecular fraction typically depends only linearly on these quantities, but
exponentially on the visual extinction. In conditions typical of local GMCs, the transition from
atomic to molecular hydrogen occurs at a visual extinction $A_{\rm V} \sim 0.1$--0.2  
\citep{kmt08} and that from C$^{+}$ to CO occurs at $A_{\rm V} \sim 1$ \citep{wolf10}. 
Large differences in either the density of the gas or the strength of the ISRF are required in order 
to significantly  alter these values.
In solar metallicity gas, the total hydrogen column densities corresponding to the two
transitions are $N_{\rm H, tot} \approx 2 \times 10^{20} \: {\rm cm^{-2}}$ for the 
H--H$_{2}$ transition and $N_{\rm H, tot} \approx 2 \times 10^{21} \: {\rm cm^{-2}}$ for the C$^{+}$--CO transition.
However, in lower metallicity environments, such as the Magellanic Clouds, the
lower dust-to-gas ratio means that a  higher column density is required.

The transition from unshielded gas to gas with a significant visual extinction also has
an important influence on the thermal state of the gas. As we have already discussed,
photoelectric heating is the dominant form of radiative heating in the diffuse ISM (Section \ref{subsec:photoelectric-heating}), but 
its effectiveness falls off rapidly with increasing extinction for $A_{\rm V} > 1$.
Consequently, the equilibrium gas temperature drops significantly as we move from
unshielded to shielded gas, as illustrated in Figure~\ref{T_AV}.

\begin{figure}
\includegraphics[scale=.65]{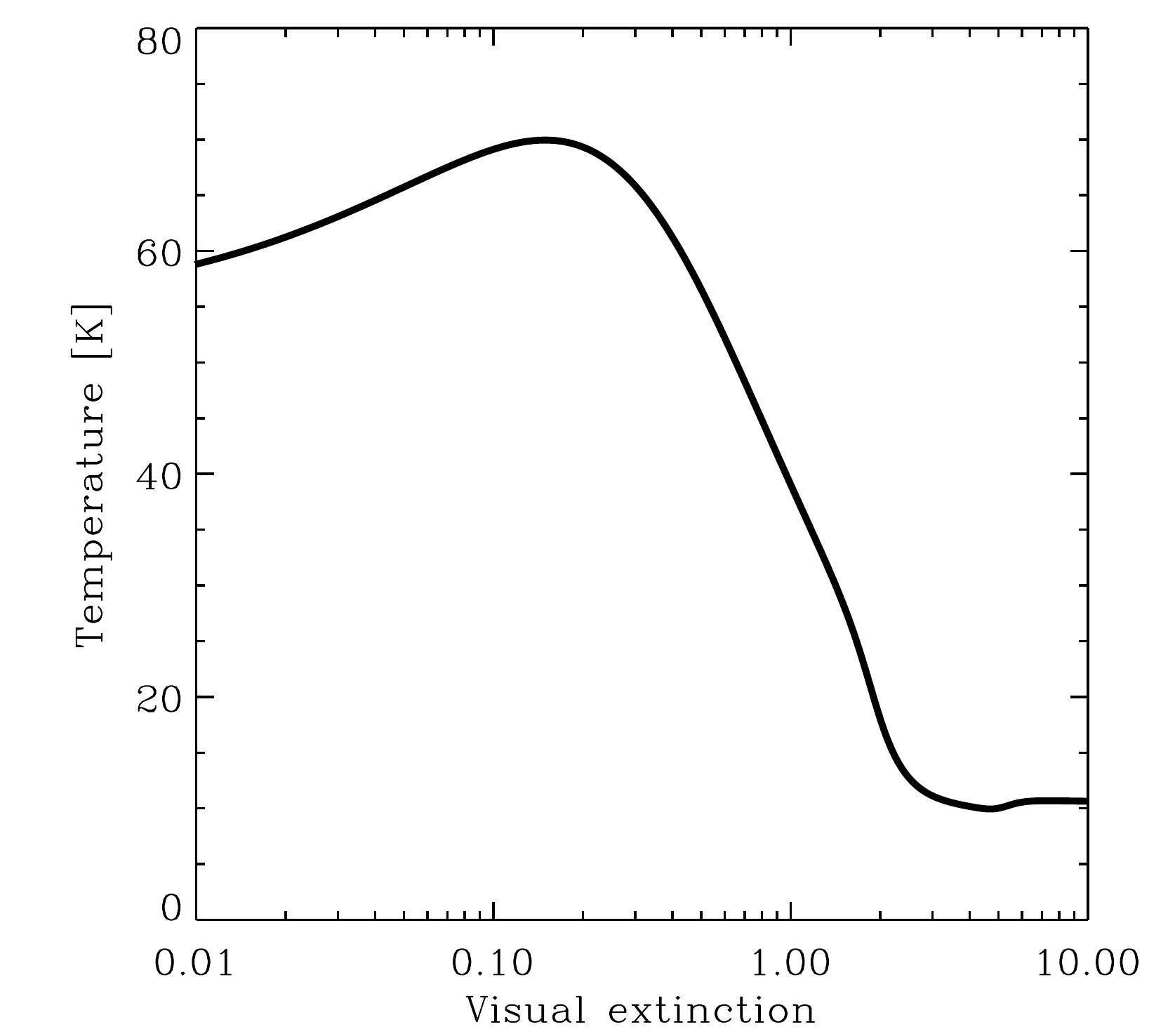}
\caption{Equilibrium temperature as a function of visual extinction within a uniform
density, semi-infinite slab of solar metallicity gas. The slab had a fixed hydrogen nuclei 
number density $n = 300 \: {\rm cm^{-3}}$, a velocity gradient of 
$3 \: {\rm km \, s^{-1} \, pc^{-1}}$, and was illuminated with a model of the ISRF 
based on \citet{dr78} at UV wavelengths and \citet{mmp83} at longer wavelengths.
The cosmic ray ionization rate of atomic hydrogen was taken to be $\zeta_{\rm H}
= 3 \times 10^{-17} \: {\rm s^{-1}}$. The rise in the temperature at very low $A_{\rm V}$
is a consequence of the transition from H to H$_{2}$: the cooling in this regime is
dominated by C$^{+}$ fine structure cooling, and at these densities and at fixed 
temperature, the C$^{+}$ cooling rate in fully atomic gas is two to three times larger 
than the  C$^{+}$ cooling rate in fully molecular gas.}
\label{T_AV}      
\end{figure}
The chemical and thermal changes that occur as we move from unshielded
to shielded regions have important implications for the gravitational stability
of the gas. The Jeans mass -- the critical mass scale above which quasi-spherical 
overdensities become unstable to their own self-gravity -- is related to the gas 
density and the temperature by
\begin{equation}
M_{\rm J} \approx 60 \; {\rm M_{\odot}}\;\mu^{-2} T^{3/2} n_{\rm tot}^{-1/2} \:,
\label{eqn:Jeans}
\end{equation}
where $\mu$ is the mean particle mass and $n_{\rm tot}$ is the total particle
number density. The factor of six decrease in $T$ that we find as we move from
low extinction to high extinction gas therefore results in a drop in $M_{\rm J}$
of roughly a factor of fifteen. The associated chemical transition from gas which is
primarily atomic to gas which is primarily molecular results in a further decrease
in $M_{\rm J}$ by roughly a factor of 2.5, and so the overall effect of increasing
the extinction is to decrease the Jeans mass by more than an order of magnitude.

The decrease in $T$ and increase in $\mu$ that occur as we move from low
$A_{\rm V}$ to high $A_{\rm V}$ gas are also responsible for a drop in the
sound speed of the gas. If the turbulent kinetic energy remains fixed, the result
is an increase in the Mach number of the turbulence. This makes it easier for
the turbulence to create high density regions (see also Section \ref{subsubsec:turb-IMF}).
The high density gas is more likely to be 
gravitationally bound, since $M_{\rm J} \propto n^{-1/2}$, and so the result of 
an increase in the Mach number of the turbulence will typically be an increase
in the star formation rate of the gas \citep{krumholz05c,pn11,hc11,fk12}.

We therefore see that it is much easier to form stars in gas clouds with high
visual extinctions than in those with low visual extinctions. In high $A_{\rm V}$
clouds, the gas temperature is lower and the gas is more likely to be molecular,
and both of these effects make star formation more likely. We therefore expect
to find a correlation between high $A_{\rm V}$ clouds and star formation. 
Moreover, since these high $A_{\rm V}$ clouds are dominated by molecular
gas, we also expect there to be a correlation between molecular gas and
star formation \citep{klm11}.

A correlation of just this kind is seen when we examine how stars form in our own
Milky Way or in nearby spiral galaxies. Work by a number of groups has shown that
on large scales, there is a tight correlation between the surface density of molecular
gas and the surface density of star formation in spiral galaxies 
\citep{wong02,leroy08,bigiel08,bigiel11,schruba11}. This
correlation is close to linear, although arguments continue as to whether it is 
truly linear \citep{leroy13}, or is actually slightly sub-linear \citep{shetty13,shetty14}. 

This correlation is often interpreted as being a consequence of molecular
cooling. It is argued that only CO cooling can lower
the gas temperature to the value of $\sim 10$~K characteristic of prestellar cores
within molecular clouds \citep{bergin07}, and that these low temperatures
are required for star formation. Although it is true that gas cannot reach 10~K
purely due to C$^{+}$ or O fine structure cooling, it can quite easily reach temperatures
as low as 20~K in well-shielded gas, as demonstrated in the detailed numerical simulations of 
\citet{gc12a}. These simulations also show that
the star formation rate of the gas is relatively insensitive to whether the gas is dominated
by atomic or molecular cooling \citep[see also][]{gc12c,krum12}, further supporting the idea that
the observed correlation between molecular gas and star formation is a consequence
of the fact that both are associated with regions of cold, dense gas. In short, molecular gas is a tracer of star formation but not its cause. 

Dust shielding may also play an important role in determining which sub-regions within
molecular clouds can successfully form stars. We know from observations of local
star-forming GMCs that stellar birth is not a completely random
process. Instead, there is a clear relationship between the observed column density of
the gas and the star formation rate. The process occurs predominantly in regions 
with column densities $N_{\rm H_{2}} > 7.5 \times 10^{21} \: {\rm cm^{-2}}$, corresponding to 
visual extinctions $A_{\rm V} > 8$ \citep[see e.g.\ the discussion
in][]{molinari14}. It remains an open question as to whether this relationship is best described in 
terms of a column density {\em threshold} \citep{onishi98,john04,lada10} or simply a steep dependence 
of the star formation rate on the column density \citep{hatch05,enoch08,heid10,guter11,burk13,lada13,evans14}.

A complete theoretical understanding of why this correlation exists remains lacking, but
in a recent study, \citet{cg13} argue that it is a further consequence of dust shielding. 
They point out that in a turbulent cloud, the angle-averaged extinction seen by an arbitrarily chosen
point along a high extinction line of sight will frequently be much lower than the extinction along
that line of sight. In their simulations, the dense structures traced by line of sight extinctions 
$A_{\rm V} > 8$ typically have much smaller mean extinctions, $\langle{A}_{\rm V}\rangle \sim 1$--2.
This roughly corresponds to the point at which dust shielding renders photoelectric heating of the 
gas ineffective, and so \citet{cg13} argue that the observed column density threshold merely
reflects the extinction required for clouds to shield themselves effectively from their environment.

\subsection{Molecular cloud formation in a galactic context}
\label{subsec:MC-formation}
As we have seen above, the transition between regions of the ISM that are dominated by atomic 
gas and regions that are dominated by molecular gas is primarily driven by changes in the column density.
In regions with low column densities, photodissociation of H$_2$ and CO is very efficient and
the equilibrium molecular fraction is small. On the other hand, in regions with high column 
densities, molecular self-shielding and dust shielding dramatically reduce the photodissociation
rates of H$_{2}$ and CO, allowing the equilibrium molecular fraction to become large. From
one point of view, then, the question of how molecular clouds form in the ISM has a simple 
answer. This happens whenever sufficient gas is brought together in one place to raise the
column density above the value needed to provide effective shielding against the ISRF, as long as the gas remains in this configuration for longer than the H$_{2}$ formation timescale,
$t_{\rm form}$. The real question, therefore, is what are the most important processes responsible 
for gathering together sufficient gas out of the diffuse ISM to make a dense molecular cloud.
This is a highly complex topic, and in these lecture notes we will do little more than to give a brief outline
of the main models that have been proposed to explain molecular cloud formation. We refer
readers in search of a more in-depth treatment of these issues to the recent reviews by 
\citet{Hennebelle:2012p72333}, \citet{dobbs14} and \citet{molinari14}.

One of the simplest models for molecular cloud formation is the  coagulation model, originally proposed 
by \citet{oort54} and subsequently elaborated by many other authors 
\citep[see e.g.][]{field65,kwan79,tomisaka84,tt09}. This model is based on a picture of the
ISM in which the cold atomic and molecular gas is organized into a series of discrete clouds
with a range of different masses. Small atomic clouds are formed directly from warmer atomic
gas by thermal instability \citep{Field:1965p8403}. Collisions between these small clouds efficiently 
dissipate energy, and so colliding clouds tend to coagulate, forming successively larger clouds. 
Once the clouds have grown large enough, they become able to shield themselves from the effects
of the ISRF, at which point they become dominated by molecular gas. Even once they have become 
molecular, however, the clouds continue to undergo regular collisions, and can potentially grow to 
very large masses. This process is terminated for a particular cloud once the feedback from the
stars forming within it becomes strong enough to disrupt the cloud.

This model has a number of appealing features. The stochasticity of the process of cloud-cloud
collisions is thought to naturally lead to a power law cloud mass function \citep[e.g.][]{field65}, and the fact
that collisions occur more frequently in denser regions of the galactic disk also provides a simple
explanation for the enhanced concentrations of molecular gas and ongoing star formation found
within most spiral arms. In addition, the coagulation model also can easily produce clouds that
are counter-rotating compared to the galactic disk \citep[e.g.][]{dobbs08coag,tt09}, 
explaining why clouds with retrograde rotation appear to be common within the ISM 
\citep[see e.g.][]{phil99,ib11}.

Unfortunately, this model also suffers from a major problem. Small molecular clouds can be
built by coagulation relatively rapidly, but large molecular clouds with masses of 
$10^{5}$--$10^{6} \: {\rm M_{\odot}}$ require of the order of 100~Myr or more to form by this
method \citep{blitz80}. Since this is an order of magnitude larger than most estimates for 
typical GMC lifetimes \citep{blitz07a}, it seems to be impossible to form massive GMCs in
low density environments in this model. In the dense environments of spiral arms, the much
higher cloud collision rate alleviates this problem to a large extent \citep{cc82,kv83,dobbs08coag}, 
but this does not provide an explanation for the existence of very massive clouds in inter-arm regions, 
as observed in galaxies such as M51 \citep{hughes13}.

A more fundamental issue with the coagulation model is that it is not clear that the picture of
the ISM on which it is based, in which GMCs are discrete objects that evolve in equilibrium between collisions and that have well-defined masses and
clear edges, is a good description of the real ISM. As we discuss in Section \ref{subsubsec:turb-obs}, observations show that GMCs are ubiquitously
surrounded by extended envelopes of atomic gas \citep[see e.g.][]{wann83,ee87,lee12,hvs13,motte14} 
and so the observational ``edge'' of a GMC -- the point at which we cease to be able to detect CO emission 
-- more likely represents a chemical transition in the gas (see Section \ref{subsec:transition-H-to-H2}), rather than any sudden change in the density.
In addition, a considerable fraction of the molecular gas of a galaxy seems to be in an extended diffuse component, rather than in discrete clouds \cite[see e.g.][]{pety2013, shetty2014, smith2014}. This finding casts further doubts on any astrophysical conclusion derived from the cloud collision picture. 

Altogether, it is highly  plausible that rather than being discrete objects with identities that persist over
long periods of time, molecular clouds are instead merely the highest density regions within
a far more extended turbulent flow of gas. This picture motivates an alternative way of thinking
about molecular cloud formation, known as the converging (or colliding) flow model for cloud
formation. The basic idea in this case is that molecular clouds form in dense, post-shock regions
formed when converging flows of lower density gas collide and interact. If the flows initially
consist of warm atomic hydrogen, then their collision can trigger a thermal instability, leading
to the rapid production of a cloud of much denser, cooler gas \citep[see e.g.][]{hp99,hp00,ki02,
ah05,heitsch05,vaz06,Hennebelle:2007p12522,hh08,banerjeeetal09}.
The mean density of the cold gas clouds produced in this way is typically of the order of
$100 \: {\rm cm^{-3}}$, high enough to allow H$_{2}$ formation to occur on a timescale
shorter than the duration of the collision \citep[see e.g.][]{clark12b}. CO will also form in
these cold clouds in regions where the column density is high enough to provide effective
shielding from the ISRF, although simulations have shown that the production of these 
high column density regions generally requires at least some part of the cold cloud to undergo
gravitational collapse \citep{hh08,clark12b}.

The converging flow model for GMC formation naturally explains why we see so few molecular
clouds that are not associated with ongoing star formation. CO observations are blind to the
inflow during its early evolution, since at this stage, the molecular abundance in the gas is very
small \citep{Hartmann01}. High molecular abundances and detectable CO luminosities are produced only 
during relatively late evolutionary phases, and work by \citet{clark12b} has shown that the time 
lag between the appearance of detectable CO emission and the onset of star formation is typically only
1--2$\,$Myr. This picture is supported by growing observational evidence that molecular clouds (as traced by CO) are continuously gaining mass during their evolution. For example, \citet{fukui09} and \citet{kawamura09} report in their analysis of GMCs in the Large Magellanic Cloud mass growth rates of several $10^{-2}\,$M$_{\odot}\,$yr$^{-1}$. It is a very appealing feature that this continuous accretion process provides a simple explanation for the presence of turbulence within GMCs. The kinetic energy associated with the convergent flow that forms the cloud in the first place is also able to drive its internal turbulence and explain many of its internal properties \cite[see ][]{klessen10,goldbaum11}.  As a consequence, the turbulent cascade extends from global galactic scales all the way down to the dissipation regime on sub-parsec scales (Section \ref{subsubsec:ISM-scales}).

Much of the work that has been done to model the formation of molecular clouds in converging 
flows has focused on the case where the flow is essentially one-dimensional, with two streams
of gas colliding head-on. However, in this scenario, it is difficult to form very massive clouds, as
a simple calculation demonstrates. Suppose we have two flows of convergent gas, each of which
has a cross-sectional area $A$, an initial number density $n_{0}$, and a length $L_{\rm flow} / 2$. The total 
mass of the cloud that can be formed by the collision of these flows is given approximately by
\begin{equation}
M_{\rm cloud} \sim \mu n_{0} A L_{\rm flow},
\end{equation}
where we have assumed that all of the gas in the flows becomes part of the cold cloud, and $\mu = 1.26\, m_{\rm H} = 2.11\times 10^{-24}\;$g typical for atomic gas. If the
gas in the flows is initially part of the warm neutral medium, then the number density is $n_{0} \sim 0.5\,$cm$^{-3}$ (see Table \ref{ism-phases}), and 
\begin{equation}
M_{\rm cloud} \sim 2300 \, {\rm M}_{\odot} \, \left(\frac{A}{1000 \: {\rm pc^{2}}}\right) 
\left(\frac{L_{\rm flow}}{150 \: {\rm pc}}\right).
\end{equation}
If the flows together have a total length $L_{\rm flow} \sim 150 \: {\rm pc}$ that is comparable 
to the molecular gas scale height of the Galactic disk (see Table~\ref{tab:MW}), and a cross-sectional area 
typical of a reasonably large GMC \citep{solomon87}, then the total mass of the resulting cloud 
is only a few thousand solar masses, much smaller than the mass of most GMCs.

There are several ways in which we might try to avoid this problem. First, we can make 
$L_{\rm flow}$ larger. However, even if we make it comparable to the atomic gas scale
height, so that $L_{\rm flow} \sim 1000 \: {\rm pc}$, the resulting cloud mass is still small,
$M_{\rm cloud} \sim 15000 \: {\rm M_{\odot}}$. Second, we can make $n_{0}$ larger. 
The value that we have adopted above is typical of the stable WNM, but thermally unstable 
diffuse atomic gas could have a density that is an order of magnitude higher
\citep[see e.g.][]{dobbs12}.
However, once again this does not increase $M_{\rm cloud}$ by a large enough amount to explain
how the most massive GMCs form. Finally, we could make $A$
larger. Simulations show that clouds formed in one-dimensional flows tend to collapse gravitationally 
in the directions perpendicular to the flow \citep[see e.g.][]{burk04,heitsch08,vazsemetal09}.
Therefore, it is reasonable to suppose that the 
cross-sectional area of the flows involved in forming the cloud may be much larger than the cross-sectional 
area of the final GMC. However, even if we increase $A$ by a factor of 20, so that the width and
height of the flow are comparable to its length, we again only increase $M_{\rm cloud}$ by an order
of magnitude. In addition, if all of the dimensions of the flow are similar, it is unclear whether we
should really think of it as a one-dimensional flow any longer. In the end, what is needed in order to explain 
the formation of the most massive GMCs in this model is a combination of these points. The flow must
consist of gas that is denser than is typical for the WNM, that has a coherent velocity over a relatively
large distance, and that either has a large cross-sectional area or is actually inflowing from multiple
directions simultaneously. How often these conditions are realized in the real ISM remains an open
question. 

Another issue that is not yet completely settled is which of several different possible physical 
processes is primarily responsible for driving these convergent flows of gas. One obvious possibility
is that these flows are driven by large-scale gravitational instability. Analysis of the behavior of
small perturbations in a thin rotating gas disk shows that the key parameter that determines whether
or not they grow exponentially is the so-called Toomre parameter \citep{toomre64},
\begin{equation}
Q = \frac{c_{\rm s, eff} \kappa}{\pi G \Sigma}.
\end{equation}
Here, $c_{\rm s, eff}$ is the effective sound-speed of the gas, which accounts not only for the thermal
sound speed, but also for the influence of the small-scale turbulent velocity dispersion, $\kappa$ is
the epicyclic frequency of the disk, and $\Sigma$ is the surface density of the gas. A pure gas disk is unstable
whenever $Q < 1$. In the case of a disk that contains a mix of gas and stars, the analysis is more
complex \citep[see e.g.][]{rafikov01, elmegreen11}, but the required value of $Q$ remains close to unity.
Measurements of $Q$ in nearby spirals and dwarf galaxies suggest that in most of these systems,
the gas is marginally Toomre stable, even when the gravity of the stellar component is taken into
account \citep{Leroy:2008p4217}. However, this does not mean that gravitational instability is unimportant
in these systems, as simulations show that star formation in disk galaxies tends to self-regulate so that
$Q \sim 1$ \citep[e.g.][]{krum10,fg13}. Briefly, the reason for this is that if $Q \ll 1$, the disk will be highly unstable and
will form stars rapidly \citep[see e.g.][]{li05,li06}. This will both deplete the gas surface density, and also increase 
$c_{\rm s, eff}$, due to the injection of thermal and turbulent energy into the gas by the various stellar 
feedback processes discussed in Section~\ref{subsub:ISM-driving-feedback}. These effects combine to 
increase $Q$ until the disk becomes marginally stable. 

Another mechanism that can drive large-scale convergent flows of gas in spiral galaxies is the
Parker instability \citep{parker66}. This is a magnetic instability which causes a  field that
is stratified horizontally in the disk to buckle due to the influence of magnetic buoyancy. Gas
then flows down the buckled magnetic field lines, accumulating near the midplane of the disk. The
characteristic length scale associated with this instability is a factor of a few larger than the
disk scale height. It therefore allows gas to be accumulated from within a large volume, and is
hence capable of producing even the most massive GMCs \citep{mous74a,mous74b}. However, the
density contrasts produced by the Parker instability are relatively small 
\citep[see e.g.][]{kim98,kim01,kim02} and so, although this instability may play a role in triggering thermal instability 
in the galactic midplane \citep{mous09}, it seems unlikely to be the main mechanism responsible
for GMC formation.

Finally, stellar feedback in the form of expanding H{\textsc{ii}} regions, stellar wind bubbles, supernova remnants 
and super-bubbles may also drive converging flows of gas in the ISM 
\citep[see e.g.][for some recent examples]{nt11,dobbs12,henne14}.
The idea that stellar feedback may trigger cloud formation, and hence also star formation, has a long
history \citep[see e.g.][for a seminal early study]{elmegreen77}. At first sight it has considerable observational support, since examples of spatial associations between molecular clouds
and feedback-driven bubbles are widespread \citep[e.g.][]{beau10,deh10,hou14}. However, this is a case in 
which the observations are somewhat misleading. The fact that a molecular cloud is associated with the 
edge of a feedback-driven bubble does not necessarily imply that the bubble is responsible for creating the cloud, since the
expanding bubble may simply have swept up some dense, pre-existing structure \cite[e.g.][]{pringle01}. Models of
cloud formation in a supernova-driven turbulent ISM without self-gravity find that although some
cold, dense clouds are formed in the expanding shells bounding the supernova remnants, the
total star formation rate expected for these regions is only $\sim 10$\% of the rate required to produce
the assumed supernova driving \citep{joung06}. Recent efforts to quantify the
effectiveness of triggering in the LMC also find that no more than about 5--10\% of the total molecular
gas mass budget can be ascribed to the direct effect of stellar feedback \citep{dawson13}. Therefore, although 
stellar feedback clearly plays an important role in structuring the ISM on small scales and contributes
significantly to the energy budget for interstellar turbulence (Section~\ref{subsub:ISM-driving-feedback}), 
it does not appear to be the main process responsible for the formation of molecular clouds.

\pagebreak
\section{Star formation}
\label{sec:collapse-SF}

\subsection{Molecular cloud cores as sites of star formation}
\label{subsec:cores}

In this Section, we focus on the small-scale characteristics of molecular clouds and discuss the properties of the low-mass cores that are the immediate progenitors of individual stars or binary systems. We begin with a discussion of the core mass spectrum, and then turn our attention to the density, thermal, chemical, kinematic, and magnetic field structure of individual cores. We distinguish between prestellar cores, which are dense cloud cores that are about to form stars in their interior, but have not yet done so (or at least show no detectable sign of stellar activity), and protostellar cores, for which we can infer the presence of  embedded protostars in the main accretion phase.

\subsubsection{Mass spectrum of molecular cloud cores}
\label{par:statistics}

In Section \ref{par:prop-MC}, we discussed the global statistical properties of molecular clouds. While a complete structural decomposition of an entire cloud leads to a power-law mass spectrum (equation \ref{eqn:mass-spectrum}), focusing on the densest parts of the clouds, on the pre- and protostellar cores, yields a different picture. As one probes smaller and smaller scales and more strongly bound objects, the inferred mass distribution becomes closer to the stellar IMF. The first large study of this kind was published by \citet{motte98}, for a population of submillimeter cores in $\rho$ Oph. Using data obtained with the IRAM 30m-telescope\footnote{http://www.iram-institute.org/EN/30-meter-telescope.php}, they discovered a total of 58 starless clumps, ranging in mass from $0.05\ $M$_{\odot}$ to  $3\ $M$_{\odot}$. Similar results have been obtained for the Serpens cloud \citep{testi98}, for Orion B North  \citep{johnstone01} and Orion B South \citep{Johnstoneetal2006}, and for the Pipe Nebula  \citep{Ladaetal2006}. Currently all observational data \citep[e.g.][]{motte98,testi98,johnstone00,johnstone01,Johnstoneetal2006,NutterWardThompson2007,alves07, DiFrancescoetal2007,WardThompsonetal2007,lada08a,kon10}  reveal a striking similarity to the IMF. To reach complete overlap one is required to introduce a mass scaling or efficiency factor of 0.2 to 0.5, depending on the considered region.  An exciting interpretation of these observations is that we are witnessing the direct formation of the IMF via fragmentation of the parent cloud. However, we note that the observational data also indicate that a considerable fraction of the prestellar cores do not exceed the critical mass for gravitational collapse, much like the clumps on larger scales. The evidence for a one-to-one mapping between prestellar cores and the stellar mass, thus, is by no means conclusive. For an extended discussion of potential caveats, see \citet{clark07} or consult the {\em Protostars and Planets VI}  review by \citet{Offner14}.

\subsubsection{Density structure} 

The density structure of prestellar cores is typically inferred through the analysis of dust emission or absorption using near-infrared extinction mapping of background starlight, millimeter/submillimeter dust continuum emission, or dust absorption against the bright mid-infrared background emission \citep{bergin07}. A main characteristic of the density profiles derived with the above techniques is that they require a central flattening within radii smaller than $2500 - 5000\,$AU, with typical central densities of $10^{5} -10^{6}\,$cm$^{-3}$ \citep{motte98,ward-thompson99}.  A popular approach is to describe these cores as truncated isothermal (Bonnor-Ebert) spheres \citep{ebert55,bonnor56} that often (but not always) provide a good fit to the data \citep{bacman01,Alves01,kandori05}. These are equilibrium solutions for the density structure of self-gravitating gas spheres bounded by external pressure. However, this density structure is not unique. Numerical calculations of the dynamical evolution of supersonically turbulent clouds show that the transient cores forming at the stagnation points of convergent flows exhibit similar morphology despite not being in dynamical equilibrium \citep{ballesteros03}.

\subsubsection{Kinematic stucture}
In contrast to the supersonic velocity fields observed in molecular clouds, dense cores have low internal velocities. Starless cores in clouds like Taurus, Perseus, and Ophiuchus systematically exhibit spectra with close to thermal linewidths, even when observed at low angular resolution \citep{myers83,jijina99}. This indicates that the gas motions inside the cores are subsonic or at best transsonic, with Mach numbers less than $\sim$2  \citep{kirk07, Andre:2007p25886, rosolowsky08a}. In addition, in some cores, inward motions have also been detected. They are inferred from the observation of optically thick, self-absorbed lines of species like CS, H$_2$CO, or HCO$^+$, in which low-excitation foreground gas absorbs part of the background emission. Typical inflow velocities are of order of $0.05-0.1\,$km$\,$s$^{-1}$  and are observed on scales of  $0.05 - 0.15\,$pc, comparable to the total size of the cores \citep{lee99}. The overall velocity structure of starless cores appears broadly consistent with the structure predicted by models in which protostellar cores form at the stagnation points of convergent flows, but the agreement is not perfect \citep{klessen05, offner08b}. Clearly more theoretical and numerical work is needed. In particular, the comparison should be based on synthetic line emission maps, requiring one to account for the chemical evolution of the gas in the core and the effects of radiative transfer \cite[e.g.][]{smith2012a, smith2013a, chira2014a}. In addition, it is also plausible that the discrepancy occurs because the simulations do not include all the necessary physics such as radiative feedback and magnetic fields. 

Subsonic turbulence contributes less to the energy budget of the cloud than thermal pressure and so cannot provide sufficient support against gravitational collapse \citep{myers83,goodman98,tafalla06}. If cores are longer lasting entities there must be other mechanisms to provide stability. Obvious candidates are magnetic fields \citep{shu87}. However, they are usually not strong enough to provide sufficient support \citep{crutcher99a, crutcher00, bourke01, crutcher09b, 2010ApJ...725..466C}. It seem reasonable to conclude that most observed cores are continuously evolving transient objects rather than long-lived equilibrium structures.  

\subsubsection{Thermal stucture}

The kinetic temperature of  dust and gas  in a core is  regulated by the interplay between various heating and cooling processes. At densities above $10^5$ cm$^{-3}$ in the inner part of the cores,  gas and dust are coupled thermally via collisions \cite[][and see also Section \ref{subsec:gas-grain-transfer}]{goldsmith78,burke83,Goldsmith01}. At lower densities,  corresponding to the outer parts of the cores, the two temperatures are not necessarily expected to be the same. Thus, the dust and gas temperature distributions need to be independently inferred from the observations. Large-scale studies of the dust temperature show that the grains in starless cores are colder than in the surrounding lower-density medium. Far-infrared observations toward the vicinity of a number of dense cores provide evidence for flat or decreasing temperature gradients with cloud temperatures of $15 -20\,$K and core values of $8 -12\,$K \citep{Ward02,toth04,laun13}. These observations are consistent with dust radiative transfer modeling in cores illuminated by interstellar radiation field \citep{langer05,keto05,stamatellos07a}.  The gas temperature in molecular clouds and cores is commonly infered from the level excitation of simple molecules like CO and NH$_3$ \citep{evans99,walmsley83}. One finds gas temperatures of $10 - 15\,$K, with a possible increase toward the lower density gas near the cloud edges. However, these measurements are difficult, since as the density drops, the molecular emission can become sub-thermal, in which case its excitation temperature no longer traces the kinetic temperature of the gas (see the discussion in Section \ref{subsec:opt-thin-2-level-atom}). In static prestellar cores (if such things exist), the main heat source is cosmic ray ionization, while in gravitationally collapsing cores, compressional heating and the dissipation of turbulence can also make significant contributions to the total heating rate \citep{gc12a}. Cooling in dense cores is dominated by molecular line emission, particularly from CO, and by heat transfer from the gas to the grains \citep{goldsmith78}. 

\begin{figure*}[tbp]
\begin{center}
\includegraphics[width=0.95\textwidth]{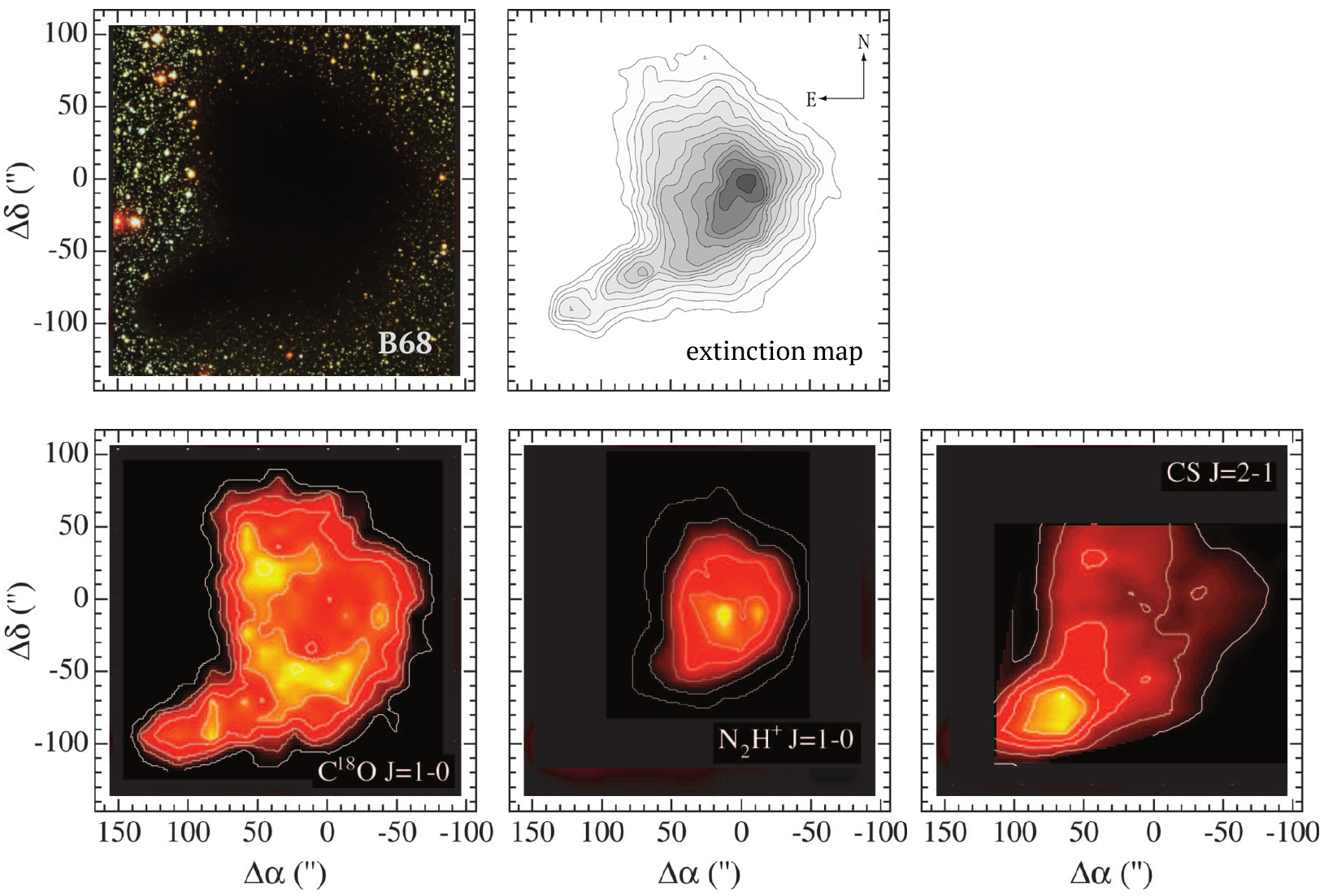}
\caption{Isolated prestellar molecular cloud core Barnard 68. {\em Upper panel:} Optical image and extinction map of the object  (see \citealp{Alves01};  images from ESO website). {\em Lower panel:} Maps of molecular line emission from  C$^{18}$O, N$_2$H$^+$, and CS (images adopted from  \citealp{lada03b}; see also see \citealt{bergin02}). The lower images illustrate the effects of depletion onto grains in the high-density central region of the core. C$^{18}$O  is clearly underabundant in the central, high-density regions of Barnard 68, while N$_2$H$^+$ traces this region very well. CS is brightest in the tail structure in the south-east corner and is highly depleted in the core center  \cite[see also][]{bergin07}.
}
\label{fig:B68}
\end{center}
\end{figure*}

\subsubsection{Chemical stucture}

Maps of integrated line intensity  can look very different for different molecular tracers. This is illustrated in Figure \ref{fig:B68}. It shows that the emission from nitrogen-bearing species, such as N$_2$H$^+$, more closely follows the dust emission, while emission from carbon-bearing molecules, such as C$^{18}$O or CS, often appears as a ``ring-like''\, structure around the dust emission peak \citep{bergin02,tafalla02,lada03b,maret07a}. The common theoretical interpretation of these data is that carbon-bearing species freeze-out on the surfaces of cold dust grains in dense portions of the cloud, while nitrogen-bearing molecules largely remain in the gas phase.  At the same time, chemical models of prestellar cores predict that molecules in the envelope of the core are destroyed by the interstellar UV field \citep{pavlyuchenkov06,aikawa08}. The resulting chemical stratification significantly complicates the interpretation of molecular line observations, and again requires the use of sophisticated chemical models which have to be coupled to the dynamical evolution \citep[e.g.][]{aikawa08,vw09,furuya12}. From the observational side, the freeze-out of many molecules makes it difficult to use their emission lines for probing the physical conditions in the inner regions of the cores.  Nevertheless, modeling of the chemical evolution  of the gas can provide us with important information on the cores. For example, the level of CS depletion can be used to constrain the age of the prestellar cores, while the deficit of CS in the envelope can indicate the strength of the external UV field \citep{bergin07}. In any case, any physical interpretation of the molecular lines in prestellar cores has to be based on chemical models and should do justice to the underlying density and velocity pattern of the gas.

\begin{figure*}[htbp]
\begin{center}
\includegraphics[width=0.90\textwidth]{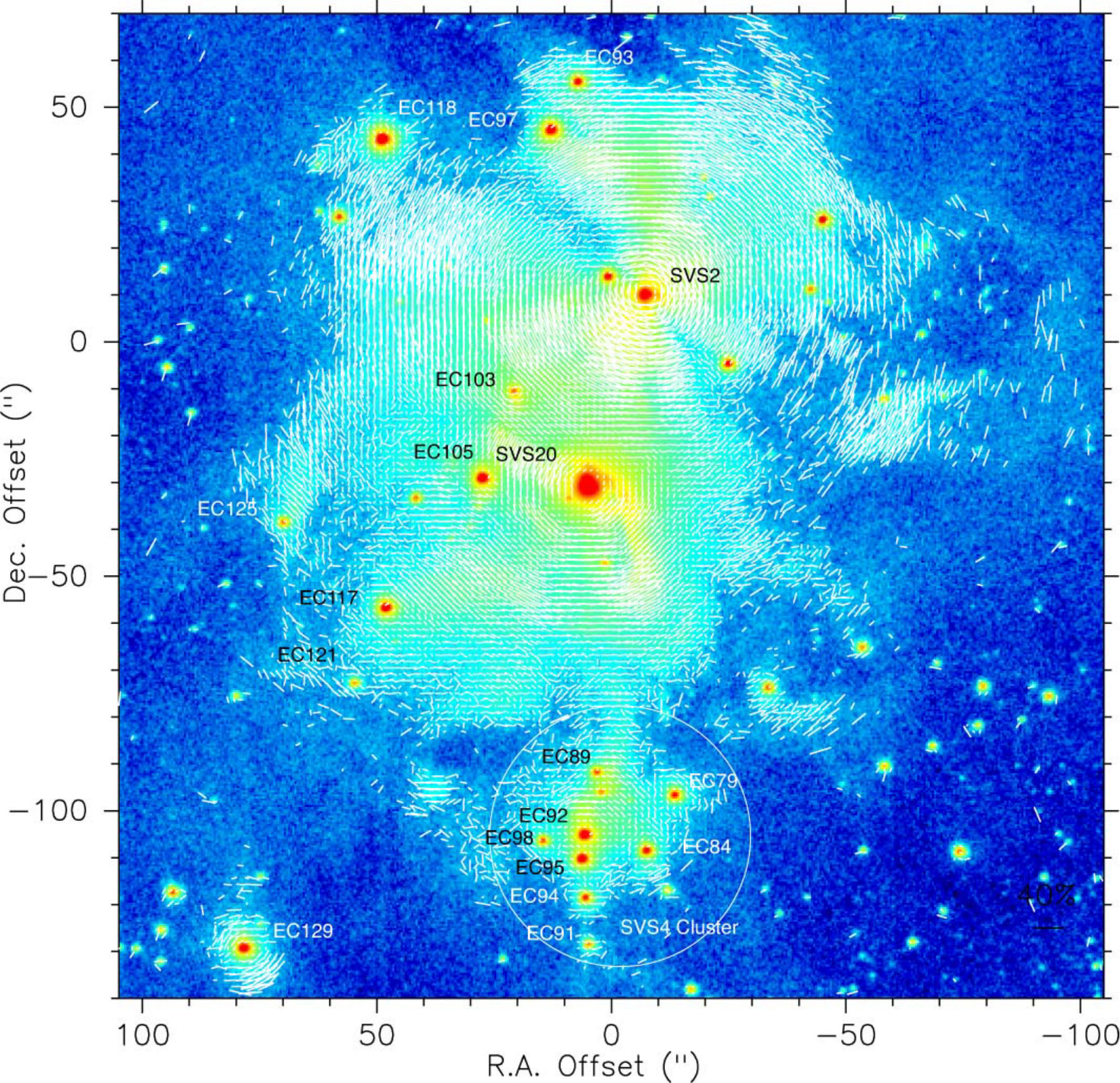}
\caption{Polarization vector map of the central region of the Serpens cloud core, superimposed on the total intensity images in logarithmic scaling.  The area of the image is 220'' $\times$ 220''. {From \citet{sugitani10}.}
}
\label{fig:mag-serpens}
\end{center}
\end{figure*}

\subsubsection{Magnetic field structure}

Magnetic fields are ubiquitously observed in the interstellar gas on all scales \citep{Crutcher03,heiles05}. However, their importance for star formation and for the morphology and evolution of molecular cloud cores remains a source of considerable controversy. A crucial parameter in this debate is the ratio between core mass and magnetic flux. 
In supercritical cores, this ratio exceeds a threshold value and collapse can proceed. In subcritical ones, magnetic fields provide  stability \citep{spitzer78,mouschovias91b,mouschovias91a}. 
Measurements of the Zeeman splitting of molecular lines in nearby cloud cores indicate mass-to-flux ratios that lie above the critical value, in some cases only by a small margin, but very often by factors of many  if non-detections are included \citep{crutcher99, bourke01, crutcher09b, 2010MNRAS.402L..64C}. The polarization of dust emission offers an alternative pathway to studying the magnetic field structure of molecular cloud cores. MHD simulations of turbulent clouds predict degrees of polarization between 1\% and 10\%, regardless of whether turbulent energy dominates over the magnetic energy (i.e.\ the turbulence is super-Alfv\'enic) or not \citep{padoan99,padoan01}. However, converting polarization into magnetic field strength is very difficult \citep{heitsch01b}.  Altogether, the current observational findings imply that magnetic fields must be considered when studying stellar birth, but also that they are not the dominant agent that determines when and where star formation sets in within a cloud. It seems fair to conclude that magnetic fields appear to be too weak to prevent gravitational collapse from occurring.

This  means that in many cases and to a reasonable approximation purely hydrodynamic simulations are sufficient to model ISM dynamics and stellar birth. However, when more precise and quantitative predictions are desired, e.g.\ when attempting to predict star formation timescales or binary properties, it is necessary to perform magnetohydrodynamic (MHD) simulations or even to consider non-ideal MHD. The latter means to take ambipolar diffusion (drift between charged and neutral particles) or Ohmic dissipation into account. Recent numerical simulations have shown that even a weak magnetic field can have noticeable dynamical effects. It can alter how cores fragment \citep{price07a, price08a, hennebelle08a, hennebelle08c, hennebelle11, Peters2011}, change the coupling between stellar feedback processes and their parent clouds \citep{Nakamura07, krumholz07f}, influence the properties of protostellar disks due to magnetic braking \citep{price07b, mellon08b, hennebelle09, seifried2011a, seifried2012a,  seifried2012b, seifried2013}, or slow down the overall evolution \citep{Heitschetal2001}.

\subsection{Statistical properties of stars and star clusters}
\label{subsec:stars}

In order to better understand how gas turns into stars, we also need to discuss here some of the key properties of young stellar systems. We restrict ourselves to a discussion of the star formation timescale, the spatial distribution of young stars, and the stellar initial mass function (IMF). We note, however, that other statistical characteristics, such as the binary fraction, its relation to the stellar mass, and the orbital parameters of binary stars are equally important for distinguishing between different star formation models. As the study of stars and star clusters is central to many areas of astronomy and astrophysics, there are a large number of excellent reviews that cover various aspects of this wide field. For further reading on embedded star clusters, we refer to \cite{ladalada03}. For the early evolution of young star clusters, we point to \citet{Krumholz:2014p85764} and \citet{Longmore:2014p85628}, as well as to \citet{kroupa05} and  \citet{pozwetal10}. More information on the stellar IMF can be found in the seminal papers by  \citet{scalo86}, \citet{kroupa02}, and \citet{chabrier03}, as well as in the reviews by \citet{kroupa13a} and \citet{Offner14}. General reviews of star formation are provided by  \citet{maclow04},  \citet{ballesteros07b}, \citet{mckee07}, \citet{Krumholz:2014p85740}, or \citet{zinnecker07}, with the latter focusing specifically on the formation of high-mass stars.

\subsubsection{Star formation timescales}
\label{par:SF-timescales}

The star formation process in molecular clouds appears to be fast \citep{ Hartmann01, Elmegreen07}. Once the collapse of a cloud region sets in, it rapidly forms an entire cluster of stars within $10^6$ years or less. This is indicated by the young stars associated with star-forming regions, typically T~Tauri stars with ages less than $10^6$ years \citep{gomez92,green95, carpenter97}, and by the small age spread in more evolved stellar clusters \citep{Hillenbrand97, palla99}.  Star clusters in the Milky Way also exhibit an amazing degree of chemical homogeneity \cite[in the case of the Pleiades, see][]{wilden02}, implying that the gas out of which these stars formed must have been chemically well-mixed initially, which could provide interesting pathways to better understand turbulent mixing in the ISM  \cite[see also][]{deavillez02, klessen03b, feng14}.   

\subsubsection{Spatial distribution} 
\label{par:spatial-dist}

The advent of sensitive infrared detectors in the last decade or so allowed us to conduct wide-area surveys. These have revealed that most stars form in clusters and aggregates of various size and mass scales, and that isolated or widely distributed star formation is the exception rather than the rule \citep{ladalada03}. The complex hierarchical structure of molecular clouds (see e.g. Figure \ref{fig:mon-orion-total-intensity})
provides a natural explanation for this finding. 

Star-forming molecular cloud cores can vary strongly in size and mass. In small, low-density clouds, stars form with low efficiency, more or less in isolation or scattered around in small groups of up to a few dozen members. Denser and more massive clouds may build up stars in associations and clusters of a few hundred members.  This appears to be the most common mode of star formation in the solar neighborhood \citep{adams01}. Examples of star formation in small groups and associations are found in the Taurus-Aurigae molecular cloud \citep{hartmann02}. Young stellar groups with a few hundred members form in the Chamaeleon I \citep{persi00} or $\rho$-Ophiuchi \citep{Bontempsetal2001} dark clouds. Each of these clouds is at a distance of about 130 to 160$\,$pc from the Sun.  Like most of the nearby young star forming regions they appear to be associated with a ring-like structure in the Galactic disk called Gould's Belt \citep{poeppel97}.

\begin{figure}
\center{\includegraphics[width=0.90\textwidth]{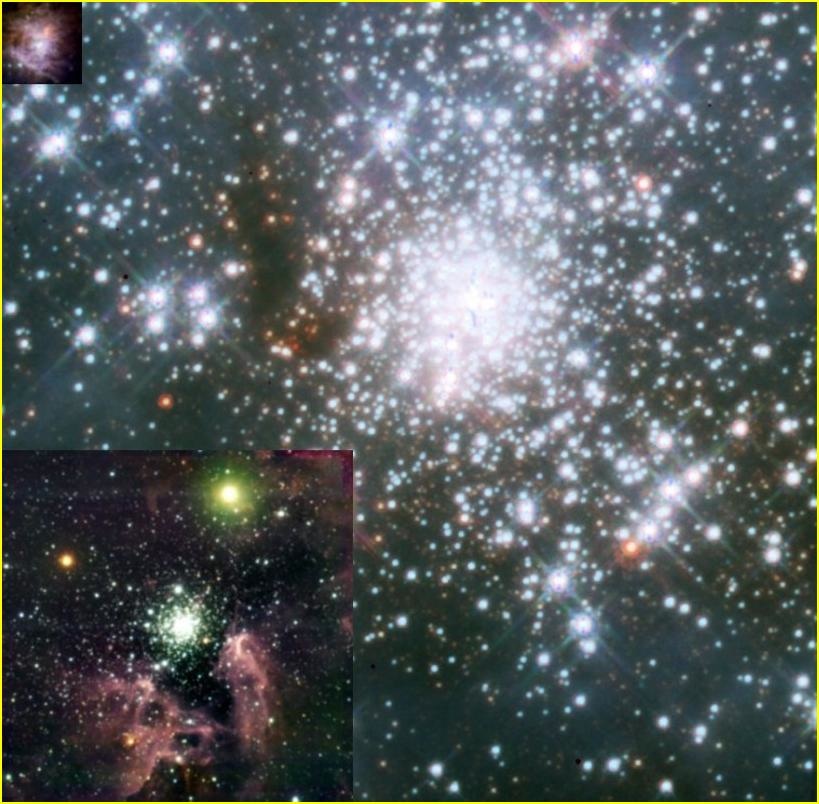}}
\caption{\label{fig:clusters}
Comparison of clusters of different masses scaled to same relative distance. 
The cluster in the upper left corner is the Orion Nebula Cluster \citep{mccaughrean01} and the one at the lower left is NGC~3603 \citep{brandl99}, both observed with the Very Large Telescope at infrared wavelength. The large cluster in the center is 30 Doradus in the LMC observed with the Hubble Space Telescope (courtesy of M.\ J.\ McCaughrean). The total mass increases roughly by a factor of ten from one cluster to the other. Image from \citet{zinnecker07}.
}
\end{figure}

The formation of dense rich clusters with thousands of stars is rare. The closest region where this happens  is the Orion Nebula Cluster  \citep{Hillenbrand97,hillenbrand98}. It lies at a distance of $410\,$pc \citep{sandstrom07, menten07, hirota07, caballero08}.  A rich cluster somewhat further away is associated with the Monoceros R2 cloud \citep{carpenter97} at a distance of $\sim 830\,$pc.  The cluster NGC~3603 is roughly ten times more massive than the Orion Nebula Cluster.  It lies in the Carina region, at about $7\,$kpc distance. It contains about a dozen O stars, and is the nearest object analogous to a starburst knot \citep{brandl99,moffat02}. To find star-forming regions building up hundreds of O stars one has to look towards giant extragalactic H\textsc{ii} regions, the nearest of which is 30 Doradus in the Large Magellanic Cloud, a satellite galaxy of our Milky Way at a distance at 55$\,$kpc. The giant star-forming region 30 Doradus is thought to contain up to a hundred thousand young stars, including more than 400 O stars \citep{hunter95,walborn97,townsley06}. Figure \ref{fig:clusters} shows that the star formation process spans many orders of magnitude in spatial scale and mass, ranging from stellar groups with no or only a few high-mass stars to massive clusters with several tens of thousands of stars and dozens if not hundreds of O stars. This variety of star-forming regions appears to be controlled by the competition between self-gravity and opposing agents such as the turbulence in the parental gas clouds, its gas pressure and magnetic field content.

\subsubsection{Observations of the stellar IMF}
\label{par:IMF}

Mass is the most important parameter determining the evolution of individual stars. The luminosity $L$ of a star scales as a very steep function of the mass $M$. The relation $L\propto M^{3.5}$ provides a reasonable estimate except for very low-mass stars and very massive ones  \citep{kippenhahn12}. Stars on the main sequence generate energy by nuclear fusion. The total energy available is some fraction of $M c^2$, with $c$ being the speed of light. Consequently, we can estimate the main sequence lifetime as $t \propto M/L$ or $t \propto M^{-2.5}$. Massive stars with high pressures and temperatures in their centers convert hydrogen into helium very efficiently. This makes them very luminous but also short-lived. Conversely, low-mass stars are much fainter but long-lived. 

Explaining the distribution of stellar masses at birth, the so-called initial mass function (IMF), is a key prerequisite to any theory of star formation. The IMF has three properties that appear to be relatively robust in diverse environments (see Figure \ref{fig:IMF}). These are the power law behavior $dN/dM \propto M^{-\alpha}$ with slope $\alpha \approx 2.3$ for masses $M$  above about  $1\,$M$_\odot$, originally determined by \citet{Salpeter1955}, the lower mass limit for the power law and the broad plateau below it before the brown dwarf regime \citep{miller79, scalo86}, and the maximum mass of stars at around $100\,$M$_\odot$  \citep{weidner04,weidner06,oey05}. Comprehensive reviews of cluster and field IMFs may be found in \citet{scalo86}, \citet{kroupa02},  \citet{chabrier03}, \citet{Bastian10}, \citet{kroupa13a}, and \citet{Offner14}.

\begin{figure}[t]
\center{\includegraphics[width=0.60\textwidth]{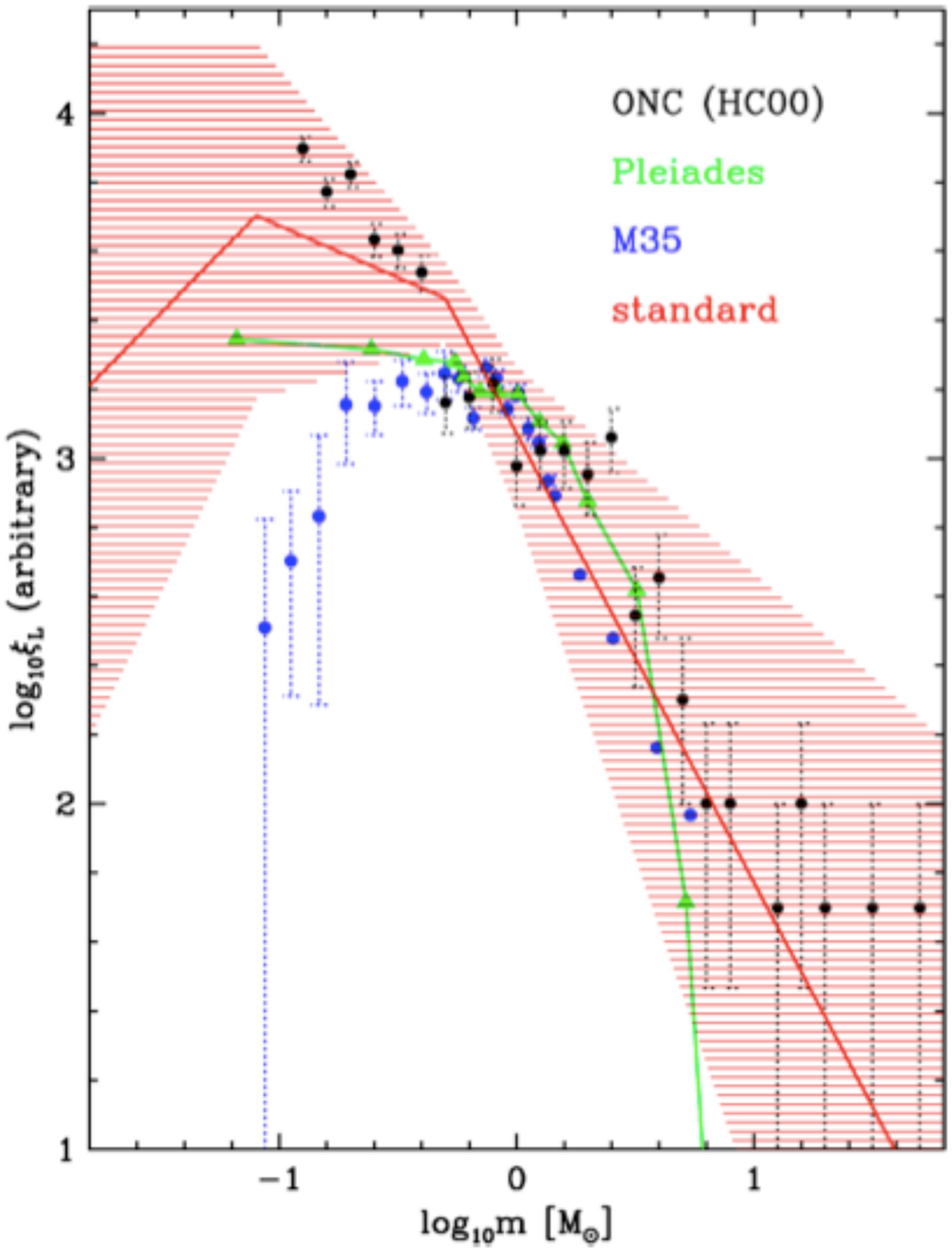}}
\parbox[b]{0.36\textwidth}{\vspace{2.0cm}
\caption{\label{fig:IMF}
Stellar mass spectrum in different nearby clusters (black symbols: Orion Nebula Cluster, green: Pleiades, blue: M35) and its description by a three-component power law (red lines with overall uncertainties indicated by the hatched region). {From \citet{kroupa02}.}\vspace*{1.2cm}}} 
\end{figure}

There are two widely accepted functional parameterizations of the IMF. The first one is based on the continuous combination of multiple power-law segments. It was proposed by \citeauthor{kroupa01} (\citeyear{kroupa01}, \citeyear{kroupa02}), and introducing the dimensionless mass $m = M / 1{\rm M}_{\odot}$, it reads
\begin{eqnarray} 
f(m ) & = &  \begin{cases}
\begin{array}{l@{~~~~{\rm for}~~}r@{~ }c@{ ~}c@{~ }r@{}l}
A k_0 \,m^{- 0.3} &  0.01 & < m &<& 0&.08\;,  \\
A k_1 \,m^{- 1.3} &  0.08  &< m& < & 0&.5\;,   \\
A k_2 \,m^{- 2.3} &  0.5 &< m\;, & \multicolumn{3}{c}{ ~}
\end{array}
\end{cases}
\label{eqn:Kroupa-IMF}
\end{eqnarray}
where $A$ is a global normalization factor, and $k_0 = 1$, $k_1 = k_0 m_1^{-0.3 + 1.3}$, and $k_2 = k_1 m_2^{-1.3 + 2.3}$ are chosen to provide a continuous transition between the power-law segments at $m_1=0.08$ and $m_2=0.5$. The quantity $f(m)dm$ denotes the number of stars in the mass interval $[m, m+dm]$. 
 A method to calculate $k_i$ is provided by \citet{PflammAltenburg2006}; see also \citet{maschberger2013a}. 

Another parameterization is suggested by  \citet{chabrier03}. It combines a log-normal with a power-law, 
\begin{eqnarray} 
f(m) =  \begin{cases}
\begin{array}{l@{~~~{\rm for}~~}r@{\ }c@{\ }l}
A k_1 {m^{-1}} \exp\left[-  \dfrac{1}{2} \left(  \dfrac{ \log_{10} m - \log_{10} 0.079}{0.69}\right)^2 \right]
 &  m  & <  1\;,  \\
A  k_2 m^{-2.3} & m & >    1\;.
\end{array}
\end{cases}
\label{eqn:Chabrier-IMF}
\end{eqnarray}
Again $A$ is a global normalization factor, and $k_1 = 0.158$ and $k_2 = 0.0443$ provide a continuous connection at $m=1$ \citep{chabrier03b, chabrier05, maschberger2013a}. Equation (\ref{eqn:Kroupa-IMF}) is easier to integrate than  (\ref{eqn:Chabrier-IMF}), as this does not involve special functions. On the other hand it has several kinks. Both converge to the \citet{Salpeter1955} power law with a slope of  $-2.3$ for large masses. They differ by about a factor of 2 at low masses. However, within the observational errors, both functional forms are more or less equivalent. 

We need to point out that the observational knowledge of the IMF is quite limited at the extreme ends of the stellar mass spectrum. Because massive stars are very rare and short-lived, only very few are sufficiently near to study them in detail and with very high spatial resolution, for example to determine  multiplicity \citep{zinnecker07}. We do not even know what is the upper mass limit for stability, both in terms of observations as well as theoretical models \citep{massey03, vink2013}. In addition, there is evidence that the upper mass end of the IMF depends on the properties of the cluster where it is measured. The upper mass limit in more massive clusters seems to be higher than in lower-mass  clusters, an effect that goes beyond the statistical fluctuations expected for purely random sampling from a universal distribution \cite[see e.g.][]{weidner04, weidner06, weidetal10}. 

At the other end of the IMF, low-mass stars and brown dwarfs are faint, so they too are difficult to study in detail \citep{burrows01}. Such studies, however, are in great demand, because secondary indicators such as the fraction of binaries and higher-order multiples as a function of mass, or the distribution of disks around very young stars and possible signatures of accretion during their formation are probably better suited to distinguish between different star formation models than just looking at the IMF \cite[e.g.][]{goodwin2005a, marks12}.

\subsection{Gravoturbulent star formation}
\label{subsec:gravoturb-SF}

The past decade has seen a paradigm shift in low-mass star formation theory \citep{maclow04, mckee07, Offner14}. The general belief since the 1980's was that prestellar cores in low-mass star-forming regions evolve quasi-statically in magnetically subcritical clouds \citep{shu87}. In this picture, gravitational contraction is mediated by ambipolar diffusion \citep{mouschovias76, mouschovias79, mouschovias81,mouschovias91b} causing a redistribution of magnetic flux until the inner regions of the core become supercritical and go into dynamical collapse.  This process was originally thought to be slow, because in highly subcritical clouds the ambipolar diffusion timescale is about 10 times larger than the dynamical one. However, for cores close to the critical value, as is suggested by observations, both timescales are comparable. Numerical simulations furthermore indicate that the ambipolar diffusion timescale becomes significantly shorter for turbulent velocities similar to the values observed in nearby star-forming region  \citep{FatuzzoAdams2002,heitsch04,zli04}. The fact that ambipolar diffusion may not be a slow process under realistic cloud conditions, as well as the fact that most cloud cores are magnetically supercritical \citep{crutcher99a,crutcher00,bourke01, crutcher09b}  has cast significant doubts on any magnetically-dominated quasi-static models of stellar birth. For a more detailed account of the shortcomings of the quasi-static, magnetically dominated star formation model, see \cite{maclow04}.

For this reason, star formation research has turned to considering supersonic turbulence as being one of the primary physical agents regulating stellar birth. The presence of turbulence, in particular of supersonic turbulence, has important consequences for molecular cloud evolution \cite[see e.g.][]{Padoan:2013p88872, dobbs14}. On large scales it can support clouds against contraction, while on small scales it can provoke localized collapse. Turbulence establishes a complex network of interacting shocks, where dense cores form at the stagnation points of convergent flows. The density can be large enough for gravitational collapse to set in. However, the fluctuations in turbulent velocity fields are highly transient.  The random flow that creates local density enhancements can disperse them again.  For local collapse to actually result in the formation of stars, high density fluctuations must collapse on timescales shorter than the typical time interval between two successive shock passages.  Only then are they able to decouple from the ambient flow and survive subsequent shock interactions.  The shorter the time between shock passages, the less likely these fluctuations are to survive. Hence, the timescale and efficiency of protostellar core formation depend strongly on the wavelength and strength of the driving source \citep{klessen00b,Heitschetal2001,Vazquez03,maclow04,krumholz05c,ballesteros07b,mckee07}, and accretion histories of individual protostars are strongly time-varying \citep{Klessen2001b,SchmejaKlessen2004}.

\begin{figure}
\center{\includegraphics[width=1.0\textwidth]{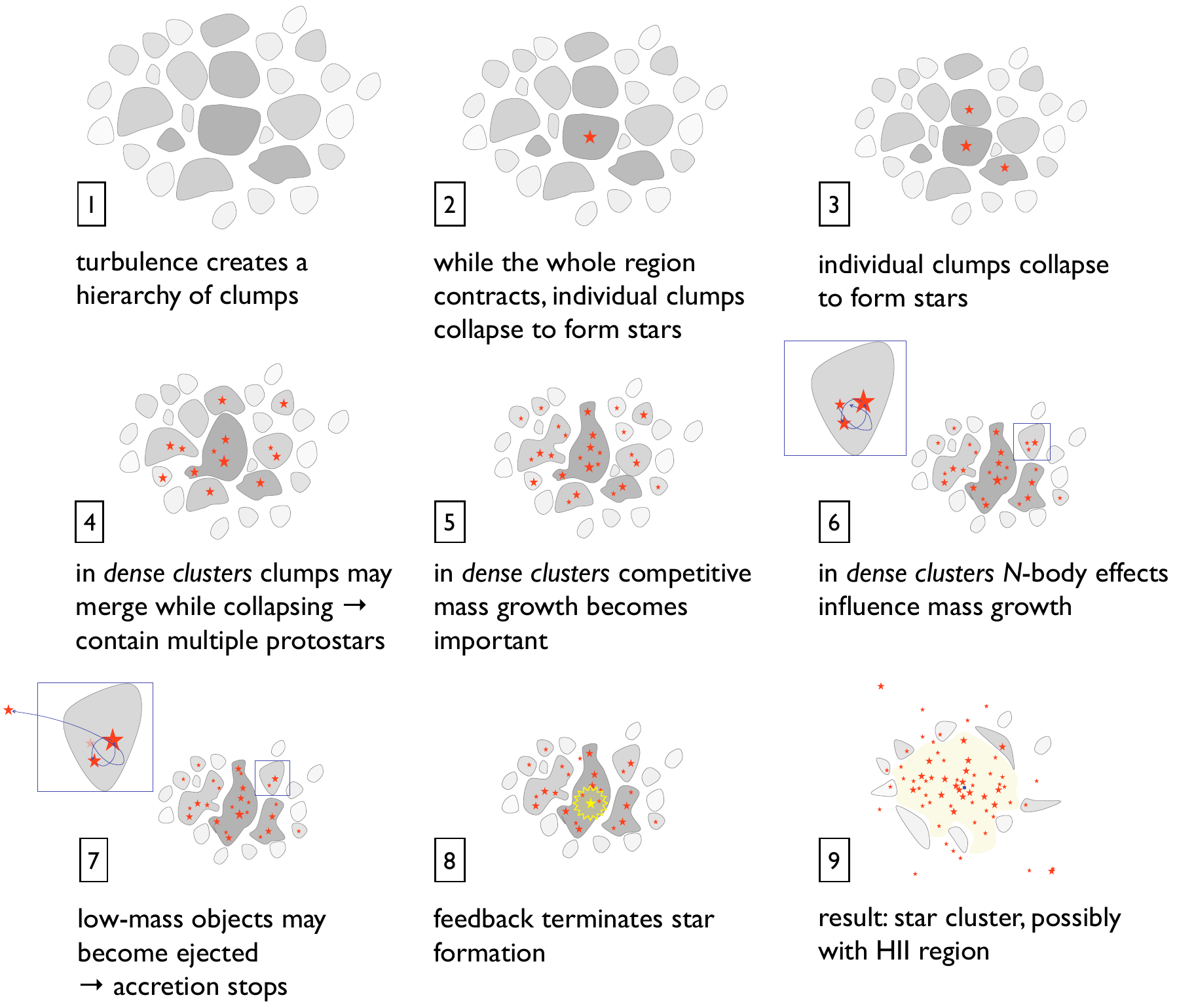}}
\caption{\label{fig:sequence}
Cartoon picture of star cluster formation in a molecular cloud core.  From \cite{klessen11b}.}
\end{figure}

Altogether, we propose an evolutionary  sequence as outlined in Figure \ref{fig:sequence}. Star cluster formation takes place in massive cloud cores of several $10^2$ to $10^3$ solar masses with sizes of a few parsecs and a velocity dispersion of about $1\,$km$\,$s$^{-1}$ (see also Table \ref{tab:mol-clouds}).  In order to form a bound cluster, the potential energy must dominate the energy budget, meaning that the entire region is contracting. The cluster-forming massive cloud cores are still in the supersonic range of the turbulent cascade (see Section \ref{subsubsec:ISM-scales}), and as a consequence they exhibit a high degree of internal substructure with large density contrasts. Some of these density fluctuations are gravitationally unstable in their own right and begin to collapse on timescales much shorter than the global contraction time, as the free-fall time $\tau_{\rm ff}$ scales with the density $\rho$ as $\tau_{\rm ff} \propto \rho^{-1/2}$.

Typically, the most massive fluctuations have the highest density and form a protostar in their center first. This nascent star can accrete from the immediate environment, but because it is located in a minimum of the cloud core's gravitational potential more gas flows towards it, and it can maintain a high accretion rate for a longer time. In contrast, stars that form in lower-mass gas clumps typically can only accrete material from their immediate surrounding and not much beyond that \cite[see e.g.][]{klessen00a,klessen00b,klessen01,bonnell04}. Because this preferentially happens in the cluster outskirts, these processes naturally lead to mass segregation as we often observe in young clusters \cite[see e.g.][for the Orion Nebula Cluster]{Hillenbrand97,hillenbrand98}. In very dense clusters, there is the possibility that clumps merge while still accreting onto their central protostars. These protostars now compete for further mass growth from a common gas reservoir. This gives rise to collective phenomena which can strongly modify the accretion behavior and hence influence the resulting mass spectrum (see Section \ref{subsec:IMF-models} below).

Once a star has reached a mass of about $10\,$M$_{\odot}$, it begins to ionize its environment. It carves out a bubble of hot and tenuous gas, which eventually will expand and enclose the entire stellar cluster. At this point no new stars can form and stellar birth has come to an end. We can observe the young cluster at infrared or even optical wavelengths, as illustrated in Figure \ref{fig:clusters}.

\subsection{Theoretical models for the origin of the IMF}
\label{subsec:IMF-models}

There are three principal pathways towards better understanding the origin of the IMF, depending on which aspects of gravitational collapse in the turbulent ISM one decides to focus on. We begin by introducing the underlying physical concepts behind the three different models in a qualitative way. Then we follow a more rigorous approach and discuss the most popular theoretical models for the IMF in some mathematical detail. We point out that the boundaries between these approaches are not clearly defined and that numerous hybrid models have been proposed in the current literature. 

It turns out that essentially all theoretical models that are able to reproduce the IMF rely on two basic ingredients. They propose that the stellar mass spectrum is determined by a sequence of stochastic processes (such as turbulence or the probabilistic nature of stellar encounters in dense clusters) combined with scale-free physics (again, as provided by the power-law nature of the turbulent energy cascade or by the simple distance dependence of gravitational interactions). The former leads to a log-normal mass spectrum, the latter to a power-law contribution. Put together, they constitute one of the most popular parameterizations of the IMF \cite[e.g.][]{chabrier03}. 

\subsubsection{Basic concepts and caveats}

Here we introduce the three basic physical concepts behind the IMF.

\paragraph{\bf Core accretion} 
\label{subsubsec:core-accretion}

This model takes as its starting point the striking similarity between the shape of the observed core mass distribution and the IMF. It is based on the assumption of a one-to-one relation between the two distributions, such that the observed cores are the direct progenitors of individual stars or binary systems. The factor of $\sim\,$3 decrease in mass between cores and stars is thought to be the result of feedback processes, mostly protostellar outflows, that eject a fixed fraction of the mass in a core rather than letting it accrete onto the star \citep{matzner00}. This model reduces the problem of the origin of the IMF to understanding the mass spectrum of bound cores. Arguments to explain the core mass distribution generally rely on the statistical properties of turbulence. Its scale-free nature leads to a power-law behavior for high masses. The thermal Jeans mass in the cloud then imposes the flattening and turn-down in the observed mass spectrum. For further discussion, see Section \ref{subsubsec:turb-IMF}.

\paragraph{\bf Collective models}
\label{subsubsec:coll-models}

A second line of reasoning accounts for the fact that stars almost always form in clusters, where the interaction between protostars, as well as between a protostellar population and the gas cloud around it may become important. In these collective models, the origin of the peak in the IMF is much the same as in the core accretion model: it is set by the Jeans mass in the prestellar gas cloud. However, rather than fragmentation in the gas phase producing a spectrum of core masses, each of which collapses down to a single star or star system, the final stellar mass spectrum  in the collective accretion model is the result of the mutual interaction between the protostars in a cluster during their main accretion phase. 

In the original competitive accretion picture \citep{bonnell01a, bonnell01b, bonnell02, bate03} protostars start out with roughly the same small mass close to the opacity limit of fragmentation \citep{rees76}. These protostars then compete with each other for mass accretion from the same reservoir of gas. In a simple Bondi-Hoyle-Lyttleton accretion scenario \cite[e.g.][]{Bondi:1952p13808}, the accretion rate scales as the square of the protostellar mass ($dM/dt \propto M^2$). That means small initial differences in mass quickly amplify and leads to a run-away growth of a few selected objects. The original idea of putting roughly equal-mass protostellar seeds into an pre-existing gas reservoir was later extended by taking the original cloud fragmentation process into account \citep{klessen1998, klessen00a, klessen01, bate05, bonnell06d, bonnell08a}. The fragmentation down to the local Jeans scale creates a mass function that lacks the power law tail at high masses that we observe in the stellar mass function. This part of the distribution forms via a second phase in which protostars with initial masses close to the Jeans mass compete for gas in the center of a dense cluster. The cluster potential channels mass toward the center, so stars that remain in the center grow to large masses, while those that are ejected from the cluster center by $N$-body interactions remain low mass \citep{bonnell04}. 

The fact that fragmentation and the formation of multiple protostars strongly influence the subsequent accretion flow in the entire cluster also allows for a different interpretation. \citet{Peters2010b} and \citet{Girichidis:2012p55106} point out that the processes described above limit the accretion of gas onto the central protostars in a cluster. They find that the gas flowing towards the potential minimum at the center of the cluster is efficiently accreted by protostars that are located at larger radii. As a consequence, the central region is effectively shielded from further accretion and none of the central objects can sustain its initially high accretion rate for a very long time. The fact that the gas fragments into a cluster of stars limits the mass growth onto the central object, which would otherwise have the available gas reservoir all for itself (as in the core accretion model described before). The gas flow towards the cluster center is reduced due to the efficient shielding by secondary protostars. Consequently, this process has been termed fragmentation-induced starvation \cite[][]{petersetal10a, Girichidis:2011p44985, Girichidis:2012p55333}. In these collective models, the apparent similarity between the core and stellar mass functions is an illusion, because the observed cores do not map the gas reservoir that is accreted by the stars \citep{clark06, smith08a}.

\paragraph{\bf Importance of the thermodynamic behavior of the gas}
\label{subsubsec:thermodynamics}

One potential drawback to both the core accretion and collective  models is that they rely on the Jeans mass to determine the peak of the IMF, but do not answer the question of how to compute it. This is subtle, because molecular clouds are nearly isothermal but at the same time contain a very wide range of densities. At a fixed temperature, the Jeans mass scales as $M_{\rm J} \propto \rho^{-1/2}$, and it is not obvious what value of the density should be used to calculate $M_{\rm J}$. A promising idea to resolve this problem forms the basis for a third model of the IMF. It  focuses on the thermodynamic properties of the gas.  The amount of fragmentation occurring during gravitational collapse depends on the compressibility of the gas \citep{li03}.  For a polytropic equation of state (\ref{eqn:EOS}) with an index $\gamma < 1$, the gas reacts very strongly to pressure gradients. Turbulent compression can thus lead to large density contrasts. The local Jeans mass (equation \ref{eqn:Jeans}) drops rapidly and many high-density fluctuations in the turbulent flow become gravitationally unstable and collapse. On the other hand, when $\gamma > 1$, compression leads to heating and turbulence can only induce small density variations. As the gas heats up, the decrease in the Jeans mass in the compressed gas is much smaller. Indeed, for $\gamma > 4/3$, compression actually results in an increasing Jeans mass. In addition, \citet{larson05} argues that $\gamma = 1$ is a critical value, because filaments in which $\gamma < 1$ are unstable to continued gravitational collapse, while those with $\gamma > 1$ are stabilized against collapse and hence cannot decrease their Jeans mass to very small values. In real molecular clouds, the effective polytropic index varies significantly as the gas density increases. At low densities, $\gamma \approx 0.7$ \citep{larson85,larson05,gc12a}, but once the gas and dust temperatures become thermally coupled at $n_{\rm crit} \approx 10^5\,$cm$^{-3}$ (see Figure \ref{fig:dust-gas-coupling} and the discussion in Section \ref{subsec:gas-grain-transfer}), one expects this value to increase, reaching $\gamma \approx 1.1$ at densities $n \gg n_{\rm crit}$ \citep{ban06}. This suggests that fragmentation will tend to occur at densities $n \approx n_{\rm crit}$, and that the Jeans mass evaluated at this point sets the mass scale for the peak of the IMF. In this model, the apparent universality of the IMF in the Milky Way and nearby galaxies is then a result of the insensitivity of the dust temperature to the intensity of the interstellar radiation field \citep{elmegreen08}.  Not only does this mechanism set the peak mass, but it also appears to produce a power-law distribution of masses at the high-mass end comparable to the observed distribution \citep{Jappsen05}.

\paragraph{\bf Caveats} 
\label{subsubsec:caveats}

Each of these models has potential problems. In the core accretion picture, hydrodynamic simulations seem to indicate that massive cores should fragment into many stars rather than collapsing monolithically \citep{dobbs05, clark06, Bonnell06}. The hydrodynamic simulations often suffer from over-fragmentation because they do not include radiative feedback from embedded massive stars \citep{krumholz06b,krumholz07a,krumholz08a}. The suggestion of a one-to-one mapping between the observed clumps and the final IMF is subject to strong debate, too. Many of the prestellar cores discussed in Section \ref{subsec:cores} appear to be stable entities \citep{johnstone00,johnstone01,Johnstoneetal2006,lada08a}, and thus are unlikely to be in a state of active star formation. In addition, the simple interpretation that one core forms on average one star, and that all cores contain the same number of thermal Jeans masses, leads to a timescale problem (\citealt{clark07}; see also the discussion in the last paragraph of Section \ref{subsubsec:turb-IMF}). Its solution actually requires a difference between the core mass function and the stellar IMF. We also note that the problems associated with neglecting radiative feedback effects also apply to the gas thermodynamic idea. The assumed cooling curves typically ignore the influence of protostellar radiation on the temperature of the gas, which simulations show can reduce fragmentation \citep{krumholz07a, commercon2011, Peters2010b,  Peters2011}. The collective accretion picture has also been challenged, on the grounds that the kinematic structure observed in star-forming regions sometimes appears  inconsistent with the idea that protostars have time to interact with one another strongly before they completely accrete their parent cores \citep{Andre:2007p25886}. For a comprehensive overview of the big open questions in star formation theory, see \citet{Krumholz:2014p85740}.

\subsubsection{IMF from simple statistical theory}

In the previous section, we discussed various models for the origin of the stellar mass function based on a range of different physical processes. Here we approach the problem from a purely statistical point of view without specifying up front which of these processes will become dominant. We consider the distribution of stellar masses as the result of a sequence of independent stochastic processes. Invoking the central limit theorem then naturally leads to a log-normal IMF \cite[for early discussions, see][]{zinnecker84, adams96}. The key assumption is that the mass $M$ of a star can be expressed as the product of $N$ independent variables $x_j$. At this point it is not necessary to specify these variables, as long as they are statistically independent and their values are determined by stochastic processes. We introduce again the dimensionless mass variable $m = M/(1\;\!{\rm M}_{\odot})$ and write
\begin{equation}
\label{eqn:mass-function}
  m = \prod_{j=1}^N x_j\;.
\end{equation}
Taking the logarithm of this equation, the logarithm of the mass is a sum of the random variables,
\begin{equation}
  \ln m = \sum_{j=1}^N \ln x_j +{\rm constant}\;,  
\end{equation}
where the constant term includes all quantities that are truly constant, e.g.\ the gravitational constant $G$ or Boltzmann's constant $k_{\rm B}$ or others. The central limit theorem shows that the distribution of the composite variable $\ln m$ always approaches a normal distribution as the number $N$ of variables approaches infinity \citep{bronstein}. For the application of the theorem, a transformation into normalized variables $\xi_j$ is useful, which are given by
\begin{equation}
  \label{eqn:new-variables}
  \xi_j \equiv \ln x_j -\left< \ln x_j \right> \equiv \ln \left( \frac{x_j}{\bar{x}_j}\right)\;.
\end{equation}
The angle brackets denote averages taken over the logarithm of the  variables,
\begin{equation}
  \ln \bar{x}_j = \left< \ln x_j \right> = \int_{-\infty}^{\infty}
  \ln x_j f_j(\ln x_j) d\ln x_j\;.
\end{equation}
Here, $f_j$ is the distribution function of the variable $x_j$.The normalized variables $\xi_j$ have zero mean and their dispersions $\sigma_j$  are given by
\begin{equation}
  \sigma^2_j = \int_{-\infty}^{\infty} \xi^2_jf_j({\xi_j})d\xi_j\;.
\end{equation}
We can define the new composite variable $\Xi$ as
\begin{equation}
  \Xi \equiv \sum_{j=1}^N \xi_j = \sum_{j=1}^N  \ln \left( \frac{x_j}{\bar{x}_j}\right)\;.
\end{equation}
It also has zero mean and, since the variables are assumed to be independent, it follows that
\begin{equation}
  \Sigma^2 = \sum_{j=1}^N  \sigma^2_j\;.
\end{equation}
For $N\rightarrow \infty$, the central limit theorem describes its distribution function as being  Gaussian with
\begin{equation}
  \label{eqn:central-limit-theorem}
  f({\Xi}) = (2 \pi \Sigma^2)^{-1/2} \exp\left(-\frac{1}{2}\frac{{\Xi}^2}{\Sigma^2}\right)\;,
\end{equation}
independent of the distribution $f_j$ of the individual variables $x_j$. The mass function
(\ref{eqn:mass-function}) then can be expressed as  
\begin{equation}
  \label{eqn:mass-formula}
  \ln m = \ln m_0 + {\Xi}\;,
\end{equation}
with $m_0$ being a characteristic mass scale defined by
\begin{equation}
  \ln m_0 \equiv \sum_{j=1}^N \left< \ln \bar{x}_j \right>\;.
\end{equation}
Combining the two equations (\ref{eqn:central-limit-theorem}) and (\ref{eqn:mass-formula}), we can write the distribution $f$ of stellar masses
  in the form
  \begin{equation}
    \label{eqn:IMF}
    \ln f(\ln m) = A - \frac{1}{2\Sigma^2} \left[\ln
    \left(\frac{m}{m_0}\right) \right]^2\;,
  \end{equation}
where $A$ is a constant. This is the log-normal form of the IMF first introduced by \citet{miller79}. It fits very well the mass distribution of multiple stellar systems in the solar vicinity with masses less than few solar masses \cite[e.g.][]{kroupa90, kroupa91} and it is often used to describe the peak of the single star IMF (Section \ref{par:IMF}).

\subsubsection{IMF from stochastically varying accretion rates}
\label{subsubsec:IMF-accretion}
To obtain the observed power-law behavior at the high-mass end of the IMF, we need to add  complexity to the model and extend this simple statistical approach. As a highly illustrative example, we follow the discussion provided by \citet{maschberger2013} and consider the case where the stellar mass is determined by accretion in a stochastically fluctuating medium. If we disregard the fluctuating part for the time being, and if we assume that the accretion rate depends on the mass to some power of $\alpha$, then the growth of an individual star can be described by the simple differential equation, 
\begin{equation}
dm = m^\alpha A dt\;,
\label{eqn:mass-growth}
\end{equation}
where the constant $A>0$ and the exponent $\alpha$ account for all physical processes involved. For example, if the protostars move with constant velocity $v$ through isothermal gas with temperature $T$ and sound speed $c_{\rm s} = (k_{\rm B} T / \mu)^{1/2}$ with Boltzmann constant $k_{\rm B}$ and mean particle mass $\mu$ (in grams), we can apply the Bondi-Hoyle-Lyttleton accretion formula \cite[e.g.][]{Bondi:1952p13808} to obtain
\begin{eqnarray}
 A & = & \frac{2 \pi G^2 \rho}{(v^2 + c_{\rm s}^2)^{3/2}}\;,\nonumber \\
 \alpha &=& 2\;. \nonumber 
\end{eqnarray}
Note that this is an approximate formula. Replacing the factor $2\pi$ by $4\pi$ gives a better fit to the Hoyle-Lyttleton rate \citep{Hoyle:1939p89066}, where the object moves highly supersonically and we can neglect the contribution of the sound speed. Detailed numerical simulations  yield  a more complex parameterization of $A$, depending on the physical parameters of the system \cite[e.g.][]{Ruffert:1994p89070, krumholz06a}. What remains, however, is the quadratic power-law dependence of the accretion rate on the mass. 

Now assume that the star grows with a statistically fluctuating mass accretion rate. This could be due to the stochastic nature of gas flows in turbulent media, and/or due to $N$-body dynamics in dense embedded clusters  \cite[e.g.][]{bonnell01a, bonnell01b, Klessen2001b}, or due to  other processes that lead to stochastic protostellar mass growth. In this case, equation (\ref{eqn:mass-growth}) turns into a stochastic differential equation, 
\begin{equation}
dm = m^\alpha (A dt + B dW)\;,
\label{eqn:stochastic-mass-growth}
\end{equation}
where $A dt$ describes the mean growth rate and $B dW$ the fluctuations around this mean. Depending on the statistical properties of $B dW$ the sum $Adt + B dW$  could become negative, which would imply mass loss and could potentially lead to negative masses. In order to avoid that, it is often sensible to restrict the stochastic variable $B dW$ to positive values or to very small amplitudes. For  $\alpha \ne 1$ (as well as for $\alpha \ne 0$) we obtain the formal solution,
\begin{equation}
m(t) = \left[ (1-\alpha) \left( \frac{m_0^{1-\alpha}}{1-\alpha} + At + BW\!(t)
\right) \right]^{\frac{1}{1-\alpha}}\;,
\label{eqn:mass-solution}
\end{equation}
where the integration constant $m_0$ is the initial mass and $W\!(t) = \int_0^t dW$ is the  integral of the stochastic variable. For Gaussian fluctuations, its distribution has zero mean and variance $t$, as is well known from the random walk problem. \citep[For a more detailed discussion, see e.g.][]{oeksendal00}. To get a mass spectrum, many realizations of the random variable need to be considered. For $B=0$, equation (\ref{eqn:mass-solution}) reduces to the solution of the deterministic growth problem, which for $\alpha >1$  reaches infinite mass in the finite time 
\begin{equation}
t_{\infty} = \frac{m_0^{1-\alpha}}{A(\alpha-1)}\;.
\end{equation}
For $B\ne 0$, the time $t_{\infty}$ no longer takes on a single value, but instead depends on the stochastic path $W\!(t)$. In reality, this solution is not desired, because the mass of a cloud core is limited. In addition, once  feedback from massive stars sets in, the local reservoir of gas available for star formation is reduced even further. Consequently, the solution (\ref{eqn:mass-solution}) makes  sense only for  $t \ll t_{\infty}$. 

If we know the statistical properties of the random process $W\!(t)$, we can calculate the mass spectrum for an ensemble of stars. For Gaussian fluctuations with zero mean and variance $t$, we obtain
\begin{equation}
f(m,t) = \frac{1}{(2 \pi)^{1/2}}\frac{1}{m^\alpha}\frac{n_{\infty}(t)}{Bt^{1/2}} \exp \left[ -\frac{1}{2 B^2 t} \left( \frac{m^{1-\alpha} - m_0^{1-\alpha} }{1-\alpha} - At\right)^2 \right]\;,
\label{eqn:maschberger-mass-spectrum}
\end{equation}
where $f(m,t)dm$ gives the fraction of stars in the mass range $m$ to $m+dm$. The factor $n_\infty(t)$ corrects for the possible contributions of stars with masses approaching infinity for $\alpha > 1$. It needs to be introduced to ensure the normalization of $f(m,t)dm$ as a probability distribution function at any time $t$ \cite[for further details, see][]{maschberger2013}. For $\alpha < 1$, we set $n_\infty(t)$  to unity. For $\alpha \rightarrow 1$, i.e.\ for average exponential growth, we obtain the log-normal distribution function motivated before (see equation \ref{eqn:IMF}),
\begin{equation}
f(m,t) = \frac{1}{(2 \pi)^{1/2}}\frac{1}{m}\frac{1}{Bt^{1/2}} \exp \left[ -\frac{1}{2 B^2 t} \left( \log m - \log m_0 - At\right)^2 \right]\;,
\label{eqn:log-normal-mass-spectrum}
\end{equation}
where the factor $1/m$ is due to the conversion from $\log m$ to $m$, as $d \log m = dm/m$. We plot the two cases $\alpha = 1$ and $\alpha = 2.3$ in Figure \ref{fig:stochastic-IMF}. The log-normal distribution function (left) peaks at the mass $m_0$. For all values $\alpha >1$ the function $f(m,t)dm$ (at the right) develops a power-law tail at large masses $m$ with slope $\alpha$, reaches a maximum slightly below $m_0$,  and exhibits a sharp decline for small masses. The case $\alpha = 2.3$ is therefore very similar to the observed stellar IMF, as discussed in Section \ref{par:IMF}. In short, the power-law tail traces the accretion behavior, while the log-normal part of the spectrum comes from the intrinsic stochasticity of the process. 
\begin{figure}[t]
\unitlength1.0cm
\begin{picture}(14,4)
\put(0.2,0.0){\includegraphics[width=0.98\textwidth]{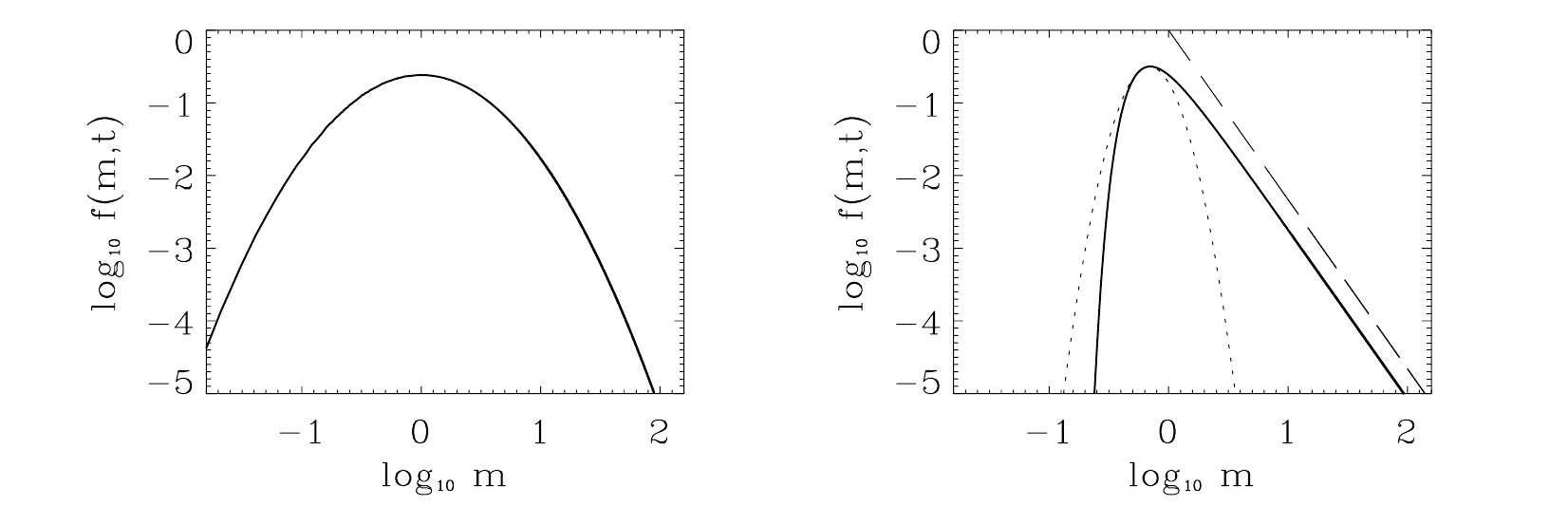}}
\put(1.9,3.15){\footnotesize{$\alpha \rightarrow 1$}}
\put(7.3,3.15){\footnotesize $\alpha = 2.3$}
\end{picture}
\caption{\label{fig:stochastic-IMF}
Mass distribution function (\ref{eqn:maschberger-mass-spectrum}) from stochastic accretion for $\alpha \rightarrow 1$ (left) and $\alpha = 2.3$ (right). The parameters $A$, $B$, and $m_0$ are all set to unity. As $\alpha$ approaches unity, the mass function becomes log-normal. The case $\alpha = 2.3$ develops a power law tail at large $m$ with slope $\alpha=2.3$ (as indicted by the long dashed line) and is very similar to the observed IMF (Section \ref{par:IMF}). The dotted curved indicates a log-normal fit to the peak of the mass distribution. {Adopted from \citet{maschberger2013}.}
 }
\end{figure}

\subsubsection{IMF from turbulence statistics}
\label{subsubsec:turb-IMF}

Besides leading to stochastic variations in the protostellar accretion rate (as discussed before in Section \ref{subsubsec:IMF-accretion}), interstellar turbulence can  influence the IMF by producing the clump structure within molecular clouds. These clumps or cores define the mass reservoir available for the formation of individual stars and small-multiple stellar systems. As a start, let us assume  -- most likely wrongly (see Section \ref{subsubsec:caveats}) --  that each core forms exactly one star with some fixed efficiency factor. If we furthermore assume that there are no other stochastic processes at play, such as competitive accretion or fragmentation-induced starvation (Section \ref{subsubsec:coll-models}), then understanding the origin of the stellar IMF boils down to identifying the physical processes that determine the  clump mass function (CMF) in star-forming molecular clouds. 

ISM turbulence is intrinsically a scale-free process as long as one stays within the inertial range (Section \ref{subsubsec:ISM-scales}). It is therefore conceivable that it could play a key role in producing the power-law tail at the high-mass end of the stellar mass distribution. Most analytic models that attempt to do so involve the following four steps. First, they come up with a model that relates key parameters of the turbulent ISM to the probability distribution function (PDF) of gas density. Second, they relate the density PDF to the clump mass spectrum. Third, they identify a set of criteria by which some of these clumps go into gravitational collapse and begin to form stars. Typically, these involve some kind of Jeans argument and give preference to the most massive and densest clumps in the cloud. Fourth, they involve a mapping procedure, which converts a certain fraction of the clump mass into the final stellar mass. 

\paragraph{\bf Density distribution function}

As discussed at the end of Section \ref{par:prop-MC}, the PDF of column densities in tenuous, non star-forming clouds is well approximated by a log-normal function. However, it develops a power-law tail at high column densities in more massive and star-forming cloud complexes. This is a signpost of gravitational contraction. A typical example for this case is the Orion A cloud. A map of its integrated CO emission is shown in  the top panel of Figure \ref{fig:orion-A-vel-disp}, and the corresponding distribution function of column densities (derived from dust emission measurements) is plotted at the bottom right  of Figure \ref{fig:column-density-PDF}.

In order to obtain an estimate of the three-dimensional density distribution we need to convert the projected column density PDFs. Numerical simulations show that the column density PDFs have a smaller width than the density PDFs and can exhibit different shapes in the high- and low-density regimes \citep{osg01,federrath10}. However, both generally show very similar statistical properties \citep{fk12}. This can be used to derive an estimate of the three-dimensional density PDF from the two-dimensional column density PDF \cite[for further details, see][]{brunt10}. The shape and width of the density PDF are governed by the presence of compressive motions in the turbulent ISM. The medium is highly compressive and locally convergent flows lead to spatially and temporally confined regions of increased density. By the same token, expansion creates lower-density voids. Consequently, the overall distribution of density in the ISM is a sensitive function of the statistical properties of the underlying turbulent flow, with key parameters being the effective Mach number, the turbulent forcing scheme (i.e.\ the ratio between compressional and rotational modes), the magnetic field strength, and the thermodynamic properties of the gas. Magnetic field lines resist compression and distortion and therefore reduce the compressibility of the gas. The competition between heating and cooling processes in the ISM (see Section \ref{sec:heating-cooling}) can act both ways. This is best seen when adopting a effective polytropic equation of state of the form 
\begin{equation}
P \propto \rho^{\gamma}\;.
\label{eqn:EOS}
\end{equation}
If the gas heats up when being compressed (for $\gamma > 1$), then pressure differences lead only to moderate density increase. However, when the gas gets colder when compressed (in the case $\gamma < 1$), the same pressure gradient can result in large density excursions. 

Analytical theory as well as numerical simulations show that the distribution of the gas density in isothermal ($\gamma = 1$), non self-gravitating, and well sampled turbulent media follows a log-normal distribution, 
\begin{equation}
{\mathrm{PDF}}(s)=\frac{1}{\sqrt{2\pi\sigma_s^2}}\exp\left(-\frac{(s-s_0)^2}{2\sigma_s^2}\right)\,.
\label{eqn:s-pdf}
\end{equation}
Here, we introduce the logarithmic density,
\begin{equation} \label{eqn:s}
s = \ln{(\rho/\rho_0)}\,,
\end{equation}
and  $\rho_0 = \langle \rho\rangle$ as well as  $s_0 = \langle s \rangle$ denote the corresponding mean values.  For a purely Gaussian distribution, the mean $s_0$ is related to the variance $\sigma_{s}^2$ of the logarithmic density $s$ via the equation
\begin{equation}
s_0=-\frac{1}{2}\,\sigma_s^2\;.
\end{equation}
This results from the normalization and mass-conservation constraints of the PDF \citep{Vazquez1994, FederrathKlessenSchmidt2008}. In turn, we can relate $\sigma_s$ to the Mach number $\cal M$, to the forcing parameter $b$, and to the ratio of the thermal energy density to the magnetic energy density $\beta$,
\begin{equation}
\sigma_s^2=\ln\left(1+b^2{\cal M}^2\frac{\beta}{\beta+1}\right)\,.
\label{eqn:sigma}
\end{equation}
For further discussions, see \citet{pn11} or \citet{Molina12}. The forcing parameter $b$ varies from a value of approximately 0.3 for turbulence that is purely driven by solenoidal (divergence-free) modes to $b\approx 1$ for purely compressive (curl-free) schemes. A natural mix of forcing modes results in  $b\approx 0.4$ \cite[see e.g.][]{FederrathKlessenSchmidt2008, federrath10, schmidt09, ssn11, Konstandin12}.  The parameter $\beta$ describes the ratio between thermal energy density and magnetic energy density,
\begin{equation}
\beta = \frac{c_{\rm s}^2}{B^2/8\pi \rho} = 2\frac{c_{\rm s}^2}{v_{\rm A}^2}\;,
\end{equation}
and can be expressed as the ratio between sound speed $c_{\rm s}$ and Alfv\'{e}n velocity $v_{\rm A} = B / \sqrt{4 \pi \rho}$ (equation \ref{eqn:Alfven-velocity}).

As indicated by the column density PDF in nearby molecular clouds depicted in Figure \ref{fig:column-density-PDF}, deviations from the pure log-normal behavior occur when parts of the gas undergo gravitational collapse and form stars. The velocity field, and as a consequence the density distribution, are no longer solely governed by turbulence statistics, but are also influenced by varying degrees of self-gravity \citep{Klessen2000, dib05, Collins2011, Kritsuk2011}. Furthermore, \citet{passot98} found deviations of the log-normal behavior in simulations of non-isothermal gas. Depending on the polytropic exponent $\gamma$ in the equation of state (\ref{eqn:EOS}), the PDF develops a power-law tail at low densities for $\gamma > 1$ and at high densities for $\gamma < 1$ \cite[see e.g.][]{li03}. The latter effect is very similar to gravitational collapse. In addition, the PDFs from hydrodynamic simulations typically change with time as the overall cloud evolution progresses \citep{BP2011, cho2011}. For example, \citet{fk13} quantify how the slope of the high-density tail of the PDF in their numerical models  flattens with increasing star-formation efficiency.  \citet{Girichidis2014} demonstrate analytically that free-fall contraction of a single core or an ensemble of collapsing spheres forms a power- law tail similar to the observed PDFs.

\paragraph{\bf Clump mass function}
The next step in this sequence is to relate the density PDF to the clump mass function (CMF). A first attempt to analytically derive the CMF from turbulence properties goes back to \citet{elmegreen93} and was then refined by \citet{pnj97} and \citet{Padoan02}. They argue that high-density clumps are simply the shock-compressed regions that are the natural outcome of supersonic turbulence. They then invoke the shock jump conditions to calculate the achievable density contrast from the distribution of Mach numbers in the flow. Because they consider magnetized media, they base their considerations on the Alfv\'{e}nic Mach number (${\cal M}_{\rm A} = v_{\rm A}/c_{\rm s}$), but similar conclusions follow for purely hydrodynamic flows \citep{elmegreen02}. To get the number of cores at a given density and length scale they argue that the flow is self-similar and that this quantity is simply determined by the available volume compressed to the density under consideration. This method has a number of shortcomings and we  refer the reader  to  \citet{Krumholz:2014p85740} for a more detailed account. 

More refined statistical approaches are based on the Press-Schechter (\citeyear{ps74}) and excursion set \citep{bond91} formalisms. These were originally introduced to describe the stochastic properties of cosmological fluctuations and quantify the behavior of random fields with structure over a wide range of scales. In particular, they can be used to count the number of objects above a certain density threshold for a given distribution function. This is exactly what is needed to determine the CMF. The first to realize this and to employ the Press-Schechter formalism to construct a model of the mass distribution of clumps from ISM turbulence was \citet{inutsuka01}. This was later extended by \citet{HennebelleChabrier2008, hennebelle09c, hennebelle13a}. The Press-Schechter method, however, gives rise to the so-called `cloud-in-cloud' problem, which occurs because the same object may be counted several times at different spatial scales. It can be resolved by introducing an appropriate correction factor  \cite[][]{jedamzik94}, but a better a approach is to resort to the excursion set formalism. In essence, one computes a large number of Monte Carlo realizations of the stochastic variable by performing a random walk in the available parameter space. This allows one to determine the expected number of cores for a given length scale and density (and thus mass) with high precision \cite[for a more detailed discussion, see][]{hopkins12a, hopkins12b, hopkins13b}. 

We would also like to mention that alternative statistical models have been proposed that are based on stochastic sampling in fractal media \citep[see][]{elmegreen96, elmegreen97b, elmegreen97, elmegreen99, elmegreen00b, elmegreen02b}.

\paragraph{\bf Collapsing cores}
Once the CMF is obtained, the next step towards the stellar mass function is to select the subset of clumps that are gravitationally unstable and that begin to collapse in order to form stars. The most simple approach is to base the selection of bound clumps on a thermal Jeans argument. \citet{jeans1902} studied the stability of isothermal gas spheres. He found that the competition between thermal pressure gradients and potential gradients introduces a critical mass, $M_{\rm J} \propto \rho^{-1/2} (T/\mu)^{3/2}$, as expressed by equation (\ref{eqn:Jeans}). The Jeans mass only depends on the density $\rho$ and the temperature $T$, as well as on the chemical composition of the gas through the mean particle mass $\mu$. If a clump is more massive than $M_{\rm J}$, then it will collapse; otherwise, it will expand. 

This approach can be extended by including the effects of micro-turbulence \citep{vonweizsaecker51, chandrasekhar51} and by considering the presence of magnetic fields \citep{mestel56, mouschovias76b, shu87}. For a more detailed account of the historic development of star formation criteria, see \citet{maclow04}. Probably the most intuitive approach to assess the stability of molecular cloud clumps is based on the virial theorem, which relates the time evolution of the moment of inertia tensor of an object to its volumetric energy densities and surface terms, and allows us to take all physical processes into account that influence the dynamical evolution of the system \cite[e.g.][]{MZ92, BP2006}. For an application to star formation, see for example \citet{krumholz05c}.

The problem that arises from the turbulent compression \citep{pnj97, Padoan02} or Press-Schechter approach is that it provides an estimate of the mass of a clump, but not of the density and of other physical properties that allow us to calculate the stability of the clump. To solve this problem, \citet{Padoan02}, for example, take a typical cloud temperature $T$ (see Table \ref{tab:mol-clouds}) and pick a random density  $\rho$ from the assumed density PDF (see above)  for each clump $M$ in the CMF. With these values they calculate the Jeans mass (Eq.~\ref{eqn:Jeans}). If the clump mass exceeds the Jeans mass, $M>M_{\rm J}$, the clump is considered to be gravitationally bound and forming stars. If it is less massive than the Jeans mass, it is disregarded. The probability for massive clumps to be unstable for randomly picked $\rho$ and $T$ values is very high, and  the mass spectrum of bound clumps is similar to the CMF at high masses. However, for lower-mass clumps, the likelihood of picking a combination of $\rho$ and $T$ such that the clump mass exceeds the Jeans mass gets smaller and smaller. As a consequence, the mass spectrum of bound clumps turns over towards smaller masses. This  calculation becomes somewhat easier in the excursion set approach. This is because the random walk through parameter space provides both the length scale $\ell$ and  the density $\rho$ for each clump. If one picks a temperature $T$, one can calculate at each step in the process whether the mass $M \sim \rho \ell^3$ of the clump under consideration exceeds the Jeans mass. Objects on the largest scales $\ell$ with $M > M_{\rm J}$ are identified as giant molecular clouds and objects on the smallest scales $\ell$ with $M > M_{\rm J}$ as star-forming clumps or prestellar cores \citep{hopkins12a, hopkins12b}. This approach can readily be extended to include the stabilizing effects of turbulence and magnetic fields \citep{krumholz05c, hennebelle09c, hennebelle13a, hopkins13a, fk12} or the influence of changes in the equation of state \citep{guszejnov14}. The overall peak of the mass spectrum is most likely determined by the balance between heating and cooling processes in the star-forming gas which sets a characteristic range of values for $M_{\rm J}$ and its variants (Section \ref{subsubsec:thermodynamics}).

\paragraph{\bf Stellar IMF }
Once an ensemble of collapsing cloud clumps is selected, as outlined above, the stellar IMF is often determined by simply mapping the clump mass to the stellar mass with some given fixed efficiency. Typical values are around 30\% (see Section \ref{subsubsec:core-accretion}). As a result the IMF has the same functional form as the mass function of bound cores (see Section \ref{par:statistics}). However, as outlined in Section \ref{subsubsec:caveats}, this simple approach has its problems. If indeed each core only forms one star (or maybe a binary system), then it needs to have about one Jeans mass. Otherwise the core is likely to  fragment \citep{goodwin04a, goodwin04b, goodwin06, holman13}. Because $M_{\rm J} \propto \rho^{-1/2}$ for a given temperature $T$, high-mass clumps should be less dense than low mass ones. This immediately leads to a timescale problem \citep{clark07}. Because $\tau_{\rm ff} \propto \rho^{-1/2}$, the collapse time scales linearly with the clump mass. The time it takes to build up a star with $10\,$M$_{\odot}$ is sufficient to form ten stars with $1\,$M$_{\odot}$. As a consequence the resulting stellar IMF should be considerably steeper than the CMF. In addition, high-mass clumps are not observed to be less dense than low-mass ones. If anything, they tend to be denser, and they are typically highly Jeans-unstable \citep{battersby10, ragan2012, ragan2013, marsh14}. 

A potential way out of this dilemma is to assume that high-mass clumps are hotter. High mass stars can indeed heat up their surroundings quite considerably (Section \ref{subsec:massive-stars}), since their luminosity $L$ scales with stellar mass $M$ as $L\propto M^{3.5}$   \cite[e.g.][]{Hansen:1994p85789}. However, there are many low-mass star-forming regions which show no signs of massive star formation (e.g.\ Taurus or $\rho$-Ophiuchi) and where the temperatures inferred for prestellar cores are uniformly low \citep{bergin07}.

In general there is thus no good reason to believe in a one-to-one mapping between the core mass function and the stellar IMF.  None of the current analytic models for the IMF includes processes such as stellar feedback in form of radiation or outflows, or fragmentation during core collapse and during the accretion disk phase, in a realistic and consistent way. The same holds for numerical simulations of star cluster formation. Altogether, it is likely that the transition from core to stars follows a complicated and stochastic pathway that may change with varying environmental conditions. For a simple cartoon picture,  see Section \ref{subsec:collapse}.

\subsection{Massive star formation}
\label{subsec:massive-stars}

Because their formation time is short, of the order of $10^5\,$yr, and because they grow while deeply embedded in
massive cloud cores, very little is known about the initial and environmental conditions of
high-mass stellar birth. In general, regions forming high-mass stars are characterized by more
extreme physical conditions than regions forming only low-mass stars,
containing cores of size, mass, and velocity dispersion roughly an order of magnitude larger than those of cores in 
regions without high-mass star formation
\citep[e.g.][]{beutheretal07, motteetal08, Krumholz:2014p85764}. 
Typical sizes of cluster-forming clumps are about $1\,$pc.
They have mean densities of $n \approx 10^5$ cm$^{-3}$, masses of $\sim$$10^3\,$M$_\odot$ and
above, and velocity dispersions ranging between $1.5$ and $4 \,$km$\,$s$^{-1}$. Whenever
observed with high resolution, these clumps break up into even denser cores that are believed
to be the immediate precursors of single or gravitationally bound multiple massive protostars. 

Massive stars usually form as members of multiple stellar systems \citep{hohaschik81, lada06, zinnecker07, repurth2014} which themselves are parts of larger clusters \citep{ladalada03, dewitetal04, testietal97, Longmore:2014p85628}. This fact    adds
additional challenges   to
the interpretation of observational data from high-mass
star forming regions as it is difficult to disentangle mutual dynamical interactions from the
influence of individual stars \citep[e.g.][]{gotoetal06,linzetal05}.
Furthermore, high-mass stars  reach the main sequence while still accreting. Their Kelvin-Helmholtz
pre-main sequence contraction   time is considerably shorter than
   their accretion time. Once  a star has reached a mass of about $10\,$M$_\odot$, its spectrum becomes UV-dominated and it begins
to ionize its environment. This means that accretion as well as ionizing and non-ionizing radiation
needs to be considered in concert \citep{keto02b,keto03,keto07, petersetal10a, Peters2010b}. It was realized
decades ago that in simple one-dimensional collapse models, the outward radiation force on the accreting
material should be significantly stronger than the inward pull of
gravity \citep{larsstarr71,kahn74},
in particular if one accounts for dust opacity. Since we see stars with $100\,$M$_\odot$ or even more \citep{bonanosetal04, figer05, rauwetal05, bestenlehner2011, borisova2012, doran2013}, a simple spherically symmetric approach to high-mass star formation must fail. 

Consequently, two different models for massive star formation have been proposed. The first one takes
advantage of the fact that high-mass stars always form as members of stellar clusters. If the central
density in the cluster is high enough, there is a
         chance
that low-mass protostars collide and so
successively build up more massive objects \citep{bonbatezin98}. As the radii of protostars
usually are considerably larger than the radii of main sequence stars in  the same mass range \citep{hosokawa2009},  this
could be a viable option. However, the stellar densities required to produce massive stars by collisions are
extremely high \citep{baumgardt11}. They seem inconsistent with the observed stellar densities of most Galactic star clusters 
\cite[e.g.][and references therein]{pozwetal10}, but could be reached in the central regions of the most extreme and massive clusters in the Local Group (such as 30 Doradus in the LMC as shown in Figure \ref{fig:clusters}; see e.g.\ \citealt{banerjee12a}). 

 An alternative approach is to argue that high-mass stars form like low-mass stars by accretion
of ambient gas that goes through a rotationally supported disk formed by angular momentum conservation.
Indeed, such disk structures are observed around a number of high-mass protostars
\citep{chini04,chini06,jiangetal08,daviesetal10}. Their presence
breaks any spherical symmetry that might have been present in the initial cloud and thus solves the
opacity problem. Radiation tends to escape along the polar axis, while matter is transported inwards
through parts of the equatorial plane shielded by the disk. Hydrodynamic simulations in two and three
dimensions focusing on the transport of non-ionizing radiation strongly
support this picture \citep{yorke02,krumholz09a, kuiper10, kuiper11}.
The same holds when taking the  effects of ionizing radiation into account \citep{petersetal10a, Peters2010b, Peters2011,commercon2011}. Once the disk becomes  gravitationally unstable, material flows along dense, opaque filaments, whereas the radiation
escapes through optically thin channels in and above the disk. Even
ionized material can be accreted, if the accretion flow is strong enough. \hii\ regions are gravitationally
trapped at this stage, but soon begin to rapidly fluctuate between trapped and extended states, as seen in some Galactic massive star-formation regions  \citep{petersetal10a, GalvanMadrid:2011p55089, depree2014}. Over time, the same
ultracompact \hii\ region can expand anisotropically, contract again, and take on any of the observed
morphological classes \citep{woodchurch89,kurtzetal94, Peters2010c}. In their
extended phases, expanding \hii\ regions drive bipolar neutral outflows characteristic of high-mass
star formation \citep{petersetal10a}. 

Another key fact that any theory of massive star formation must account for is the apparent presence of an upper mass limit at around $100 - 150\,$M$_\odot$ \citep{massey03}.
It holds for the Galactic field, but in dense clusters, apparently higher-mass stars have been reported \cite[e.g.][]{crowther10, doran2013}. If this mass limit holds, then purely random sampling of the initial mass function (IMF)
\citep{kroupa02,chabrier03} without an upper mass limit should have yielded stars above
$150\,$M$_\odot$ (\citealt{weidner04,figer05,oey05,weidetal10}; see however, \citealt{selmel08}). Altogether, the situation is not fully conclusive. If indeed there is an upper mass limit, it raises the question of its physical origin.  It has been speculated before that radiative
stellar feedback might be responsible for this limit \cite[for a detailed discussion see e.g.][]{zinnecker07} or
alternatively that the internal stability limit of stars with non-zero metallicity lies in this
mass regime \citep{appen70a,appen70b,appen87,baraetal01}.
However,  fragmentation could also limit protostellar mass growth, as suggested by the numerical simulations of \citet{Peters2010b}. The likelihood of fragmentation to occur and the number of fragments
to form depends sensitively on the physical conditions in the star-forming cloud and its initial and
environmental parameters \citep[see e.g.][]{Girichidis:2012p55106}. Understanding the build-up of massive
stars therefore requires detailed knowledge about the physical processes that initiate and regulate the
formation and dynamical evolution of the molecular clouds these stars form in 
\citep{vazsemetal09}.

\citet{Peters2010b, Peters2011}, \citet{kuiper11}, and  \citet{commercon2011} argue that ionizing radiation \cite[see also][]{Krumholz:2014p85764}, just like its non-ionizing, lower-energy counterpart,
         cannot
shut off the accretion flow onto massive stars. Instead it is the dynamical processes
in the gravitationally unstable accretion
         flow
that inevitably
      occurs
during the collapse of
high-mass cloud cores that
         control
the mass growth of individual protostars.  Accretion onto the
central star is shut off by the fragmentation of the disk and the formation of  lower-mass companions
which intercept inward-moving material. \citet{Peters2010b, Peters2011} call this process fragmentation-induced starvation and show that it occurs unavoidably in regions of high-mass star formation where the mass flow onto the
disk exceeds the inward transport of matter due to viscosity only and thus renders the disk unstable
to fragmentation (see also Section \ref{subsubsec:coll-models}).

As a side note, it is interesting to speculate that  fragmentation-induced starvation is important not only for present-day star formation but also in the primordial universe during the formation of metal-free Population III stars. Consequently, we expect these stars to be in binary or small number multiple systems and to be of lower mass than usually inferred \citep{abeletal02,brommetal09}. Indeed, current numerical simulations provide the first hints that this might be the case \citep[e.g.][]{clark11a, greif11b, stacy2013}.

\subsection{Final stages of star and planet formation}
\label{subsec:collapse}

\begin{figure*}[tbp]
\begin{center}
\includegraphics[height=4.52cm]{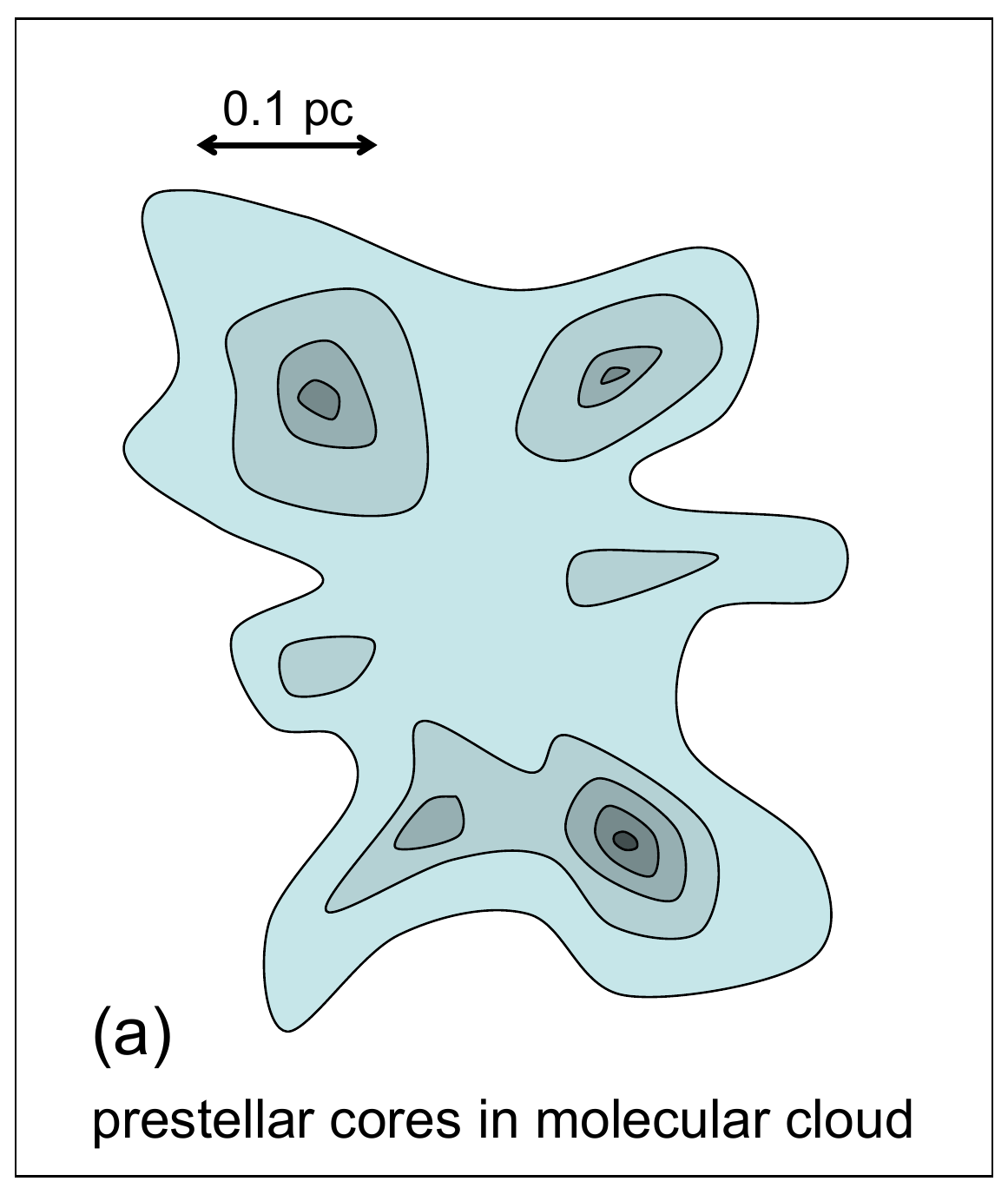}\hspace*{0.01cm}
\includegraphics[height=4.5cm]{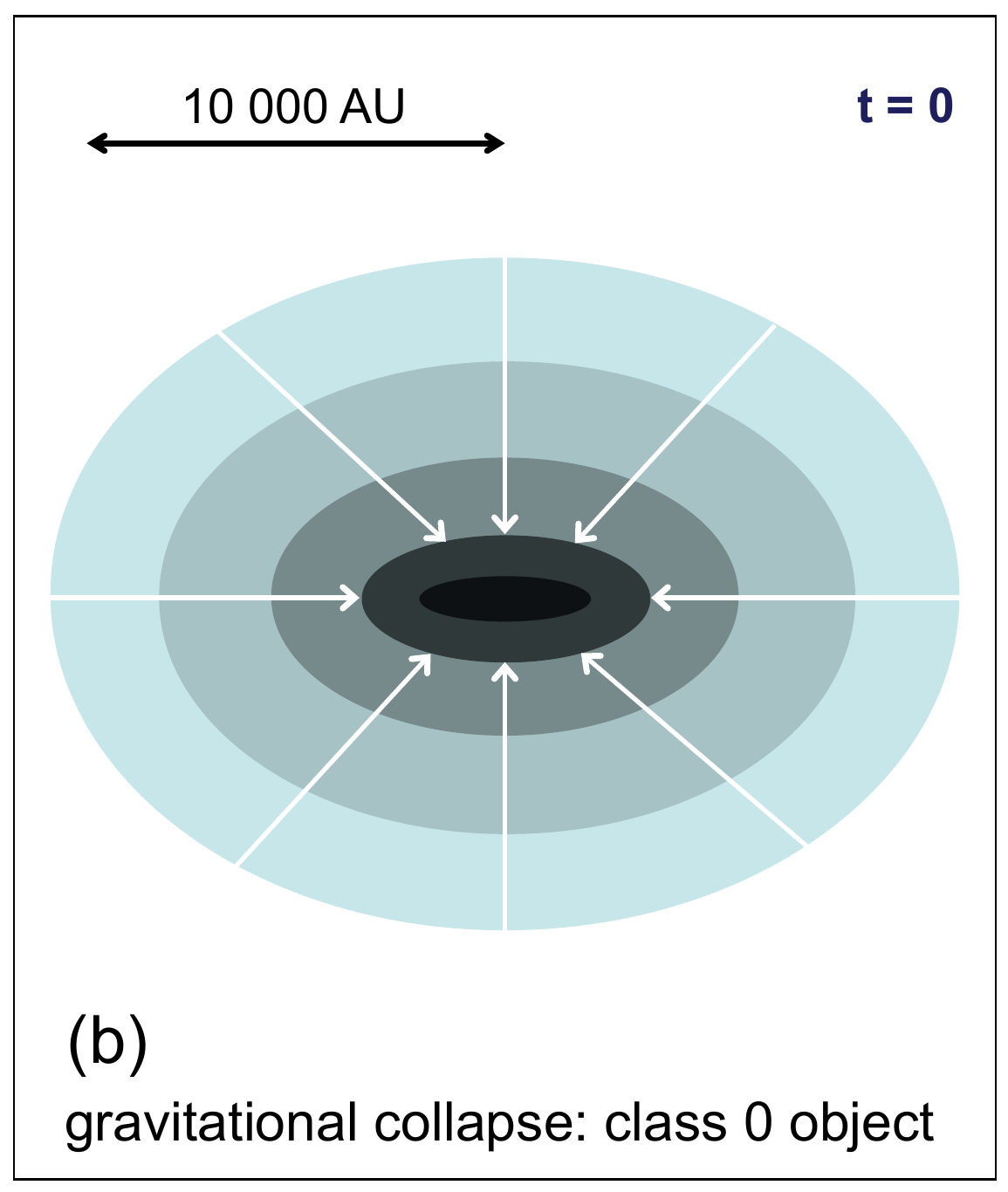}
\includegraphics[height=4.5cm]{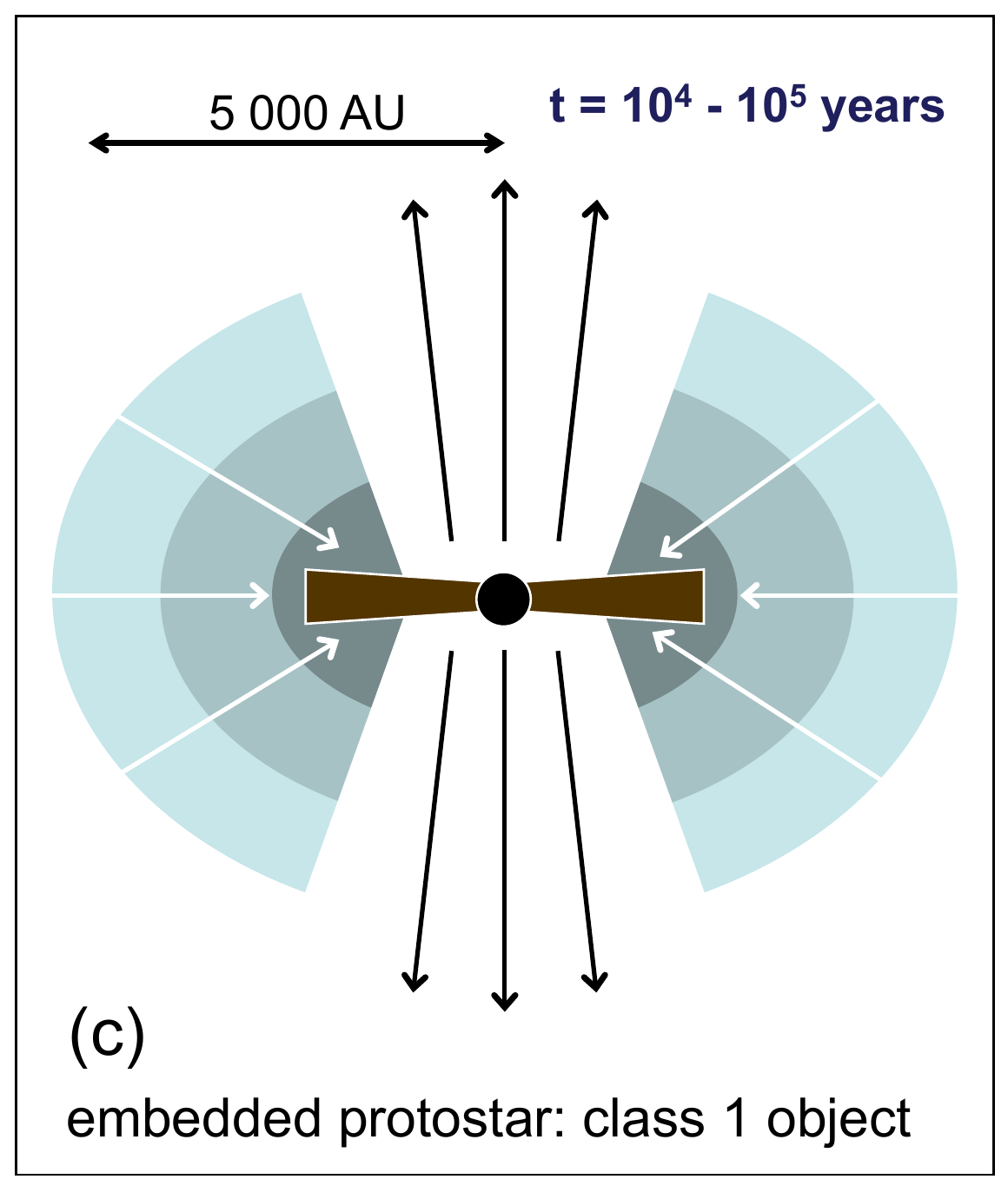}
\includegraphics[height=4.5cm]{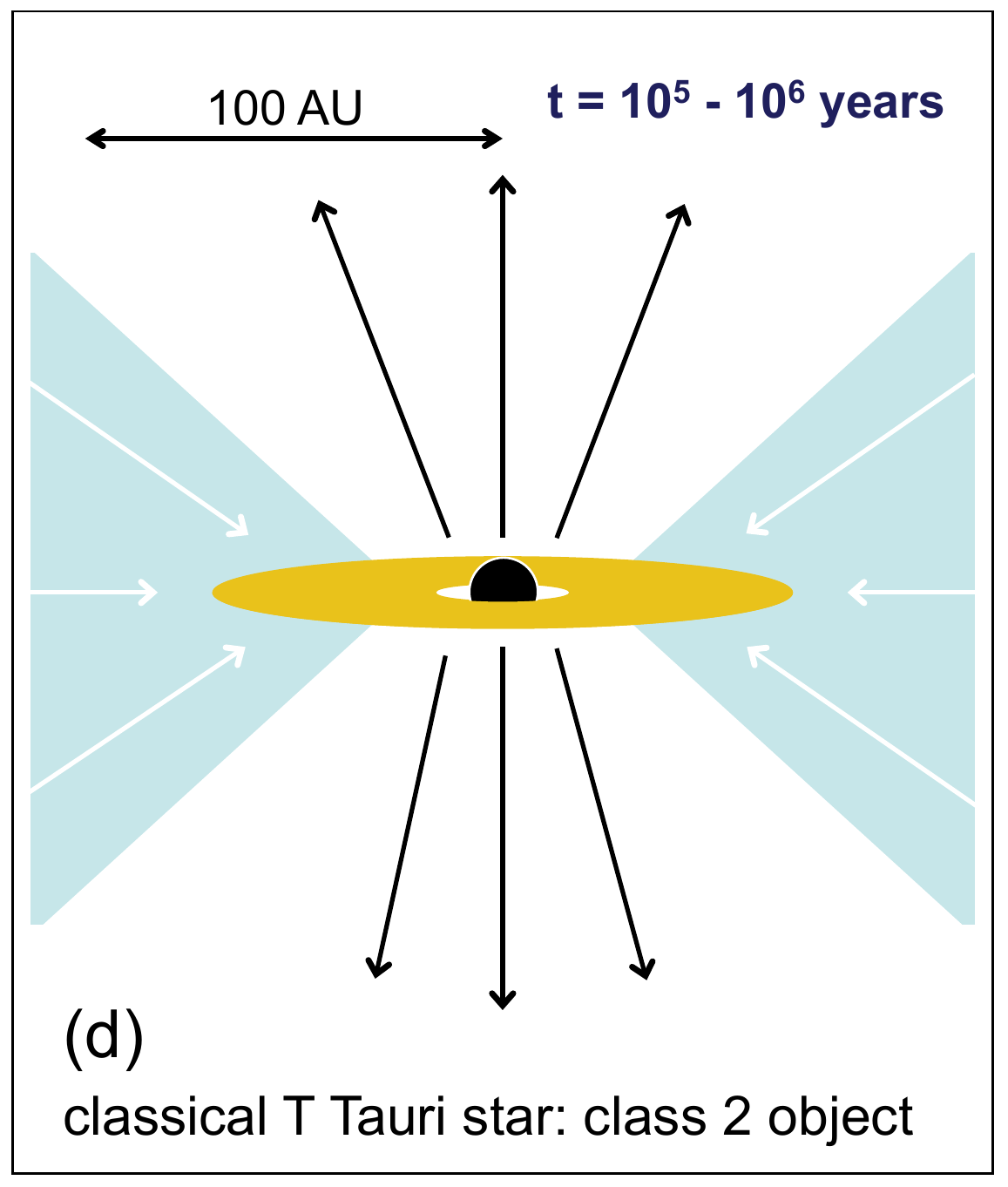}
\includegraphics[height=4.5cm]{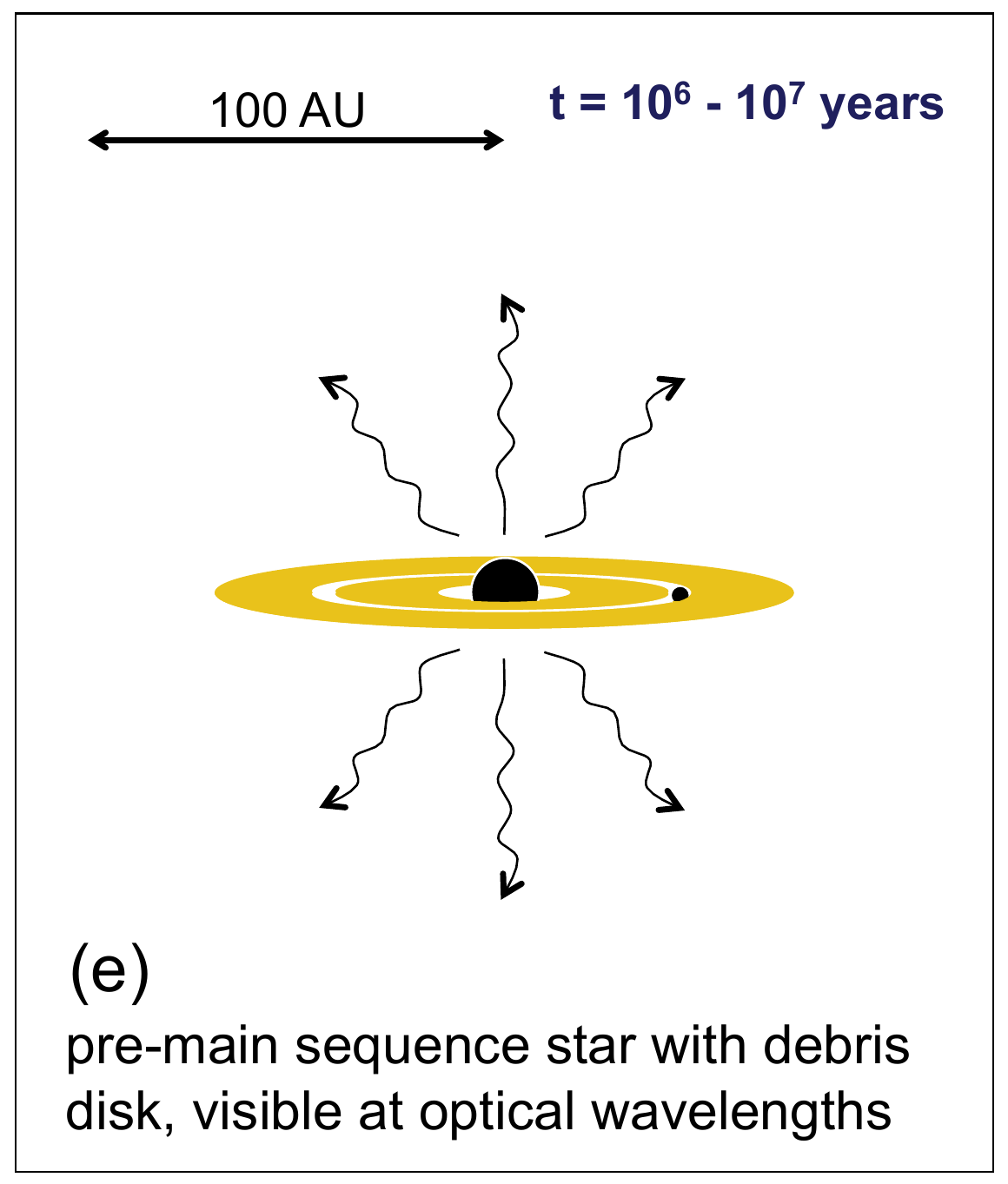}
\includegraphics[height=4.5cm]{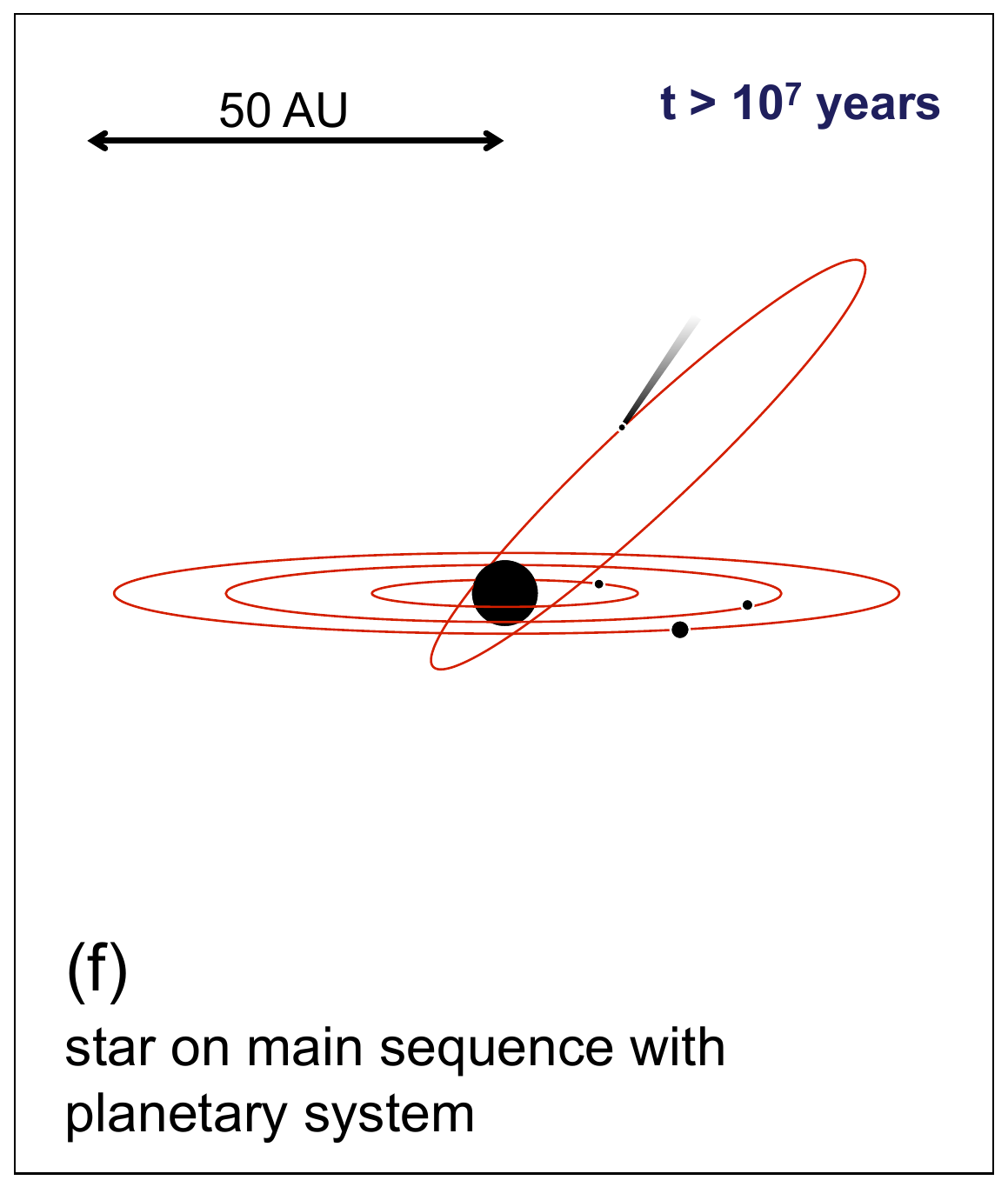}
\caption{Schematic view of the star and planet formation process. (a) Prestellar cores form by turbulent compression inside larger molecular clouds. (b) Some become gravitational unstable and begin to collapse. During the main accretion phase, the young protostar in the center experiences rapid mass growth. This is the class 0 phase of protostellar evolution. (c) Because of angular momentum conservation, the infalling material settles into a protostellar / protoplanetary disk. Magnetically launched outflows begin to disperse the infalling envelope. This is the class 1 phase. (d) The central protostar becomes visible as more and more of the envelope is removed. This is the class 2 phase. (e) The envelope is removed, and the central star becomes fully visible. Only a remnant disk remains in which planet formation continues. Low mass stars are still on the Kelvin-Helmholtz pre-main sequence contraction phase. (f) Finally, the original gas and dust disk is cleared and what remains is a central star with a planetary system, such as we observe in our solar system. Note that the cartoon picture describes the situation for isolated low-mass stars. For high-mass stars, the situation is more complicated, because the disk is likely to fragment into  a binary or higher-order stellar system during the main accretion phase. 
\label{fig:SF-sequence}
}
\end{center}
\end{figure*}

Here, we summarize again the main phases of the star and planet formation process. The entire sequence is illustrated in Figure \ref{fig:SF-sequence}. It begins with the formation of molecular cloud complexes in the turbulent multi-phase ISM of the Galaxy (as we discuss in Section \ref{par:prop-MC}), and continues {\em (a)} with supersonic turbulence generating high density clumps with a wide range of densities and sizes (Section \ref{subsec:cores}). {\em (b)} Some of these density fluctuations may become gravitationally unstable and begin to collapse. The central density increases until the compressional heat generated by the contraction can no longer be radiated away efficiently. A quasi-hydrostatic object then forms in the center of the core. This protostar continues to grows in mass by accreting material from the infalling envelope. In the class 0 phase of protostellar evolution, the mass budget is still dominated by the enclosing envelope. It is optically think and absorbs the accretion luminosity generated as the infalling material comes to a halt at the protostellar surface. The spectral energy distribution (SED) of the system is thus dominated by the reprocessed emission from the cold envelope radiating mainly at sub-mm wavelengths. 

Due to the conservation of angular momentum, most of the infalling matter will not directly end up in the central protostar,  {\em (c)} but instead it will build up a rotationally supported accretion disk. If the mass load onto this disk during the main accretion phase exceeds its capability to transport material inwards by gravitoviscous processes, then the disk becomes unstable and will fragment into a binary or higher-order multiple stellar system. This is very likely to happen for high-mass stars, but occurs less frequently for low-mass objects, as indicated by the strong mass dependence of the stellar multiplicity fraction \citep{lada06}. Molecular cloud cores are magnetized. The magnetic field is compressed and amplified by dynamo processes during the contraction, and eventually the accretion disk is able to launch a magnetically driven outflow along the rotational axis of the system \cite[for a summary, see][]{Pudritz07}. The outflow begins to disperse the remaining envelope material. First this happens along the polar axis, but later on  larger and larger volumes are affected. This affects the observed SED. As more and more of the inner regions of the disk become visible, the peak of the emission moves towards infrared wavelengths. Favorable viewing angles even permit us to look down onto the protostellar photosphere, which contributes to the emission at near infrared and optical wavelengths. {\em (d)} Clearing the infalling envelope and the corresponding changes in SED mark the transition from class 1 to class 2 objects. In this T Tauri phase of protostellar evolution, most of the mass is already assembled in the central star with the remaining accretion disk contributing only a few percent to the overall matter budget. Nevertheless, this is where planets begin to form. 

At some point, {\em (e)} the envelope is completely removed (or accreted by the disk), and the central star becomes fully visible. Only a debris disk remains in which planet formation continues. The central protostar is expected to be fully convective and the energy loss due to the emission of radiation at its surface is compensated by the release of gravitational energy. It slowly contracts and by doing so becomes hotter and denser. This is the classical Kelvin-Helmholtz contraction phase of pre-main sequence evolution. For solar-type stars it lasts about $20 - 30$ million years. Finally, the central conditions are right for nuclear fusion to set in. The star enters the main sequence and settles into a quasi-equillibrium state, where its radiative energy losses are compensated by nuclear burning processes converting hydrogen into helium.  {\em (f)} After some time, the remaining debris disk is also cleared away and we are left with the central star, most likely being surrounded by a planetary system, such as we see in our own solar system or as we observe around other nearby stars.\footnote{The latest updates and findings of the research activities on extrasolar planets can be found at the following websites: {\sl www.exoplanet.eu} and {\sl www.exoplanets.org}.}

\section{Summary}
\label{sec:summary}

In these lecture notes, we have made an attempt to identify and characterize the key astrophysical processes that provide the link between the dynamical behavior of the interstellar medium and the formation of stars. We hope that we have made it  clear that one part cannot be understood without solid knowledge of the other. Both are connected via a number of competing feedback loops. We have argued that the evolution of the galactic ISM on large scales depends on the detailed microphysics in very complicated and often counter-intuitive ways. Conversely,  global dynamical processes set the initial and boundary conditions for the formation of dense clouds on small scales and the birth of stars in their interior. Altogether, ISM dynamics spans an enormous dynamic range, from the extent of the galaxy as a whole down to the local blobs of gas that collapse to form individual stars or binary systems. Similarly, it covers many orders of magnitude in time, from the hundreds of millions of years it takes to complete one galactic rotation down to the hundreds of years it takes an ionization front to travel through a star-forming cloud. Improving our understanding of the interstellar medium and its ability to give birth to new stars is a complex multi-scale and multi-physics problem. It requires insights from quantum physics and chemistry, and knowledge of magnetohydrodynamics, plasma physics as well as gravitational dynamics. It also demands a deep understanding of the coupling between matter and radiation, together with input from high-resolution multi-frequency and multi-messenger astronomical observations. 

After a brief introduction to the field in Section \ref{sec:introduction}, we began our discussion  in Section \ref{sec:comp-ISM} with a detailed account of the main constituents of the interstellar medium. These are gas, dust, the interstellar radiation field, and cosmic rays. Next, we turned our attention to the various heating and cooling processes that govern the thermodynamic behavior of the ISM. In Section \ref{sec:heating-cooling}, we introduced the microphysical processes that regulate the coupling between matter and radiation as well as between the different matter components. We identified the observed interstellar turbulence as a key agent driving the dynamical evolution of the Galactic ISM. These turbulent flows play a dual role. As we discussed in Section  \ref{sec:turbulence}, turbulence can prevent or delay collapse on large scales, but on small scales it may actually trigger local compression and  star formation. In addition, we showed that ISM turbulence dissipates quickly and needs to be continuously replenished for a galaxy to reach an approximate steady state. This led to a critical comparison of the various astrophysical processes that have been proposed to drive interstellar turbulence in galaxies such as our Milky Way. Because star formation is always found to be associated with molecular clouds, we discussed the physical (and chemical) processes that govern the formation of these densest and coldest components of the ISM in Section \ref{sec:cloud-form}. We paid special attention to the chemical reactions that lead to the formation of H$_2$ and to its most important tracer CO. We found that dust attenuation plays a key role in this process, and we discussed molecular cloud formation in the context of global ISM dynamics on galactic scales. Finally, in Section \ref{sec:collapse-SF} we zoomed in on smaller and smaller scales, and summarized the properties of molecular cloud cores as the direct progenitors of individual stars and stellar systems. Furthermore, we motivated and described the current statistical and theoretical models of stellar birth and tried to explain the seemingly universal observed distribution of stellar masses at birth, the initial mass function (IMF), as the result of a sequence of stochastic events mostly governed by the interplay between turbulence and self-gravity in the star-forming gas. 

We hope that we have illustrated in these lecture notes that the question of stellar birth in the multi-phase ISM of our Milky Way and elsewhere in the universe is far from being solved. On the contrary,  the field of ISM dynamics and star formation is rapidly evolving and has gone through a significant transformation in recent years. We acknowledge that scientific progress in this area requires the concerted and combined efforts of theory, observations, as well as laboratory experiments. We notice a general trend away from only taking isolated processes and phenomena into account and towards a more integrated multi-scale and multi-physics approach in today's theoretical models and computer simulations. Observational studies now regularly attempt to accumulate and combine information from as many different wavebands as possible, and to cover as large an area on the sky with as much detail and resolution as possible. New large facilities such as ALMA on the ground or Gaia in space have the potential for real scientific breakthroughs.\footnote{Information about the Atacama Large Millimeter/Submillimeter Array (ALMA) and about the Gaia satellite can be found at {\sl www.almaobservatory.org} and {\sl sci.esa.int/gaia/}.} All our theoretical and observational efforts would be in vain without complementary laboratory studies that provide fundamental information and cross sections for molecular and ionic reactions as well as transition frequencies and data on dust physics, that constitute the physical and chemical basis of our understanding of the ISM.

We end this summary with a list of questions, which we think are amongst the most important  open problems in the field of ISM dynamics and star formation studies. We note that  these questions are closely related to each other, and that the answer to one question may hold the key to resolving another. 

\vspace*{0.1cm}\noindent{\em What drives interstellar turbulence?} Observations show that turbulence in molecular clouds is ubiquitous. With the exception of the dense cores discussed in Section \ref{subsec:cores}, it seems to follow a universal relationship between velocity dispersion and size (Larson's relation, see Section \ref{subsubsec:turb-obs}). Even extragalactic molecular clouds exhibit similar behavior.  In addition, there are few variations in the turbulent properties between molecular cloud regions with ongoing  star formation and those without. This seems to argue in favor of a galaxy-scale driving process (Section~\ref{subsub:ISM-driving-external}). On the other hand, there are also no systematic variations in GMC properties within a galaxy or between galaxies, which would seem to argue that internal processes must be important as well (Section~\ref{subsub:ISM-driving-feedback}). What is the relative importance of internal and external forcing mechanisms in driving ISM turbulence? Does the answer depend on the length scales that one examines, or on the place where one looks? So far, the `smoking gun' to answer these questions, both observationally or theoretically, remains to be found. 

\vspace*{0.1cm}\noindent{\em How is the star formation process correlated with galaxy properties? And how can we best study that problem?}  On large scales, star formation appears to follow a fairly universal scaling behavior. This holds for galaxies that range from being mildly dominated by atomic hydrogen (such as the Milky Way) to those that are strongly dominated by molecular hydrogen (such as local starbursts). Does the presence or absence of a significant atomic phase play an important role in regulating star formation, either directly (e.g.\ by limiting the amount of molecular gas available for star formation) or indirectly (e.g.\ by driving turbulent motions via thermal instability)? How does the star formation process change, if at all, in galaxies such as dwarfs that contain very little molecular gas? On the observational side, one of the key questions is whether and to what extent commonly used observational tracers of the star formation rate (SFR), such as H$\alpha$, 24$\,\umu$m dust emission, or [C{\textsc{ii}}] fine structure emission, can reliably recover the true rates? Accurate measurements of the SFR in galaxies are of great importance for many different fields of astrophysical research, and yet remain difficult to carry out. In nearby molecular clouds, counts of young stellar objects can give a direct measurement of the SFR, but this technique cannot be used in extragalactic systems where individual objects cannot be detected and resolved. Instead, indirect indicators of the SFR must be used, such as the H$\alpha$ luminosity or the total far-infrared emission. A central assumption underpinning these methods is that the energy radiated by these tracers comes primarily from newly-formed massive stars. If this is not the case, then these tracers will give a misleading view of the SFR. Answering these questions requires both dedicated observations and numerical models that allow us to explore the conditions in which the different tracers of the SFR can be used safely, and to understand when and why they fail. These can then also address the question of the scale over which the above correlations hold. Do we still see a good correlation between the tracers and the SFR on small scales (tens of parsecs or less), or only when we average on scales of hundreds of parsecs?

\vspace*{0.1cm}\noindent{\em What are the best observational tracers to study ISM dynamics and molecular cloud assembly? } We know that dense molecular clouds must be assembled from gas that is initially in a more diffuse state (Section \ref{sec:cloud-form}), but whether this process is driven primarily by turbulence or by gravity is unclear. At the present time, we are not even sure what we should observe in order to best distinguish between these two models. It seems likely that CO forms in significant quantities only once a large fraction of the cloud mass is already assembled, since it resists photodissociation only in regions with relatively high extinctions  (Section \ref{dust-shield-import}), and so CO observations are unlikely to provide strong constraints on the assembly process. H{\textsc{i}} 21~cm observations may be better suited for this purpose, but only if the inflowing gas is primarily atomic. If, instead, it is largely composed of H$_2$, then chemical tracers of this phase (e.g.\ HD or HF) may be more useful. Fine structure emission from [C{\textsc{ii}}] or [O{\textsc{i}}] may also trace the inflowing gas, but only if it is warm enough to excite the lines. Addressing this issue requires us to perform dedicated numerical simulations coupled with a time-dependent chemical network and to produce synthetic observations in the tracers of interest in a post-processing step, which then can be compared one-to-one with real observational data.

\vspace*{0.1cm}\noindent{\em Which observational diagnostics are best suited to recover the true physical properties of the ISM?} Our knowledge of physical cloud properties (such as mass or spatial extent) often relies on indirect measurements, particularly in extragalactic systems. For example, it is often assumed that all molecular clouds are in virial equilibrium, which allows us to estimate their masses from the observed CO linewidths. In the Milky Way, we can hope to benchmark this approach by using more direct measurements of cloud masses, as derived from e.g.\ dust extinction. However, this can be done easily only for nearby clouds, meaning that the range of environments in which these estimates can be directly tested is quite limited. Besides better and more detailed observations, progress will require us to examine the performance of the different available measures of cloud mass in dedicated numerical simulations for a wide variety of different physical environments. Only these calculations provide full access to the six-dimensional phase space, and by doing so enable us to figure out which measures are the least biased and potentially also to derive correction factors to improve the observational estimates.

\vspace*{0.1cm}\noindent{\em What physical processes determine the distribution of stellar masses at birth?} 
How reliable are observations that suggest that the stellar IMF and binary distribution at the present day are similar in different galactic environments? In particular, in rich (and more distant) clusters, our observational basis needs to be extended to lower masses. The same holds for variations with metallicity as can be traced in the Local Group. Is the IMF in the Large Magellanic Cloud (with a metal abundance of about half of the solar value) and the Small Magellanic Cloud (with a metal abundance of about one fifth solar or less) really similar to the Milky Way? On the theoretical side, what processes are responsible for the (non-)variation of the IMF? The critical mass for gravitational collapse can vary enormously between different environments. Yet the IMF in globular clusters, for example, appears to be the same as in regions of distributed star formation such as Taurus. How can the statistical theoretical models introduced in Section \ref{subsec:IMF-models} be extended to address these questions? Better understanding the physical origin of the IMF will remain a key driver of star formation research for a long time to come.

{\acknowledgement{\mbox{~}

Writing these lecture notes would have been impossible without the help and input from many collaborators and colleagues. In particular, we want to thank Christian Baczynski, Javier Ballesteros-Paredes, Robi Banerjee, Erik Bertram, Henrik Beuther, Frank Bigiel, Peter Bodenheimer, Ian A.\ Bonnell, Andreas Burkert, Paul C.\ Clark, Cornelis P.\ Dullemond, Edith Falgarone, Christoph Federrath, Philipp Girichidis, Alyssa Goodman, Dimitrios Gouliermis, Fabian Heitsch, Patrick Hennebelle, Thomas Henning, Mark H.\ Heyer, Philip F.\ Hopkins, Juan Iba\~{n}ez Mejia, Eric R.\ Keto, Lukas Konstandin, Pavel Kroupa, Mark R.\ Krumholz, Mordecai-Mark Mac~Low, Faviola Molina, Volker Ossenkopf, Thomas Peters, Ralph E.\ Pudritz, Sarah Ragan, Julia Roman-Duval, Daniel Seifried, Dominik R.\ G.\ Schleicher, Wolfram Schmidt, Nicola Schneider, Jennifer Schober, Rahul Shetty, Rowan J.\ Smith, J\"{u}rgen Stutzki, Laszlo Sz\H{u}cs, Enrique Vazquez-Semadeni, Antony P.\ Whitworth, and Hans Zinnecker for many stimulating and encouraging discussions.

We acknowledge support from the Deutsche Forschungsgemeinschaft (DFG) via the SFB 881 {\em The Milky Way System} (subprojects B1, B2, B5 and B8), and the SPP (priority program) 1573 {\em Physics of the ISM}. We also acknowledge substantial support from the European Research Council under the European Community's Seventh Framework Program (FP7/2007-2013) via the ERC Advanced Grant {\em STARLIGHT: Formation of the First Stars} (project number 339177). 

}}



\end{document}